\documentclass{aa}  

\usepackage{graphicx}
\usepackage{txfonts}
\usepackage{grffile}
\usepackage{lscape}
\def\nh3{NH$_{3}$}
\def\kms{km~s$^{-1}$}

\def\Vlsr{$V_{\rm LSR}$}

\def\G24{G24.78$+$0.08}
\def\HII{H{\sc ii}}

\newcommand{\ms}{$M_{\odot}$}
\newcommand{\ls}{$L_{\odot}$}
\newcommand{\msyr}{$M_{\odot}$~yr$^{-1}$}

\newcommand{\degree}{$^{\circ}$}

\usepackage{color}

\begin{document} 

   \title{Protostellar Outflows at the EarliesT Stages (POETS). III. H$_2$O masers tracing disk-winds and jets near luminous YSOs.}
    
  \subtitle{}

  \titlerunning{H$2$O maser kinematics.}
   
   \author{L. Moscadelli\inst{1}
          \and
          A. Sanna\inst{2}
          \and
          C. Goddi\inst{3,4}
          \and
          V. Krishnan\inst{1,5}
          \and
          F. Massi\inst{1}
          \and
          F. Bacciotti\inst{1}}
  
   \institute{INAF-Osservatorio Astrofisico di Arcetri, Largo E. Fermi 5, 50125 Firenze, Italy \\
              \email{mosca@arcetri.astro.it}
             \and
              Max-Planck-Institut f\"{u}r Radioastronomie, Auf dem H\"{u}gel 69, 53121 Bonn, Germany 
             \and 
              Leiden Observatory, Leiden University, PO Box 9513, 2300 RA Leiden, The Netherlands 
             \and
             Department of Astrophysics/IMAPP, Radboud University, PO Box 9010, 6500 GL, Nijmegen, The Netherlands  
             \and
             South African Radio Astronomy Observatory (SARAO), 2 Fir street, Black River Park, Observatory, Cape Town, 7925, South Africa            
             }

   \date{}

 
  \abstract
   {Although recent observations and theoretical simulations have pointed out that accretion disks and jets can be essential for the formation of stars with a mass of up to at least 20~\ms, the processes regulating mass accretion and ejection are still uncertain.}
   {The goal of the  Protostellar Outflows at the EarliesT Stages (POETS) survey is to image the disk-outflow interface on scales of \ 10-100~au \  in a statistically significant sample (36) of luminous young stellar objects (YSO), targeting both the molecular and ionized components of the outflows.}
   {The outflow kinematics is studied at milliarcsecond scales through very long baseline interferometry (VLBI) observations of the 22~GHz water masers, which are ideal test particles to measure the three-dimensional (3D) motion of shocks owing to the interaction of winds and jets with ambient gas. We employed the Jansky Very Large Array (JVLA) at 6,~13,~and~22~GHz in the A-~and~B-Array configurations to determine the spatial structure and the spectral index of the radio continuum emission, and address its nature.}
   {In about half of the targets, the water masers observed at separation \ $\le$1000~au from the YSOs trace either or both of these kinematic structures: 1)~a spatially elongated distribution oriented at close angle with the direction of collimation of the maser proper motions (PM), and \ 2)~a linear local standard of rest (LSR) velocity (\Vlsr) gradient across the YSO position. The kinematic structure~(1) is readily interpreted in terms of a protostellar jet, as confirmed in some targets via the comparison with independent observations of the YSO jets, in thermal (continuum and line) emissions, reported in the literature. The kinematic structure~(2) is interpreted in terms of a disk-wind (DW) seen almost edge-on on the basis of several pieces of evidence: \ first, it is invariably directed perpendicular to the YSO jet; \ second, it agrees in orientation and  polarity with the \Vlsr\ gradient in thermal emissions (when reported in the literature) identifying the YSO disk at scales of \ $\le$1000~au; third, the PMs of the masers delineating the \Vlsr\ gradients hint at flow motions at a speed of \ 10--20~\kms\ directed at large angles with the disk midplane. In the remaining targets, the maser PMs are not collimated but rather tend to align along two almost perpendicular directions. To explain this peculiar PM distribution, and in light of the observational bias strongly favoring masers moving close to the plane of sky, we propose that, in these sources, the maser emission could originate in DW-jet systems slightly inclined ($\le$30\degree) with respect to edge-on. Magneto-centrifugally driven DWs could in general account for the observed velocity patterns of water masers.} 
 {}


   \maketitle
%
\section{Introduction}

Thanks to the recent advance in sensitivity and angular resolution of (sub-)millimeter interferometers, it is now possible to map the dust and molecular line emission from young stellar objects (YSO) on scales of a few 10--100~au, and get insights into the physical conditions and kinematics of the circumstellar gas. In particular, Atacama Large Millimeter Array (ALMA) observations reveal that rotating disks and jets can be associated with YSOs of all masses up to at least 20~\ms\ \citep{Joh15,Lee17,Gin18,Ile16,Mos19,San19a}. In a few cases, accretion via spiral filaments is also proposed \citep{Izq18,God18}. From a theoretical standpoint, the interplay between disks and jets is crucial for the formation of high-mass ($>$8~\ms) stars \citep[see, e.g.,][]{Kru09,Kui11,Kui15} to prevent powerful stellar radiation from sweeping away the natal cocoon before enough mass is gained by the protostar. In fact, disks are expected to channel and boost the ram pressure of the accreting gas, whereas the energetic stellar photons should leak through the polar cavities of the jet. Although these recent observational and theoretical achievements assess the general importance of disks and jets in the formation of high-mass stars, the detailed spatial distribution and kinematics of the dusty and gaseous constituents of the disks, as well as the processes regulating mass accretion and ejection, still have yet to be constrained through observations and understood theoretically.

YSO jets are usually modeled as magneto-centrifugally driven winds, powered by the rotation and gravitational energy and launched along the magnetic field lines either from the disk inner edge (``X-wind'', \citet{shu95}), or across a much larger (up to a few 100~au) portion of the disk (magneto-centrifugal (MC) ``disk-wind'', \citet{Pud05}). Recent non-ideal magneto-hydrodynamic (MHD) simulations of the collapse of a massive prestellar core and the formation of outflows, achieving unprecedentedly high angular resolution, resolve a MC fast jet from a slow, wide-angle magnetic-pressure-driven (MP) tower flow, the latter dominating the angular-momentum transport \citep{Koe18}.
In the specific case of high-mass YSOs, several models suggest that the magnetic field could be less efficient in driving and collimating the outflows, which is due to several factors, such as the gravitational fragmentation of the accretion disk, the thermal pressure of the ionized gas at the base of the jet, and the powerful (isotropic) radiation pressure from the YSO \citep{Pet11,Vai11}. 

From an observational point of view, we can sort the protostellar winds into two large classes on the basis of their geometrical appearance: 1)~polar winds (hereafter named PW), and \ 2)~disk-winds (DW). PWs originate at radii \ $\ll$1~au \ from the star and their launching region cannot be  spatially resolved with the present observational techniques. Radiatively driven (via ultraviolet absorption lines) stellar winds \citep{Abb82} and MC X-winds are examples of PWs. DWs emerge from the circumstellar disk at radii ranging from \ $\la$10~au to a few 100~au, and their launching region can be studied through observations achieving angular resolutions of \ $\sim$1~mas, in principle. Various driving mechanisms have been proposed for DWs: \ MC DWs \citep{Pud05}, \ MP tower flows \citep{Koe18}, and \ photoevaporated DWs \citep{Hol94}.

Previous single-dish surveys \citep{Beu02b,Wu04,Zha05,Kim06,deVil15} of protostellar outflows  in high-mass star forming regions achieve angular resolutions, between \ 5\arcsec\ and 30\arcsec\ (corresponding to linear scales of \ $\ge$10$^4$~AU at distances of a few kpc of high-mass YSOs), 
that are about ten times larger than the clustering scales ($\sim$10$^3$~AU) of multiple forming stars, and they are generally insufficient in isolating the flow pattern produced by a single high-mass YSO. Besides single-dish surveys, some targeted interferometric studies of massive outflows at millimeter wavelengths in typical molecular outflow tracers ($^{12}$CO, $^{13}$CO, SiO, and HCO$^{+}$), achieving angular resolutions of \ $\sim$1\arcsec--10\arcsec\ \citep[see][for a recent review]{Arc07}, succeed in isolating the outflow of the most massive YSO in the region. However, the sampled scales still remain ten to hundreds of times larger than the model-predicted disk size ($\sim$100~AU). They are thus insufficient in investigating the flow launching region, as well as in determining the physical nature of the engine powering the larger scale molecular outflow. 

A large number of weak ($\le$1~mJy) and compact ($\le$0\farcs6) radio sources associated with luminous ($\ga$10$^3$~\ls) YSOs have been recently discovered by means of subarcsecond resolution, sensitive Jansky Very Large Array (JVLA) observations by \citet{Ros16}. Based on the analysis of the radio continuum properties of the whole sample of 70 objects, about 30\% of the sources are considered to be good jet candidates associated with high-mass protostars, while the most compact emissions could be pressure-confined \HII~regions \citep{Ros19}. These recent results indicate that radio jets can be commonly associated with high-mass YSOs similarly as for low-mass protostars \citep{Ang96}. Although the radio continuum observations by \citet{Ros16} achieve sufficient angular resolution toward the nearest targets to potentially resolve the flow launching region, they do not provide any information on the gas kinematics, which is essential to discriminate, for instance, between fast ionized winds, and confined or slow-expanding \HII~regions.   

To our knowledge, up to now, the nearest high-mass YSO Source~I in Orion-BN/KL \citep{Mat10,Iss17} is the only object in which the launching region of the protostellar outflow has been spatially resolved and the gas kinematics determined, through SiO maser very long baseline interferometry (VLBI) observations. In this unique object, these observations provide clear evidence for DW operation, although the driving mechanism of the DW is still uncertain \citep{Mat10}.


We have been carrying out the  Protostellar Outflows at the EarliesT Stages (POETS) survey, with the goal of imaging the disk-outflow interface on scales of \ 10-100~au \  in a statistically significant sample (36) of luminous YSOs, targeting both the molecular and ionized components of the outflows. The outflow kinematics is studied at milliarcsecond scales through VLBI observations of the 22~GHz water masers, which are ideal test particles to measure the three-dimensional (3D) motion of shocks owing to the interaction of winds and jets with ambient gas \citep{Tor03,Mos05,God06a,San12,Bur16,Hun18}. We have employed the JVLA at C- (6~GHz), Ku- (13~GHz), and K-band (22~GHz) in the A-~and~B-Array configurations (FWHM beams of \ 0\farcs1--0\farcs4) to determine the spatial structure and the spectral index of the radio continuum emission, and address its nature. \citet{Mos16} (hereafter Mos16) presented the first results of the POETS survey for a pilot sample of 11~targets, describing  also the target selection, the observations and the data analysis.  \citet{San18} (hereafter Paper~I) reported on and interpreted the radio continuum data of the whole sample. Here, we present the combined analysis of the radio continuum and water maser observations for the targets not included in Mos16, and focus on the water maser kinematics. In an accompanying paper (Moscadelli et~al., in prep., hereafter Paper~IV), we report on the statistics of the maser properties and examine global correlations in the observed physical quantities.

In Sect.~\ref{obs_res}, we present the information derived from the spatial and velocity distribution of the H$_2$O masers, which we use in combination with the radio continuum to characterize the outflow properties in each POETS target. In Sect.~\ref{kin-mas}, we compare the 3D velocity patterns of the H$_2$O masers in different targets, and propose a coherent picture to interpret the maser kinematics. The proposed interpretation is discussed in Sect.~\ref{discu}, and our conclusions are drawn in Sect.~\ref{conclu}.

%
\longtab{
\begin{landscape}
\begin{longtable}{l c c r r r r c c r r r r c r}       
\caption{\label{wat_kin} Water maser kinematics}\\             
\hline \hline                 
Source &     d     &  L$_{\rm bol}$         &     N$_{\rm mas}$ & $\langle R \rangle$ & $\Delta R$  &  $D_{\rm MD}$ &  MA$_{\rm MD}$ & $A_{\rm VG}$ &  $\langle$ A$_{\rm PM}$ $\rangle$ & PD$_{\rm PM}$ &  MA$_{\rm RC}$ & MA$_{\rm LD}$ & A$_{\rm LJ}$&  Ref. \\  
       &    (kpc)  &  (L$_{\odot}$)         &               &          (au)         &     (au)    &    (au)    &   (\degr)      & (\kms / au) &               (\kms)              &    (\degr)    &    (\degr)     &  (\degr) & (\degr) &  \\
\hline 
\endfirsthead
\caption{continued.}\\
\hline \hline                 
Source &     d     &  L$_{\rm bol}$         &     N$_{\rm mas}$ & $\langle R \rangle$ & $\Delta R$  &  $D_{\rm MD}$ &  MA$_{\rm MD}$ & $A_{\rm VG}$ &  $\langle$ A$_{\rm PM}$ $\rangle$ & PD$_{\rm PM}$ &  MA$_{\rm RC}$ &  MA$_{\rm LD}$ & A$_{\rm LJ}$&  Ref. \\  
       &    (kpc)  &  (L$_{\odot}$)         &               &          (au)         &     (au)    &    (au)    &   (\degr)      & (\kms / au) &               (\kms)              &    (\degr)    &    (\degr)     &  (\degr) & (\degr) &  \\
\hline 
\endhead
\hline
\endfoot
%
  \multicolumn{15}{c}{\bf Group~N: near the YSO}  \\
 \multicolumn{15}{c}{}  \\
\multicolumn{15}{c}{Subgroup~N-G: $V_{\rm LSR}$ gradient} \\
\hline
G011.92$-$0.61 &   3.37$\pm$0.35 &   1.2 $\times$ 10$^4$                  &       27 &   412.3 &   232.2 &    1019.1 & 127.8$\pm$11.9 &  1.6 $\times$ 10$^{-2}$  &     21.0 &   71       &      62.0$\pm$9.0  & 127   & 53 & 13, 2 \\
G035.02$+$0.35 &   2.33$\pm$0.22 &   1.0 $\times$ 10$^4$                  &       27 &   306.1 &   212.5 &     682.9 & 145.6$\pm$2.1  &  2.2 $\times$ 10$^{-2}$  &     16.3 &   72       &     152.0$\pm$8.0  & 157   & 25 & 14, 4 \\
G075.76$+$0.34 &   3.51$\pm$0.28 &   1.4 $\times$ 10$^4$                  &       19 &   351.3 &    96.0 &     376.4 & 142.3$\pm$1.7  &  7.0 $\times$ 10$^{-3}$  &     14.4 &   84       &      76.0$\pm$5.0  &    & 75  & 5  \\
G075.78$+$0.34-{\em IG} &   3.72$\pm$0.43 &   1.1 $\times$ 10$^4$         &       27 &   155.7 &    84.9 &     349.9 &  33.3$\pm$1.7  &  3.9 $\times$ 10$^{-2}$  &     16.2 &   44       &      82.0$\pm$4.0  &    &   &   \\
G176.52$+$0.20-{\em IG} &   0.96$\pm$0.02 &   1.5 $\times$ 10$^2$         &       39 &    23.2 &     4.5 &      23.8 & 150.2$\pm$1.6  &  1.1 $\times$ 10$^{-1}$  &      9.8 &  137       &         ...        &    & 73  &  8 \\
G236.82$+$1.98-{\em IG} &   3.36$\pm$0.20 &  {\it 2.3 $\times$ 10$^3$ }   &       23 &   154.7 &    32.8 &     241.1 &  42.2$\pm$4.5  &  8.6 $\times$ 10$^{-2}$  &     17.3 &  {\bf 42 } &      42.0$\pm$17.0 &  53\tablefootmark{a}  & 127\tablefootmark{a}  &  15 \\
\hline
 \multicolumn{15}{c}{}  \\
\multicolumn{15}{c}{Subgroup~N-C: collimated motion} \\
\hline
G005.88$-$0.39 &   2.99$\pm$0.18 &   5.8 $\times$ 10$^4$                 &       26 &  1648.3 &  1544.8 &   18047.7 &  33.3$\pm$18.8 &  ...  &     18.6 &  {\bf 55  } &     145.0$\pm$9.0  &    & 30  &  1 \\
G012.68$-$0.18 &   2.40$\pm$0.18 &   5.7 $\times$ 10$^3$                 &       21 &   254.7 &   102.3 &     870.3 & 166.2$\pm$43.7 &  ...  &     24.4 &  {\bf 65  } &      55.0$\pm$6.0  &    &   &   \\
G075.78$+$0.34-{\em OG} &   3.72$\pm$0.43 &   1.1 $\times$ 10$^4$        &       34 &  1405.3 &   876.6 &    4298.0 & 110.2$\pm$1.3  &  ...  &     24.8 &  {\bf 96  } &      82.0$\pm$4.0  &    &   &   \\
G076.38$-$0.62 &   1.30$\pm$0.09 &   {\it 1.4 $\times$ 10$^4$ }          &       50 &    70.6 &    16.3 &     144.4 &  73.1$\pm$2.2  &  ...  &     52.1 &  {\bf 53  } &         ...        &    &   &   \\
G092.69$+$3.08  &   1.63$\pm$0.05 &   ({\it 4.7 $\times$ 10$^3$})        &       50 &   205.5 &    40.9 &     144.0 &  37.9$\pm$8.3  &  ...  &     23.7 &  {\bf 49  } &      56.0$\pm$12.0 &    &   &   \\
G105.42$+$9.88\tablefootmark{b}~3B &   0.89$\pm$0.05 &   ({\it 5.8 $\times$ 10$^2$})      &       72 &    17.7 &     3.6 &      31.5 & 134.1$\pm$17.8 &  ...  &      5.9 &  {\bf 44  } &      60.0$\pm$6.0  &    & 57  & 6  \\
G108.20$+$0.59 &   4.37$\pm$0.53 &   2.1 $\times$ 10$^3$                 &       39 &   213.9 &   118.4 &     788.3 &  36.7$\pm$12.7 &  ...  &     22.5 &  {\bf 15  } &         ...        &    &   &   \\
G111.25$-$0.77 &   3.40$\pm$0.18 &   {\it 5.0 $\times$ 10$^3$}           &       57 &   425.9 &   322.2 &    1428.2 &  42.6$\pm$41.0 &  ...  &     16.7 &  {\bf 64  } &     125.0$\pm$5.0  &    &   &   \\
G176.52$+$0.20-{\em OG} &   0.96$\pm$0.02 &   1.5 $\times$ 10$^2$        &       32 &    76.3 &    38.4 &     149.8 &  66.9$\pm$22.5 &  ...  &     23.7 &  {\bf 57  } &         ...        &    & 73  &  8 \\
G236.82$+$1.98-{\em OG} &   3.36$\pm$0.20 &   {\it 2.3 $\times$ 10$^3$}  &       21 &   417.1 &   123.6 &     346.7 & 162.4$\pm$3.0  &  ...  &      7.2 &  {\bf 141 } &      42.0$\pm$17.0 &  53\tablefootmark{a}  & 127\tablefootmark{a}  & 15  \\
{\it AFGL~5142 } &   2.14$\pm$0.05 &   5.0 $\times$ 10$^3$               &       23 &   280.4 &   135.1 &     860.3 & 147.3$\pm$4.8  &  ...  &     14.9 &  {\bf 134 } &     130.0$\pm$10.0 &    & 142  & 12  \\
{\it IRAS 20126$+$4104} &   1.64$\pm$0.05 &   1.3 $\times$ 10$^4$        &       26 &   586.2 &   103.3 &    1138.0 & 117.1$\pm$6.9  &  ...  &     56.5 &  {\bf 113 } &     125.0$\pm$10.0 &    & 125  & 11  \\
\hline
 \multicolumn{15}{c}{}  \\
\multicolumn{15}{c}{Subgroup~N-UNC: undetermined or non-collimated motion} \\
\hline
G009.99$-$0.03\tablefootmark{c}  &   {\it 5.0 }    &   {\it 1.2 $\times$ 10$^4$}       &    16    &   608.2 &  174.4   &   1364.5 &   44.3$\pm$ 1.3  &  ...  &  ...  &  ... &  71.0$\pm$21.0 &    &   &   \\
G012.43$-$1.12\tablefootmark{c}  &   {\it 3.7 }    &   {\it 4.2 $\times$ 10$^4$}       &        3 &   216.2 &     7.8 &      17.1 &  ...          &  ...  &     ...   &  ...  &         ...         &    &   &   \\
G012.90$-$0.24  &   2.45$\pm$0.15 &   8.6 $\times$ 10$^2$            &        2 &   128.4 &     1.5 &       3.2 &   ...          &  ...  &     51.0  &  176  &     123.0$\pm$19.0  &    &   &   \\
G014.64$-$0.58 &   1.83$\pm$0.07 &   1.1 $\times$ 10$^3$             &        5 &   173.9 &     8.1 &      86.5 &  53.4$\pm$1.4  &  ...  &     23.4  &   45  &     178.0$\pm$4.0   &    &   &   \\
G026.42$+$1.69\tablefootmark{c} &   {\it 3.1 } &   {\it 9.0 $\times$ 10$^3$}          &        7 &   230.8 &    38.0 &     166.8 & 122.0$\pm$11.1 &  ...  &     ...   &  ...  &     152.0$\pm$3.0   &    &   &   \\
G049.19$-$0.34 &   5.29$\pm$0.20 &   6.0 $\times$ 10$^3$             &       11 &   191.0 &    33.8 &     312.1 & 108.4$\pm$4.1  &  ...  &     25.8  &   26  &         ...         &    &   &   \\
G074.04$-$1.71 &   1.59$\pm$0.05 &   {\it 3.7 $\times$ 10$^2$}       &       29 &   184.1 &   178.6 &     568.0 & 156.0$\pm$26.4 &  ...  &     18.6  &   46  &     160.0$\pm$15.0  &    & 158  &  5 \\
G079.88$+$1.18 &   1.61$\pm$0.07 &   8.6 $\times$ 10$^2$             &        6 &   138.3 &    40.5 &     148.8 &  45.0$\pm$10.0 &  ...  &     32.7  &   57  &         ...         &    &   &   \\
G090.21$+$2.32 &   0.67$\pm$0.02 &   2.7 $\times$ 10$^1$             &       15 &    30.6 &     3.7 &      13.7 &   ...          &  ...  &      4.2  &  111  &      77.0$\pm$6.0   &    &   &   \\
G097.53$+$3.18--M &   7.52$\pm$0.96 &   ({\it 8.8 $\times$ 10$^4$})     &       82 &   869.9 &   523.0 &    3272.8 &  33.2$\pm$28.9 &  ...  &     16.4  &   97  &      55.0$\pm$6.0   &    &   &   \\
G100.38$-$3.58 &   3.44$\pm$0.10 &   {\it 8.5 $\times$ 10$^3$}       &       49 &   261.2 &   119.6 &     973.7 &  18.1$\pm$9.1  &  ...  &     20.2  &  150  &      13.0$\pm$14.0  &    &   &   \\
G105.42$+$9.88\tablefootmark{b}~3A &   0.89$\pm$0.05 &  ({\it 5.8 $\times$ 10$^2$})   &        2 &    14.9 &     0.2 &       1.5 &  ...           &  ...  &     16.3  &   13  &      66.0$\pm$27.0  &    &   &   \\
G168.06$+$0.82 &   7.69$\pm$2.37 &   1.6 $\times$ 10$^4$             &       15 &   793.0 &   355.9 &    2268.8 & 117.2$\pm$21.9 &  ...  &     32.0  &   54  &     131.0$\pm$7.0   &    &   &   \\
G182.68$-$3.27 &   6.71$\pm$0.50 &  {\it 8.6 $\times$ 10$^2$}        &        3 &   302.4 &    54.3 &     134.3 &  18.2$\pm$1.3  &  ...  &     28.0  &  ...  &         ...         &    &   &   \\
G183.72$-$3.66 &   1.75$\pm$0.04 &   {\it 9.7 $\times$ 10$^2$}       &       22 &    84.7 &    29.9 &     184.0 &  33.0$\pm$34.7 &  ...  &      9.5  &   49  &      54.0$\pm$15.0  &    & 30  &  9 \\
G229.57$+$0.15\tablefootmark{d}~1 &   4.52$\pm$0.29 &   2.2 $\times$ 10$^3$           &        2 &   159.5 &     1.9 &       4.2 &  ...           &  ...  &     18.9  &  ...  &      99.0$\pm$21.0  &    &   &   \\
G229.57$+$0.15\tablefootmark{d}~2 &   4.52$\pm$0.29 &   2.2 $\times$ 10$^3$           &       26 &   273.5 &    90.6 &     615.5 &  52.1$\pm$12.5 &  ...  &     19.8  &   90  &         ...         &    &   &   \\
G359.97$-$0.46\tablefootmark{c}  &   {\it 4.0 }    &   {\it 5.7 $\times$ 10$^4$}       &  4   &     954.1  &  22.2  &    80.6    &  62.1$\pm$8.4  &  ...  &  ...      &  ...  &  28.0$\pm$17.0      &    &   &   \\
\hline 
\\
\multicolumn{15}{c}{\bf Group~F: far from the YSO} \\
\multicolumn{15}{c}{}  \\
\multicolumn{15}{c}{Subgroup~F-A: associated with the YSO} \\
\hline
G016.58$-$0.05 &   3.58$\pm$0.30 &   1.3 $\times$ 10$^4$         &       35 &  1800.8 &   420.7 &    1404.7 & 113.1$\pm$12.1 &  ...  &     42.8 &   74       &      89.0$\pm$15.0 &    & 89  &  3 \\
G031.58$+$0.08 &   4.90$\pm$0.72 &   2.0 $\times$ 10$^4$         &       11 & 11096.8 &   132.2 &     594.3 &  16.8$\pm$3.7  &  ...  &     34.3 &  143       &     146.0$\pm$46.0 &    &   &   \\
G111.24$-$1.24 &   3.47$\pm$0.53 &   {\it 1.0 $\times$ 10$^4$}   &       21 &  3526.1 &    11.2 &      53.6 & 141.5$\pm$38.7 &  ...  &     30.5 & {\bf  57 } &     110.0$\pm$20.0 &    &   &   \\
G160.14$+$3.16 &   4.10$\pm$0.10 &   {\it 8.4 $\times$ 10$^3$}   &        5 & 17272.3 &    35.3 &     144.5 &  70.0$\pm$1.4  &  ...  &     25.5 & {\bf 130 } &     142.0$\pm$14.0 &    & 125.5  & 7  \\
G240.32$+$0.07 &   4.72$\pm$0.47 &   8.3 $\times$ 10$^3$         &       10 &  8444.5 &    15.9 &     111.1 &   7.5$\pm$39.0 &  ...  &     33.0 &  109       &      48.0$\pm$11.0 &    &  144 &  10 \\
\hline
\multicolumn{15}{c}{}  \\
\multicolumn{15}{c}{Subgroup~F-NA: not associated with the YSO} \\
\hline
G012.91$-$0.26 &   2.53$\pm$0.20 &   2.7 $\times$ 10$^4$   &       18 & 16756.3 &    38.2 &     196.8 &  56.1$\pm$38.6 &  ...  &     82.0 &  ...  &         ...          &    &   &   \\
G108.59$+$0.49 &   2.51$\pm$0.20 &   2.8 $\times$ 10$^3$   &        5 &  7296.1 &    11.5 &     223.2 &  36.3$\pm$0.2  &  ...  &     44.0 &  ...  &         ...          &    &   &   \\
\end{longtable}
\tablefoot{Column~1 reports the name of the POETS targets: the two sources in italic characters have been observed prior of the POETS survey; Cols.~2~and~3 give the trigonometric parallax distance from BeSSeL (available for all but four sources, for which the kinematic distance is reported) and the evaluated bolometric luminosity, respectively: kinematic distances and more uncertain luminosities are given in italic characters, and values which might be severe upper limits are enclosed within brackets; Col.~4 lists the number of detected maser features; Cols.~5~and~6 report the average maser distance from the continuum peak and the corresponding standard deviation, respectively; Cols.~8~and~7 give the PA of the major-axis of the maser spatial distribution, and the spread of the maser positions along the major-axis, respectively: if the spread is \ $\le$20~au, the PA of the spatial distribution is not shown; Col.~9 reports the amplitude of the maser \Vlsr\ gradient for the targets in which it is detected; Cols.~10~and~11 list the average PM amplitude and the PA of the preferential direction of the maser PMs (see Sect.~\ref{obs_pm}), respectively, using boldface characters for the PA of the preferential direction if the maser PMs are collimated: targets for which these parameters are not reported have zero or just a single PM measured; Col.~12 indicates the PA of the major-axis of the radio continuum emission associated with the water masers, if the deconvolution with the observing beam was succesful; Cols.~13~and~14 give the PA of the major-axis of the circumstellar disk and the PA of the sky-projected jet axis, if a disk and/or a jet have been detected towards the YSO through high angular-resolution observations in thermal (continuum or line) emission; Col.~15 reports the reference index of the literature observations.\\
\tablefoottext{a}{The elongated chain of H$_2$ 2.2~$\mu$m knots (PA $\approx$53\degree) and the extinction lane in the $J$-band image (PA $\approx$127\degree) are associated with the bright NIR source \#1 of \citet{Var12}, which could correspond with our POETS target \ G236.82$+$1.98 \ within the uncertainties in the astrometric calibration of the UKIRT Wide Field Camera NIR images.}\\
\tablefoottext{b}{In this target, the water maser emission comes from two nearby YSOs, resolved through subarcsecond VLA observations, named VLA~3A and VLA~3B.}\\
\tablefoottext{c}{In this target, the water maser emission has faded away over the VLBI observing epochs, and nor trigonometric parallax distance neither reliable maser proper motions have been measured.}\\
\tablefoottext{d}{In this target, the water maser emission comes from two nearby YSOs, resolved through subarcsecond VLA observations, named VLA-1 and VLA-2.}
}
\tablebib{(1)~\citet{Sol04}; (2)~\citet{Cyg11}; (3)~\citet{Mos16}; (4)~\citet{San19b};
(5)~\citet{Mas19}; (6)~\citet{Tri04}; (7)~\citet{Var10}; (8)~\citet{Fon09};
(9)~\citet{Bro16}; (10) \citet{Qiu14}; (11)~\citet{Ces13}; (12)~\citet{God11a}; (13)~\citet{Ile16}; (14)~\citet{Bel14}; (15)~\citet{Var12}
}
\end{landscape}
}


\begin{figure*}
\centering
\includegraphics[width=0.7\textwidth]{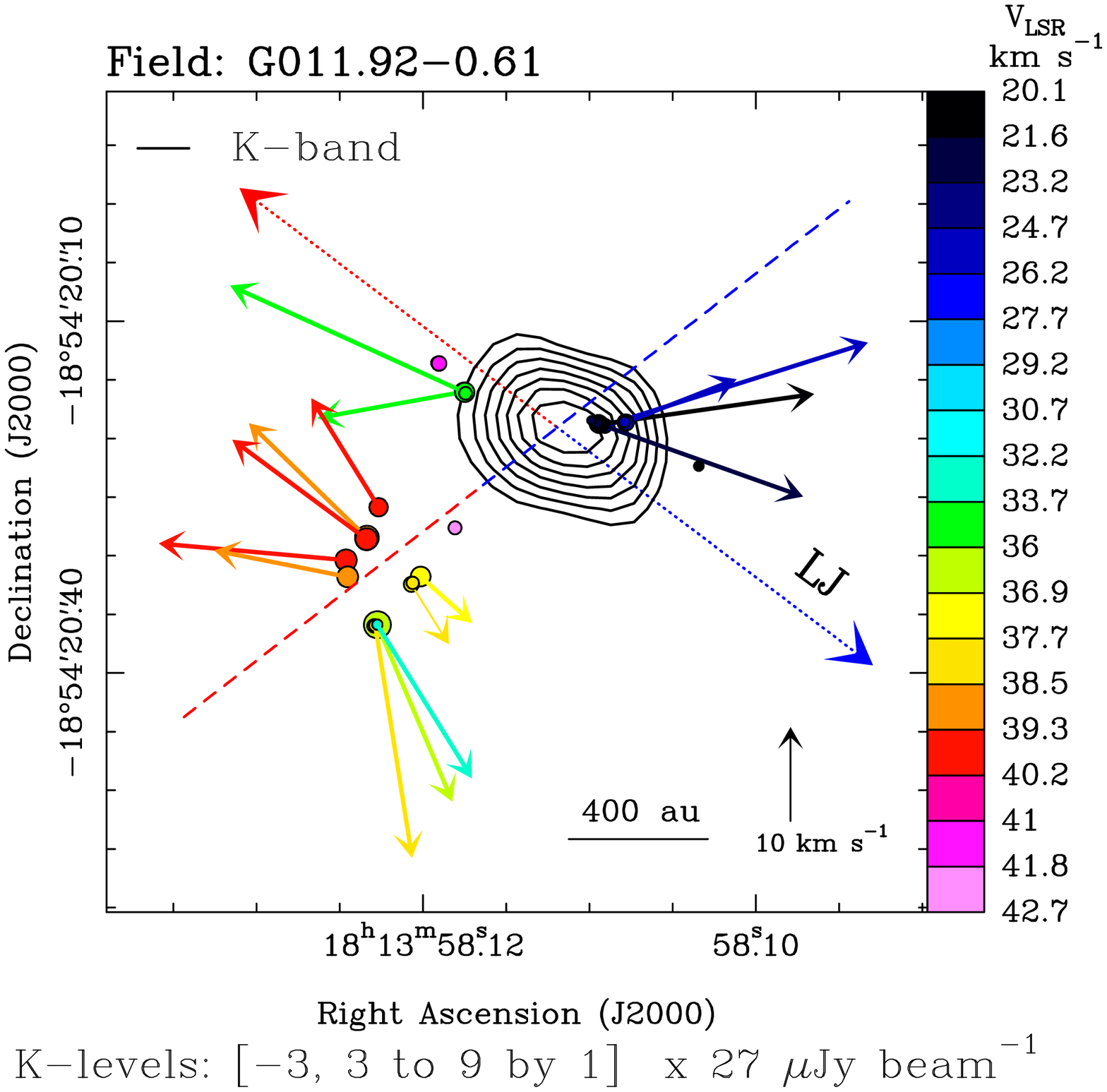}

\hspace*{-11cm} \includegraphics[width=0.46\textwidth,angle=0]{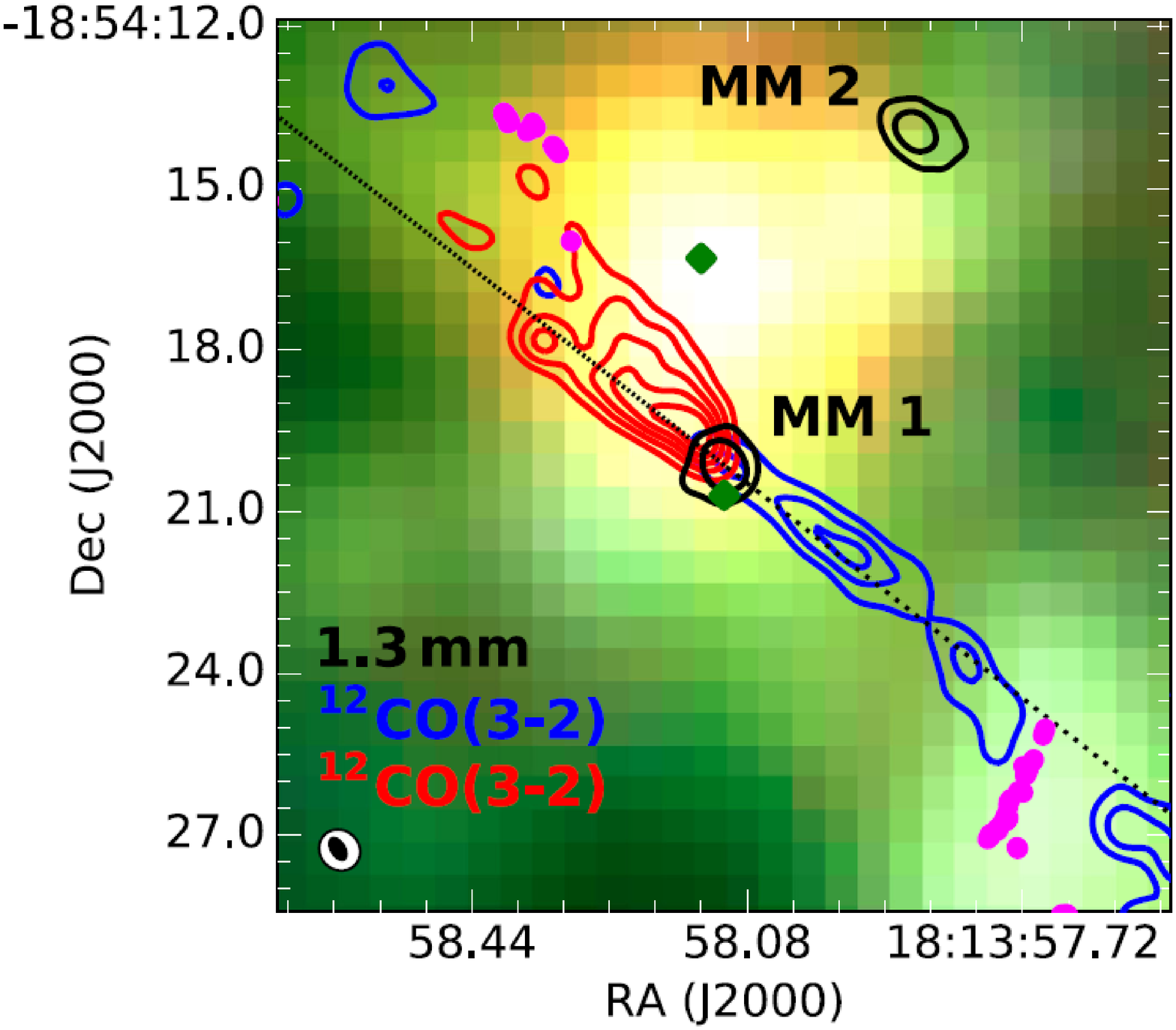}

\vspace*{-7.5cm} \hspace*{8cm} \includegraphics[width=0.6\textwidth]{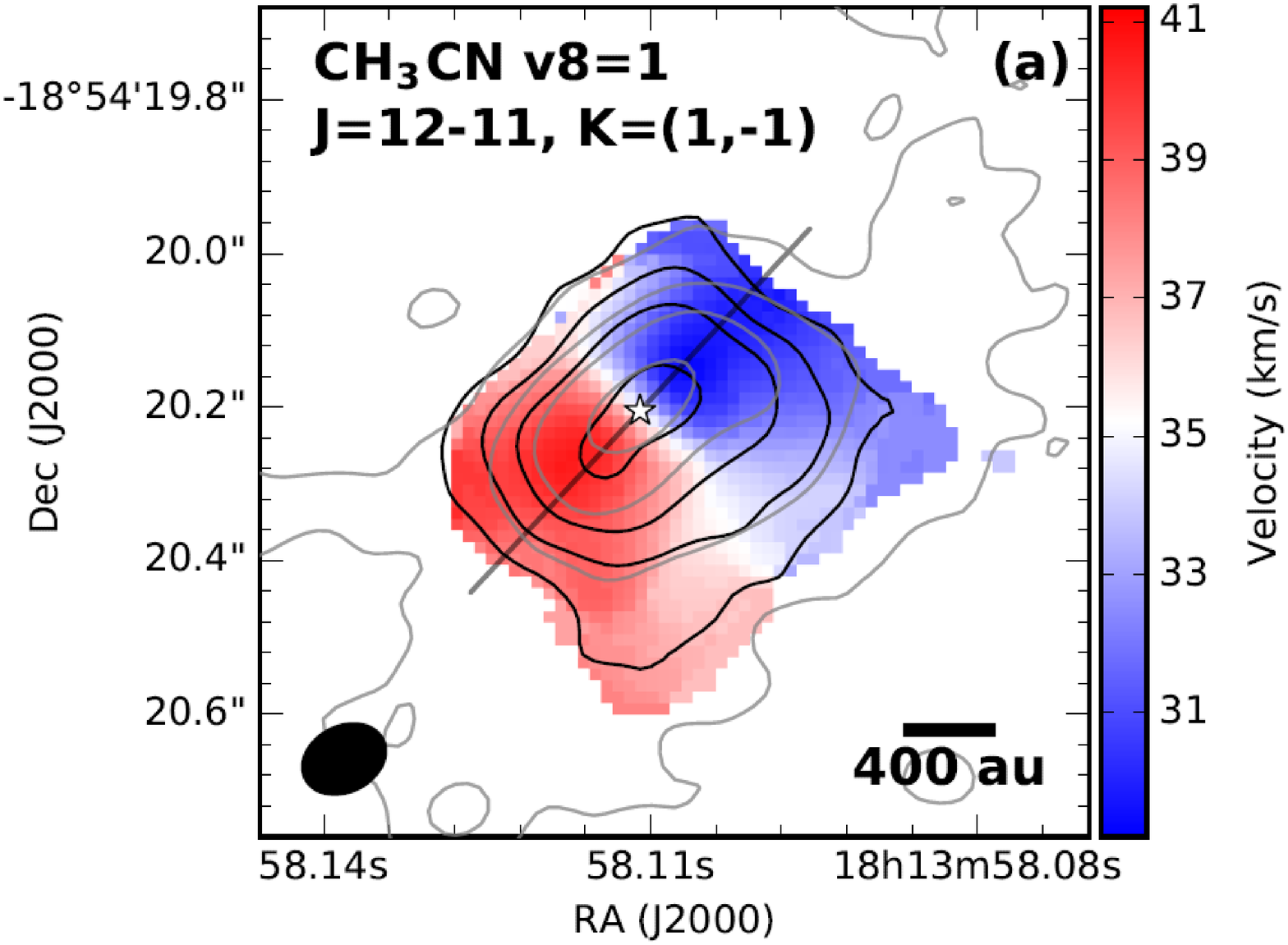}
\caption{Comparison of gas kinematics traced with water masers and thermal emission for source \  G011.92$-$0.61. \ {\it Upper~panel:}~water maser absolute positions ({\it colored dots}) and PMs ({\it colored arrows}) overlaid on the JVLA Ku-~or~K-band (as indicated in the top-left corner) continuum map ({\it black contours}). Colors denote the maser \Vlsr\ as coded in the wedge on the right side. The dot size scales logarithmically with the maser intensity. All the plotted PMs have \ $S/N$~$\ge$~3, and their amplitude scale is given on the bottom. The plotted contours of the JVLA continuum are reported below the plot. The \Vlsr\ gradient traced by the water masers is indicated by a {\it dashed line}, using {\it colors} to denote the {\it red}-~and~{\it blue}-shifted side of the gradient. The {\it dotted arrows} labeled~LJ give the axis of the collimated \ $^{12}$CO outflow observed with the SMA by \citet{Cyg11}, using {\it colors} to distinguish the {\it red}-~and~{\it blue}-shifted flow lobe. 
\ {\it Bottom-left~panel:}~adapted from Fig.~1 by \citet{Ile16}. SMA 1.3~mm continuum ({\it black~contours}) and {\it blue/red}-shifted $^{12}$CO~(3-2) emission. The outflow PA is indicated by the {\it dotted black} line. \ {\it Bottom-right~panel:}~reproduced from Fig.~2 of \citet{Ile18} by permission of the American Astronomical Society. Integrated intensity ({\it black contours}) and intensity-weighted velocity ({\it colorscale}) for the \ CH$_3$CN~v8=1 \ J~=~12-11 (K=1,-1) line observed with ALMA by \citet{Ile18}. The {\it continuous black segment} denotes the PA of the \Vlsr\ gradient in the \ CH$_3$CN~v8=1 \ line.}
\label{F:G11.92}
\end{figure*}

\begin{figure*}
\centering
\includegraphics[width=0.7\textwidth]{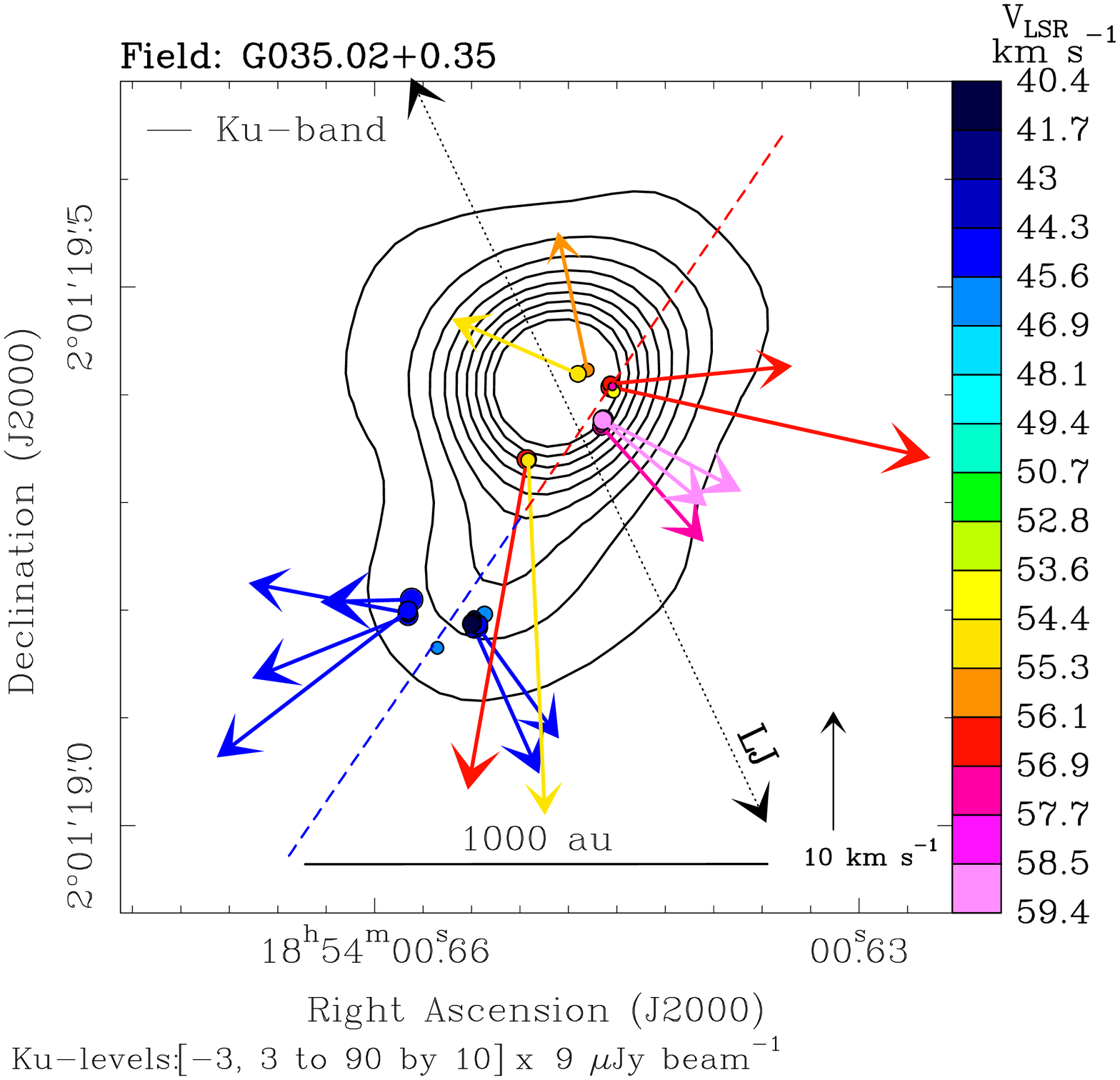}

\vspace*{-2.0cm}\includegraphics[width=0.64\textwidth]{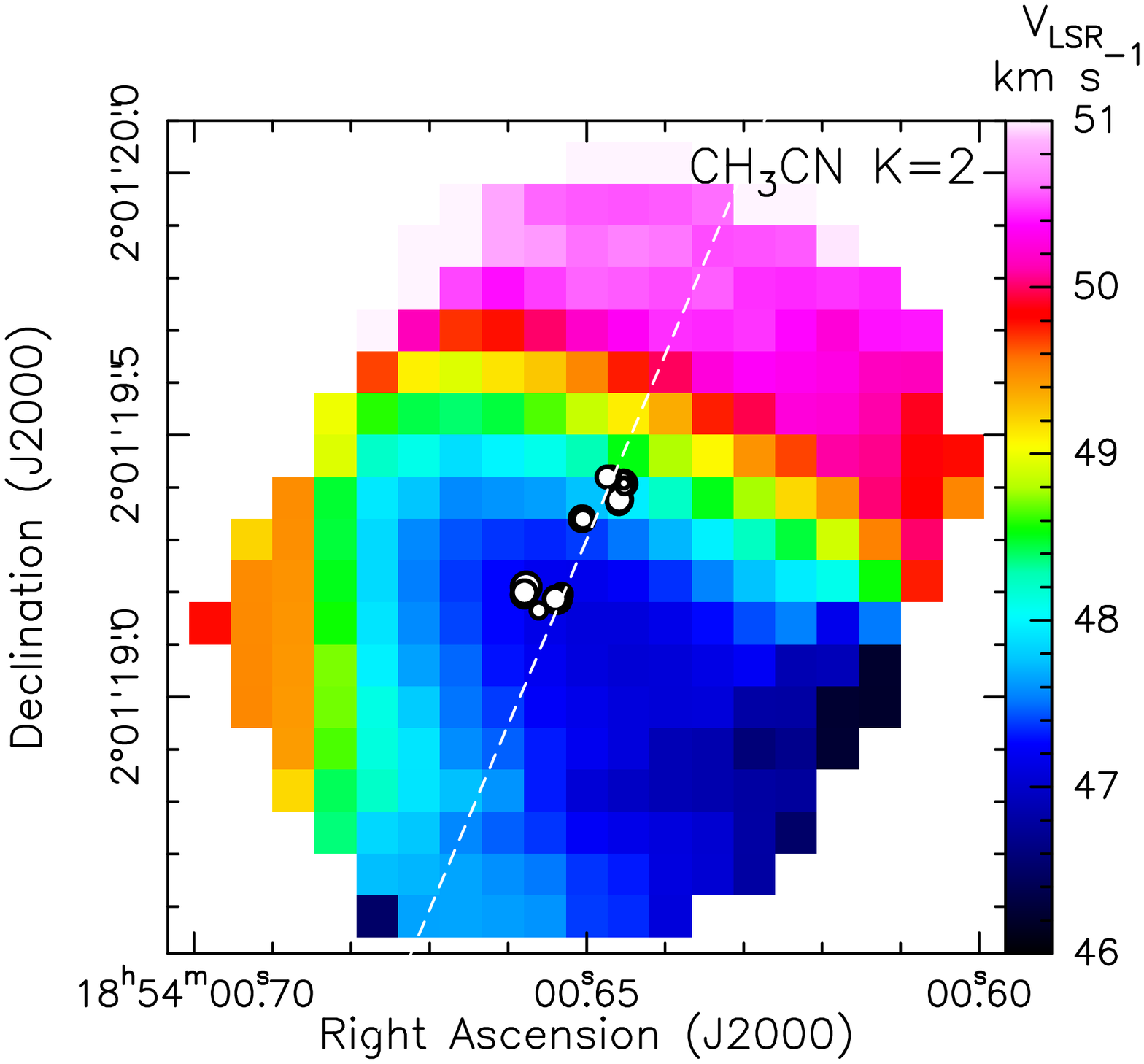}
\caption{Comparison of gas kinematics traced with water masers and thermal emission for source \  G035.02$+$0.35. \ {\it Upper~panel:}~{\it black contours}, {\it colored dots} and {\it arrows}, and the {\it colored dashed line} have the same meaning as in the upper~panel of Fig.~\ref{F:G11.92}. The {\it black dotted arrows} labeled~LJ give the axis of the ionized jet imaged with the JVLA by \citet{San19b}. \ 
{\it Lower~panel:}~Intensity-weighted velocity ({\it colorscale}) of the \ CH$_3$CN~v=0 \ J~=~19-18 (K=2) emission observed with ALMA by \citet{Bel14}. The {\it white dots} mark the position of the water masers plotted in the upper~panel. The {\it dashed white line} indicates the PA of the \Vlsr\ gradient in the \ CH$_3$CN~v=0 \ line.}
\label{F:G35.02}
\end{figure*}

\section{Data analysis and observational results}
\label{obs_res}

In Mos16 we combined the JVLA continuum images with the water maser positions and proper motions (PM) for 11 targets. 
Here, we extend the analysis to the remaining 25 POETS sources, by comparing the water maser observations with the continuum maps at the highest
angular-resolution from Paper~I (corresponding to Ku or, if missing, K~band). The combined radio continuum and water maser observations are shown in \ Figs.~\ref{F:G11.92},~\ref{F:G35.02},~\ref{F:G176.52-G236.82},~and~\ref{F:G9.99-G12.43}-\ref{F:G240.32-G359.97}. The analysis of the BeSSeL\footnote{The Bar and Spiral Structure Legacy (BeSSeL) survey is a VLBA key project, whose main goal is to derive the structure and kinematics of the Milky Way by measuring accurate positions, distances (via trigonometric parallaxes) and proper motions of methanol and water masers in hundreds of high-mass star forming regions distributed over the Galactic Disk \citep{Rei14}.} Very Long Baseline Array (VLBA) data to determine water maser absolute positions and motions, and the method to correct the maser velocities for the apparent proper motion, are described in Mos16. We include the sources IRAS~20126$+$4104 and AFGL~5142, as well, whose water maser emission was studied via multi-epoch VLBI observations by our group in the past \citep{Mos05, God06a}.  Table~\ref{wat_kin} summarizes the main characteristics of the maser spatial and velocity distribution for all the POETS targets.
In the following, we describe the maser parameters and the procedure employed to derive them.

\subsection{H$_2$O maser distance from the radio continuum}
\label{obs_rad}

As already discussed in Mos16, the peak of the radio continuum at Ku~or~K~band is a good proxy for the YSO position, within the absolute position accuracy, $\approx$20~mas, of the JVLA observations.
In most of our targets, the continuum emission is dominated by a single (unresolved or slightly resolved) source, whose peak position is determined by fitting a 2D Gaussian profile\footnote{In a few cases where two nearby peaks are observed, we choose the peak closer to most of the water masers. In three objects, G005.88$-$0.39, IRAS~20126$+$4104, and AFGL~5142, previously studied at subarcsecond resolution with millimeter and near-infrared interferometers, we have used the position of the YSO identified through continuum or molecular line emissions.}.
Columns~5~and~6 of Table~\ref{wat_kin} list the average maser distance, $\langle R \rangle$, from the continuum peak and the corresponding standard deviation, $\Delta R$, respectively, for each POETS target.  Based on the maser distance from the radio continuum peak, the targets are classified in two groups (see Table~\ref{wat_kin}): \ 1)~``near the YSO'' (hereafter named group~``N''), if the water maser emission is mainly found within \ $\la$1000~au from the continuum peak (see, for instance, Figs.~\ref{F:G11.92}~and~\ref{F:G35.02}, upper~panels), and \ 2)~``far from the YSO'' (group~``F''), if the water masers have distances of several 1000~au. The PMs of the more distant masers can be oriented along directions forming small or large angles with the line connecting the continuum peak with the masers (compare Fig.~\ref{F:G111.24-G160.14}, lower~panel, with Fig.~\ref{F:G12.90-G12.91}, lower~panel). In the latter case, occurring only for the  sources \ G012.91$-$0.26 \ and \ G108.59$+$0.49 (see Fig.~\ref{F:G108.20-G108.59}, lower~panel), taking into account also the large maser-continuum separation \ $\ga$10$^4$~au, it is possible that the water masers are excited by a nearby low-mass YSO undetected in the radio, rather than being associated with the YSO responsible for the radio continuum. The far masers are further divided in two subgroups, separating the sources associated with the YSO (subgroup~``F-A'') from those possibly not associated (subgroup~``F-NA''), which are excluded from the following analysis.
 
\subsection{H$_2$O maser internal spatial distribution}
\label{obs_spd}
By fitting a line to the maser positions in each POETS target, we have derived the PA of the major-axis, MA$_{\rm MD}$, of the maser spatial distribution, and the spread, $D_{\rm MD}$, of the maser positions along the major-axis, reported in Cols.~8~and~7, respectively, of Table~\ref{wat_kin}. It is clear that in most targets the maser distribution is significantly elongated, as indicated by the relatively small error in \ MA$_{\rm MD}$, $\la$10\degree. In most cases, the masers extend over distances from a few 10~au to $\sim$1000~au. In five sources, G012.43$-$1.12 (see Fig.~\ref{F:G9.99-G12.43}, lower~panel), G012.90$-$0.24 (see Fig.~\ref{F:G12.90-G12.91}, upper~panel), G090.21$+$2.32 (see Fig.~\ref{F:G90.21-G105.42}, upper~panel), G105.42$+$9.88 VLA~3A (see Fig.~\ref{F:G90.21-G105.42}, bottom-left~panel), and \ G229.57$+$0.15~VLA-1 (see Fig.~\ref{F:G183.72-G229.57}, bottom-left~panel), the masers cluster inside too small a region ($\le$20~au in size) to consider their spatial elongation significant, and we do not report the corresponding value of \ MA$_{\rm MD}$. 

\subsection{Linear velocity gradients}
\label{obs_grad}
A noticeable result that we first report in this work is that in six targets the water masers present \ 1)~an elongated distribution on top of the YSO (pinpointed by the radio continuum peak), together with \ 2)~a clear local standard of rest (LSR) velocity (\Vlsr) gradient along the major-axis of the distribution. In three cases, G011.92$-$0.61 (Fig.~\ref{F:G11.92}, upper-panel), G035.02$+$0.35 (Fig.~\ref{F:G35.02}, upper-panel), and G075.76$+$0.34 (Fig.~\ref{F:G75.76-G75.78}, upper~panel), the whole set of observed masers trace the \Vlsr\ gradient. In the other three sources,  G176.52$+$0.20 (Fig.~\ref{F:G176.52-G236.82}, top-right~panel), G236.82$+$1.98 (Fig.~\ref{F:G176.52-G236.82}, bottom-right~panel), and G075.78$+$0.34 (Fig.~\ref{F:G75.76-G75.78}, bottom-right~panel), only a subset of masers closer to the continuum peak trace the  \Vlsr\ gradient, while the masers at larger separation from the YSO present a remarkably different spatial and 3D velocity distribution. Accordingly, in these latter cases, we have divided the masers in two subsets: \ 1)~in the \Vlsr\ gradient (hereafter named ``{\em IG}''), and \ 2)~out of the \Vlsr\ gradient (named ``{\em OG}''), and we have analyzed the two subsets independently. The six targets (or subsets of masers) exhibiting a \Vlsr\ gradient are collected within the subgroup ``\Vlsr\ gradient'' in Table~\ref{wat_kin} (named subgroup~``N-G''). Col.~9 of Table~\ref{wat_kin} reports the gradient amplitude, $A_{\rm VG}$, derived by a linear fit of the \Vlsr\ vs positions projected along the major-axis of the spatial distribution.

\subsection{Distribution of proper motions}
\label{obs_pm}

A high degree of collimation of the water maser PMs provides evidence for a jet and the collimation direction of the maser PMs identifies the sky-projected jet axis. Figures~\ref{histo_PA_1}-\ref{histo_PA_3} show the histogram of the PM directions for all the targets with at least two measured PMs. In each source, only the more accurate PMs are considered, whose amplitude is determined with a signal-to-noise ratio ($S/N$) better than 3. The range of PA (N to E) to evaluate the PM directions is narrowed to \ 0\degree $\le$ PA $<$ 180\degree, by adding \ 180\degree\ to the PA of the PMs pointing toward W. On the basis of the histograms of PM~PA, we have set a criterion to assess the degree of likelihood that the water masers are tracing a collimated flow. The PMs of a given source are considered to be collimated if: \ 1)~over 70\% of the PMs are within a PA range \ $\le$60\degree, and \ 2)~at least \ 4 PMs fall in that specific range. The choice of the 60\degree\ range is based on the average PM~PA error \ $\le$10\degree, and the assumption that a collimated flow has an opening angle \ $\le$30\degree. The 70\%-threshold is selected to include sources where the association between water masers and a known protostellar jet is clearly established (for instance, IRAS~20126$+$4104 and AFGL5142), and to reject more ambiguous cases.  Condition~2 filters out sources with a very poor sampling of the PM distribution. 

Adopting the above criterion,  excluding the sources already cataloged as ``\Vlsr\ gradient'', the maser targets near the YSO are split in two other subgroups in Table~\ref{wat_kin}: \ 1)~collimated motion (subgroup~``N-C''), and \ 2)~undetermined or non-collimated motion (subgroup~``N-UNC''). An example of a collimated flow traced by the water masers is the target \ G105.42$+$9.88 VLA~3B (see Fig.~\ref{F:G90.21-G105.42}, bottom-right~panel), where five, almost parallel PMs are measured; on the other hand, the maser kinematics in the nearby source \ G105.42$+$9.88 VLA~3A (Fig.~\ref{F:G90.21-G105.42}, bottom-left~panel) \ is undetermined, because only two PMs are observed here. The subgroup~N-UNC includes also the sources \ G009.99$-$0.03 \ and 
\ G012.43$-$1.12 (Fig.~\ref{F:G9.99-G12.43}), G026.42$+$1.69 (Fig.~\ref{F:G14.64-G26.42}, lower~panel), G182.68$-$3.27 (Fig.~\ref{F:G168.06-G182.68}, lower~panel),  G229.57$+$0.15~VLA-1 (Fig.~\ref{F:G183.72-G229.57}, bottom-left~panel), and \ G359.97$-$0.46 (Fig.~\ref{F:G240.32-G359.97}, lower~panel), with zero or only a single PM measured. Finally, an example of a well-sampled, but non-collimated motion is provided by the source \ G097.53$+$3.18--M (see Mos16, Fig.~9, upper~panel), where the water maser PMs are clearly not aligned around one particular direction. 

Employing the PM~PA histograms in Figs.~\ref{histo_PA_1}-\ref{histo_PA_3}, for each target we have defined the PA of the preferential direction of the maser PMs, PD$_{\rm PM}$, in two steps. First, we have identified the  \ 60\degree\ range in PA that includes the largest fraction of measured PMs. Then, considering only the PMs falling within that specific 60\degree-range, the value of \ PD$_{\rm PM}$ \ is calculated as the PA of the direction along which the average projection of the normalized\footnote{We normalize the PMs such as to make our estimate of \ PD$_{\rm PM}$ \ dependent on the PM PA only, and not on the PM amplitude.} PMs is maximum. Columns~10~and~11 of Table~\ref{wat_kin} report the average PM amplitude, A$_{\rm PM}$, and the value of \ PD$_{\rm PM}$. 
It is interesting to note that, in each of the three targets \ G075.78$+$0.34 (Fig.~\ref{F:G75.76-G75.78}, bottom-left~panel), G176.52$+$0.20 (Fig.~\ref{F:G176.52-G236.82} top-left~panel), and \ G236.82$+$1.98 (Fig.~\ref{F:G176.52-G236.82}, bottom-left~panel), the subset of masers out of the \Vlsr\ gradient show well collimated PMs, and the preferential direction of the PMs, PD$_{\rm PM}$, differ less than \ 20\degree\ from the major-axis, MA$_{\rm MD}$, of the spatial distribution of the masers (see Table~\ref{wat_kin}).

\subsection{Radio continuum morphology}
\label{obs_con}
The radio continuum emission associated with the water masers has been fully analyzed in Paper~I. Based on the positive spectral index, between \ $-$0.1 \ and \ 1.3, of the integrated fluxes and, in a few cases, on the resolved spatial structure, most of the radio sources are interpreted in terms of (proto)stellar winds or jets. We are interested in comparing the continuum morphology with the velocity pattern of the water masers on scales of a few 100~au. To this purpose, in each target, we select the continuum map observed at the highest angular-resolution (the one used to produce the plots in combination with the water masers), and, if the deconvolution with the observing beam was successful, in Table~\ref{wat_kin} (Col.~12) we report the Gaussian-fit PA of the major-axis of the continuum emission, MA$_{\rm RC}$. Following Paper~I (see Table~A.1), the radio emission at the three observing bands (C, Ku, and K) is only slightly resolved in several targets, resulting in large uncertainties in the value of \ MA$_{\rm RC}$.  Besides, in some sources, the direction of elongation of the radio emission can vary significantly across the three observing bands (see, for instance, the targets \ G092.69$+$3.08 (Mos16, Fig.~8) and G097.53$+$3.18--M (Mos16, Fig.~9, upper~panel), where the  C-~and~K-band continua are stretched along two almost perpendicular directions), which complicates the interpretation of the nature of the radio continuum. Despite these limitations, the available information on the continuum structure proves to be very useful to interpret the kinematic structures traced by the H$_2$O masers, as discussed in Sect.~\ref{rad_win}.

\subsection{Jets and disks from the literature}
For each POETS target, we have been searching the literature to find independent evidence of the presence of circumstellar disks and jets. The unambiguous association with the POETS target is ensured if the detection of disks and jets is based on observations with angular resolution \ $\la$1\arcsec. In most cases, we rely on (sub-)mm interferometric (ALMA; Submillimeter~Array, SMA; Plateau~de~Bure~Interferometer, PdBI; Combined Array for Research in Millimeter Astronomy, CARMA) observations of thermal (dust) continuum or (molecular) line emissions, in fewer cases on subarcsecond, radio continuum (JVLA) or near-infrared (NIR) (Hubble Space Telescope, HST; Large Binocular Telescope, LBT; United Kingdom InfraRed Telescope, UKIRT) data. We report the PA of the major-axis of the disk, MA$_{\rm LD}$, and the sky-projected axis of the jet, A$_{\rm LJ}$, in Cols.~13~and~14, respectively, of Table~\ref{wat_kin}.

\begin{figure*}
\includegraphics[width=0.48\textwidth]{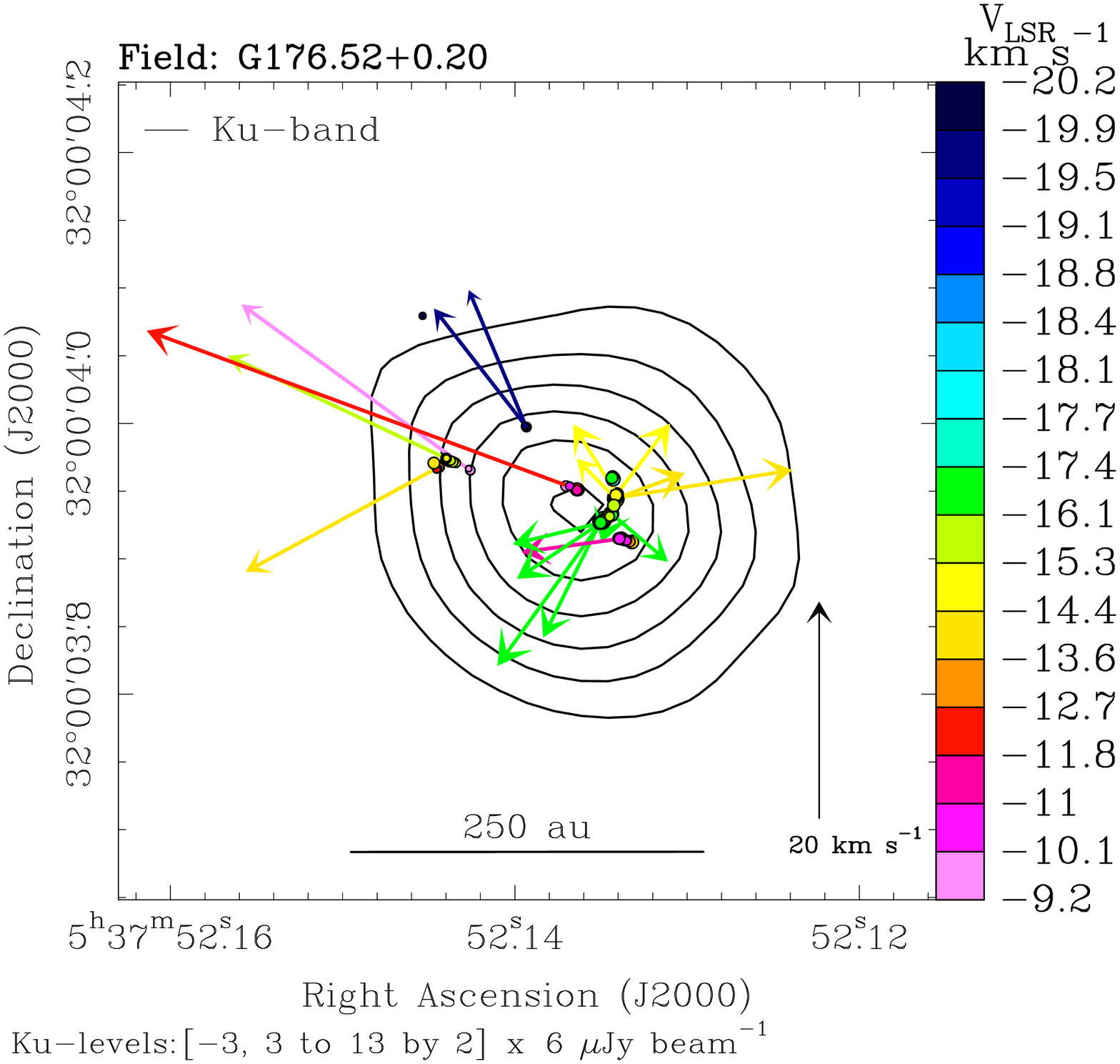}
\hspace*{0.4cm}\includegraphics[width=0.47\textwidth]{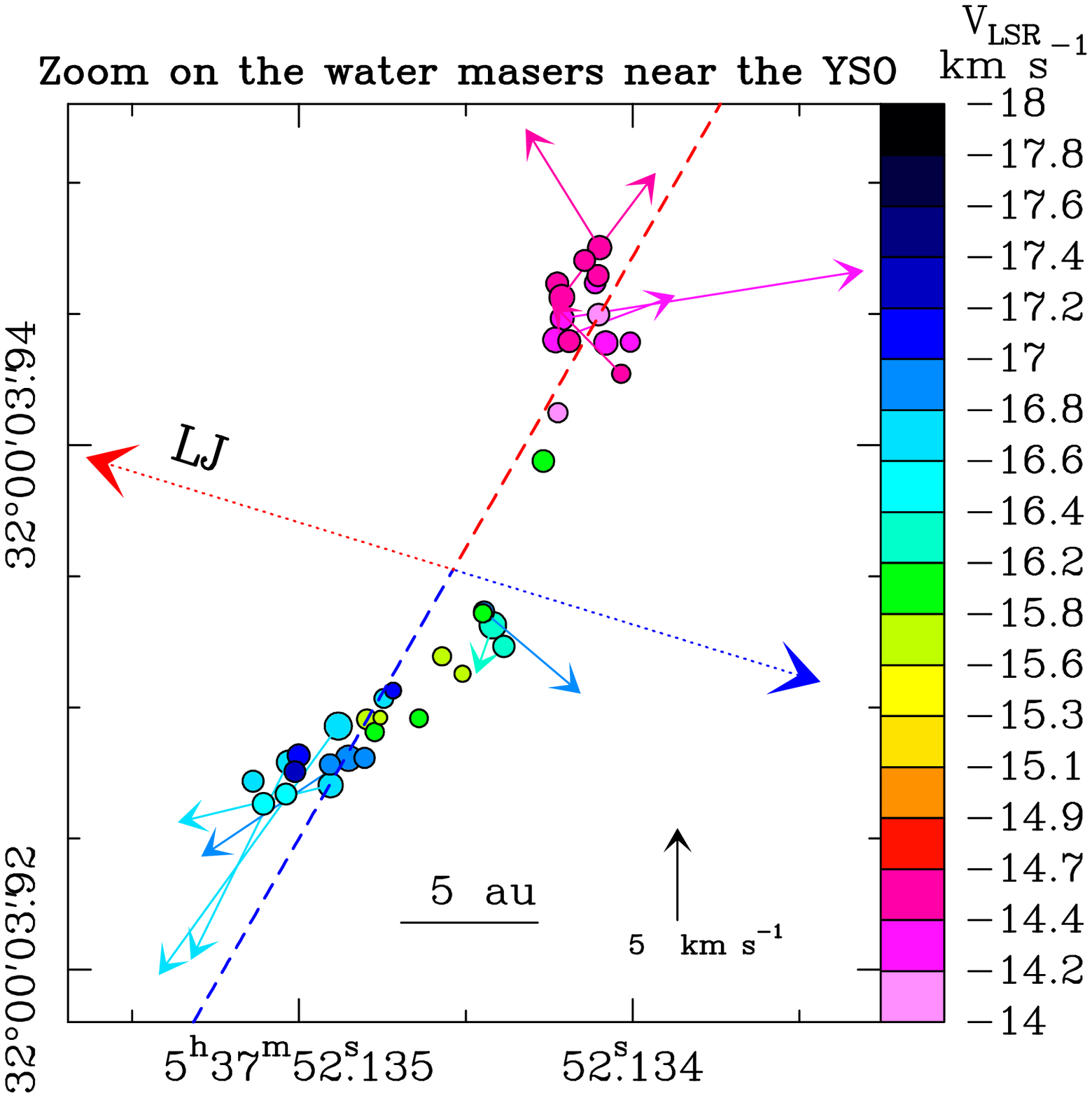}

\vspace*{0.75cm}\includegraphics[width=0.49\textwidth]{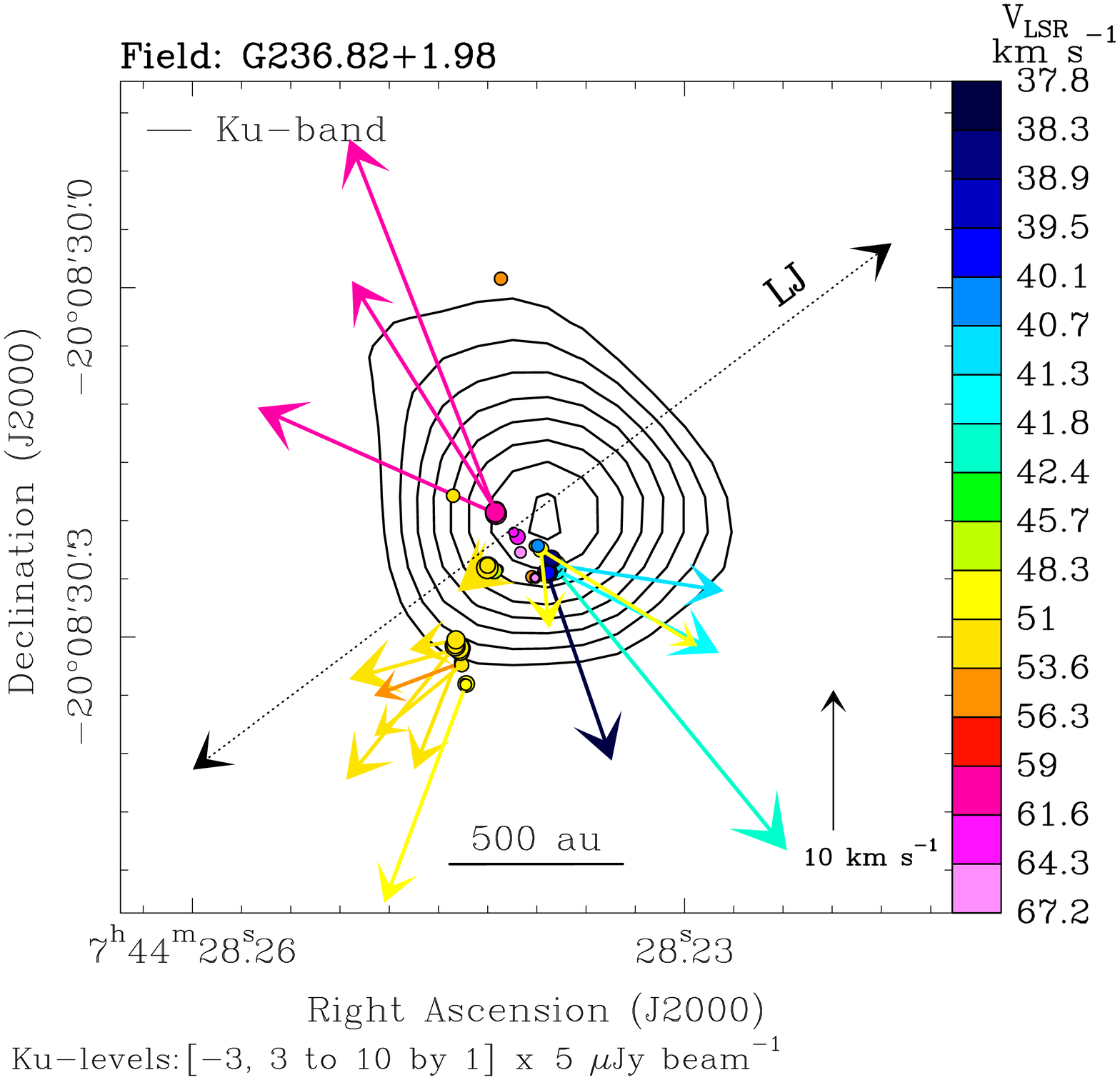}
\includegraphics[width=0.55\textwidth]{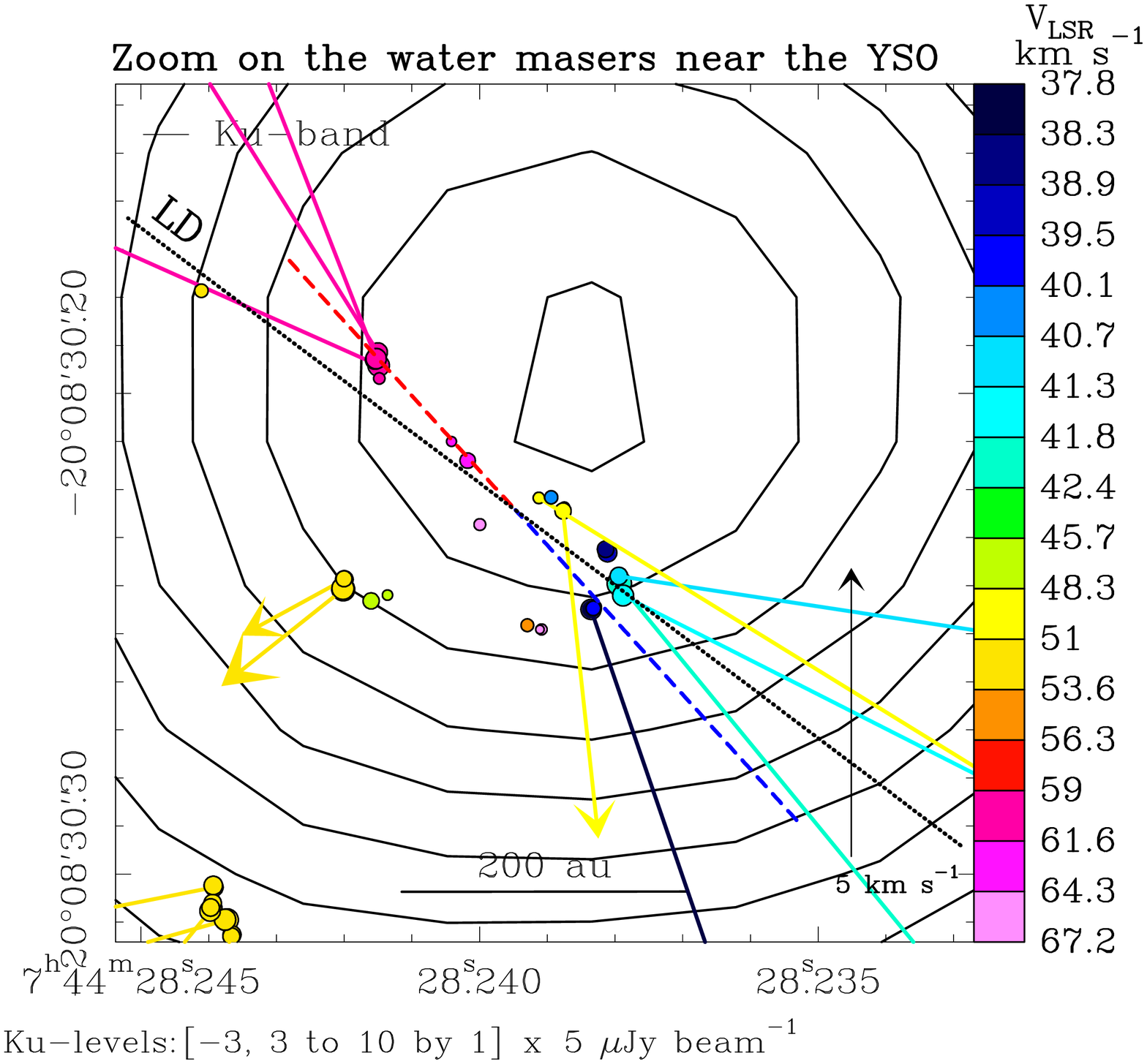}
\caption{In each panel, {\it black contours}, and {\it colored dots} and {\it arrows} have the same meaning as in the upper~panel of Fig.~\ref{F:G11.92}.  \ {\it Upper~panels:}~large view ({\it left}) and zoom on the water maser kinematics near the YSO ({\it right}) for the source \ G176.52$+$0.20.  The \Vlsr\ scale of the left and right panels is different. The \Vlsr\ gradient traced by the water masers is indicated by a {\it dashed line}, using {\it colors} to denote the {\it red}-~and~{\it blue}-shifted side of the gradient. The {\it dotted arrows} labeled~LJ give the axis of the collimated \ $^{12}$CO outflow observed with the SMA by \citet{Fon09}, using {\it colors} to distinguish the {\it red}-~and~{\it blue}-shifted flow lobe. \ {\it Lower~panels:}~large view ({\it left}) and zoom on the water masers near the YSO ({\it right}) for the source \ G236.82$+$1.98. The \Vlsr\ gradient traced by the water masers is indicated by a {\it dashed line}, using {\it colors} to denote the {\it red}-~and~{\it blue}-shifted side of the gradient. The {\it black dotted arrows} labeled~LJ ({\it left}) and the {\it black dotted line} labeled~LD ({\it right}) denote the orientations of the elongated chain of H$_2$ 2.2~$\mu$m knots and the extinction lane in the $J$-band image, respectively, observed in proximity of \ G236.82$+$1.98 by \citet{Var12}, using the Wide Field Camera mounted on the 3.8~m UKIRT.}
\label{F:G176.52-G236.82}
\end{figure*}

\begin{figure}
\centering
\includegraphics[width=0.36\textwidth]{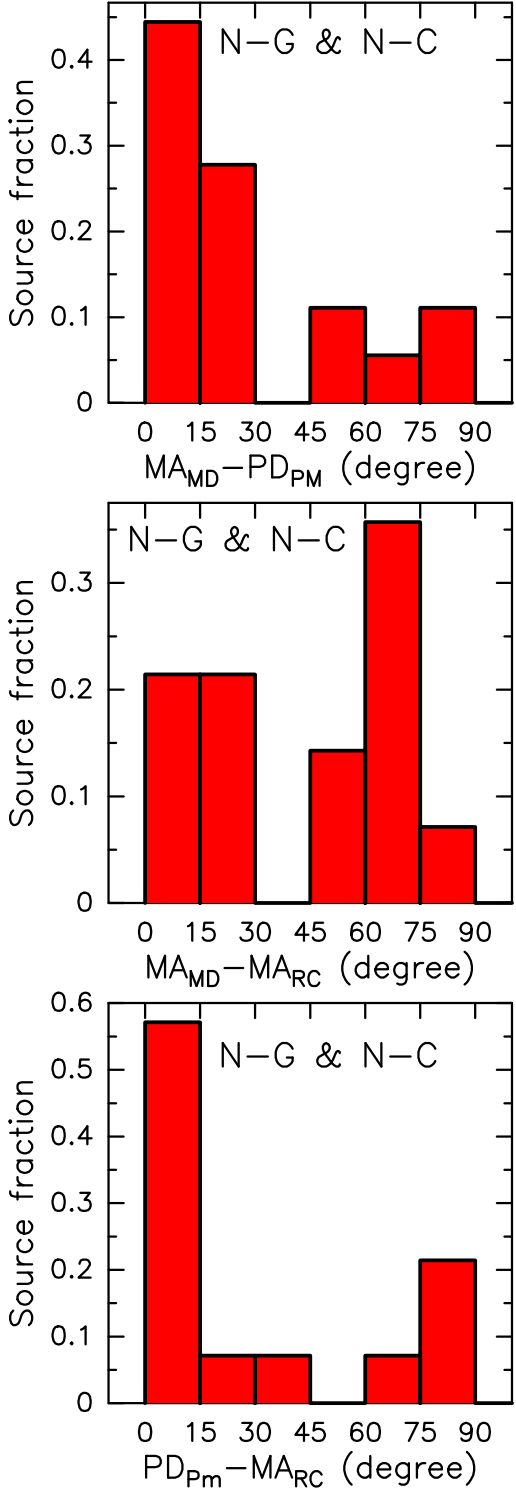}
\caption{Histograms of the difference between the PA of the major-axis of the maser distribution, MA$_{\rm MD}$, and the PA of the preferential direction of the maser PMs, PD$_{\rm PM}$ ({\it upper~panel}), between \ MA$_{\rm MD}$ \ and the PA of the major-axis of the radio continuum emission, MA$_{\rm RC}$ ({\it middle~panel}), and between \ PD$_{\rm PM}$ \ and \ MA$_{\rm RC}$ ({\it lower~panel}). To produce these histograms we have selected all the POETS targets belonging to the combined subgroups~N-C~and~N-G (see Table~\ref{wat_kin}). The histogram bin is 15\degree\ and the histogram values are normalized by the total number of considered targets. The source-average errors on the differences \ MA$_{\rm MD}$$-$PD$_{\rm PM}$, MA$_{\rm MD}$$-$MA$_{\rm RC}$, and \ PD$_{\rm PM}$$-$MA$_{\rm RC}$ \ are \ 26\degree, 21\degree, and 24\degree, respectively.}
\label{histo_90}
\end{figure}

\begin{figure}
\centering
\includegraphics[width=0.5\textwidth]{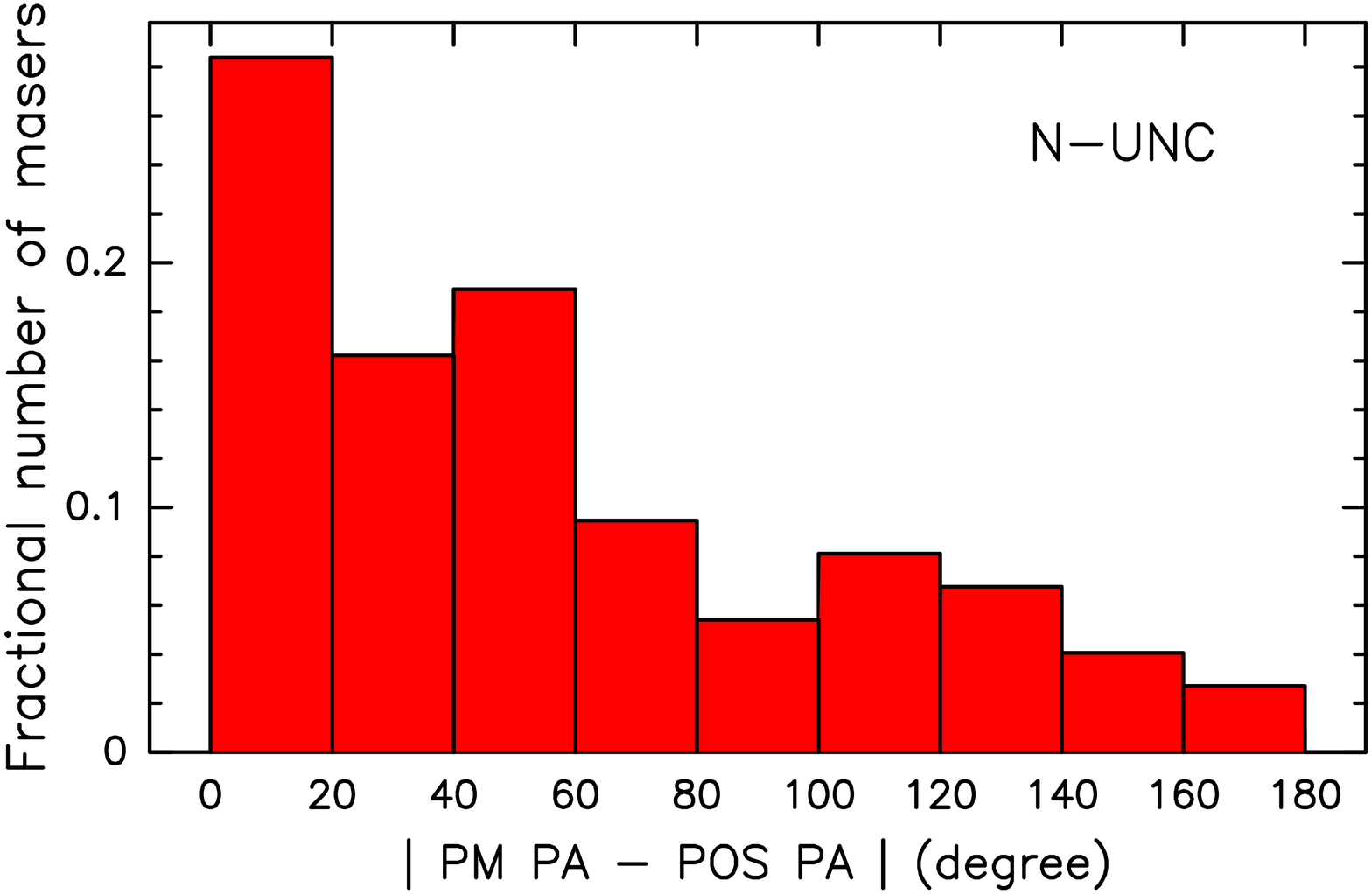}

\vspace*{0.4cm}\includegraphics[width=0.5\textwidth]{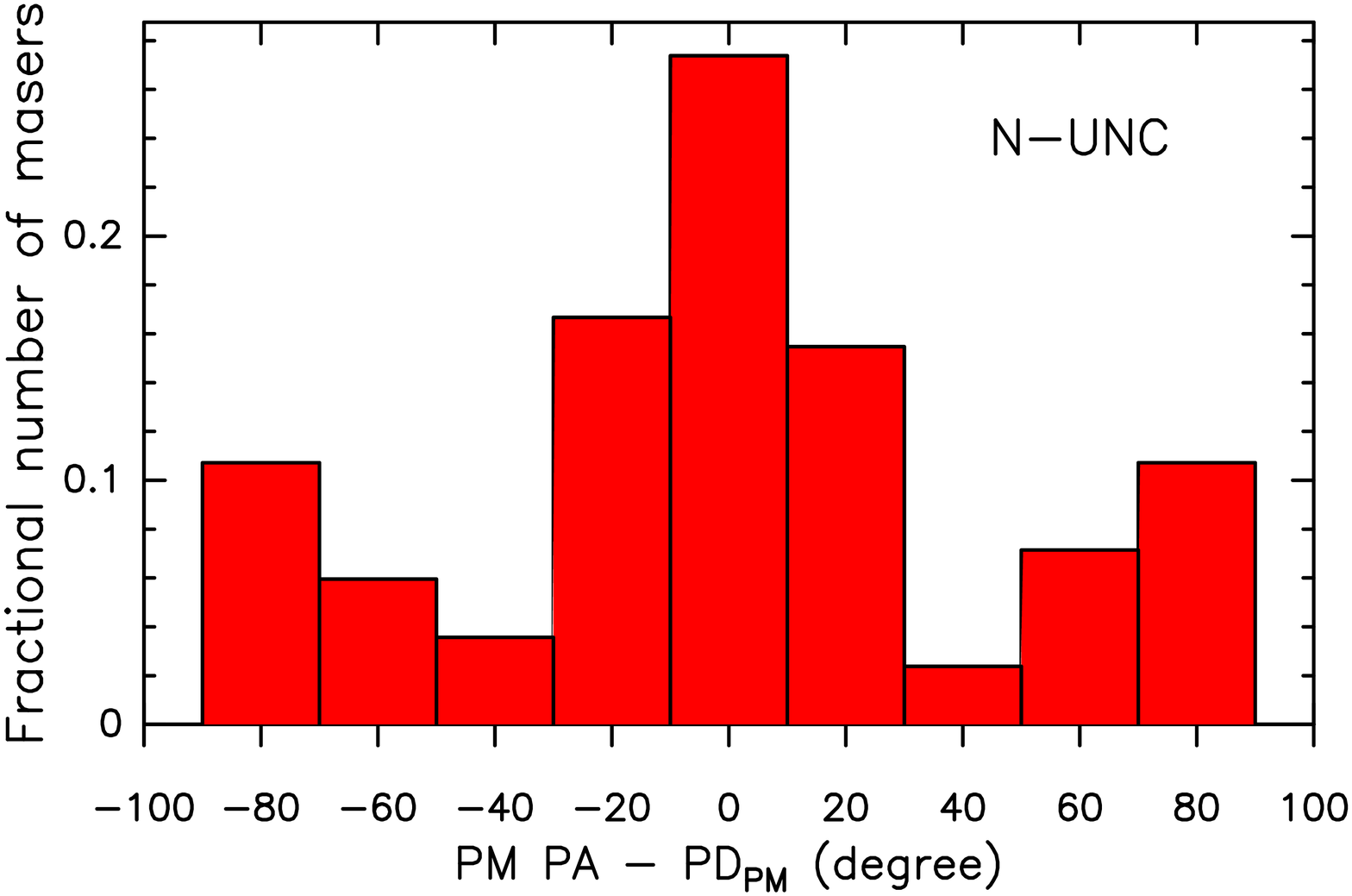}
\caption{Histograms of the absolute value of the difference between the PA of the maser PM, PM~PA, and the PA of the maser position vector from the YSO, POS~PA ({\it upper~panel}), and of the difference between the PM~PA and the PA of the preferential direction of the water maser PMs, PD$_{\rm PM}$ ({\it lower~panel}). To produce these histograms we have selected the POETS targets belonging to the subgroup~N-UNC (see Table~\ref{wat_kin}) with at least two measured PMs. The histogram bin is 20\degree\ and the histogram values are normalized by the total number of considered PMs. The source-average error on the difference \ PM~PA $-$ POS~PA \ and corresponding standard deviation are \ 32\degree~and~6\degree. The source-average error on the PM~PA and corresponding standard deviation are \ 9\degree~and~3\degree.}
\label{PM_POS-CD}
\end{figure}


\section{The kinematic structures of the water masers}
\label{kin-mas}

In the following, to interpret the water maser kinematics, we make the basic assumption that a single YSO is responsible for the continuum emission and the water maser motion. The case of multiple YSOs is considered in Sect.~\ref{mul-YSO}. A detailed discussion of the maser kinematics of each POETS target is beyond the scope of this paper. Rather, we aim to highlight common kinematic features in the various sources, and organize them in a coherent picture.
In Sect.~\ref{obs_res}, the group of POETS targets with water masers near the YSO (group~N) has been divided into different subgroups by noting the presence of linear \Vlsr\ gradients (subgroup~N-G), collimated motion (N-C), and undetermined or non-colllimated motion (N-UNC). Assuming that a single YSO accelerates the water masers, hereafter we propose a scenario that could explain the different maser configurations.

\subsection{Subgroup~N-G: linear \Vlsr\ gradients}

\label{LVG}

An interesting result of the POETS survey is the observation of a linear \Vlsr\ gradient in the water masers on top of the YSO position. For the sources \ G011.92$-$0.61 (Fig.~\ref{F:G11.92}) \ and \ G035.02$+$0.35 (Fig.~\ref{F:G35.02}), we can compare the maser 3D velocity pattern with the kinematics of the thermal emission studied at an angular resolution of a few 0\farcs1 through recent sensitive ALMA observations. In both sources, it is observed a \Vlsr\ gradient in high-density molecular tracers centered on the water maser position and with size of \ $\sim$1000~au \citep{Ile18,Bel14}. The orientations of the \Vlsr\ gradients in the maser and thermal emissions coincide within the errors and show the same polarity, red- and blue-shifted to SE and NW, respectively, for \ G011.92$-$0.61 (see Fig.~\ref{F:G11.92}), and blue- and red-shifted to SE and NW, respectively, for \ G035.02$+$0.35 (see Fig.~\ref{F:G35.02}). Following the interpretation of the thermal \Vlsr\ gradients in terms of rotating disks seen almost edge-on, the \Vlsr\ pattern of the water masers should also trace the disk rotation closer to the YSO. However, in \ G011.92$-$0.61 \ and \ G035.02$+$0.35, the measured water maser PMs appear to be inconsistent with both edge-on and face-on rotation. The former case is in conflict with finding most of the PMs directed at large angles from the disk mid-plane (indicated with a colored dashed line in the upper~panels of Figs.~\ref{F:G11.92}~and~\ref{F:G35.02}). The latter case can be also excluded because PM amplitudes of \ 10--20~\kms\ at distances of \ $>$~500~au from the YSO would imply a too large ($>$150~\ms) central mass, if interpreted in terms of rotation velocities under the effect of the YSO gravity. 

In \ G011.92$-$0.61 \ the protostellar jet has been clearly characterized via SMA $^{12}$CO observations by \citet{Cyg11}. It is very well collimated along a SW-NE direction, with the blue-~and~red-shifted lobes extending to SW and NE, respectively (see Fig.~\ref{F:G11.92}, bottom-left~panel). Inspecting Fig.~\ref{F:G11.92}~(upper~panel), we note that the water masers moving away from the disk mid-plane toward SW are significantly more blue-shifted than the nearby masers moving toward NE. Therefore, the maser \Vlsr\ pattern appears to reflect the outflow bipolarity as well. In \ G035.02$+$0.35, the presence of a radio jet (directed along SW-NE and approximately perpendicular to the direction of the maser and thermal \Vlsr\ gradient) has recently been reported by \citet{San19b}, but we have no information on the flow velocity. Looking at Fig.~\ref{F:G35.02}~(upper~panel), one can actually note circumstantial evidence that the water masers moving toward SW are more red-shifted than those moving toward NE.

The simplest interpretation to account for \ 1)~the agreement in the \Vlsr\ distributions of the maser and thermal emissions, tracing both edge-on rotation and the bipolar outflow, \ 2)~the large angles between the direction of many maser PMs and the disk major-axis, and \ 3)~the relatively high maser speeds,  is in terms of a DW, seen almost edge-on. In fact, the 3D velocities of the water masers hint at a flow originating from a relatively large (a few 100~au) portion of the YSO disk.

The prototype for an edge-on DW is the kinematic structure observed around the high-mass YSO Orion-BN/KL Source~I \citep{Mat10}. The surface of the edge-on disk and the emergence of the wind are traced with the vibrationally-excited SiO masers, which draw two symmetric, flat parabolic arms (separated by \ $\approx$~20~au, total size \ $\le$~100~au) on the sky. The PMs (with average speed of \ 14~\kms) of the SiO masers are tangent locally to the parabolic arms and point away from the YSO, whereas their \Vlsr\ are red- and blue-shifted (by up to 25~\kms) on either sides of the symmetry axis of the parabolas. Ground-state SiO and water masers are observed at larger distances from the YSO (100--1000~au) and trace a jet, showing PMs well collimated along the DW axis \citep{Gre13}. 

The gas kinematics in the POETS targets \ G011.92$-$0.61 \ and \ G035.02$+$0.35 \ is sampled by our water maser observations much less homogeneously than in Source~I \ by the SiO masers. Accounting for the larger distance of our targets, we cannot expect to properly resolve the DW from the base of the jet, which can explain why the PMs of the water masers are directed at varying angles, from parallel to perpendicular, with respect to the disk major-axis. Although the \Vlsr\ gradient in the water masers mainly reflects a nearly edge-on rotation, a slight inclination of the disk axis with respect to the plane of the sky can cause a systematic shift of the average maser \Vlsr\ toward red or blue on either sides of the disk mid-plane. 

As noted in Sect.~\ref{obs_pm}, in each of the three targets \ G075.78$+$0.34 (Fig.~\ref{F:G75.76-G75.78}, bottom-left~panel), G176.52$+$0.20 (Fig.~\ref{F:G176.52-G236.82}, top-left~panel), and \ G236.82$+$1.98 (Fig.~\ref{F:G176.52-G236.82}, bottom-left~panel), the masers in the subset~{\em OG} show an elongated spatial distribution, oriented at close angle with the direction of collimation of the maser PMs. These are the kinematic properties expected for jets, and, for \ G176.52$+$0.20 \ and \ G236.82$+$1.98, this interpretation is supported by the observation on scales of 10$^4$~au of a collimated molecular flow \citep{Fon09} and an elongated chain of H$_2$ 2.2~$\mu$m knots \citep{Var12}, respectively, about parallel to the maser jet (see Fig.~\ref{F:G176.52-G236.82}).
Looking at Figs.~\ref{F:G176.52-G236.82}~and~\ref{F:G75.76-G75.78}~(lower~panels) and comparing the values of  \ MA$_{\rm MD}$ \ and \ PD$_{PM}$ \ of the maser subset~{\em OG} \ with the value of \  MA$_{\rm MD}$ \ of the corresponding maser subset~{\em IG} \ in Table~\ref{wat_kin}, it is also evident that, in each of the three sources, the \Vlsr\ gradient traced by the subset~{\em IG} is about perpendicular (within 10\degree\--20\degree) to the axis of the jet identified by the subset~{\em OG}. The finding of a \Vlsr\ gradient directed perpendicular to a jet hints at a rotating disk seen almost edge-on, and, for \ G236.82$+$1.98, this interpretation is corroborated by the detection of an extinction lane in the $J$-band UKIRT image of \citet{Var12} about parallel to the maser \Vlsr\ gradient (see Fig.~\ref{F:G176.52-G236.82}, bottom-right~panel). However, repeating the same considerations made above for \ G011.92$-$0.61 \ and \ G035.02$+$0.35, the PMs of the masers in the subsets~{\em IG} are not consistent with edge-on rotation, and we are led to conclude that, in \ G075.78$+$0.34, G176.52$+$0.20, and \ G236.82$+$1.98, this subset of masers could trace a DW, too.
Similarly to the case of Orion-BN/KL Source~I, in these three targets the water masers could trace both the wide-angle DW near the YSO and the jet at larger separation. Regarding the last member of the subgroup~N-G \ G075.76$+$0.34 \ (see Table~\ref{wat_kin}), the supposition that the maser \Vlsr\ gradient originates in a rotating disk close to edge-on is supported by the detection of a jet in H$_2$ 2.2~$\mu$m emission from our recent LBT survey \citep{Mas19}, directed about perpendicular to the \Vlsr\ gradient (see Fig.~\ref{F:G75.76-G75.78}, upper~panel). 

\subsection{Subgroup~N-C: collimated motion}
\label{col_PM}

Inspecting the subgroup~N-C in Table~\ref{wat_kin}, for ten out of the twelve sources  (excluding \ G012.68$-$0.18  and \ G105.42$+$9.88~VLA~3B) the values of \ MA$_{\rm MD}$ \ and \ PD$_{PM}$ \ are quite similar, differing by \ $\la$20\degree.  If the spatial distribution of the masers is elongated at close angle with the direction of collimation of the PMs, it is straightforward to interpret the kinematic structures in terms of jets. 
The literature data, when available, fully support this interpretation. 
In fact, in five of these ten targets (G005.88$-$0.39, G176.52$+$0.20-{\em OG}, G236.82$+$1.98-{\em OG}, AFGL5142, and IRAS~20126$+$4104), the PA of the sky-projected axis, A$_{\rm LJ}$, of the collimated outflow traced in thermal emission at larger scales differs by \ $\le$ 15\degree \ from the PA of spatial elongation, MA$_{\rm MD}$,  or motion, PD$_{PM}$, \ of the water masers (see Table~\ref{wat_kin}).

For the remaining two targets \ G012.68$-$0.18 \ and \ G105.42$+$9.88 VLA~3B \ of the subgroup~N-C, the difference between \ MA$_{\rm MD}$ \ and \ PD$_{PM}$ \ is \ 79\degree \ and \ 90\degree, respectively, indicating that the maser positions are stretched along a direction almost perpendicular (within \ $\la$10\degree) to the maser PMs. Particularly instructive is the case 
of \ G105.42$+$9.88 VLA~3B (see Fig.~\ref{F:G90.21-G105.42}, bottom-right~panel), where, although a clear \Vlsr\ gradient is not seen, the water masers have a flattened distribution with size \ $\sim$30~au, oriented perpendicularly to the radio jet observed with the VLA by \citet{Tri04} (see Table~\ref{wat_kin}). The spatial and velocity maser configurations suggest that the maser emission emerges relatively close to the YSO disk surface, and that we are seeing the base of the jet, that is, the transition from the DW to the jet, where the motion of the gas is already collimated along the jet axis, but its spatial distribution keeps track of the disk geometry. 

In G092.69$+$3.08 (see Mos16, Fig.~8), another clear case where the maser motions are well collimated and trace the YSO jet, one notes a maser \Vlsr\ gradient transversal to the jet direction, hinting at rotation around the jet axis. Clues to a gradient in the  maser \Vlsr\ perpendicular to the jet axis are also discernible in the sources \ G012.68$-$0.18 \ (see Mos16, Fig.~3) \ and \ IRAS~20126$+$4104 \ \citep[see][Fig.~2]{Mos05}. The quest for jet rotation is crucial as it supports the idea that  magnetically accelerated jets extract the excess angular momentum from the  system, and thus they are a key element for the star formation process \citep[see, e.g.,][]{Pud07}. Indications of jet rotation have been reported for low-mass stars \citep[][]{Bac02,Cof04,Lee17}.  Importantly, our water maser observations suggest that jet rotation is present also in high-mass systems.

In summary, in several targets, the properties of the jets traced with the water masers lead us to conjecture that the jets originate from DWs. In Sect.~\ref{dw-nat}, we will discuss the connection between DWs and jets more diffusely.

\subsection{Comparison of water maser and radio continuum}
\label{rad_win}
 
In Sects.~\ref{LVG}~and~\ref{col_PM}, we have interpreted the maser kinematics in the targets belonging to the subgroups~N-G and N-C in terms of DWs and jets. 
Here, we discuss the relation between the water maser properties and the radio continuum.
The upper~panel of Fig.~\ref{histo_90} shows the histogram of the difference between the values of \ MA$_{\rm MD}$ \ and \ PD$_{PM}$, for all the targets in the combined subgroups~N-G and N-C. In 89\% of the sources, the spatial elongation of the masers is either parallel or perpendicular (within 30\degree) to the (preferential) direction of motion. The middle~and~lower~panels of Fig.~\ref{histo_90} show the histograms of the difference of the PA of the major-axis of the continuum emission, MA$_{\rm RC}$, with  \ MA$_{\rm MD}$ \ and \ PD$_{PM}$, respectively. In the large majority ($\ge$ 85\%) of the sources of the combined subgroups~N-G and N-C, the radio continuum is elongated either parallel or perpendicular with respect both the maser spatial elongation and the maser motion.  Noticeable examples are the sources \ G035.02$+$0.35 (see Fig.~\ref{F:G35.02}, upper~panel) and \ G236.82$+$1.98 (see Fig.~\ref{F:G176.52-G236.82}, lower~panels), where the radio continuum extends parallel to the maser \Vlsr\ gradients, and the source \ G011.92$-$0.61 (see Fig.~\ref{F:G11.92}, upper~panel), where the continuum is elongated along a direction close to the orientation of the maser PMs.

The three histograms of Fig.~\ref{histo_90} considered above indicate that, in each target of the combined subgroups~N-G and N-C, the three directions:  \ the maser spatial elongation, \ the maser motion, and \ the radio continuum elongation, are either parallel or perpendicular each other. The proposed interpretation of the maser kinematics in terms of a DW or a jet is built on such a geometrical correspondence. In fact, in each source, using the water masers, we can identify two main, mutually perpendicular axes on the sky: \ 1)~the disk major-axis, and \ 2)~the (sky-projected) jet axis.  Regarding the radio continuum, we conjecture that the ionized emission at cm-wavelengths observed near the YSOs could result from two contributions:  \ 1)~a (partially) ionized disk, and \ 2)~a jet.  Interestingly, toward Orion-BN/KL Source~I, the 7~mm continuum emission is elongated perpendicularly to the jet and bisects the DW structure traced with the SiO masers, suggesting a likely origin from an ionized accretion disk \citep{Rei07,God11c}.  Our targets are at larger distances than Source~I, and the radio emissions from the disk and the jet could not be resolved in most cases. The observed changes in the structure of the continuum emission at different frequency bands could be related to the diverse dependency of the optical depths of the disk and jet contributions on the frequency: at C-band both the disk and the jet could be optically thick, while at K-band the disk could become partially optically thin. 

\subsection{Subgroup~N-UNC: undetermined or non-collimated motion}
\label{sca_kin}

In about half of the water masers located near the YSOs (belonging to the subgroup~N-UNC in Table~\ref{wat_kin}), there is no hint at a subset of masers tracing a linear \Vlsr\ gradient and the maser PMs do not align along a specific direction. Examples of these maser configurations are the sources \ G049.19$-$0.34 (see Fig.~\ref{F:G31.58-G49.19}, lower~panel), G097.53$+$3.18--M (Mos16, see Fig.~9, upper~panel) and \ G100.38$-$3.6 (Mos16, see Fig.~10). In these targets, the maser kinematics is more difficult to interpret. We have first considered a case in which the non-collimated directions of the maser PMs are produced by a PW launched very near or directly from the YSO. In such a scenario, the velocity of the flowing gas have to be directed radially with respect to the position of the YSO. Figure~\ref{PM_POS-CD} (upper~panel) shows the histogram of the difference between the PA of the maser PM (PM~PA) and the PA of the position vector (POS~PA) of the maser from the YSO, cumulating all the sources in the subgroup~N-UNC. We note that \ 56\% \ of the PM directions deviate more than \ 40\degree\ from the line joining the maser with the YSO, and that such a deviation is larger than the average error of \ 32\degree\ in the difference between PM~PA and POS~PA. We conclude that it is unlikely that PWs drive the maser motions in these sources, since the sky-projections of the position vector (from the YSO) and the velocity of most of the water masers are not parallel each other.

A close inspection of the distributions of PMs unveils a common feature, that is, in the majority of sources, most of the PMs are aligned along two almost perpendicular directions. This is true not only for the targets with a relatively large number of measured PMs, as \ G100.38$-$3.6 (Mos16, see Fig.~10) and \ G049.19$-$0.34 (see Fig.~\ref{F:G31.58-G49.19}, lower~panel), but also for the targets with only a few PMs, as \ G014.64$-$0.58 (Fig.~\ref{F:G14.64-G26.42}, upper~panel) and \ G079.88$+$1.18 (Fig.~\ref{F:G76.38-G79.88}, lower~panel). Figure~\ref{PM_POS-CD} (lower~panel) shows the histogram of the difference between the PM~PA and the PA of the preferential direction of the water maser PMs, PD$_{\rm PM}$, cumulating all the targets in the subgroup~N-UNC. It is clear that the PM directions in these sources tend to concentrate either within \ $\pm$30\degree\ about \ PD$_{\rm PM}$ \ or within \ $\pm$30\degree\ about the direction  perpendicular to \ PD$_{\rm PM}$. Thus, also for this subgroup of targets, similarly as found for the combined subgroups~N-G and N-C, two perpendicular directions appear to be significant for the maser kinematics. In the following, we show that a possible explanation of this peculiar distribution of PMs is in terms of a DW-jet system seen not edge-on but with the axis slightly inclined, $\la$30\degree, with respect to the plane of the sky. In this case, the preferential directions of alignment for the PMs are either the disk major-axis or the sky-projection of the DW-jet axis, which are perpendicular to each other. 

In Paper~IV, we present the statistics of the inclination of the maser 3D velocities with the plane of the sky and provide evidence for an observational bias, in obedience to which masers moving close ($\la$30\degree) to the plane of the sky are strongly favored. This agrees with the J-~and~C-shock models for the origin of the water masers \citep{Eli92, Hol13, Kau96}, where a long velocity-coherent, amplification path along the line of sight (l.o.s.) is predicted for planar shocks being accelerated and moving close to the plane of the sky. In the prototypical (edge-on) DW-jet of Source~I \citep{Mat10}, the flow velocities are preferentially tangent to the disk close to the YSO and recollimate along the jet axis at larger separation. Even if the DW is seen slightly away from edge-on, the masers which move either  \ 1)~on the disk plane along the disk major-axis, or \ 2)~within the recollimated flow, \ have velocities directed close to the plane of the sky and can be preferentially detected. These two groups of masers define the two perpendicular directions on the sky along which the observed maser PMs align. Accordingly, for the  sources of the subgroup~N-UNC, PD$_{\rm PM}$ \ should represent either the direction of the disk major-axis or that of the sky-projected axis 
of the jet. 
The observation of maser emission from the intermediate region of the DW where the gas flow is not yet recollimated is much less favored, since there, if the DW is not edge-on, the l.o.s. component of the flow velocities can vary rapidly along the l.o.s., effectively reducing the maser amplification path.

\subsection{Conclusive remarks}

To summarize the main points of Sects.~\ref{LVG}--\ref{sca_kin}, we propose a single physical scenario, the DW-jet, to interpret the kinematics of the water masers near the YSOs. In a subgroup of sources, the observation of a linear \Vlsr\ gradient in the water maser emission and maser PMs directed over a continuous range of PA (see Sect.~\ref{LVG}), are interpreted as the signature of an edge-on view. In another subgroup, the non-detection of maser \Vlsr\ gradients and preferentially perpendicular PMs, could indicate that the viewing angle is moderately different from edge-on.

We note that the targets in the subgroup~N-UNC and in the combined subgroups~N-G and N-C share the same range of bolometric and radio continuum luminosity, and present similar morphologies of the radio continuum emission.  Therefore, the two subgroups should also have comparable YSO masses and evolutionary stages. Our interpretation that the observed differences in the maser kinematics are due only to a different viewing angle of alike kinematic configurations, are in line with these considerations.

\section{Discussion}
\label{discu}

\subsection{Nature of the disk-winds traced with H$_2$O masers}
\label{dw-nat}

The main characteristics of the DW kinematics as traced with the water masers in our POETS targets are:

\begin{enumerate}

\vspace*{0.2cm}\item The flow apparently originates from a region that extends from a few 10~au (for \ G176.52$+$0.20, L$_{\rm bol}$ $\sim$ 10$^2$~\ls, see Fig.~\ref{F:G176.52-G236.82}, top-right~panel) to several 100~au (for the sources \ G011.92$-$0.61 and \ G035.02$+$0.35, L$_{\rm bol}$ $\sim$ 10$^4$~\ls, see Figs.~\ref{F:G11.92}~and~\ref{F:G35.02}, upper~panels), increasing with the YSO luminosity and mass.

\vspace*{0.2cm}\item The wind velocity ranges from $\approx$5~\kms\ (for \ G176.52$+$0.20) up to $\approx$10--20~\kms\ (for the sources \ G011.92$-$0.61 and \ G035.02$+$0.35), increasing with the YSO mass.

\vspace*{0.2cm}\item In several targets, the direction of the wind velocity changes from parallel to perpendicular to the disk across distances of \ $\sim$100~au (see discussion in Sect.~\ref{sca_kin}).
 
\end{enumerate}

In the following, we discuss whether MC or photoevaporated DW models could account for these kinematic features. We do not consider the case of MP tower flows because no analytic models for them are presently available in the literature and their properties are more uncertain. The shocks giving rise to the maser emission  may be internal to the DW or may arise in the collision of the DW with the surrounding medium.
 
\subsubsection{Magneto-centrifugal DW}
\label{mc-dw}

MC DWs are presently considered the most likely way to accelerate and collimate protostellar jets. MC winds have an important role in the formation process as they can remove angular momentum from the system through the simultaneous magnetic braking of the disk \citep{Fra14}. For these winds, the launching region on the disk is expected to scale with the YSO mass from \ $\sim$10~au to $\ga$100~au ranging from \ 1~\ms\ to \ $\ga$10~\ms\ \citep{Pud05,Sei12,Koe18}. The wind is launched along the magnetic field lines, which follow a hourglass shape and have progressively larger opening angles at increasing disk radii. The wind velocity must exceed the escape velocity, $V_{esc}$, from the YSO, approximately given by the expression \ $ V_{esc} \approx 42 \; \sqrt{M_{\rm YSO} / {R_W}}$~\kms, where \ $M_{\rm YSO}$ \ and \ $R_W$ \ are, respectively, the YSO mass in solar unit and the footpoint radius (where the wind is launched) in astronomical unit. For the targets \ G176.52$+$0.20 \ and \ G236.82$+$1.98 (see Fig.~\ref{F:G176.52-G236.82}, right~panels), the size of the maser disk is comparable with the YSO position uncertainty, and we cannot meaningfully estimate the wind footpoint. In \ G011.92$-$0.61, using \ $M_{\rm YSO} = 40$~\ms\ \citep{Ile18} and the maximum observed maser radius of \ $\approx$700~au (see Fig.~\ref{F:G11.92}), we obtain \  $ V_{esc} \approx 10$~\kms. Since this value is slightly below the observed maser velocities, the masers could effectively be generated in a MC DW. More in general, checking in Table~\ref{wat_kin} the maser targets of group~N with at least one measured PM, 26 out of 32 have average PM amplitudes in the range \ 10--25~\kms. The maser velocities are comparable or larger than the estimate of \ $ V_{esc} \approx 13$~\kms\ calculated assuming order-of-magnitude values for \ $M_{\rm YSO} \approx 10$~\ms\ and \  $ R_W \approx 100$~au. Therefore, the MC DW model could generally account for both \ (1)~the size of the wind region across the disk, and \ (2)~the measured wind speed. Concerning point~(3), we note that, in the most external parts of the wind, the flow, driven by the hourglass-shaped magnetic field, is nearly parallel to the disk surface at its base, and it is recollimated along the DW axis on short length scales. That could qualitatively explain the observation of nearby clusters of masers with (almost) perpendicular PMs in several POETS targets. 
We stress that, thanks to the magnetic collimation, MC DWs would also provide a natural justification for the jets traced by the water masers in many POETS targets.

\subsubsection{Photo-evaporated DW}
\label{pe-dw}

The upper layers of a circumstellar disk can be heated and ionized by the incident energetic (above the Lyman limit) radiation from the star. 
At sufficiently large distance from the source, the thermal energy of the heated gas exceeds the gravitational binding energy and the gas can be centrifugally launched and escapes in a photo-evaporated disk wind (PDW) \citep{Hol94,Ale14}.
The length scale of the launching region of such a wind is given by  the gravitational radius \ $r_g = GM_{\rm YSO} / {c_s}^2 $, with  $G$ the gravitational constant, $M_{\rm YSO}$ \ the YSO mass and \ $c_s$ \ the sound speed of 
the ionized gas (about \ 8--10~\kms\ in typical conditions). The peak of mass loss rate per unit disk area is 
reached at a critical radius of \ $r_g /5$ \ according to the models, but  the outflow activity extends out to the disk border.
For a 10 solar mass star, $r_g$ \ is on the order of 100~au, thus a PDW is expected to be launched from a few tens of au to the outer disk border in this stellar mass range. 

The characteristics of PDWs in individual sources depend on the stellar spectrum and on the disk properties, 
but a few  general properties can be identified. PDWs  can drive a significant mass loss 
 (up to \ $\sim$ 10$^{-6}$--10$^{-5}$~\msyr\ for high mass stars, e.g., \citet[][]{Lug04}), 
and are the dominant disk dispersal mechanism in the late  phases. 
The particles in PDWs  leave the disk carrying their own angular momentum, so the wind can be seen to rotate in the same sense of the underlying disk.  
Typical final poloidal velocities in PWDs around massive stars are \ 10--30~\kms, thus slightly above the local escape speed, 
but they are expected to be smaller than in the MC DW case, as no magnetic lever arm contributes here to the acceleration of the flow.    
Another important difference is the expected shape for the streaming surfaces of the wind.  
The hydrodynamic models elaborated for PDWs  show that as much as in the MC DW case, 
the flow is accelerated along nested surfaces of progressively larger opening angle, starting from about\ 50\degree\ with respect to the disk axis for 
the inner streamlines to close to \ 90\degree\ at the outer layers. Differently from the hourglass configuration of the MC DWs, however, 
the PDWs  lack the powerful  magnetic recollimation, and thus they  are never redirected  toward the axis \citep[see, e.g.,][]{Lug04,Ale14}.  
The material therefore leaves the disk following divergent streaming surfaces with high inclination and cannot produce a collimated jet.   

The above considerations suggest that, in a few targets, at least part of the detected maser configurations could originate in a PDW. Distinctive features of a PDW can be the simultaneous occurrence of an ionized disk and flow velocities inclined at large angles with respect the to the disk axis.
The most promising case is the maser subset~{\em IG} tracing the \Vlsr\ gradient in the source \ G236.82$+$1.98 (Fig.~\ref{F:G176.52-G236.82}, lower~panels). These masers  are aligned along the SW-NE direction, and their PMs identify a direction parallel to the  elongation of the radio continuum, and perpendicular to the direction of the PMs of the maser subset~{\em OG}, at the SE edge of the continuum, which probably trace a shocked jet. The SW-NE masers are located at about 100~au from the center of the system, and  move radially away from the symmetry axis with a well defined rotation pattern. The poloidal velocities are consistent with those expected for a PDW.  

However, PDW cannot in general account for water maser PMs forming small angles, $\le$30\degree, with the disk rotation axis, since even the innermost streamlines of a PDW are predicted to have opening angles larger than about 50\degree. Thus, considering the nine targets of the combined subgroups~N-G~and~N-C where the direction of the YSO jet is known from the literature (see Table~\ref{wat_kin}), all the water maser PMs collimated along the jet axis cannot be explained through PDWs, while they would be readily accounted for via MC DWs. PDWs alone cannot also explain the presence of maser clusters with almost perpendicular PMs, as observed in most of the targets of the combined subgroups~N-G~and~N-UNC, because the streamlines of PDWs are predicted to be almost straight and the maximum angle between the innermost and outermost streamlines should be  \ $\la$ 90\degree - 50\degree = 40\degree. In conclusion, our observations appear to indicate that only in a few cases a PDW may be the unique origin of the observed maser distribution, while in most cases a layered, hourglass-shaped MC DW is a better description of the water maser kinematics.  
This is consistent with the notion that magnetically-driven accretion-ejection activity dominates the dynamics of the disk for most of its lifetime, 
while photo-evaporation is dominant only at late stages, on timescales one order of magnitude smaller,  
in which the accretion is reduced and the mass loss by photoevaporation is responsible for the final dispersal of the disk.

\subsection{An alternative interpretation: YSO multiplicity}
\label{mul-YSO}

The DW-jet interpretation, proposed in Sect.~\ref{kin-mas}, is based on the assumption that a single YSO is responsible for the radio continuum emission and the motion of the water masers. However, in some POETS targets, the spatial and 3D velocity distribution of the water masers, together with the morphology of the radio continuum, suggest that multiple YSOs might be responsible for the excitation and the motion of the gas on scales of a few 100~au. The same sources \ G011.92$-$0.61 (Fig.~\ref{F:G11.92}) \ and \ G035.02$+$0.35 (Fig.~\ref{F:G35.02}), in which our claim for the presence of a DW is based on both maser VLBI and thermal ALMA observations (see Sect.~\ref{LVG}), could host more than one YSO. In both sources, the water masers do not present a homogeneous spatial distribution but gather in two distinct clusters, separated by \ $\approx$500~au along SE-NW, and emitting at different, average \Vlsr. In both targets, the peak of the radio continuum emission is close to the center of the maser cluster to NW, and, in \ G035.02$+$0.35, a spur of continuum emission extends toward the SE maser cluster, as well. In each water maser cluster, the directions of the PMs would also be consistent with a motion of recession from the cluster center. An alternative scenario is thus that, in both targets, two YSOs, each located at the center of one of the two maser clusters, excite the radio continuum and accelerate the water masers. In this interpretation, the SE-NW \Vlsr\ gradient observed in thermal and maser emission could be partly due to the different \Vlsr\ of the nearby YSOs.

Similar considerations apply to other POETS sources, where the water masers concentrate in tight clusters near local maxima, spurs or protuberances of the radio continuum emission, which could be the signature of multiple YSOs, rather than prominent features of (slightly resolved) continuous structures. 
However, we tend to believe that only in a minor fraction of our targets the gas kinematics can be effectively driven by multiple YSOs, since, otherwise, the distinctive features of the water maser PM distributions underlined in Sect.~\ref{kin-mas} would remain difficult to explain. In fact, the observation of PMs oriented along one specific direction (for the targets belonging to the subgroup~N-C) or two perpendicular directions (subgroup~N-UNC) is in contrast to a scenario where the water masers trace independent outflows, which, in general, should be oriented (and accelerate the masers) at any mutual angle. Thus, if most of our targets hosted multiple YSOs, the final result would be a global PM distribution uniformly scattered across the whole PA range \ [$-$90\degree, 90\degree], which is remarkably different from what we find (Fig.~\ref{PM_POS-CD}, lower~panel). Therefore, we consider the case of multiple YSOs to be less realistic than our DW-jet interpretation.
\section{Conclusions}
\label{conclu}

This paper reports on the properties of protostellar winds at scales of \ $\sim$10--100~au investigated by complementing VLBI H$_2$O maser and JVLA radio continuum observations. The POETS survey, for the first time, has observed a relatively large sample (36) of luminous YSOs with enough sensitivity ($\sim$10~$\mu$Jy in the radio continuum)  and angular resolution ($\le$10~au in the maser emission) to resolve and study the launching region of the winds and jets. The large majority of the H$_2$O masers is observed near the YSO, at separation \ $\le$1000~au. In about half of these sources, the maser spatial distribution is elongated and oriented at close angle with the direction of collimation of the maser PMs. These kinematic features are typical of jets, and this interpretation is also supported by comparison with independent observations of outflows from the same YSOs, using thermal (continuum and line) emission, reported in the literature for several targets. 

In six sources, the H$_2$O masers are found to trace a linear \Vlsr\ gradient across the YSO position. 
Based on the mutual perpendicular orientation of these gradients and the YSO jets, and on the comparison with observations of the YSO disks in thermal emission, when available in the literature, we consider that the maser \Vlsr\ gradients can arise in (quasi) edge-on rotating disks. However, the PMs of the masers delineating the \Vlsr\ gradients are non consistent with edge-on rotation, but rather suggestive of flow motions at speed of \ 10--20~\kms\ directed at large angles with the disk midplane. Thus, the 3D velocity pattern of the masers in these structures is reminiscent of DWs. In the targets where the H$_2$O masers trace DWs or jets, the radio continuum is elongated either parallel or perpendicular to the maser jet, which indicates that the radio emission can emerge from either the ionized disk or the ionized jet. 

In the remaining targets where the H$_2$O masers are observed near the YSO, the number of detected masers (and corresponding PMs) is too limited to constrain the gas kinematics (undetermined motion) or the maser PMs are clearly non-collimated. In these sources with undetermined or non-collimated motion, the maser PMs tend to align along two almost perpendicular directions. Considering that the observation of water masers is strongly biased toward masing cloudlets moving close to the plane of the sky, this peculiar PM distribution is consistent with a scenario where masers originate in a DW-jet system slightly inclined ($\le$30\degree) with respect to edge-on, whereas PMs are directed either along the disk major-axis or the sky-projected jet axis. In conclusion, a single physical scenario, the DW-jet, could explain the kinematics of all 
the observed distributions of water masers in our study. 
We argue that only a magneto-centrifugally accelerated outflow can account at the same time for the masers' motions distributed as a wide-angle DW, and for the configurations describing a collimated jet. 

\begin{acknowledgements}
We thank Maite Beltr{\' a}n for providing the velocity map of the \ CH$_3$CN \ emission in \ G035.02$+$0.35. 
We thank John Ilee for permission to reproduce two figures from his articles on G011.92$-$0.61.
We thank Zhi-Yun Li and Elisabetta Rigliaco for fruitful discussions.
 
\end{acknowledgements}

%
   \bibliographystyle{aa} 
   \bibliography{biblio} 
%

\clearpage

\begin{appendix}

\section{Overlay of radio continuum and water maser observations}
\label{over_c+m}

\begin{figure*}
\centering
\includegraphics[width=0.65\textwidth]{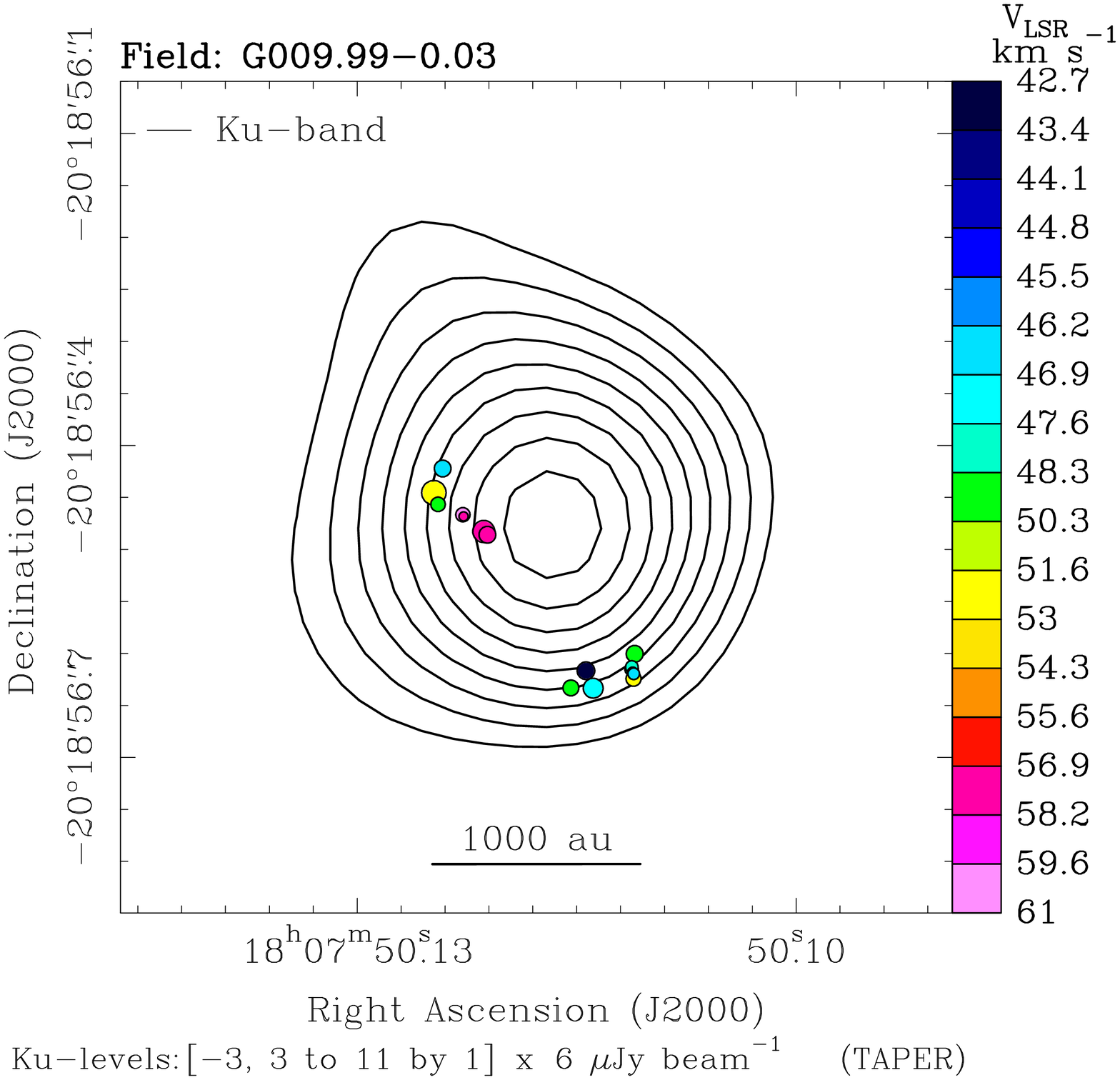}
\includegraphics[width=0.65\textwidth]{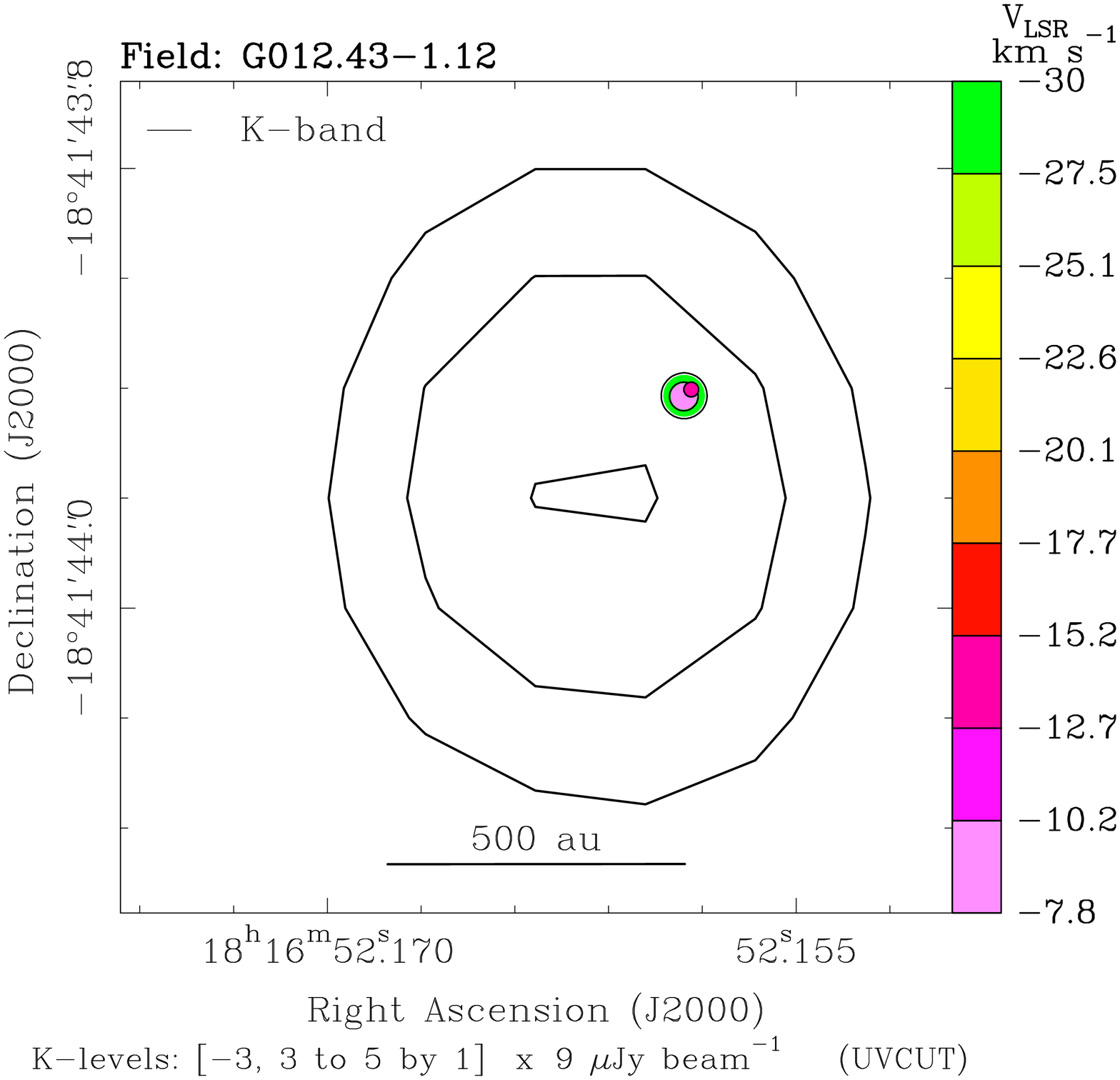}
\caption{Overlay of radio continuum and water maser positions and velocities. Each panel refers to the POETS target indicated in the observed field above the corresponding plot. In each panel, {\it black contours}, and {\it colored dots} and {\it arrows} have the same meaning as in the upper~panel of Fig.~\ref{F:G11.92}.}
\label{F:G9.99-G12.43}
\end{figure*}

\begin{figure*}
\centering
\includegraphics[width=0.65\textwidth]{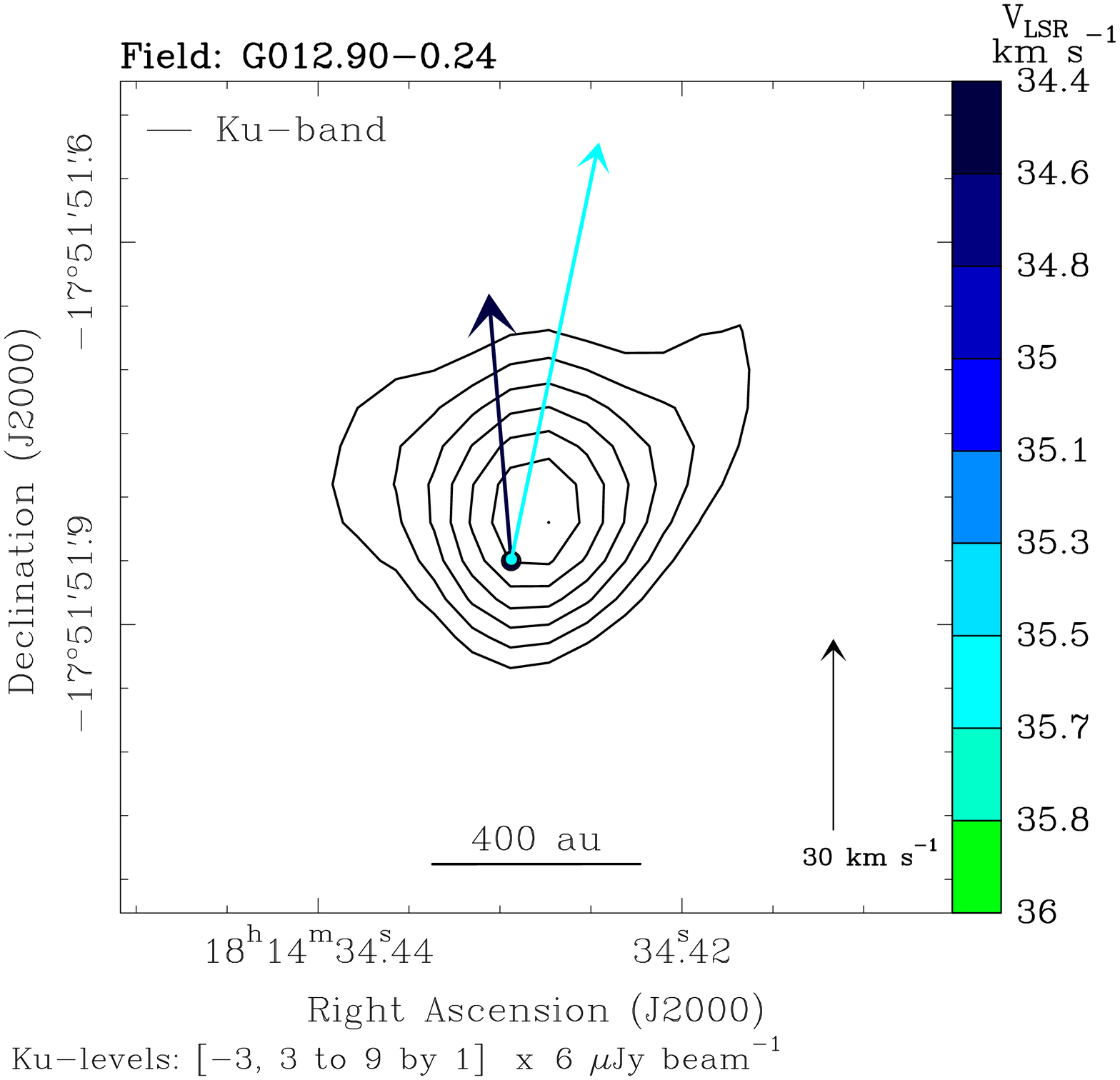}
\includegraphics[width=0.65\textwidth]{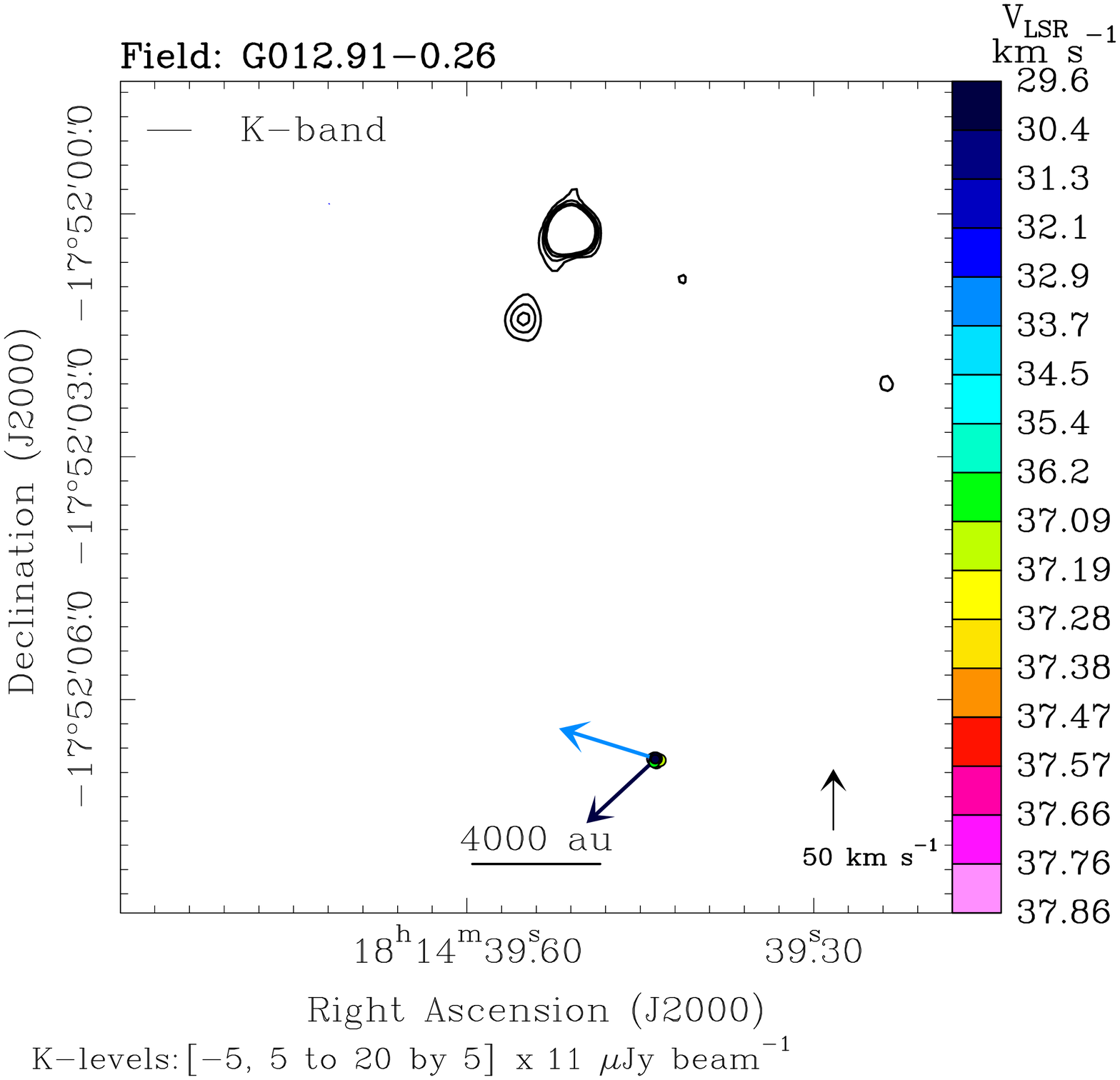}
\caption{Overlay of radio continuum and water maser positions and velocities. Each panel refers to the POETS target indicated in the observed field above the corresponding plot. In each panel, {\it black contours}, and {\it colored dots} and {\it arrows} have the same meaning as in the upper~panel of Fig.~\ref{F:G11.92}.}
\label{F:G12.90-G12.91}
\end{figure*}

\begin{figure*}
\centering
\includegraphics[width=0.65\textwidth]{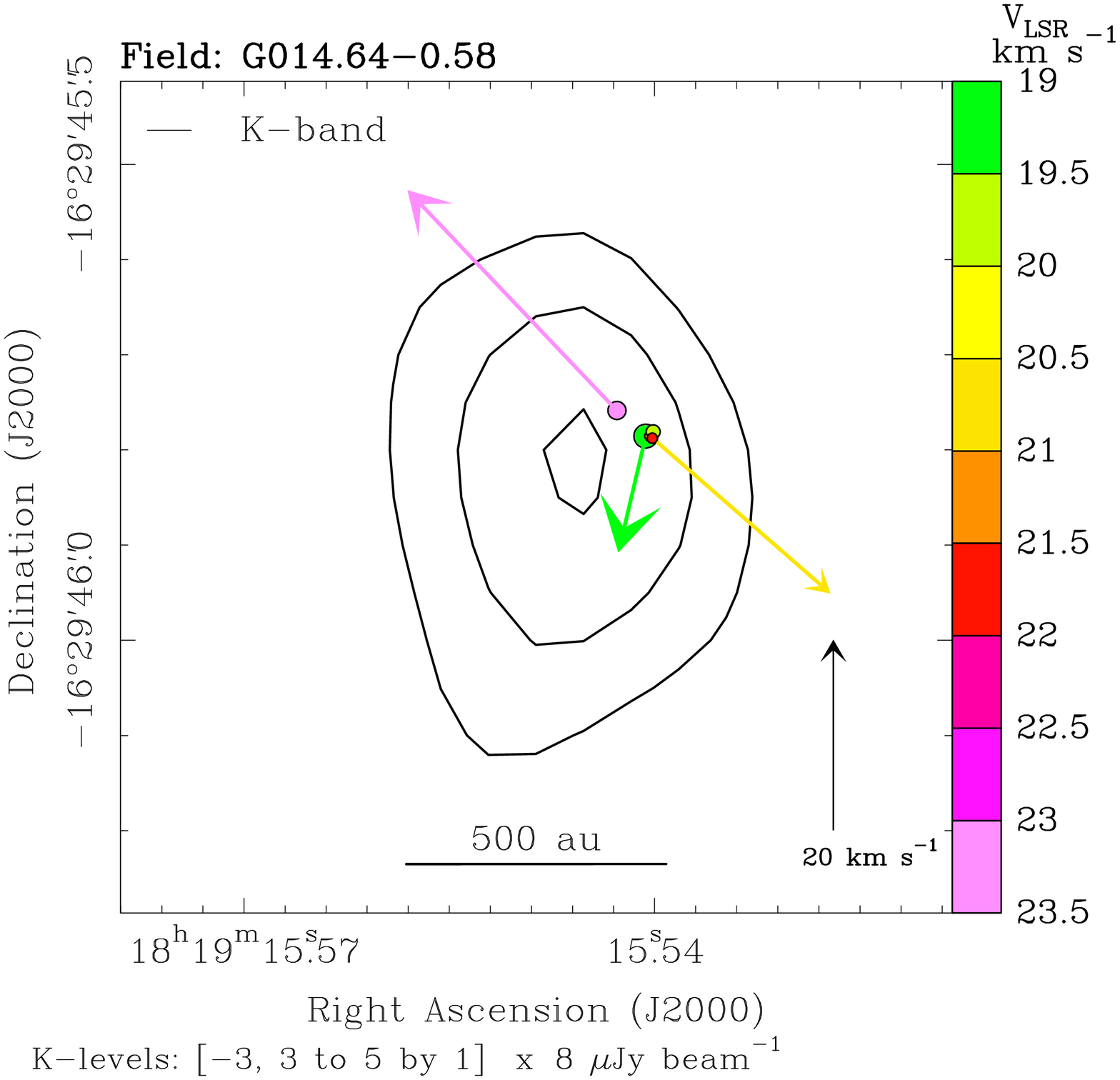}
\includegraphics[width=0.65\textwidth]{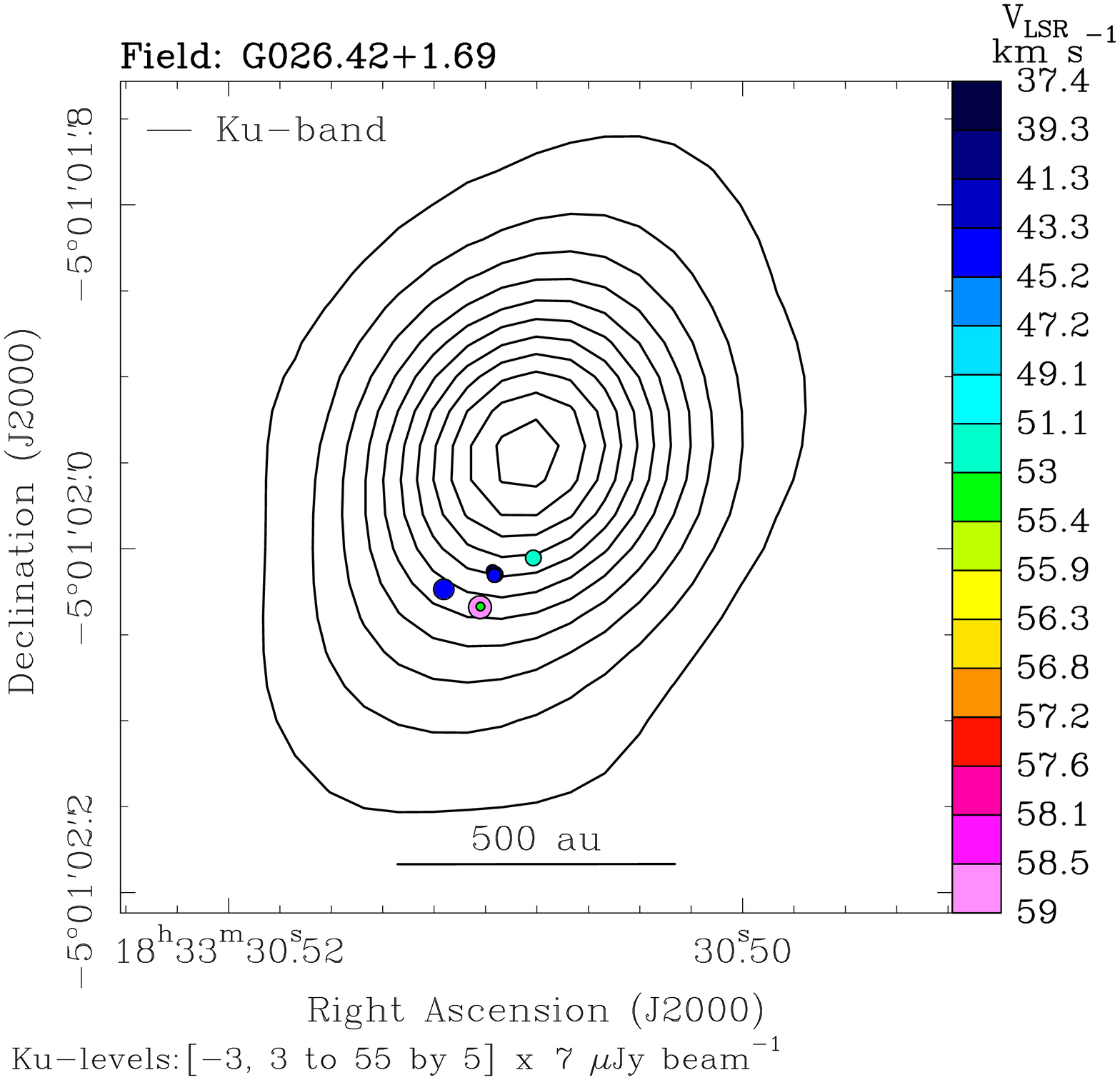}
\caption{Overlay of radio continuum and water maser positions and velocities. Each panel refers to the POETS target indicated in the observed field above the corresponding plot. In each panel, {\it black contours}, and {\it colored dots} and {\it arrows} have the same meaning as in the upper~panel of Fig.~\ref{F:G11.92}.}
\label{F:G14.64-G26.42}
\end{figure*}

\begin{figure*}
\centering
\includegraphics[width=0.63\textwidth]{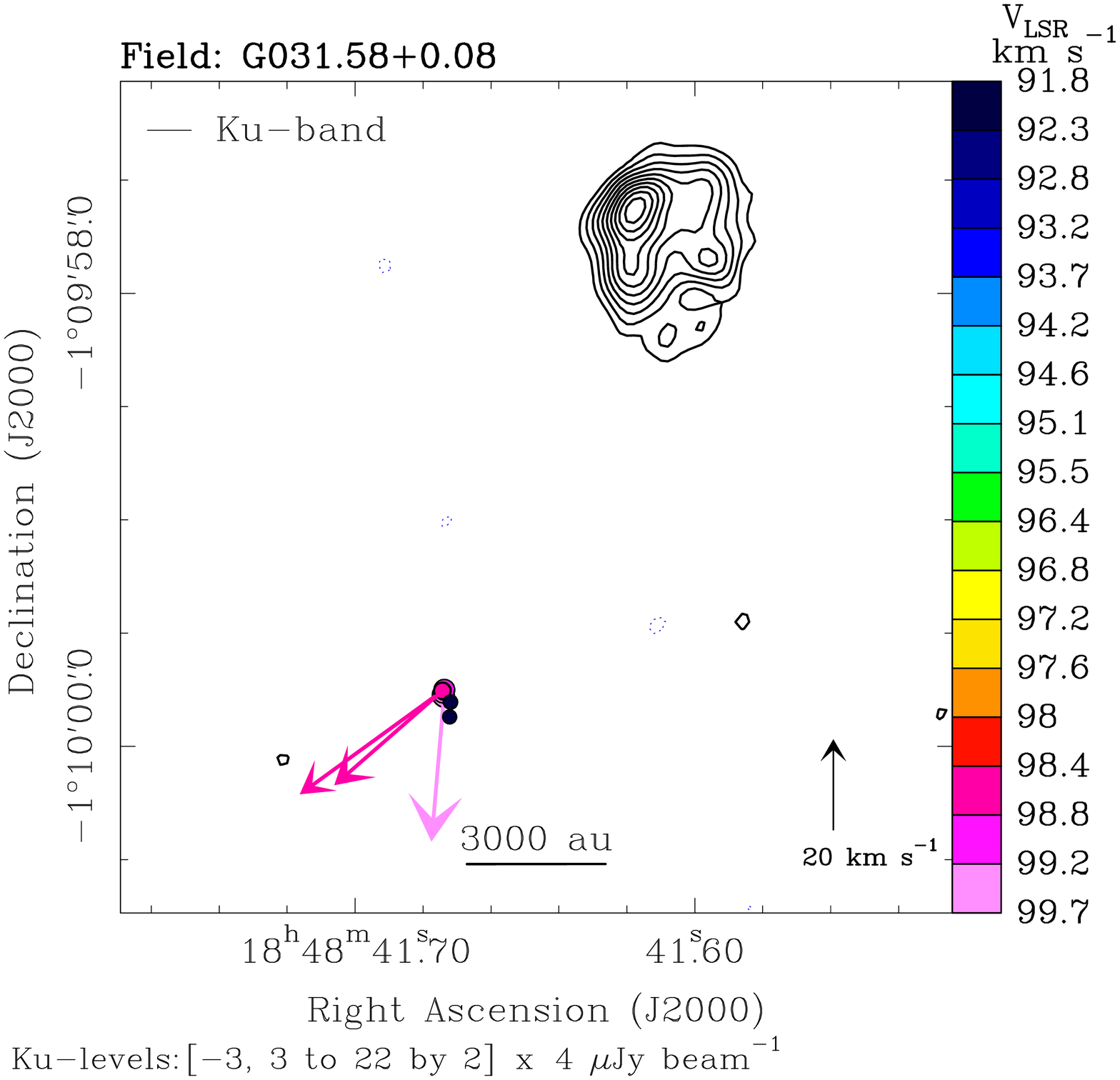}
\includegraphics[width=0.65\textwidth]{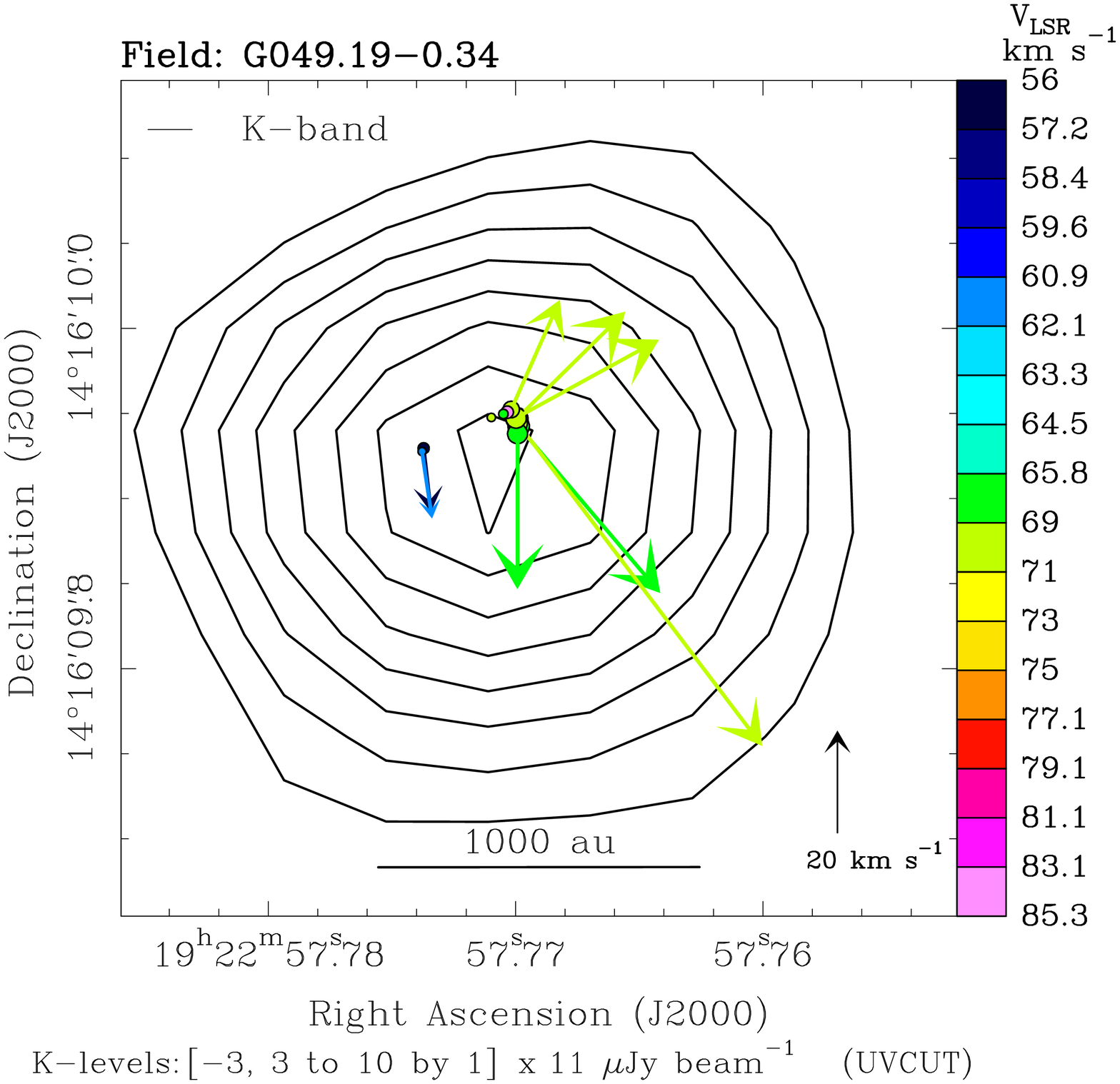}
\caption{Overlay of radio continuum and water maser positions and velocities. Each panel refers to the POETS target indicated in the observed field above the corresponding plot. In each panel, {\it black contours}, and {\it colored dots} and {\it arrows} have the same meaning as in the upper~panel of Fig.~\ref{F:G11.92}.}
\label{F:G31.58-G49.19}
\end{figure*}

\begin{figure*}
\centering
\includegraphics[width=0.62\textwidth]{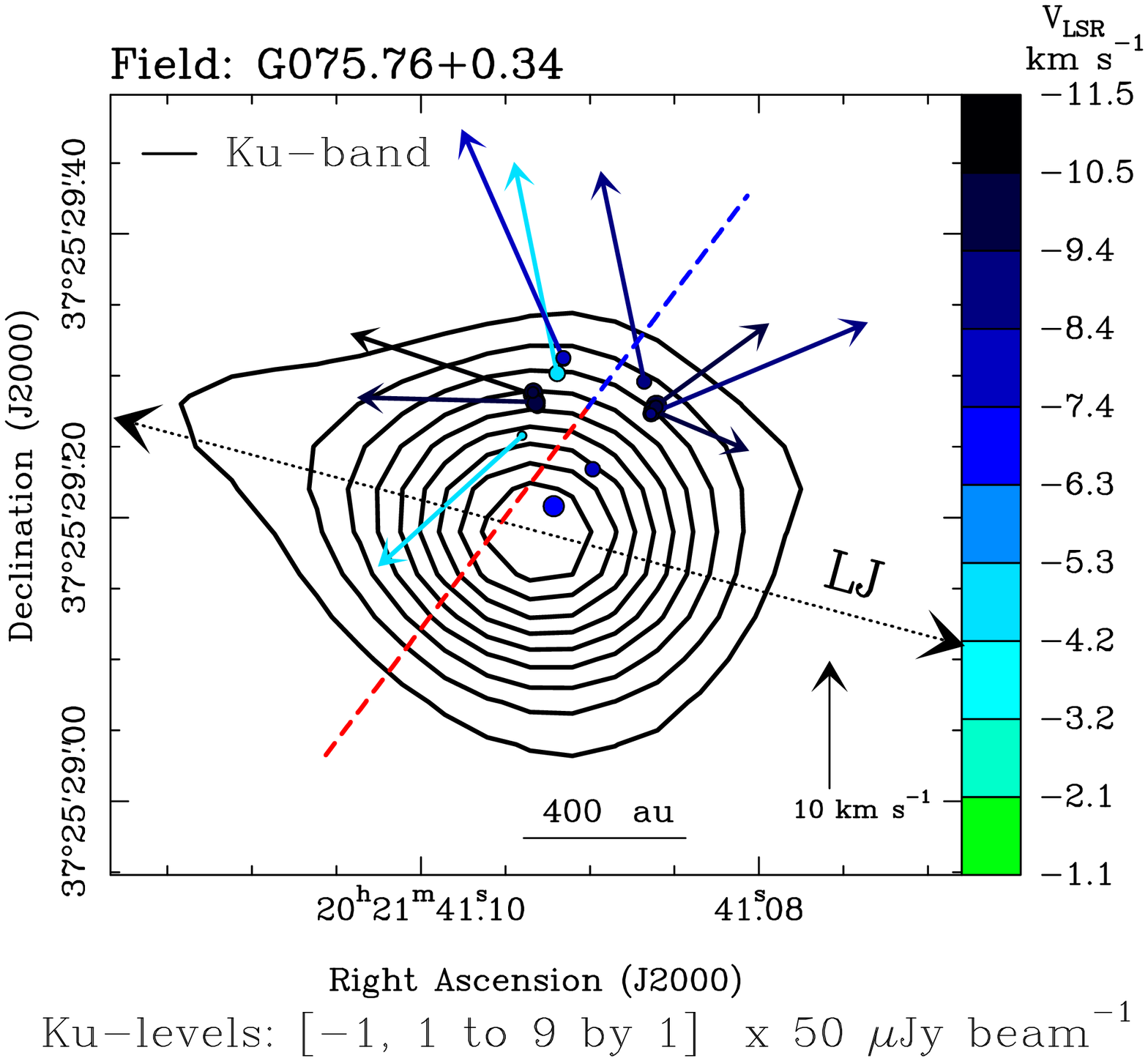}

\includegraphics[width=0.64\textwidth]{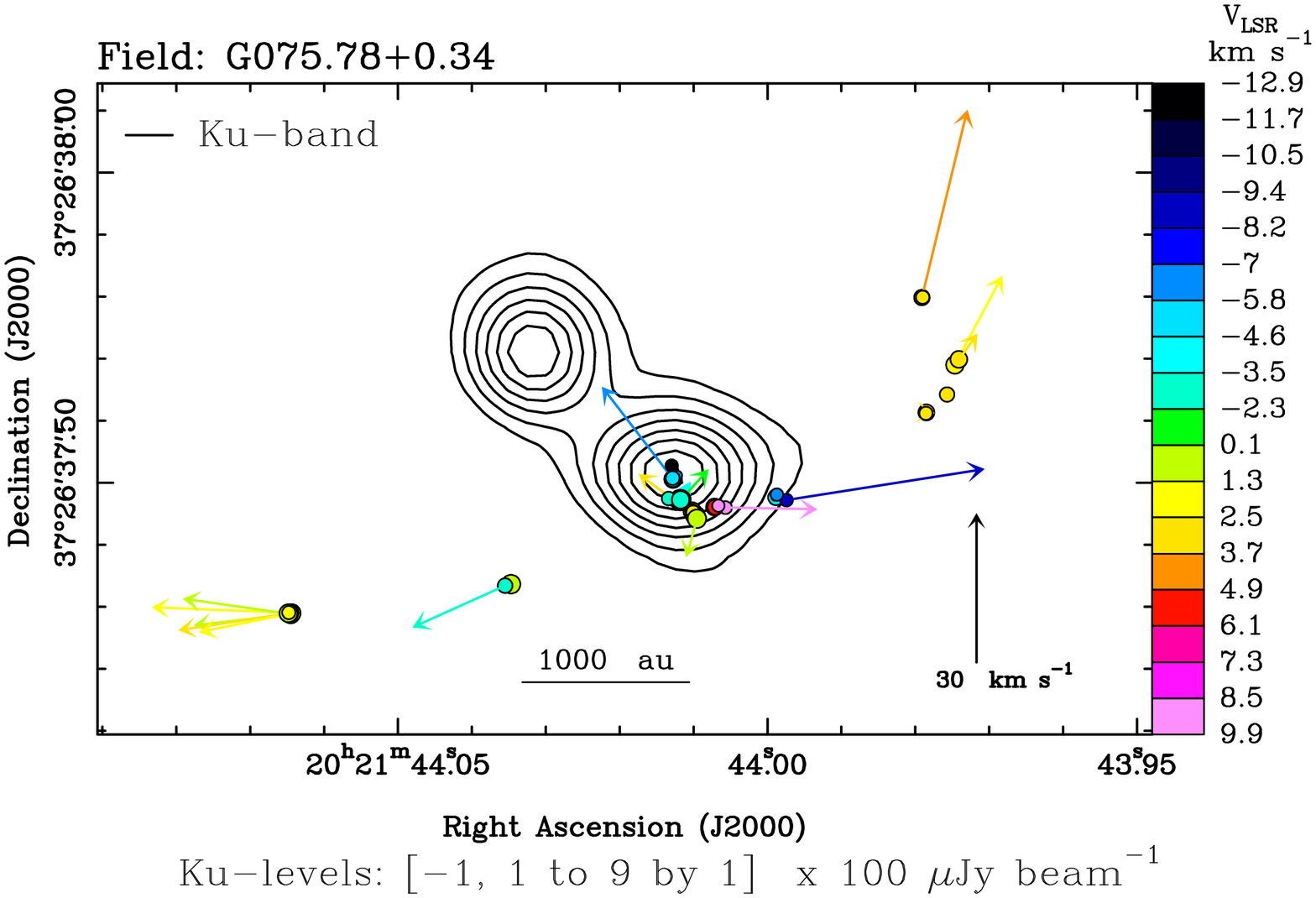}
\includegraphics[width=0.35\textwidth]{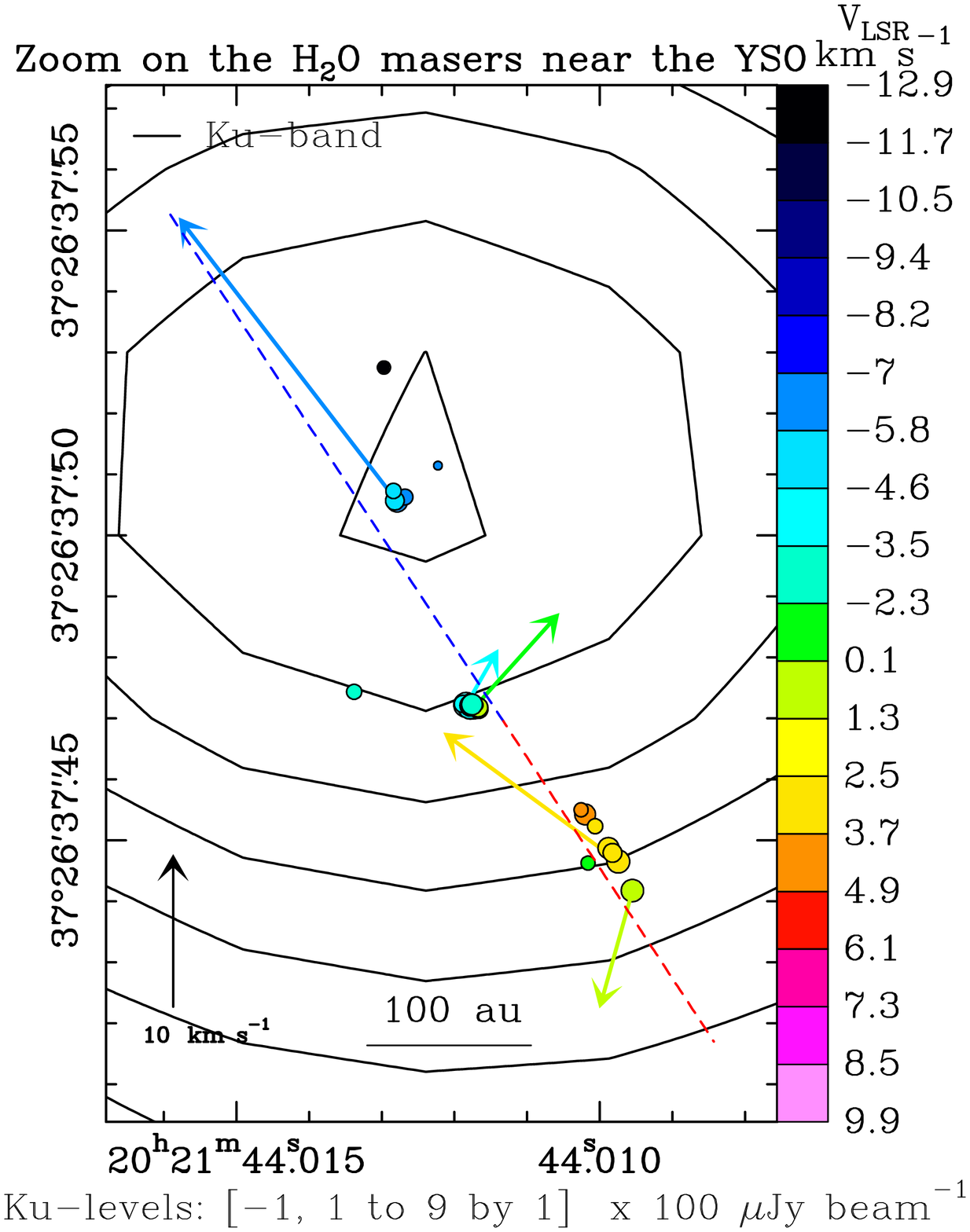}
\caption{Overlay of radio continuum and water maser positions and velocities. Each panel refers to the POETS target indicated in the observed field above the corresponding plot. In each panel, {\it black contours}, and {\it colored dots} and {\it arrows} have the same meaning as in the upper~panel of Fig.~\ref{F:G11.92}.  \ {\it Upper~panel:}~the \Vlsr\ gradient traced by the water masers is indicated by a {\it dashed line}, using {\it colors} to denote the {\it red}-~and~{\it blue}-shifted side of the gradient. The {\it black dotted arrows} labeled~LJ denote the axis of the jet in H$_2$ 2.2~$\mu$m emission discovered in our recent LBT survey \citep{Mas19}.  \ {\it Lower~panels:}~large view ({\it left}) and zoom on the water masers near the YSO ({\it right}) for the source \ G075.78$+$0.34. In the right~panel, the \Vlsr\ gradient traced by the water masers is indicated by a {\it dashed line}, using {\it colors} to denote the {\it red}-~and~{\it blue}-shifted side of the gradient.}
\label{F:G75.76-G75.78}
\end{figure*}

\begin{figure*}
\centering
\includegraphics[width=0.65\textwidth]{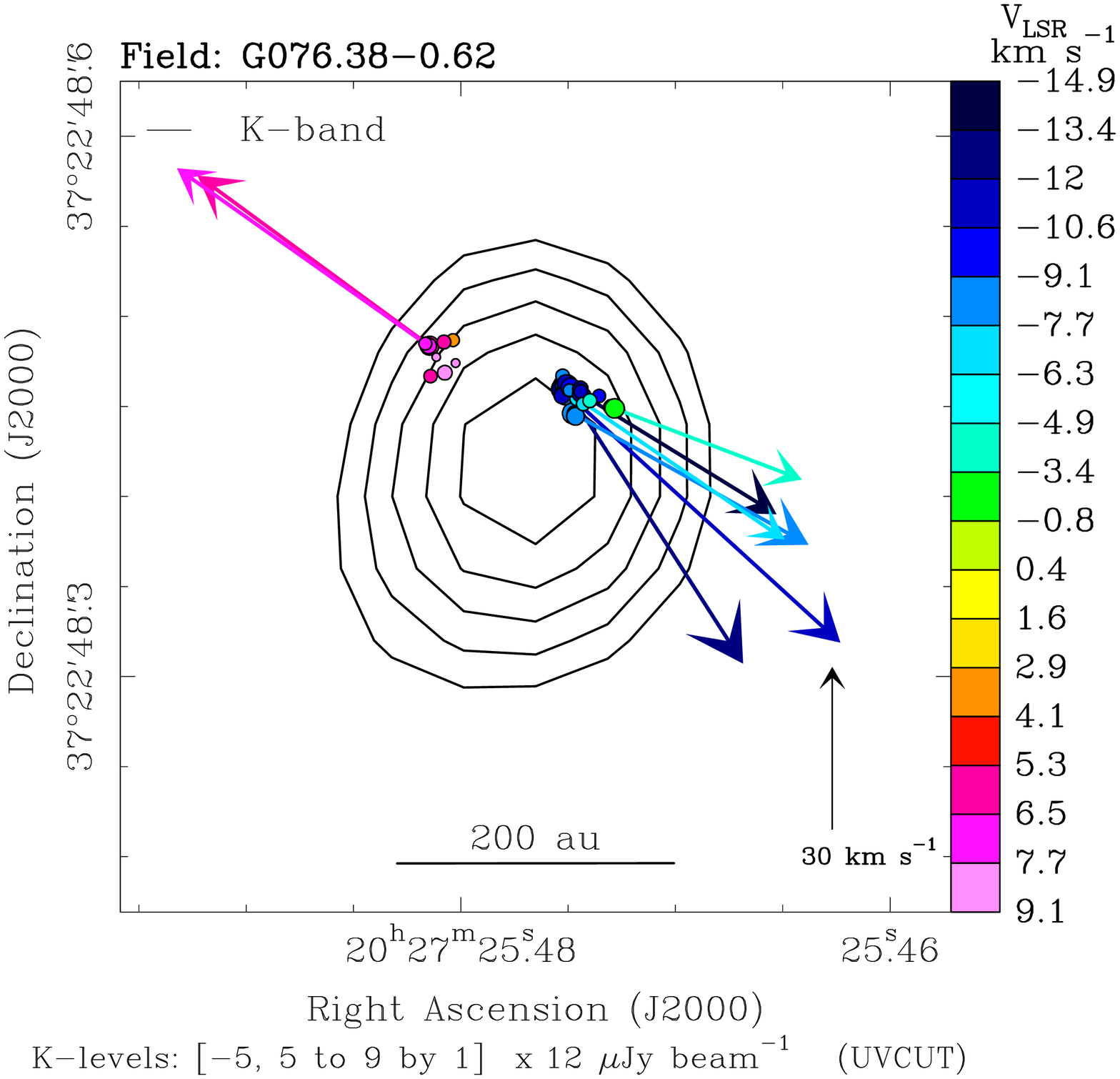}
\includegraphics[width=0.65\textwidth]{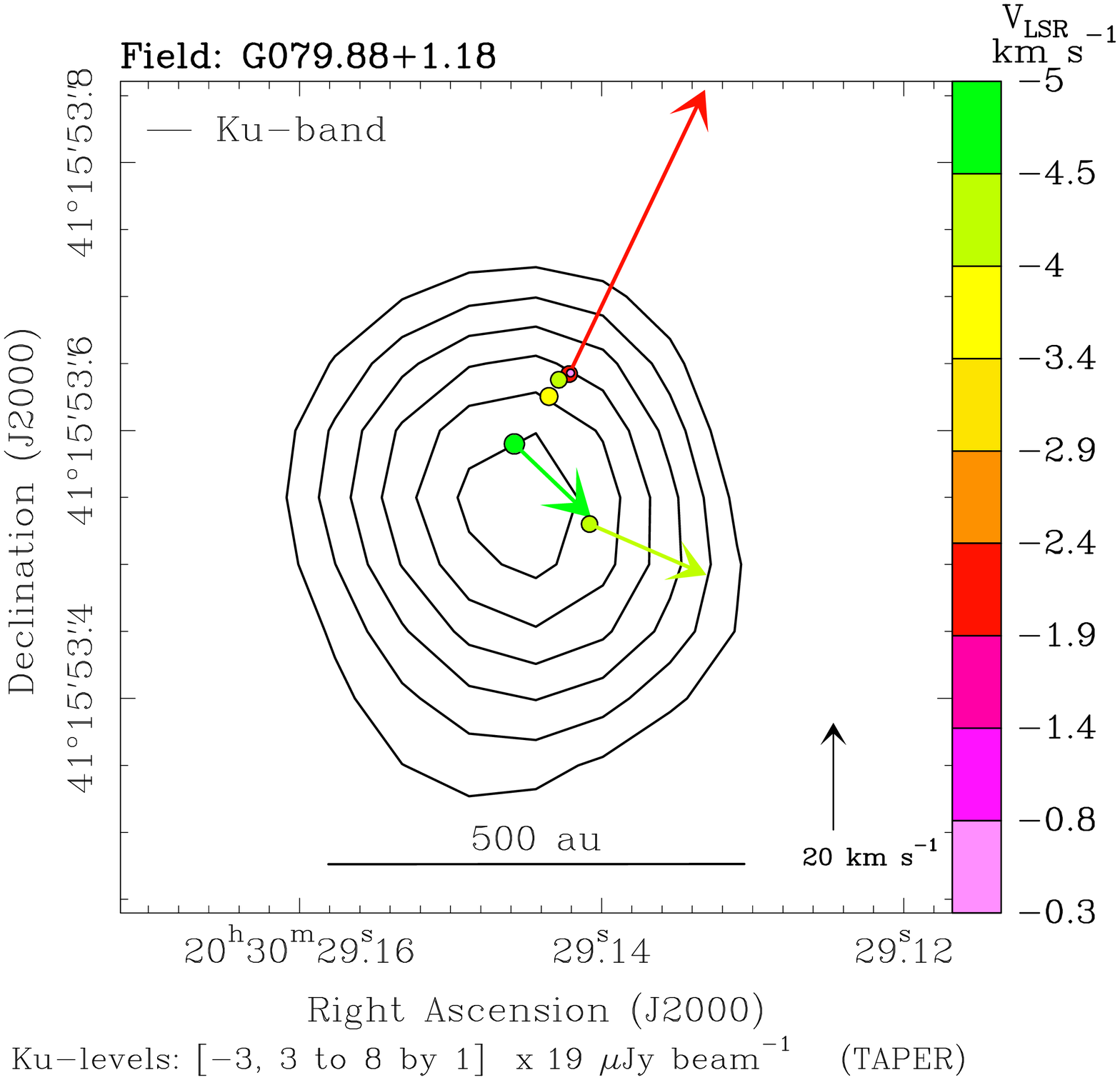}
\caption{Overlay of radio continuum and water maser positions and velocities. Each panel refers to the POETS target indicated in the observed field above the corresponding plot. In each panel, {\it black contours}, and {\it colored dots} and {\it arrows} have the same meaning as in the upper~panel of Fig.~\ref{F:G11.92}.}
\label{F:G76.38-G79.88}
\end{figure*}

\begin{figure*}
\centering
\includegraphics[width=0.65\textwidth]{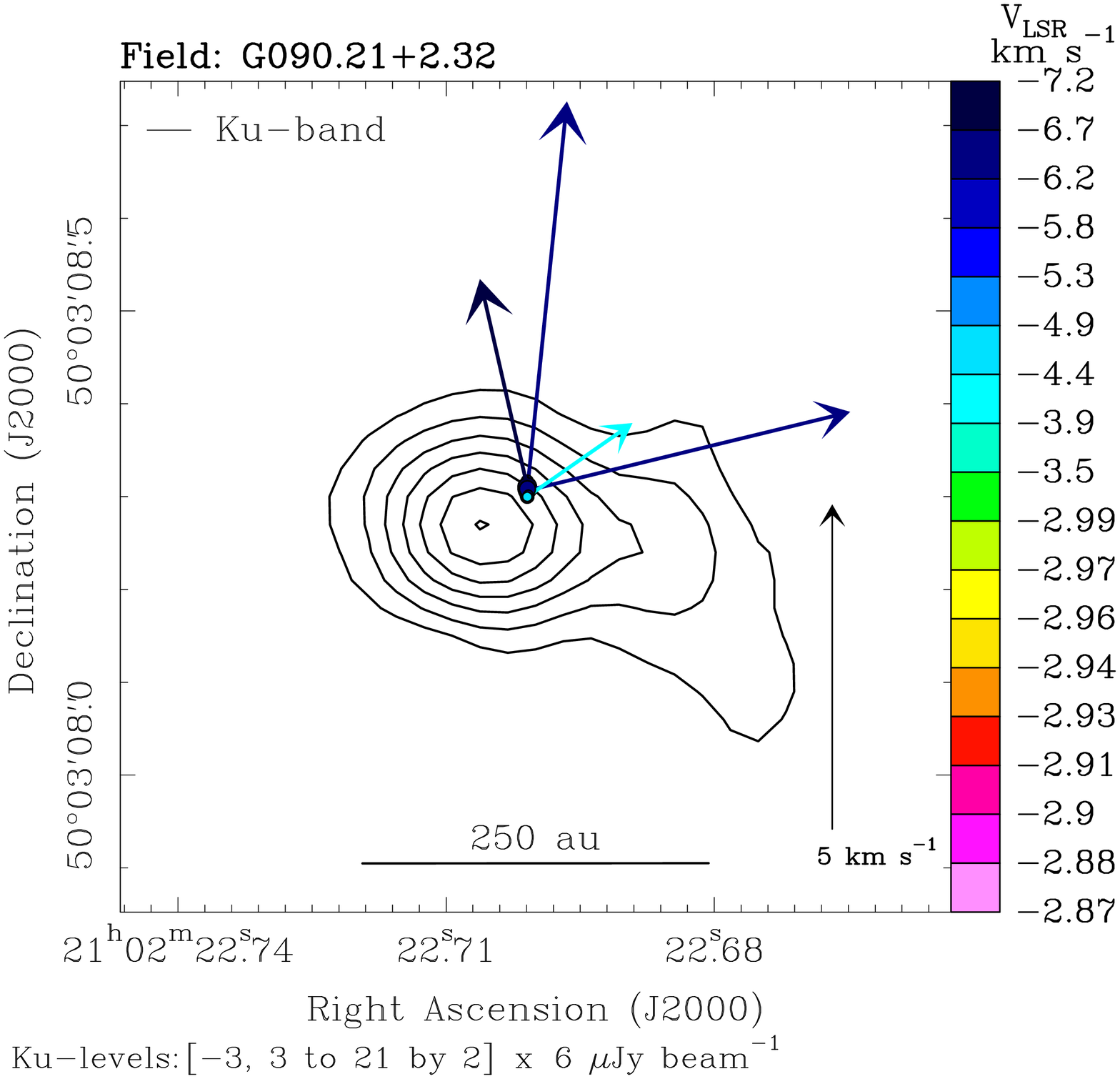}

\vspace*{0.5cm}\includegraphics[width=0.52\textwidth]{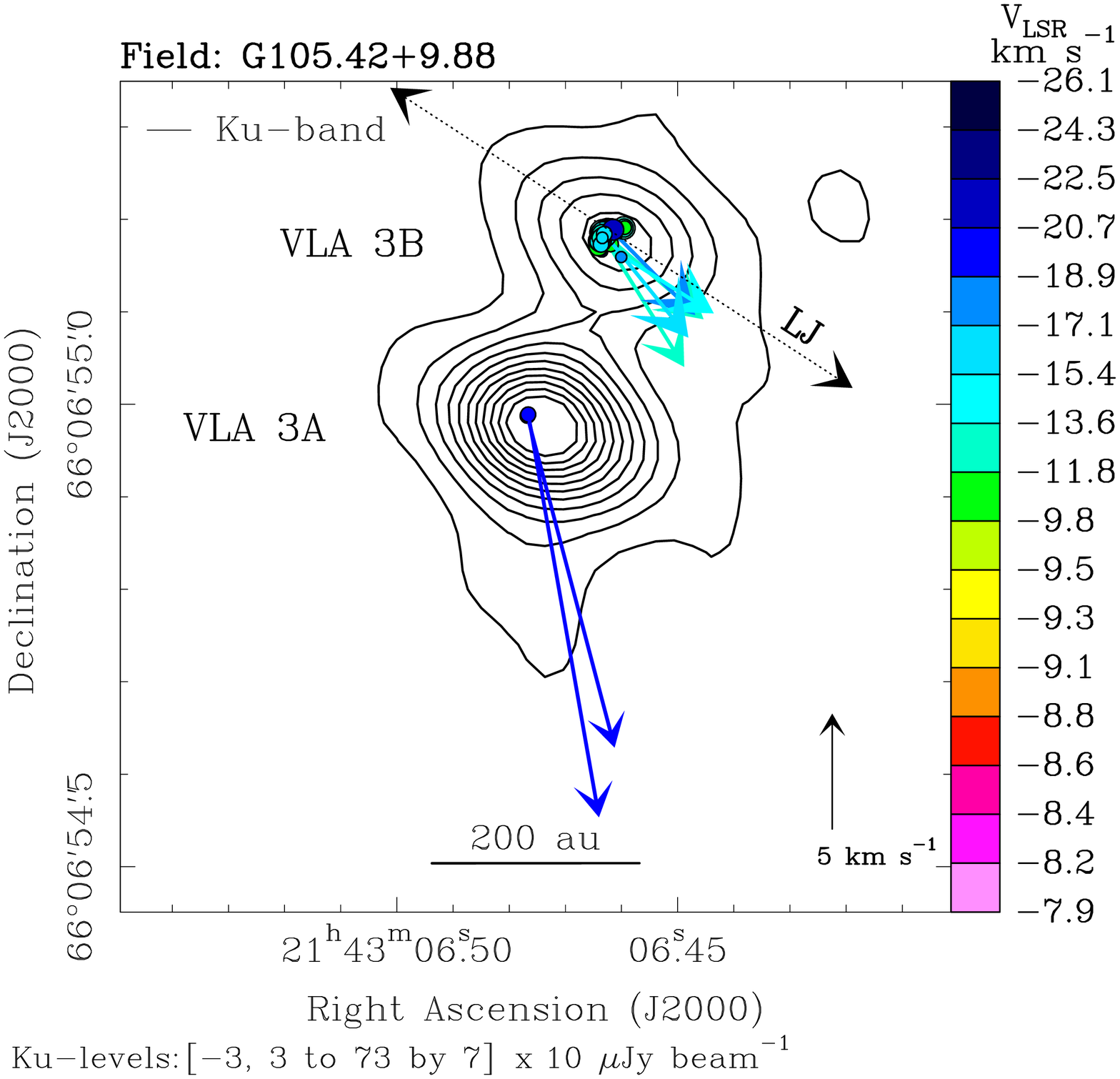}
\includegraphics[width=0.46\textwidth]{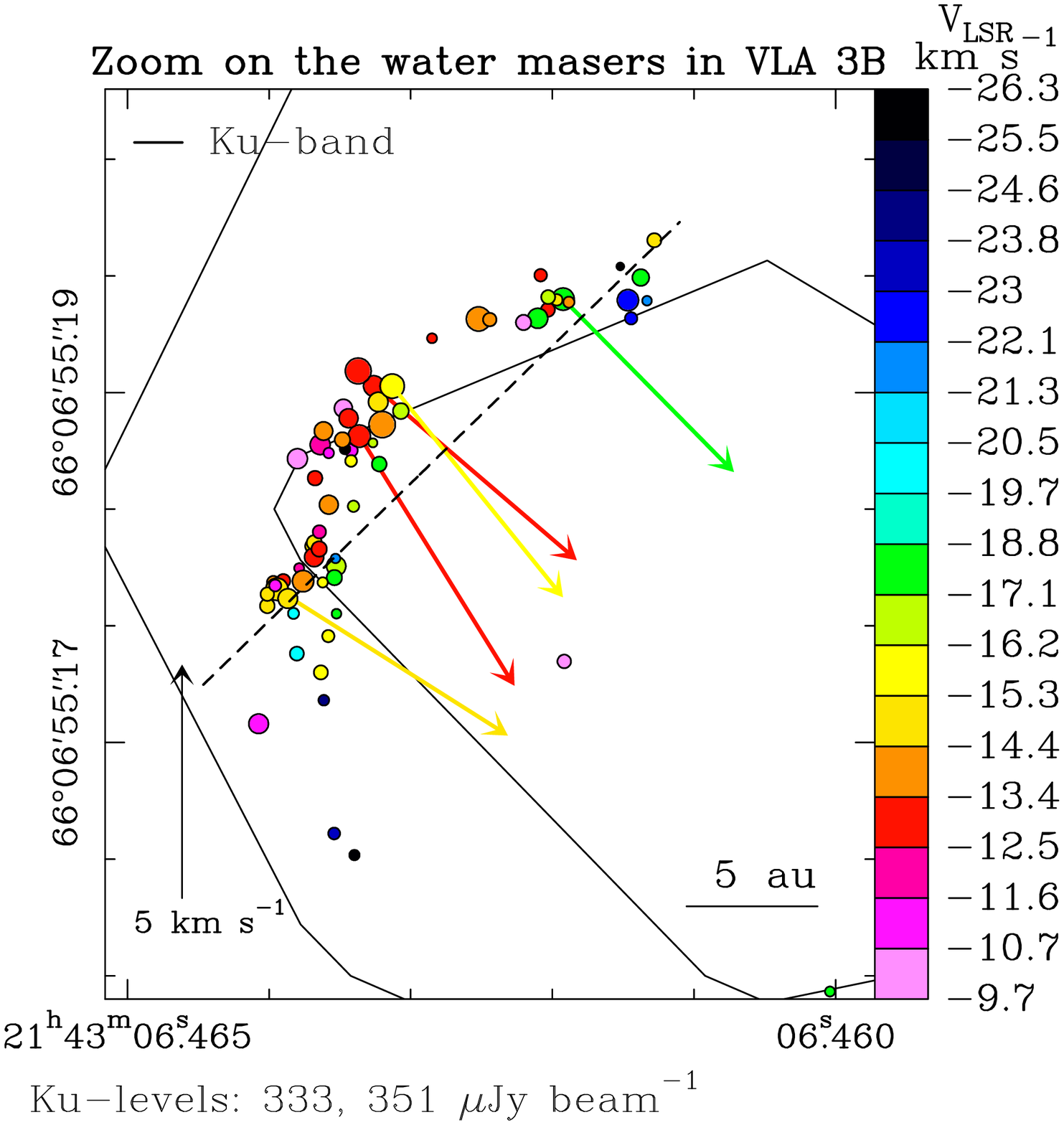}
\caption{Overlay of radio continuum and water maser positions and velocities. Each panel refers to the POETS target indicated in the observed field above the corresponding plot. In each panel, {\it black contours}, and {\it colored dots} and {\it arrows} have the same meaning as in the upper~panel of Fig.~\ref{F:G11.92}. \ {\it Lower~panels:}~large view ({\it left}) and zoom on the water maser kinematics near the YSO VLA~3B ({\it right}) for the source \ G105.42$+$9.88. The \Vlsr\ scale of the left and right panels is different. The {\it black dotted arrows} labeled~LJ in the left panel denote the axis of the radio jet observed with the VLA by \citet{Tri04}. The major-axis of the spatial distribution of the water masers is indicated by a {\it black dashed line} in the right panel. }
\label{F:G90.21-G105.42}
\end{figure*}

\begin{figure*}
\centering
\includegraphics[width=0.65\textwidth]{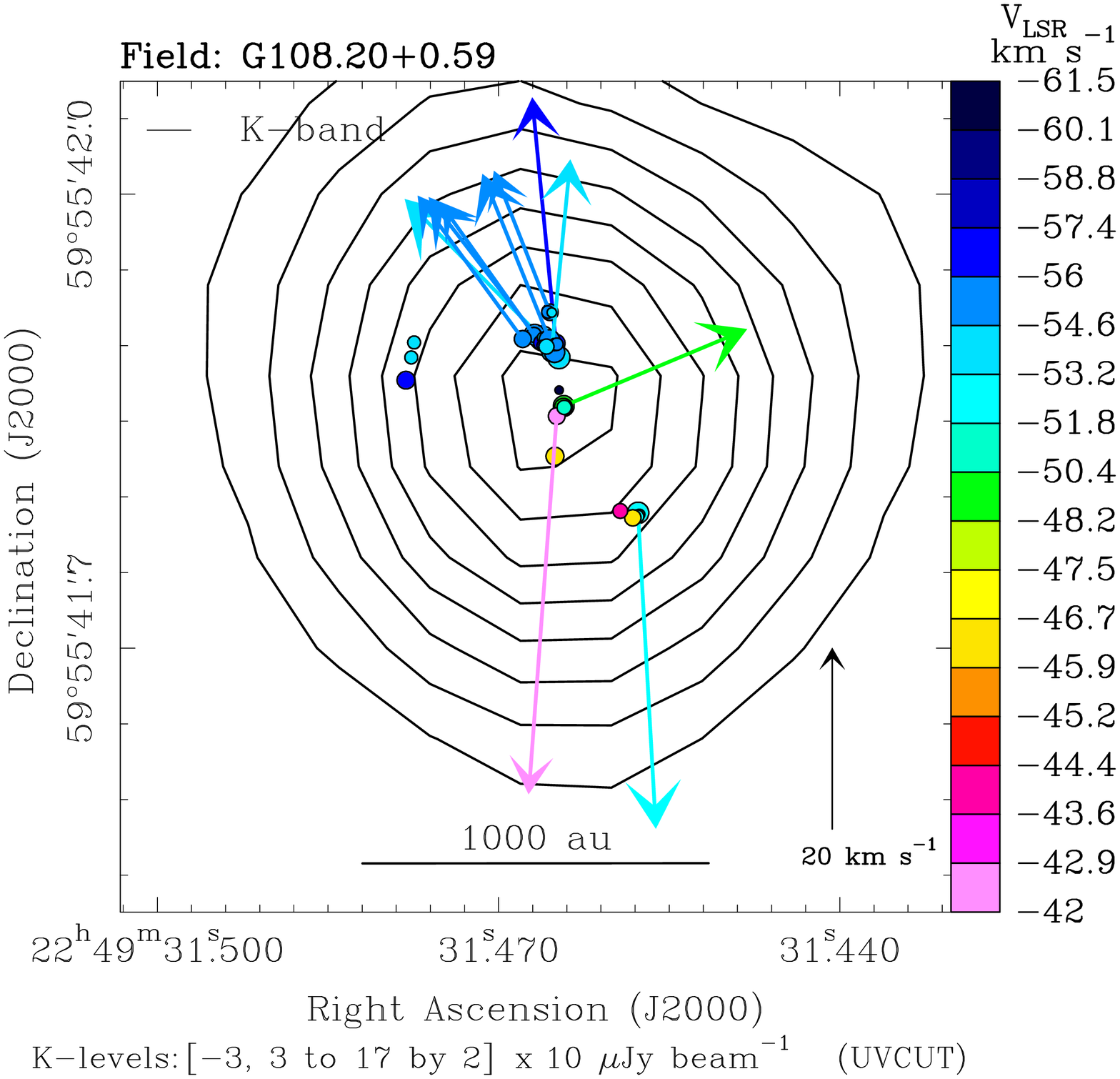}
\includegraphics[width=0.65\textwidth]{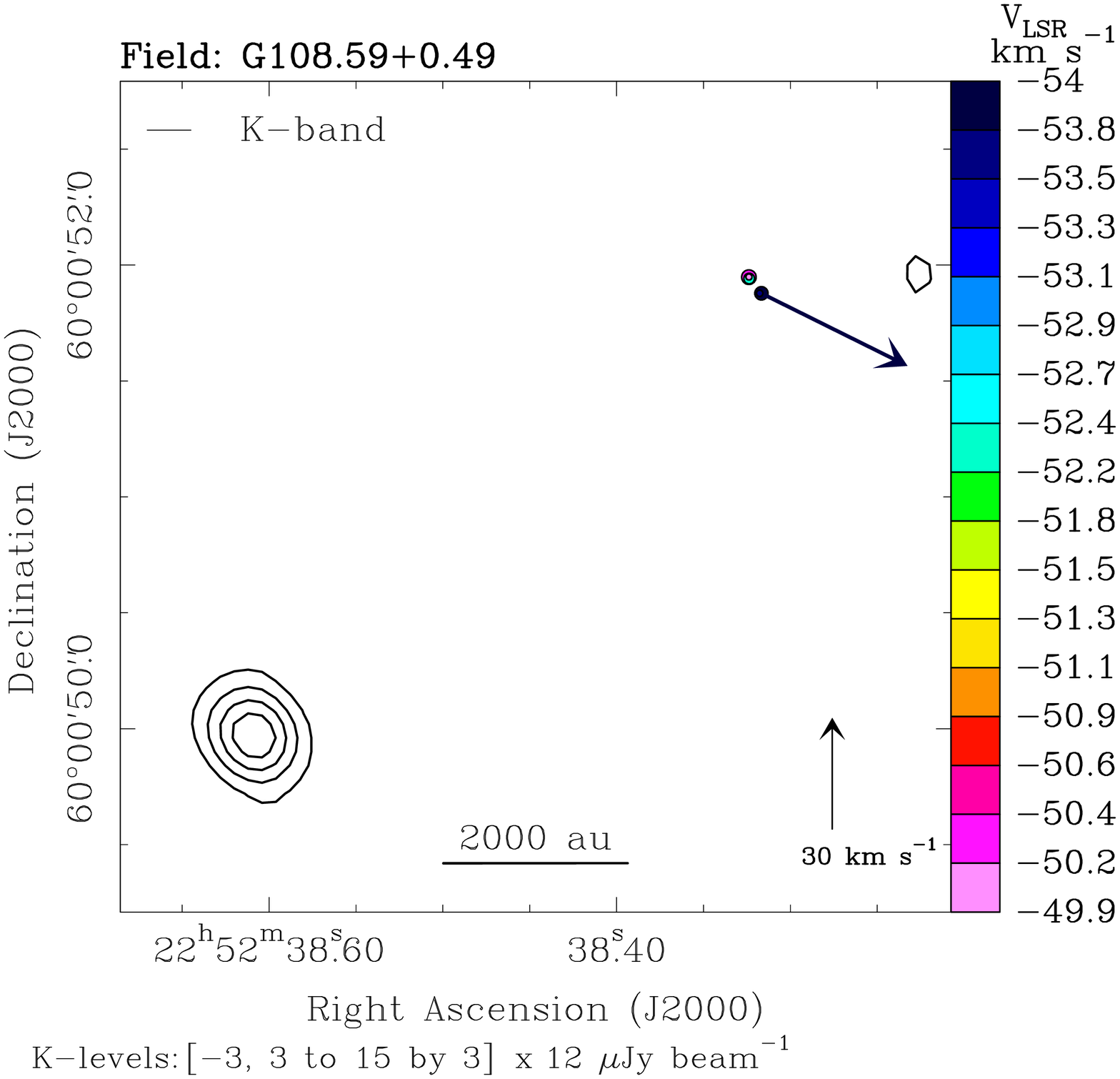}
\caption{Overlay of radio continuum and water maser positions and velocities. Each panel refers to the POETS target indicated in the observed field above the corresponding plot. In each panel, {\it black contours}, and {\it colored dots} and {\it arrows} have the same meaning as in the upper~panel of Fig.~\ref{F:G11.92}.}
\label{F:G108.20-G108.59}
\end{figure*}

\begin{figure*}
\centering
\includegraphics[width=0.65\textwidth]{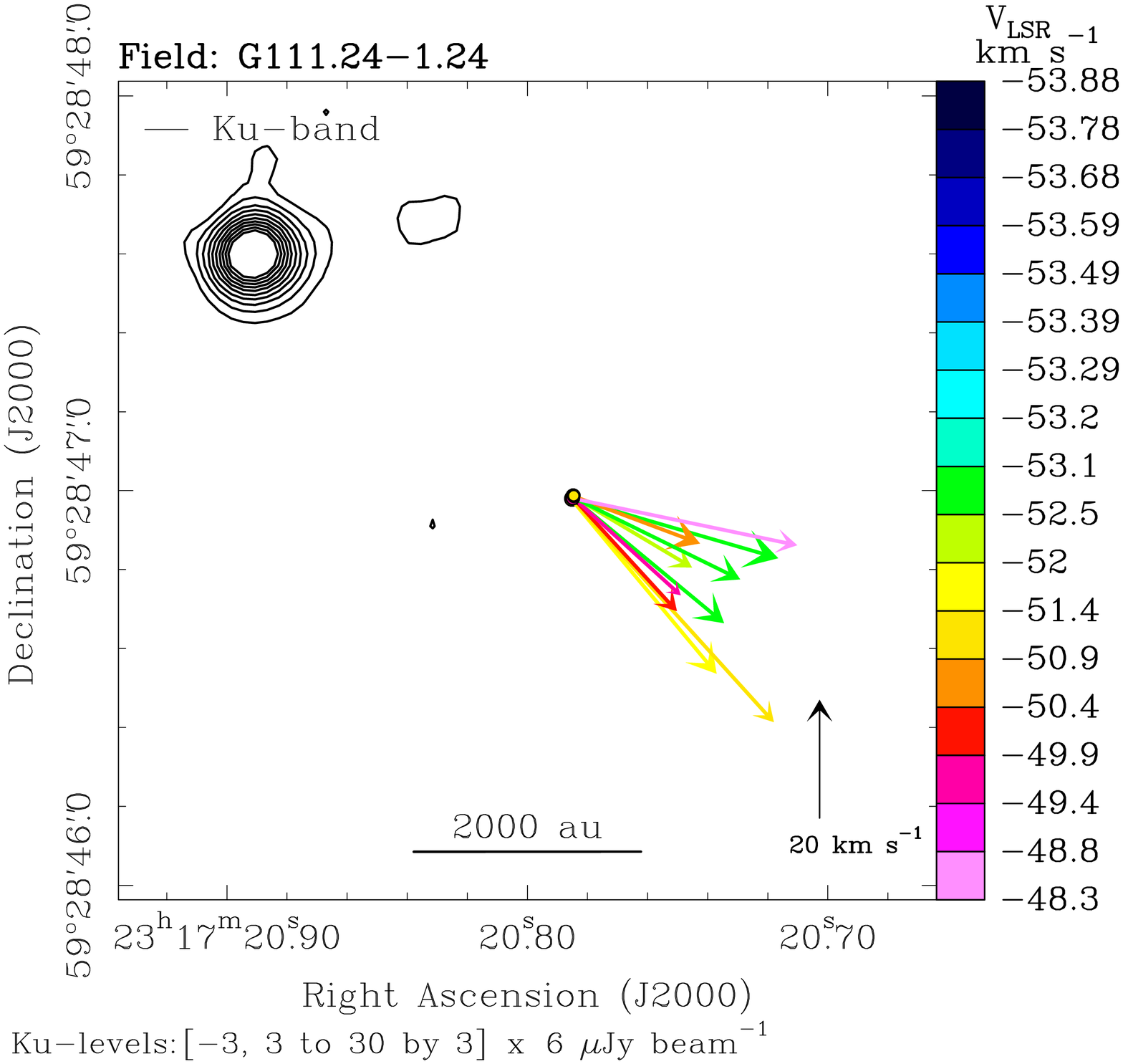}
\includegraphics[width=0.65\textwidth]{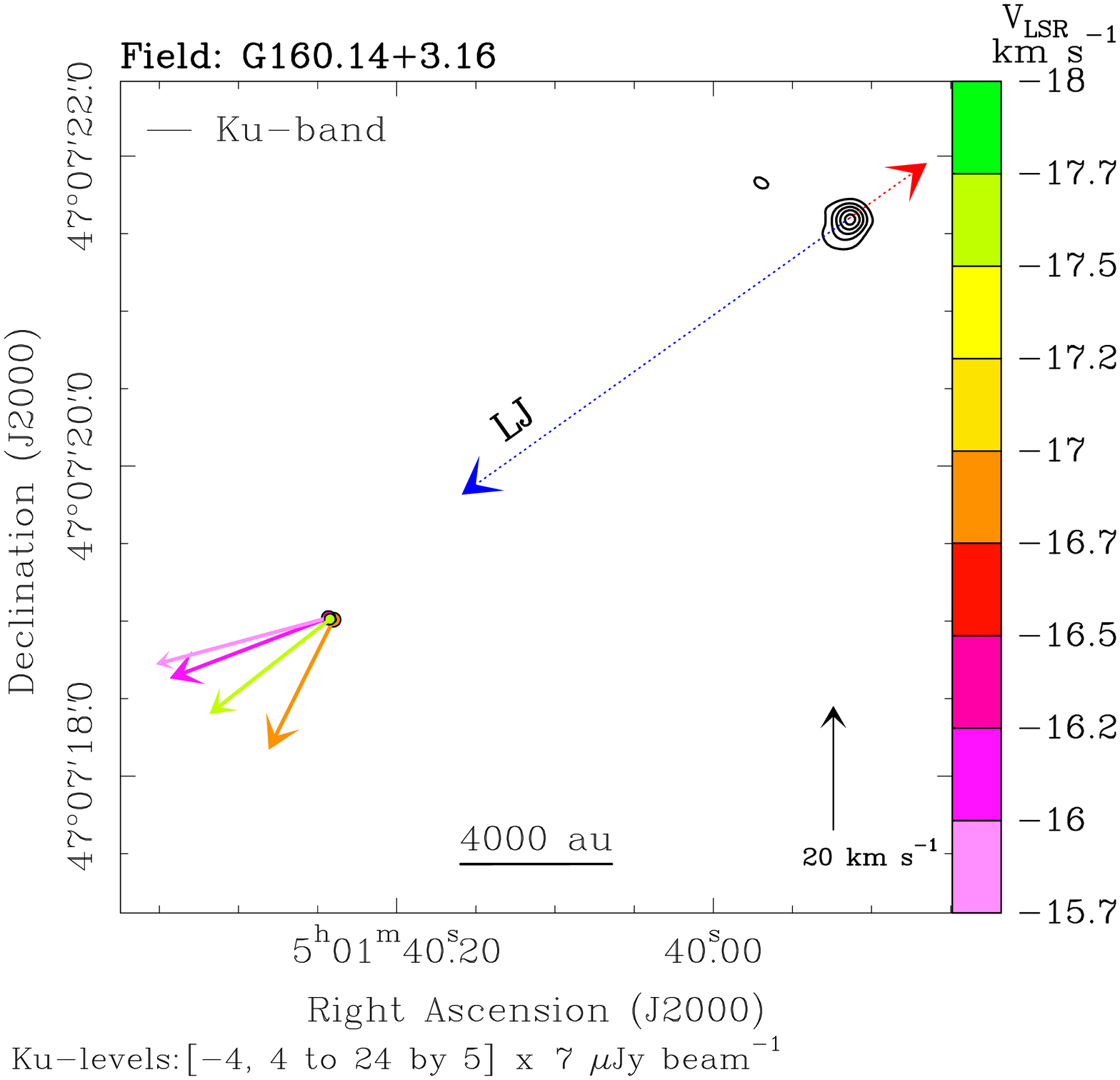}
\caption{Overlay of radio continuum and water maser positions and velocities. Each panel refers to the POETS target indicated in the observed field above the corresponding plot. In each panel, {\it black contours}, and {\it colored dots} and {\it arrows} have the same meaning as in the upper~panel of Fig.~\ref{F:G11.92}. \ {\it Lower~panel:}~The {\it dotted arrows} labeled~LJ give the axis of the jet traced in H$_2$ 2.2~$\mu$m emission by \citet{Var10}, using {\it colors} to distinguish the {\it red}-~and~{\it blue}-shifted lobe.}
\label{F:G111.24-G160.14}
\end{figure*}

\begin{figure*}
\centering
\includegraphics[width=0.65\textwidth]{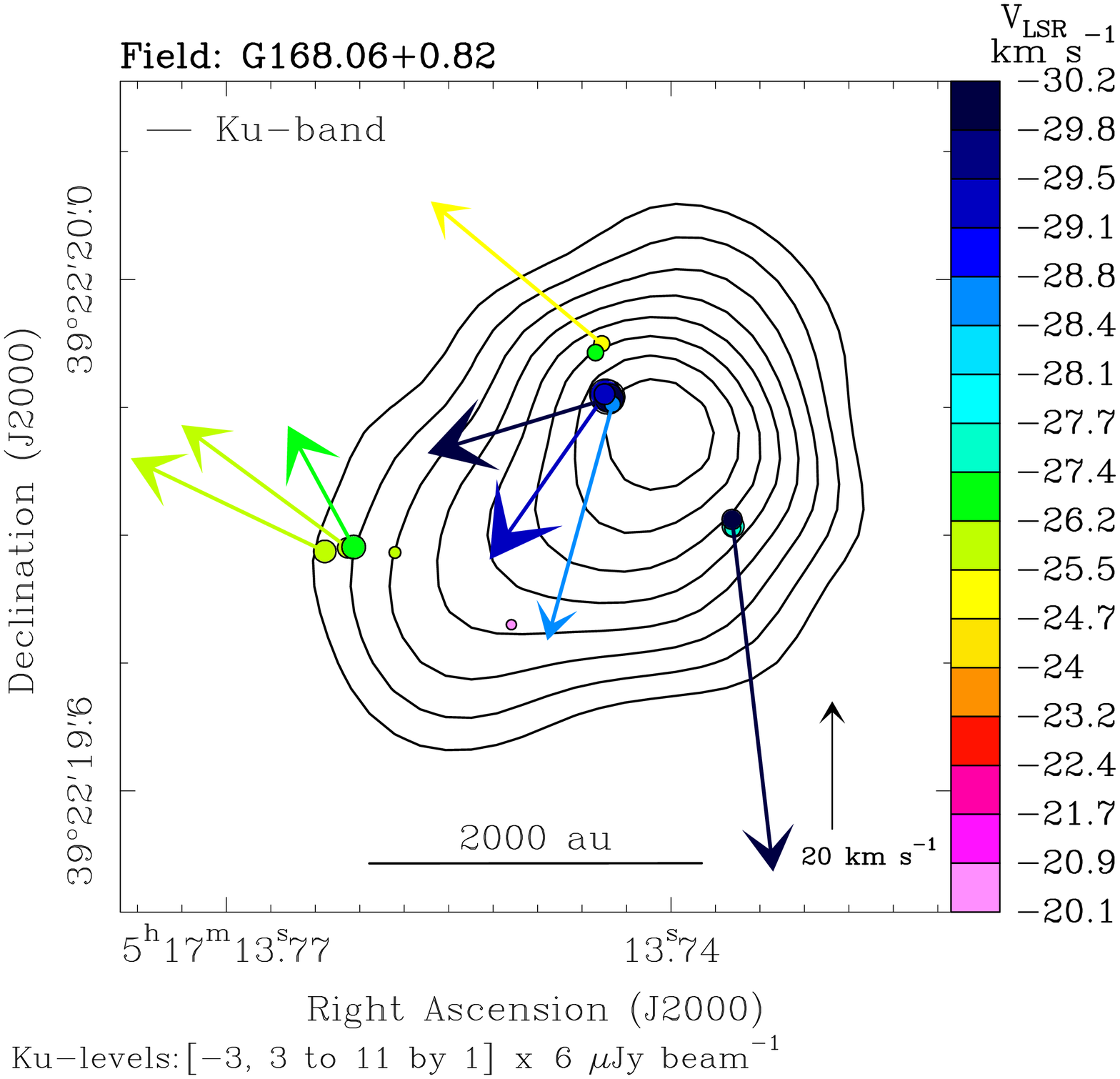}
\includegraphics[width=0.65\textwidth]{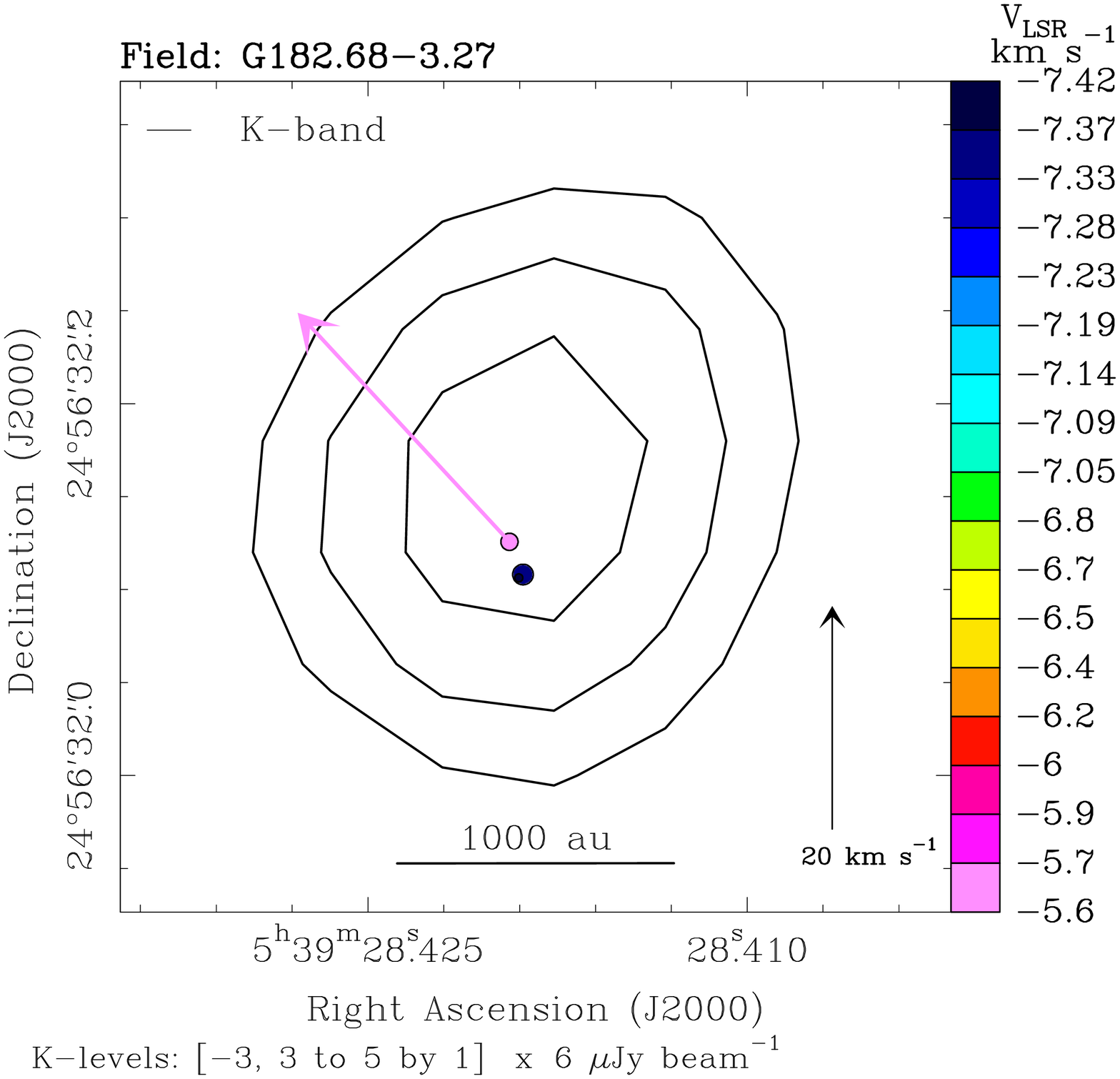}
\caption{Overlay of radio continuum and water maser positions and velocities. Each panel refers to the POETS target indicated in the observed field above the corresponding plot. In each panel, {\it black contours}, and {\it colored dots} and {\it arrows} have the same meaning as in the upper~panel of Fig.~\ref{F:G11.92}.}
\label{F:G168.06-G182.68}
\end{figure*}

\begin{figure*}
\centering
\includegraphics[width=0.65\textwidth]{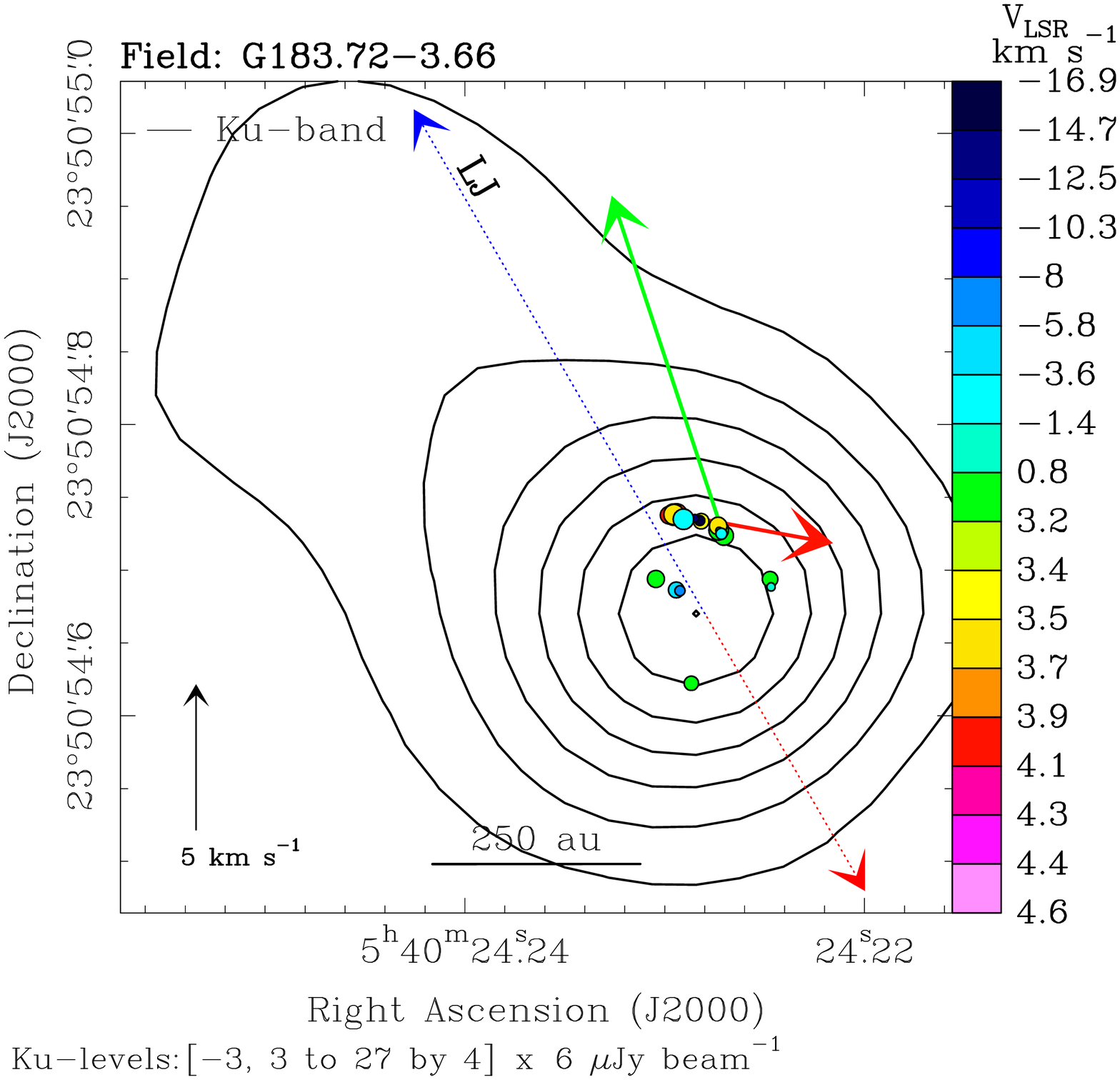}

\vspace*{0.5cm}\includegraphics[width=0.48\textwidth]{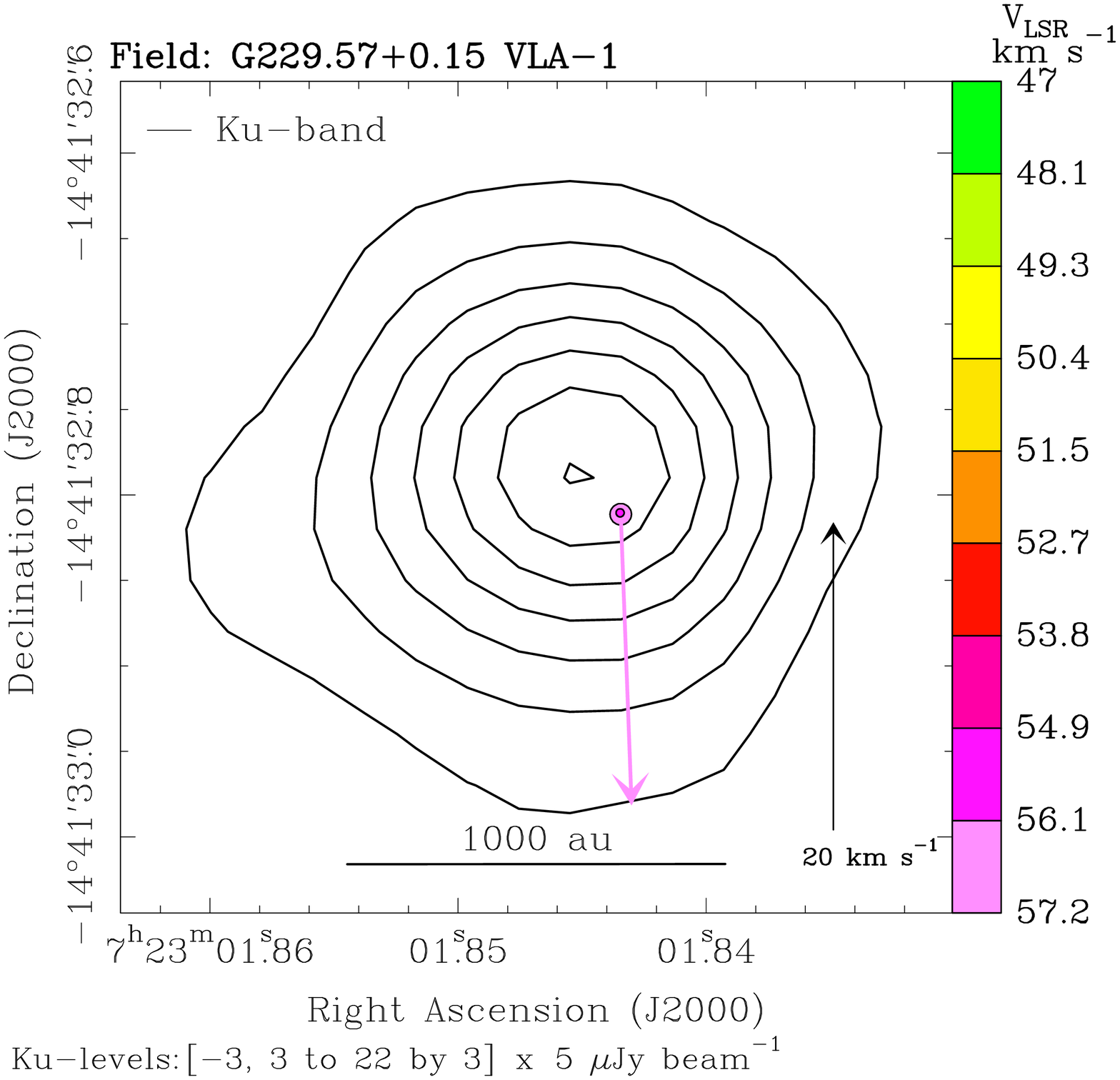}
\includegraphics[width=0.48\textwidth]{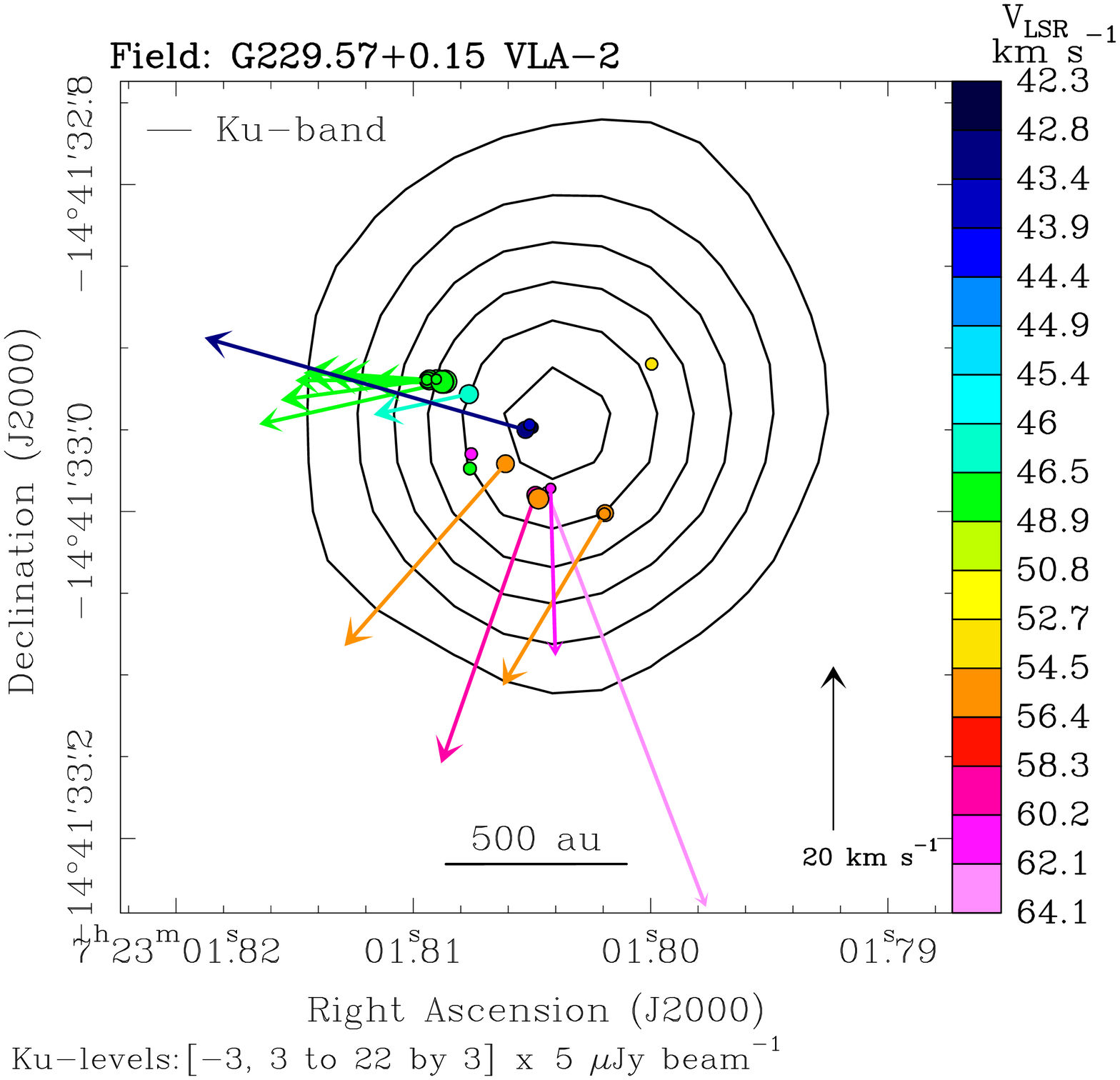}
\caption{Overlay of radio continuum and water maser positions and velocities. Each panel refers to the POETS target indicated in the observed field above the corresponding plot. In each panel, {\it black contours}, and {\it colored dots} and {\it arrows} have the same meaning as in the upper~panel of Fig.~\ref{F:G11.92}. \ {\it Upper~panel:}~The {\it dotted arrows} labeled~LJ give the axis of the collimated outflow traced in $^{12}$CO emission with CARMA by \citet{Bro16}, using {\it colors} to distinguish the {\it red}-~and~{\it blue}-shifted flow lobe.}
\label{F:G183.72-G229.57}
\end{figure*}

\begin{figure*}
\centering
\includegraphics[width=0.65\textwidth]{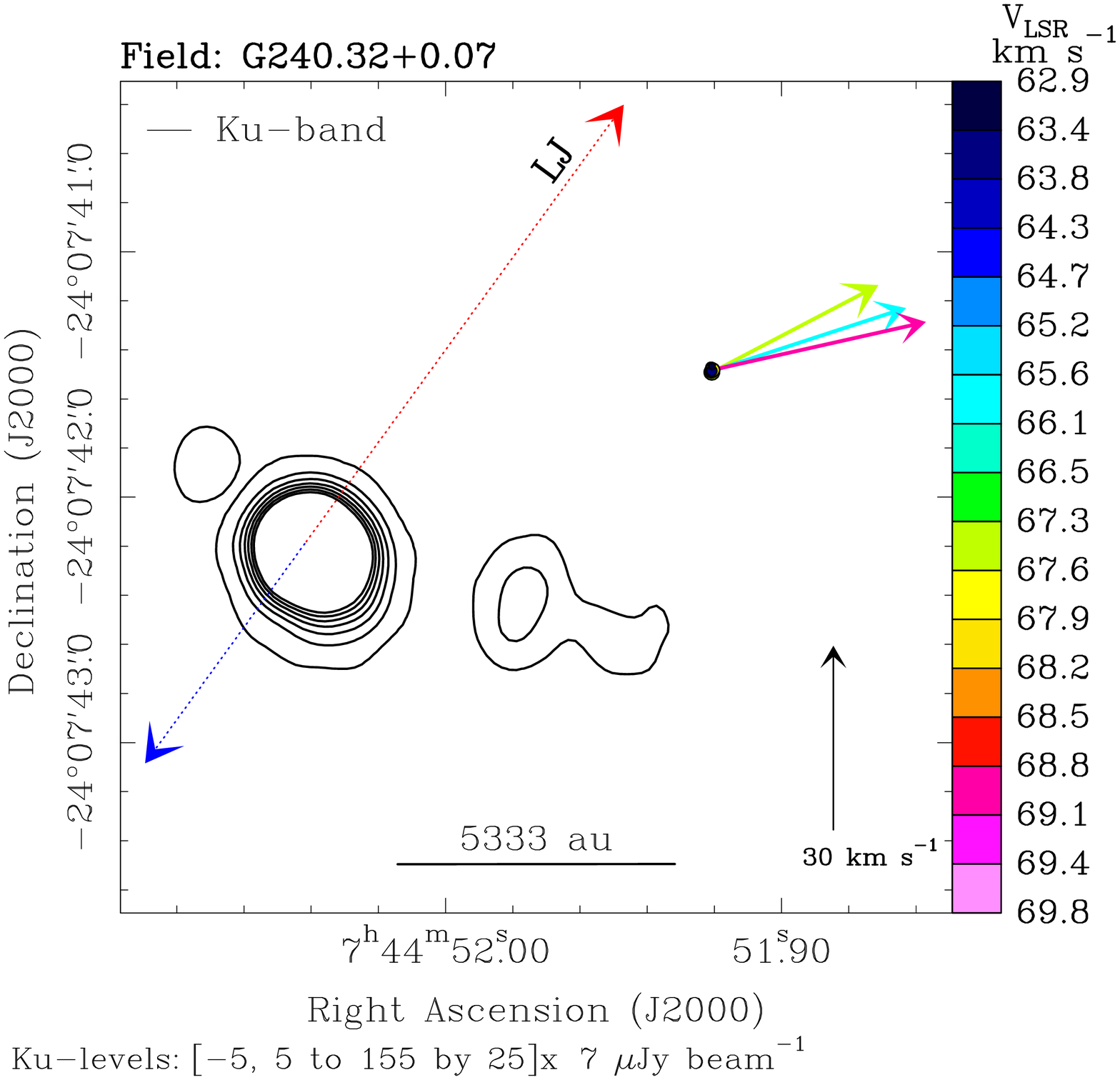}
\includegraphics[width=0.65\textwidth]{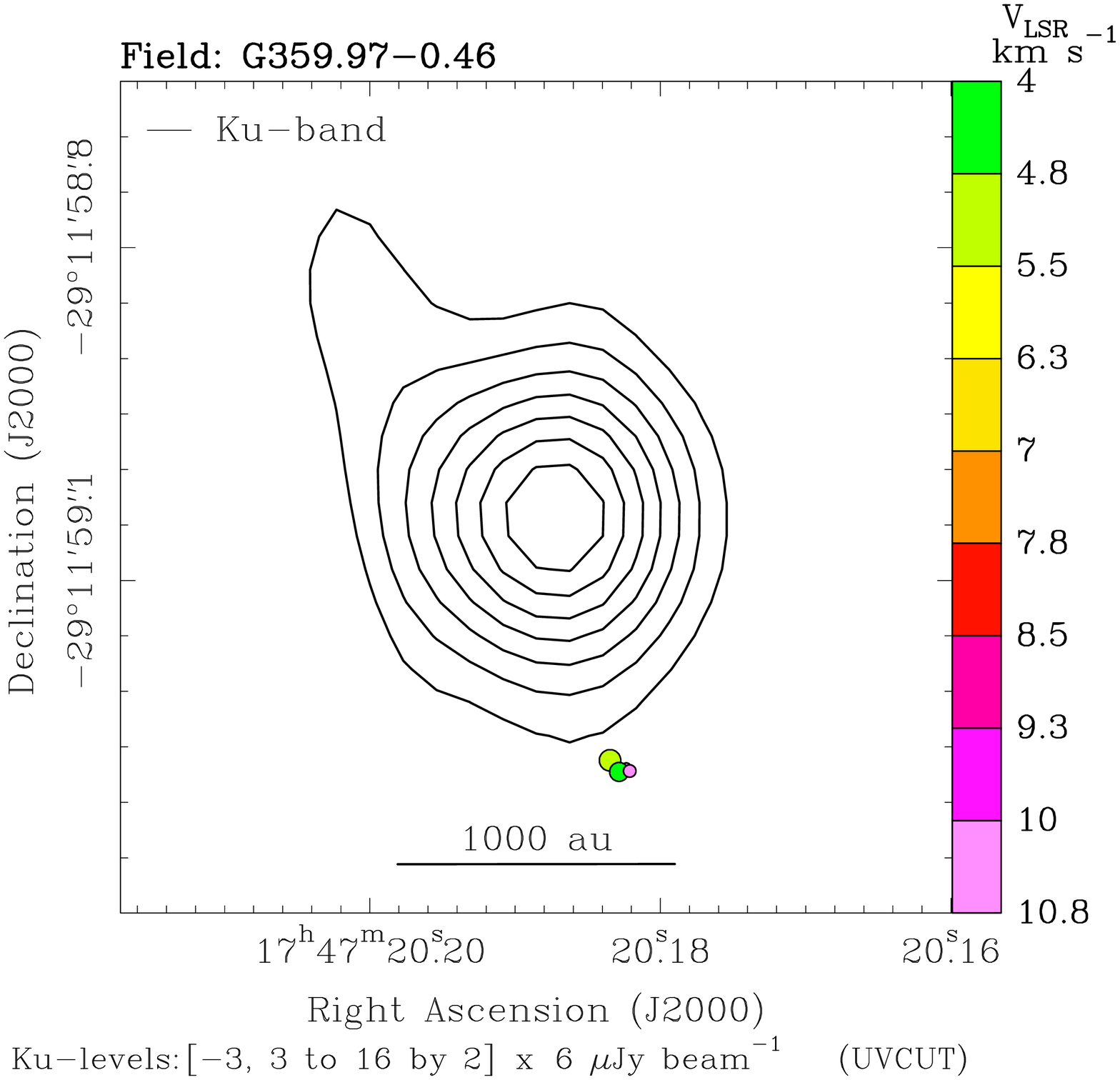}
\caption{Overlay of radio continuum and water maser positions and velocities. Each panel refers to the POETS target indicated in the observed field above the corresponding plot. In each panel, {\it black contours}, and {\it colored dots} and {\it arrows} have the same meaning as in the upper~panel of Fig.~\ref{F:G11.92}. \ {\it Upper~panel:}~The {\it dotted arrows} labeled~LJ denote the axis of the collimated outflow traced in $^{12}$CO emission with the SMA by \citet{Qiu14}, using {\it colors} to distinguish the {\it red}-~and~{\it blue}-shifted flow lobe.}
\label{F:G240.32-G359.97}
\end{figure*}

In this Section, we complete the presentation of the plots overlaying the radio continuum emission with the water maser positions and PMs for all the POETS targets. Figs.~\ref{F:G9.99-G12.43}--\ref{F:G31.58-G49.19}, and Figs.~\ref{F:G76.38-G79.88}--\ref{F:G240.32-G359.97} \ refer to the targets nor included in Mos16, and whose plots have neither entered the main body of this present paper. 
Fig.~\ref{F:G75.76-G75.78} shows the linear \Vlsr\ gradients traced by the water masers in proximity of the YSO in the two sources \ G075.76$+$0.34 \ and \ G075.78$+$0.34, whose large-scale plots were already shown in Mos16. Tables~\ref{G009_99b}-\ref{G359_97b} list the maser parameters (intensity, \Vlsr, absolute position and proper motion) of the 25 new POETS targets presented in this work.

\clearpage

\longtab[1]{ 
\begin{longtable}{rrcrrrr} 
\caption{\label{G009_99b} 22.2~GHz H$_2$O maser parameters for G009.99$-$0.03}\\ 
\hline\hline
\multicolumn{1}{c}{Feature} &  \multicolumn{1}{c}{I$_{\rm peak}$} & \multicolumn{1}{c}{$V_{\rm LSR}$} & \multicolumn{1}{c}{$\Delta~x$} & \multicolumn{1}{c}{$\Delta~y$} & \multicolumn{1}{c}{$V_{x}$} & \multicolumn{1}{c}{$V_{y}$} \\
\multicolumn{1}{c}{Number}  &  \multicolumn{1}{c}{(Jy beam$^{-1}$)} & \multicolumn{1}{c}{(km s$^{-1}$)} & \multicolumn{1}{c}{(mas)} & \multicolumn{1}{c}{(mas)} & \multicolumn{1}{c}{(km s$^{-1}$)} & \multicolumn{1}{c}{(km s$^{-1}$)} \\
\hline
\endfirsthead
\caption{continued.}\\
\hline\hline
\multicolumn{1}{c}{Feature} &  \multicolumn{1}{c}{I$_{\rm peak}$} & \multicolumn{1}{c}{$V_{\rm LSR}$} & \multicolumn{1}{c}{$\Delta~x$} & \multicolumn{1}{c}{$\Delta~y$} & \multicolumn{1}{c}{$V_{x}$} & \multicolumn{1}{c}{$V_{y}$} \\
\multicolumn{1}{c}{Number}  &  \multicolumn{1}{c}{(Jy beam$^{-1}$)} & \multicolumn{1}{c}{(km s$^{-1}$)} & \multicolumn{1}{c}{(mas)} & \multicolumn{1}{c}{(mas)} & \multicolumn{1}{c}{(km s$^{-1}$)} & \multicolumn{1}{c}{(km s$^{-1}$)} \\
\hline
\endhead
\hline
\endfoot
\hline
\endlastfoot
    1 &      8.10 &    52.2 &    $-$63.7$\pm$2.0 &    $-$41.7$\pm$2.0 & ... $ \; \; \; \, $  &  ... $ \; \; \; \, $  \\
    2 &      3.25 &    57.8 &   $-$111.7$\pm$2.0 &    $-$79.0$\pm$2.0 & ... $ \; \; \; \, $  &  ... $ \; \; \; \, $  \\
    3 &      1.41 &    47.0 &   $-$216.9$\pm$2.0 &   $-$229.7$\pm$2.0 & ... $ \; \; \; \, $  &  ... $ \; \; \; \, $  \\
    4 &      0.68 &    42.8 &   $-$210.0$\pm$2.0 &   $-$212.9$\pm$2.0 & ... $ \; \; \; \, $  &  ... $ \; \; \; \, $  \\
    5 &      0.67 &    57.9 &   $-$115.2$\pm$2.0 &    $-$82.1$\pm$2.0 & ... $ \; \; \; \, $  &  ... $ \; \; \; \, $  \\
    6 &      0.61 &    46.6 &    $-$72.2$\pm$2.0 &    $-$18.4$\pm$2.0 & ... $ \; \; \; \, $  &  ... $ \; \; \; \, $  \\
    7 &      0.60 &    48.9 &   $-$256.8$\pm$2.0 &   $-$196.7$\pm$2.0 & ... $ \; \; \; \, $  &  ... $ \; \; \; \, $  \\
    8 &      0.42 &    49.5 &   $-$195.5$\pm$2.0 &   $-$229.4$\pm$2.0 & ... $ \; \; \; \, $  &  ... $ \; \; \; \, $  \\
    9 &      0.34 &    52.4 &   $-$255.7$\pm$2.0 &   $-$220.6$\pm$2.0 & ... $ \; \; \; \, $  &  ... $ \; \; \; \, $  \\
   10 &      0.32 &    60.9 &    $-$91.6$\pm$2.0 &    $-$62.8$\pm$2.0 & ... $ \; \; \; \, $  &  ... $ \; \; \; \, $  \\
   11 &      0.30 &    49.0 &    $-$67.7$\pm$2.0 &    $-$52.9$\pm$2.0 & ... $ \; \; \; \, $  &  ... $ \; \; \; \, $  \\
   12 &      0.23 &    49.2 &   $-$253.9$\pm$2.0 &   $-$211.3$\pm$2.0 & ... $ \; \; \; \, $  &  ... $ \; \; \; \, $  \\
   13 &      0.18 &    48.2 &   $-$254.1$\pm$2.0 &   $-$209.4$\pm$2.0 & ... $ \; \; \; \, $  &  ... $ \; \; \; \, $  \\
   14 &      0.14 &    46.8 &   $-$255.9$\pm$2.0 &   $-$215.8$\pm$2.0 & ... $ \; \; \; \, $  &  ... $ \; \; \; \, $  \\
   15 &      0.13 &    57.0 &    $-$92.3$\pm$2.0 &    $-$64.5$\pm$2.0 & ... $ \; \; \; \, $  &  ... $ \; \; \; \, $  \\
   16 &      0.12 &    53.6 &   $-$254.9$\pm$2.0 &   $-$214.5$\pm$2.0 & ... $ \; \; \; \, $  &  ... $ \; \; \; \, $  \\
\end{longtable}
\tablefoot{
\\
Column~1 gives the feature label number; 
Cols.~2~and~3 provide the intensity of the strongest spot
and the intensity-weighted LSR velocity, respectively, averaged over the
observing epochs; Cols.~4~and~5 give the offsets (with
the associated errors) in absolute position along the R.A. and Dec. axes; Cols.~6~and~7 give the components of the absolute
proper motion (with the associated errors) along the R.A. and Dec. axes. \\
The absolute position to which the position offsets refer is:
R.A.~(J2000) = 18$^{\rm h}$ 7$^{\rm m}$ 50\fs1293,  
Dec~(J2000) = $-$20\degree\ 18$^{\prime}$ 56\farcs404. 
The absolute positions are evaluated at the BeSSeL/VLBA observing epoch: September 20, 2010. 
} 
} 

\clearpage

\longtab[2]{ 
\begin{longtable}{rrcrrrr} 
\caption{\label{G012_43b} 22.2~GHz H$_2$O maser parameters for G012.43$-$1.12}\\ 
\hline\hline
\multicolumn{1}{c}{Feature} &  \multicolumn{1}{c}{I$_{\rm peak}$} & \multicolumn{1}{c}{$V_{\rm LSR}$} & \multicolumn{1}{c}{$\Delta~x$} & \multicolumn{1}{c}{$\Delta~y$} & \multicolumn{1}{c}{$V_{x}$} & \multicolumn{1}{c}{$V_{y}$} \\
\multicolumn{1}{c}{Number}  &  \multicolumn{1}{c}{(Jy beam$^{-1}$)} & \multicolumn{1}{c}{(km s$^{-1}$)} & \multicolumn{1}{c}{(mas)} & \multicolumn{1}{c}{(mas)} & \multicolumn{1}{c}{(km s$^{-1}$)} & \multicolumn{1}{c}{(km s$^{-1}$)} \\
\hline
\endfirsthead
\caption{continued.}\\
\hline\hline
\multicolumn{1}{c}{Feature} &  \multicolumn{1}{c}{I$_{\rm peak}$} & \multicolumn{1}{c}{$V_{\rm LSR}$} & \multicolumn{1}{c}{$\Delta~x$} & \multicolumn{1}{c}{$\Delta~y$} & \multicolumn{1}{c}{$V_{x}$} & \multicolumn{1}{c}{$V_{y}$} \\
\multicolumn{1}{c}{Number}  &  \multicolumn{1}{c}{(Jy beam$^{-1}$)} & \multicolumn{1}{c}{(km s$^{-1}$)} & \multicolumn{1}{c}{(mas)} & \multicolumn{1}{c}{(mas)} & \multicolumn{1}{c}{(km s$^{-1}$)} & \multicolumn{1}{c}{(km s$^{-1}$)} \\
\hline
\endhead
\hline
\endfoot
\hline
\endlastfoot
    1 &     30.00 &   $-$29.4 &      2.9$\pm$2.0 &      9.2$\pm$2.0 & ... $ \; \; \; \, $  &  ... $ \; \; \; \, $  \\
    2 &      0.56 &    $-$7.8 &      3.1$\pm$2.0 &      8.8$\pm$2.0 & ... $ \; \; \; \, $  &  ... $ \; \; \; \, $  \\
    3 &      0.06 &   $-$14.7 &     $-$0.3$\pm$2.0 &     12.0$\pm$2.0 & ... $ \; \; \; \, $  &  ... $ \; \; \; \, $  \\
\end{longtable}
\tablefoot{
\\
Column~1 gives the feature label number; 
Cols.~2~and~3 provide the intensity of the strongest spot
and the intensity-weighted LSR velocity, respectively, averaged over the
observing epochs; Cols.~4~and~5 give the offsets (with
the associated errors) in absolute position along the R.A. and Dec. axes; Cols.~6~and~7 give the components of the absolute
proper motion (with the associated errors) along the R.A. and Dec. axes. \\
The absolute position to which the position offsets refer is:
R.A.~(J2000) = 18$^{\rm h}$ 16$^{\rm m}$ 52\fs1584,  
Dec~(J2000) = $-$18\degree\ 41$^{\prime}$ 43\farcs913. 
The absolute positions are evaluated at the BeSSeL/VLBA observing epoch: March 19, 2011. 
} 
} 

\clearpage

\longtab[3]{ 
\begin{longtable}{rrcrrrr} 
\caption{\label{G012_90b} 22.2~GHz H$_2$O maser parameters for G012.90$-$0.24}\\ 
\hline\hline
\multicolumn{1}{c}{Feature} &  \multicolumn{1}{c}{I$_{\rm peak}$} & \multicolumn{1}{c}{$V_{\rm LSR}$} & \multicolumn{1}{c}{$\Delta~x$} & \multicolumn{1}{c}{$\Delta~y$} & \multicolumn{1}{c}{$V_{x}$} & \multicolumn{1}{c}{$V_{y}$} \\
\multicolumn{1}{c}{Number}  &  \multicolumn{1}{c}{(Jy beam$^{-1}$)} & \multicolumn{1}{c}{(km s$^{-1}$)} & \multicolumn{1}{c}{(mas)} & \multicolumn{1}{c}{(mas)} & \multicolumn{1}{c}{(km s$^{-1}$)} & \multicolumn{1}{c}{(km s$^{-1}$)} \\
\hline
\endfirsthead
\caption{continued.}\\
\hline\hline
\multicolumn{1}{c}{Feature} &  \multicolumn{1}{c}{I$_{\rm peak}$} & \multicolumn{1}{c}{$V_{\rm LSR}$} & \multicolumn{1}{c}{$\Delta~x$} & \multicolumn{1}{c}{$\Delta~y$} & \multicolumn{1}{c}{$V_{x}$} & \multicolumn{1}{c}{$V_{y}$} \\
\multicolumn{1}{c}{Number}  &  \multicolumn{1}{c}{(Jy beam$^{-1}$)} & \multicolumn{1}{c}{(km s$^{-1}$)} & \multicolumn{1}{c}{(mas)} & \multicolumn{1}{c}{(mas)} & \multicolumn{1}{c}{(km s$^{-1}$)} & \multicolumn{1}{c}{(km s$^{-1}$)} \\
\hline
\endhead
\hline
\endfoot
\hline
\endlastfoot
    1 &     22.61 &    34.5 &    191.3$\pm$2.0 &    242.0$\pm$2.0 &    3.5$\pm$4.1 &   42.0$\pm$5.0  \\
    2 &      9.03 &    35.6 &    191.1$\pm$2.0 &    243.3$\pm$2.0 &  $-$13.7$\pm$6.3 &   65.5$\pm$9.0  \\
\end{longtable}
\tablefoot{
\\
Column~1 gives the feature label number; 
Cols.~2~and~3 provide the intensity of the strongest spot
and the intensity-weighted LSR velocity, respectively, averaged over the
observing epochs; Cols.~4~and~5 give the offsets (with
the associated errors) in absolute position along the R.A. and Dec. axes; Cols.~6~and~7 give the components of the absolute
proper motion (with the associated errors) along the R.A. and Dec. axes. \\
The absolute position to which the position offsets refer is:
R.A.~(J2000) = 18$^{\rm h}$ 14$^{\rm m}$ 34\fs416,  
Dec~(J2000) = -17\degree\ 51$^{\prime}$ 52\farcs092. 
The absolute positions are evaluated at the BeSSeL/VLBA observing epoch: December 19, 2010. 
} 
} 

\clearpage

\longtab[4]{ 
\begin{longtable}{rrcrrrr} 
\caption{\label{G012_91b} 22.2~GHz H$_2$O maser parameters for G012.91$-$0.26}\\ 
\hline\hline
\multicolumn{1}{c}{Feature} &  \multicolumn{1}{c}{I$_{\rm peak}$} & \multicolumn{1}{c}{$V_{\rm LSR}$} & \multicolumn{1}{c}{$\Delta~x$} & \multicolumn{1}{c}{$\Delta~y$} & \multicolumn{1}{c}{$V_{x}$} & \multicolumn{1}{c}{$V_{y}$} \\
\multicolumn{1}{c}{Number}  &  \multicolumn{1}{c}{(Jy beam$^{-1}$)} & \multicolumn{1}{c}{(km s$^{-1}$)} & \multicolumn{1}{c}{(mas)} & \multicolumn{1}{c}{(mas)} & \multicolumn{1}{c}{(km s$^{-1}$)} & \multicolumn{1}{c}{(km s$^{-1}$)} \\
\hline
\endfirsthead
\caption{continued.}\\
\hline\hline
\multicolumn{1}{c}{Feature} &  \multicolumn{1}{c}{I$_{\rm peak}$} & \multicolumn{1}{c}{$V_{\rm LSR}$} & \multicolumn{1}{c}{$\Delta~x$} & \multicolumn{1}{c}{$\Delta~y$} & \multicolumn{1}{c}{$V_{x}$} & \multicolumn{1}{c}{$V_{y}$} \\
\multicolumn{1}{c}{Number}  &  \multicolumn{1}{c}{(Jy beam$^{-1}$)} & \multicolumn{1}{c}{(km s$^{-1}$)} & \multicolumn{1}{c}{(mas)} & \multicolumn{1}{c}{(mas)} & \multicolumn{1}{c}{(km s$^{-1}$)} & \multicolumn{1}{c}{(km s$^{-1}$)} \\
\hline
\endhead
\hline
\endfoot
\hline
\endlastfoot
    1 &      1.09 &    37.3 &    $-$57.8$\pm$1.0 &   $-$348.0$\pm$1.5 & ... $ \; \; \; \, $  &  ... $ \; \; \; \, $  \\
    2 &      0.69 &    33.6 &    $-$48.4$\pm$1.0 &   $-$330.0$\pm$1.5 &   79.6$\pm$3.9 &   24.8$\pm$10.6  \\
    3 &      0.52 &    37.8 &    $-$13.1$\pm$1.0 &   $-$336.4$\pm$1.5 & ... $ \; \; \; \, $  &  ... $ \; \; \; \, $  \\
    4 &      0.32 &    30.0 &    $-$48.6$\pm$1.0 &   $-$336.5$\pm$1.5 &   57.1$\pm$8.0 &  $-$52.7$\pm$21.1  \\
    5 &      0.28 &    32.1 &    $-$43.0$\pm$1.0 &   $-$325.8$\pm$1.5 & ... $ \; \; \; \, $  &  ... $ \; \; \; \, $  \\
    6 &      0.24 &    36.5 &    $-$59.4$\pm$1.0 &   $-$348.2$\pm$1.5 & ... $ \; \; \; \, $  &  ... $ \; \; \; \, $  \\
    7 &      0.23 &    37.1 &    $-$56.2$\pm$1.0 &   $-$382.1$\pm$1.5 & ... $ \; \; \; \, $  &  ... $ \; \; \; \, $  \\
    8 &      0.23 &    37.1 &    $-$78.2$\pm$1.0 &   $-$345.8$\pm$1.5 & ... $ \; \; \; \, $  &  ... $ \; \; \; \, $  \\
    9 &      0.23 &    37.0 &    $-$34.0$\pm$1.0 &   $-$361.1$\pm$1.5 & ... $ \; \; \; \, $  &  ... $ \; \; \; \, $  \\
   10 &      0.22 &    37.5 &    $-$49.8$\pm$1.0 &   $-$349.6$\pm$1.5 & ... $ \; \; \; \, $  &  ... $ \; \; \; \, $  \\
   11 &      0.22 &    37.2 &    $-$97.1$\pm$1.0 &   $-$350.7$\pm$1.5 & ... $ \; \; \; \, $  &  ... $ \; \; \; \, $  \\
   12 &      0.21 &    37.1 &    $-$36.2$\pm$1.0 &   $-$323.8$\pm$1.5 & ... $ \; \; \; \, $  &  ... $ \; \; \; \, $  \\
   13 &      0.17 &    36.8 &    $-$28.1$\pm$1.0 &   $-$375.3$\pm$1.5 & ... $ \; \; \; \, $  &  ... $ \; \; \; \, $  \\
   14 &      0.13 &    32.2 &    $-$56.3$\pm$1.0 &   $-$348.0$\pm$1.5 & ... $ \; \; \; \, $  &  ... $ \; \; \; \, $  \\
   15 &      0.12 &    37.1 &    $-$77.5$\pm$1.0 &   $-$346.3$\pm$1.5 & ... $ \; \; \; \, $  &  ... $ \; \; \; \, $  \\
   16 &      0.11 &    29.6 &    $-$42.0$\pm$1.0 &   $-$336.2$\pm$1.5 & ... $ \; \; \; \, $  &  ... $ \; \; \; \, $  \\
   17 &      0.09 &    33.1 &    $-$59.7$\pm$1.0 &   $-$349.3$\pm$1.5 & ... $ \; \; \; \, $  &  ... $ \; \; \; \, $  \\
   18 &      0.08 &    33.0 &    $-$48.8$\pm$1.0 &   $-$347.3$\pm$1.5 & ... $ \; \; \; \, $  &  ... $ \; \; \; \, $  \\
\end{longtable}
\tablefoot{
\\
Column~1 gives the feature label number; 
Cols.~2~and~3 provide the intensity of the strongest spot
and the intensity-weighted LSR velocity, respectively, averaged over the
observing epochs; Cols.~4~and~5 give the offsets (with
the associated errors) in absolute position along the R.A. and Dec. axes; Cols.~6~and~7 give the components of the absolute
proper motion (with the associated errors) along the R.A. and Dec. axes. \\
The absolute position to which the position offsets refer is:
R.A.~(J2000) = 18$^{\rm h}$ 14$^{\rm m}$ 39\fs4400,  
Dec~(J2000) = $-$17\degree\ 52$^{\prime}$ 06\farcs400. 
The absolute positions are evaluated at the BeSSeL/VLBA observing epoch: September 19, 2010. 
} 
} 

\clearpage

\longtab[5]{ 
\begin{longtable}{rrcrrrr} 
\caption{\label{G014_tab} 22.2~GHz H$_2$O maser parameters for G014.64$-$0.58}\\ 
\hline\hline
\multicolumn{1}{c}{Feature} &  \multicolumn{1}{c}{I$_{\rm peak}$} & \multicolumn{1}{c}{$V_{\rm LSR}$} & \multicolumn{1}{c}{$\Delta~x$} & \multicolumn{1}{c}{$\Delta~y$} & \multicolumn{1}{c}{$V_{x}$} & \multicolumn{1}{c}{$V_{y}$} \\
\multicolumn{1}{c}{Number}  &  \multicolumn{1}{c}{(Jy beam$^{-1}$)} & \multicolumn{1}{c}{(km s$^{-1}$)} & \multicolumn{1}{c}{(mas)} & \multicolumn{1}{c}{(mas)} & \multicolumn{1}{c}{(km s$^{-1}$)} & \multicolumn{1}{c}{(km s$^{-1}$)} \\
\hline
\endfirsthead
\caption{continued.}\\
\hline\hline
\multicolumn{1}{c}{Feature} &  \multicolumn{1}{c}{I$_{\rm peak}$} & \multicolumn{1}{c}{$V_{\rm LSR}$} & \multicolumn{1}{c}{$\Delta~x$} & \multicolumn{1}{c}{$\Delta~y$} & \multicolumn{1}{c}{$V_{x}$} & \multicolumn{1}{c}{$V_{y}$} \\
\multicolumn{1}{c}{Number}  &  \multicolumn{1}{c}{(Jy beam$^{-1}$)} & \multicolumn{1}{c}{(km s$^{-1}$)} & \multicolumn{1}{c}{(mas)} & \multicolumn{1}{c}{(mas)} & \multicolumn{1}{c}{(km s$^{-1}$)} & \multicolumn{1}{c}{(km s$^{-1}$)} \\
\hline
\endhead
\hline
\endfoot
\hline
\endlastfoot
    1 &      8.84 &    19.3 &    $-$27.8$\pm$2.0 &    $-$28.8$\pm$2.0 &    2.9$\pm$0.8 &  $-$12.2$\pm$0.8  \\
    2 &      2.09 &    23.4 &      2.5$\pm$2.0 &     $-$1.8$\pm$2.0 &   22.0$\pm$0.7 &   23.2$\pm$0.7  \\
    3 &      0.88 &    19.8 &    $-$35.6$\pm$2.0 &    $-$24.2$\pm$2.0 & ... $ \; \; \; \, $  &  ... $ \; \; \; \, $  \\
    4 &      0.54 &    20.8 &    $-$32.5$\pm$2.0 &    $-$28.7$\pm$2.0 &  $-$19.0$\pm$1.8 &  $-$16.5$\pm$1.9  \\
    5 &      0.49 &    21.7 &    $-$34.7$\pm$2.0 &    $-$31.0$\pm$2.0 & ... $ \; \; \; \, $  &  ... $ \; \; \; \, $  \\
\end{longtable}
\tablefoot{
\\
Column~1 gives the feature label number; 
Cols.~2~and~3 provide the intensity of the strongest spot
and the intensity-weighted LSR velocity, respectively, averaged over the
observing epochs; Cols.~4~and~5 give the offsets (with
the associated errors) in absolute position along the R.A. and Dec. axes; Cols.~6~and~7 give the components of the absolute
proper motion (with the associated errors) along the R.A. and Dec. axes. \\
The absolute position to which the position offsets refer is:
R.A.~(J2000) = 18$^{\rm h}$ 19$^{\rm m}$ 15\fs5426,  
Dec~(J2000) = $-$16\degree\ 29$^{\prime}$ 45\farcs757. 
The absolute positions are evaluated at the BeSSeL/VLBA observing epoch: March 29, 2011. 
} 
} 

\clearpage

\longtab[6]{ 
\begin{longtable}{rrcrrrr} 
\caption{\label{G026_tab} 22.2~GHz H$_2$O maser parameters for G026.42$+$1.69}\\ 
\hline\hline
\multicolumn{1}{c}{Feature} &  \multicolumn{1}{c}{I$_{\rm peak}$} & \multicolumn{1}{c}{$V_{\rm LSR}$} & \multicolumn{1}{c}{$\Delta~x$} & \multicolumn{1}{c}{$\Delta~y$} & \multicolumn{1}{c}{$V_{x}$} & \multicolumn{1}{c}{$V_{y}$} \\
\multicolumn{1}{c}{Number}  &  \multicolumn{1}{c}{(Jy beam$^{-1}$)} & \multicolumn{1}{c}{(km s$^{-1}$)} & \multicolumn{1}{c}{(mas)} & \multicolumn{1}{c}{(mas)} & \multicolumn{1}{c}{(km s$^{-1}$)} & \multicolumn{1}{c}{(km s$^{-1}$)} \\
\hline
\endfirsthead
\caption{continued.}\\
\hline\hline
\multicolumn{1}{c}{Feature} &  \multicolumn{1}{c}{I$_{\rm peak}$} & \multicolumn{1}{c}{$V_{\rm LSR}$} & \multicolumn{1}{c}{$\Delta~x$} & \multicolumn{1}{c}{$\Delta~y$} & \multicolumn{1}{c}{$V_{x}$} & \multicolumn{1}{c}{$V_{y}$} \\
\multicolumn{1}{c}{Number}  &  \multicolumn{1}{c}{(Jy beam$^{-1}$)} & \multicolumn{1}{c}{(km s$^{-1}$)} & \multicolumn{1}{c}{(mas)} & \multicolumn{1}{c}{(mas)} & \multicolumn{1}{c}{(km s$^{-1}$)} & \multicolumn{1}{c}{(km s$^{-1}$)} \\
\hline
\endhead
\hline
\endfoot
\hline
\endlastfoot
    1 &      5.88 &    58.9 &     16.7$\pm$2.0 &    $-$32.0$\pm$2.0 & ... $ \; \; \; \, $  &  ... $ \; \; \; \, $  \\
    2 &      2.13 &    43.5 &     37.8$\pm$2.0 &    $-$21.6$\pm$2.0 & ... $ \; \; \; \, $  &  ... $ \; \; \; \, $  \\
    3 &      0.42 &    43.0 &      8.0$\pm$2.0 &    $-$12.9$\pm$2.0 & ... $ \; \; \; \, $  &  ... $ \; \; \; \, $  \\
    4 &      0.37 &    52.0 &    $-$14.3$\pm$2.0 &     $-$3.3$\pm$2.0 & ... $ \; \; \; \, $  &  ... $ \; \; \; \, $  \\
    5 &      0.19 &    37.4 &      9.4$\pm$2.0 &    $-$11.0$\pm$2.0 & ... $ \; \; \; \, $  &  ... $ \; \; \; \, $  \\
    6 &      0.19 &    45.1 &      8.5$\pm$2.0 &    $-$13.6$\pm$2.0 & ... $ \; \; \; \, $  &  ... $ \; \; \; \, $  \\
    7 &      0.09 &    53.9 &     16.5$\pm$2.0 &    $-$31.6$\pm$2.0 & ... $ \; \; \; \, $  &  ... $ \; \; \; \, $  \\
\end{longtable}
\tablefoot{
\\
Column~1 gives the feature label number; 
Cols.~2~and~3 provide the intensity of the strongest spot
and the intensity-weighted LSR velocity, respectively, averaged over the
observing epochs; Cols.~4~and~5 give the offsets (with
the associated errors) in absolute position along the R.A. and Dec. axes; Cols.~6~and~7 give the components of the absolute
proper motion (with the associated errors) along the R.A. and Dec. axes. \\
The absolute position to which the position offsets refer is:
R.A.~(J2000) = 18$^{\rm h}$ 33$^{\rm m}$ 30\fs5091,  
Dec~(J2000) = $-$05\degree\ 1$^{\prime}$ 02\farcs002. 
The absolute positions are evaluated at the BeSSeL/VLBA observing epoch: June 27, 2011. 
} 
} 

\clearpage

\longtab[7]{ 
\begin{longtable}{rrcrrrr} 
\caption{\label{G031_tab} 22.2~GHz H$_2$O maser parameters for G031.58$+$0.08}\\ 
\hline\hline
\multicolumn{1}{c}{Feature} &  \multicolumn{1}{c}{I$_{\rm peak}$} & \multicolumn{1}{c}{$V_{\rm LSR}$} & \multicolumn{1}{c}{$\Delta~x$} & \multicolumn{1}{c}{$\Delta~y$} & \multicolumn{1}{c}{$V_{x}$} & \multicolumn{1}{c}{$V_{y}$} \\
\multicolumn{1}{c}{Number}  &  \multicolumn{1}{c}{(Jy beam$^{-1}$)} & \multicolumn{1}{c}{(km s$^{-1}$)} & \multicolumn{1}{c}{(mas)} & \multicolumn{1}{c}{(mas)} & \multicolumn{1}{c}{(km s$^{-1}$)} & \multicolumn{1}{c}{(km s$^{-1}$)} \\
\hline
\endfirsthead
\caption{continued.}\\
\hline\hline
\multicolumn{1}{c}{Feature} &  \multicolumn{1}{c}{I$_{\rm peak}$} & \multicolumn{1}{c}{$V_{\rm LSR}$} & \multicolumn{1}{c}{$\Delta~x$} & \multicolumn{1}{c}{$\Delta~y$} & \multicolumn{1}{c}{$V_{x}$} & \multicolumn{1}{c}{$V_{y}$} \\
\multicolumn{1}{c}{Number}  &  \multicolumn{1}{c}{(Jy beam$^{-1}$)} & \multicolumn{1}{c}{(km s$^{-1}$)} & \multicolumn{1}{c}{(mas)} & \multicolumn{1}{c}{(mas)} & \multicolumn{1}{c}{(km s$^{-1}$)} & \multicolumn{1}{c}{(km s$^{-1}$)} \\
\hline
\endhead
\hline
\endfoot
\hline
\endlastfoot
    1 &     12.22 &    99.4 &      0.5$\pm$1.0 &      0.5$\pm$1.0 & ... $ \; \; \; \, $  &  ... $ \; \; \; \, $  \\
    2 &      6.72 &    99.0 &     $-$4.3$\pm$1.0 &     24.1$\pm$1.0 & ... $ \; \; \; \, $  &  ... $ \; \; \; \, $  \\
    3 &      6.61 &    99.6 &     $-$2.1$\pm$1.0 &     $-$6.2$\pm$1.0 &    2.7$\pm$6.4 &  $-$32.0$\pm$6.5  \\
    4 &      1.40 &    98.1 &      3.2$\pm$1.0 &     20.8$\pm$1.0 & ... $ \; \; \; \, $  &  ... $ \; \; \; \, $  \\
    5 &      1.23 &    99.3 &      4.7$\pm$1.0 &      3.5$\pm$1.0 & ... $ \; \; \; \, $  &  ... $ \; \; \; \, $  \\
    6 &      1.11 &    98.4 &      4.7$\pm$1.0 &     19.6$\pm$1.0 &   23.9$\pm$6.5 &  $-$20.8$\pm$6.5  \\
    7 &      0.61 &    98.7 &     $-$0.3$\pm$1.0 &     22.8$\pm$1.0 &   31.7$\pm$6.6 &  $-$23.0$\pm$6.8  \\
    8 &      0.22 &    92.0 &    $-$31.3$\pm$1.0 &    $-$29.8$\pm$1.0 & ... $ \; \; \; \, $  &  ... $ \; \; \; \, $  \\
    9 &      0.17 &    91.8 &    $-$27.8$\pm$1.0 &    $-$95.4$\pm$1.0 & ... $ \; \; \; \, $  &  ... $ \; \; \; \, $  \\
   10 &      0.15 &    97.2 &      0.7$\pm$1.0 &     21.7$\pm$1.0 & ... $ \; \; \; \, $  &  ... $ \; \; \; \, $  \\
   11 &      0.12 &    98.1 &     $-$0.8$\pm$1.0 &     20.1$\pm$1.0 & ... $ \; \; \; \, $  &  ... $ \; \; \; \, $  \\
\end{longtable}
\tablefoot{
\\
Column~1 gives the feature label number; 
Cols.~2~and~3 provide the intensity of the strongest spot
and the intensity-weighted LSR velocity, respectively, averaged over the
observing epochs; Cols.~4~and~5 give the offsets (with
the associated errors) in absolute position along the R.A. and Dec. axes; Cols.~6~and~7 give the components of the absolute
proper motion (with the associated errors) along the R.A. and Dec. axes. \\
The absolute position to which the position offsets refer is:
R.A.~(J2000) = 18$^{\rm h}$ 48$^{\rm m}$ 41\fs6740,  
Dec~(J2000) = $-$01\degree\ 9$^{\prime}$ 59\farcs775. 
The absolute positions are evaluated at the BeSSeL/VLBA observing epoch: April 3, 2011. 
} 
} 

\clearpage

\longtab[8]{ 
\begin{longtable}{rrcrrrr} 
\caption{\label{G035_tab} 22.2~GHz H$_2$O maser parameters for G035.02$+$0.35}\\ 
\hline\hline
\multicolumn{1}{c}{Feature} &  \multicolumn{1}{c}{I$_{\rm peak}$} & \multicolumn{1}{c}{$V_{\rm LSR}$} & \multicolumn{1}{c}{$\Delta~x$} & \multicolumn{1}{c}{$\Delta~y$} & \multicolumn{1}{c}{$V_{x}$} & \multicolumn{1}{c}{$V_{y}$} \\
\multicolumn{1}{c}{Number}  &  \multicolumn{1}{c}{(Jy beam$^{-1}$)} & \multicolumn{1}{c}{(km s$^{-1}$)} & \multicolumn{1}{c}{(mas)} & \multicolumn{1}{c}{(mas)} & \multicolumn{1}{c}{(km s$^{-1}$)} & \multicolumn{1}{c}{(km s$^{-1}$)} \\
\hline
\endfirsthead
\caption{continued.}\\
\hline\hline
\multicolumn{1}{c}{Feature} &  \multicolumn{1}{c}{I$_{\rm peak}$} & \multicolumn{1}{c}{$V_{\rm LSR}$} & \multicolumn{1}{c}{$\Delta~x$} & \multicolumn{1}{c}{$\Delta~y$} & \multicolumn{1}{c}{$V_{x}$} & \multicolumn{1}{c}{$V_{y}$} \\
\multicolumn{1}{c}{Number}  &  \multicolumn{1}{c}{(Jy beam$^{-1}$)} & \multicolumn{1}{c}{(km s$^{-1}$)} & \multicolumn{1}{c}{(mas)} & \multicolumn{1}{c}{(mas)} & \multicolumn{1}{c}{(km s$^{-1}$)} & \multicolumn{1}{c}{(km s$^{-1}$)} \\
\hline
\endhead
\hline
\endfoot
\hline
\endlastfoot
    1 &     12.03 &    44.6 &   $-$214.5$\pm$2.0 &     10.0$\pm$2.0 &    7.7$\pm$1.3 &   $-$0.2$\pm$1.4  \\
    2 &      6.32 &    56.6 &   $-$399.9$\pm$2.0 &    207.5$\pm$2.0 &  $-$26.9$\pm$1.2 &   $-$6.0$\pm$1.4  \\
    3 &      4.65 &    45.2 &   $-$211.2$\pm$2.0 &     $-$4.9$\pm$2.0 &   13.2$\pm$1.0 &   $-$5.3$\pm$2.4  \\
    4 &      4.11 &    56.6 &   $-$321.4$\pm$2.0 &    140.2$\pm$2.0 &    5.0$\pm$1.6 &  $-$27.9$\pm$1.8  \\
    5 &      3.20 &    58.8 &   $-$392.0$\pm$2.0 &    176.5$\pm$2.0 &   $-$8.8$\pm$1.9 &   $-$7.3$\pm$2.1  \\
    6 &      3.04 &    44.7 &   $-$276.2$\pm$2.0 &    $-$13.2$\pm$2.0 & ... $ \; \; \; \, $  &  ... $ \; \; \; \, $  \\
    7 &      3.01 &    59.3 &   $-$392.7$\pm$2.0 &    177.1$\pm$2.0 &  $-$11.6$\pm$1.4 &   $-$6.1$\pm$1.3  \\
    8 &      2.67 &    44.4 &   $-$211.1$\pm$2.0 &     $-$0.2$\pm$2.0 &   16.1$\pm$2.1 &  $-$12.4$\pm$4.2  \\
    9 &      2.64 &    45.1 &   $-$272.7$\pm$2.0 &    $-$17.2$\pm$2.0 &   $-$5.5$\pm$1.9 &  $-$12.2$\pm$2.2  \\
   10 &      2.63 &    40.5 &   $-$270.7$\pm$2.0 &    $-$11.9$\pm$2.0 & ... $ \; \; \; \, $  &  ... $ \; \; \; \, $  \\
   11 &      2.01 &    45.0 &   $-$276.6$\pm$2.0 &    $-$16.0$\pm$2.0 &   $-$6.8$\pm$1.2 &   $-$9.4$\pm$1.1  \\
   12 &      1.83 &    41.7 &   $-$270.1$\pm$2.0 &    $-$12.8$\pm$2.0 & ... $ \; \; \; \, $  &  ... $ \; \; \; \, $  \\
   13 &      1.60 &    57.7 &   $-$391.0$\pm$2.0 &    170.9$\pm$2.0 &   $-$8.6$\pm$1.3 &   $-$9.8$\pm$1.5  \\
   14 &      1.50 &    45.4 &   $-$210.9$\pm$2.0 &     $-$2.5$\pm$2.0 &   13.4$\pm$1.6 &    2.5$\pm$1.9  \\
   15 &      1.08 &    55.3 &   $-$369.0$\pm$2.0 &    219.6$\pm$2.0 &   10.6$\pm$2.3 &    4.7$\pm$3.0  \\
   16 &      0.95 &    45.6 &   $-$282.2$\pm$2.0 &     $-$3.7$\pm$2.0 & ... $ \; \; \; \, $  &  ... $ \; \; \; \, $  \\
   17 &      0.76 &    56.6 &   $-$399.0$\pm$2.0 &    210.3$\pm$2.0 &  $-$15.4$\pm$1.1 &    1.5$\pm$1.3  \\
   18 &      0.51 &    54.7 &   $-$322.9$\pm$2.0 &    139.3$\pm$2.0 &   $-$1.4$\pm$1.2 &  $-$30.0$\pm$1.3  \\
   19 &      0.39 &    40.9 &   $-$272.6$\pm$2.0 &     $-$6.9$\pm$2.0 & ... $ \; \; \; \, $  &  ... $ \; \; \; \, $  \\
   20 &      0.33 &    55.7 &   $-$377.6$\pm$2.0 &    223.1$\pm$2.0 &    2.5$\pm$1.1 &   11.7$\pm$1.2  \\
   21 &      0.25 &    53.9 &   $-$402.1$\pm$2.0 &    203.1$\pm$2.0 & ... $ \; \; \; \, $  &  ... $ \; \; \; \, $  \\
   22 &      0.25 &    46.2 &   $-$238.4$\pm$2.0 &    $-$34.9$\pm$2.0 & ... $ \; \; \; \, $  &  ... $ \; \; \; \, $  \\
   23 &      0.22 &    44.1 &   $-$270.6$\pm$2.0 &    $-$19.7$\pm$2.0 & ... $ \; \; \; \, $  &  ... $ \; \; \; \, $  \\
   24 &      0.20 &    59.1 &   $-$393.2$\pm$2.0 &    178.1$\pm$2.0 & ... $ \; \; \; \, $  &  ... $ \; \; \; \, $  \\
   25 &      0.15 &    57.5 &   $-$390.3$\pm$2.0 &    170.0$\pm$2.0 & ... $ \; \; \; \, $  &  ... $ \; \; \; \, $  \\
   26 &      0.06 &    54.6 &   $-$388.2$\pm$2.0 &    166.6$\pm$2.0 & ... $ \; \; \; \, $  &  ... $ \; \; \; \, $  \\
   27 &      0.06 &    57.6 &   $-$401.0$\pm$2.0 &    207.9$\pm$2.0 & ... $ \; \; \; \, $  &  ... $ \; \; \; \, $  \\
\end{longtable}
\tablefoot{
\\
Column~1 gives the feature label number; 
Cols.~2~and~3 provide the intensity of the strongest spot
and the intensity-weighted LSR velocity, respectively, averaged over the
observing epochs; Cols.~4~and~5 give the offsets (with
the associated errors) in absolute position along the R.A. and Dec. axes; Cols.~6~and~7 give the components of the absolute
proper motion (with the associated errors) along the R.A. and Dec. axes. \\
The absolute position to which the position offsets refer is:
R.A.~(J2000) = 18$^{\rm h}$ 54$^{\rm m}$ 00\fs6720,  
Dec~(J2000) = 02\degree\ 1$^{\prime}$ 19\farcs200. 
The absolute positions are evaluated at the BeSSeL/VLBA observing epoch: April 15, 2011. 
} 
} 

\clearpage

\longtab[9]{ 
\begin{longtable}{rrcrrrr} 
\caption{\label{G049_tab} 22.2~GHz H$_2$O maser parameters for G049.19$-$0.34}\\ 
\hline\hline
\multicolumn{1}{c}{Feature} &  \multicolumn{1}{c}{I$_{\rm peak}$} & \multicolumn{1}{c}{$V_{\rm LSR}$} & \multicolumn{1}{c}{$\Delta~x$} & \multicolumn{1}{c}{$\Delta~y$} & \multicolumn{1}{c}{$V_{x}$} & \multicolumn{1}{c}{$V_{y}$} \\
\multicolumn{1}{c}{Number}  &  \multicolumn{1}{c}{(Jy beam$^{-1}$)} & \multicolumn{1}{c}{(km s$^{-1}$)} & \multicolumn{1}{c}{(mas)} & \multicolumn{1}{c}{(mas)} & \multicolumn{1}{c}{(km s$^{-1}$)} & \multicolumn{1}{c}{(km s$^{-1}$)} \\
\hline
\endfirsthead
\caption{continued.}\\
\hline\hline
\multicolumn{1}{c}{Feature} &  \multicolumn{1}{c}{I$_{\rm peak}$} & \multicolumn{1}{c}{$V_{\rm LSR}$} & \multicolumn{1}{c}{$\Delta~x$} & \multicolumn{1}{c}{$\Delta~y$} & \multicolumn{1}{c}{$V_{x}$} & \multicolumn{1}{c}{$V_{y}$} \\
\multicolumn{1}{c}{Number}  &  \multicolumn{1}{c}{(Jy beam$^{-1}$)} & \multicolumn{1}{c}{(km s$^{-1}$)} & \multicolumn{1}{c}{(mas)} & \multicolumn{1}{c}{(mas)} & \multicolumn{1}{c}{(km s$^{-1}$)} & \multicolumn{1}{c}{(km s$^{-1}$)} \\
\hline
\endhead
\hline
\endfoot
\hline
\endlastfoot
    1 &     55.17 &    69.1 &     $-$1.0$\pm$1.0 &    $-$37.1$\pm$1.0 &  $-$21.2$\pm$1.8 &   21.0$\pm$2.0  \\
    2 &     31.18 &    68.6 &     $-$2.0$\pm$1.0 &    $-$41.6$\pm$1.0 &  $-$27.8$\pm$1.8 &  $-$32.7$\pm$5.4  \\
    3 &     27.25 &    70.9 &     $-$0.5$\pm$1.0 &    $-$36.7$\pm$1.0 &  $-$28.0$\pm$3.3 &   15.4$\pm$3.0  \\
    4 &     12.88 &    68.3 &     $-$1.3$\pm$1.0 &    $-$45.9$\pm$1.0 &    0.0$\pm$3.9 &  $-$30.4$\pm$5.5  \\
    5 &     10.12 &    69.4 &     $-$1.5$\pm$1.0 &    $-$39.8$\pm$1.0 &  $-$47.9$\pm$3.4 &  $-$63.2$\pm$9.0  \\
    6 &      5.91 &    70.6 &      2.4$\pm$1.0 &    $-$31.7$\pm$1.0 &   $-$9.5$\pm$3.5 &   21.5$\pm$1.8  \\
    7 &      1.80 &    85.1 &      4.7$\pm$1.0 &    $-$33.3$\pm$1.0 & ... $ \; \; \; \, $  &  ... $ \; \; \; \, $  \\
    8 &      1.26 &    56.0 &     53.9$\pm$1.0 &    $-$54.5$\pm$1.0 &   $-$1.4$\pm$1.2 &  $-$12.5$\pm$1.1  \\
    9 &      0.91 &    69.3 &     14.1$\pm$1.0 &    $-$36.4$\pm$1.0 & ... $ \; \; \; \, $  &  ... $ \; \; \; \, $  \\
   10 &      0.80 &    68.6 &      7.0$\pm$1.0 &    $-$34.3$\pm$1.0 & ... $ \; \; \; \, $  &  ... $ \; \; \; \, $  \\
   11 &      0.69 &    61.6 &     54.8$\pm$1.0 &    $-$56.4$\pm$1.0 &   $-$2.0$\pm$2.8 &  $-$13.1$\pm$4.0  \\
\end{longtable}
\tablefoot{
\\
Column~1 gives the feature label number; 
Cols.~2~and~3 provide the intensity of the strongest spot
and the intensity-weighted LSR velocity, respectively, averaged over the
observing epochs; Cols.~4~and~5 give the offsets (with
the associated errors) in absolute position along the R.A. and Dec. axes; Cols.~6~and~7 give the components of the absolute
proper motion (with the associated errors) along the R.A. and Dec. axes. \\
The absolute position to which the position offsets refer is:
R.A.~(J2000) = 19$^{\rm h}$ 22$^{\rm m}$ 57\fs770,  
Dec~(J2000) = 14\degree\ 16$^{\prime}$ 09\farcs984. 
The absolute positions are evaluated at the BeSSeL/VLBA observing epoch: June 27, 2011. 
} 
} 

\clearpage

\longtab[10]{ 
\begin{longtable}{rrcrrrr} 
\caption{\label{G076_tab} 22.2~GHz H$_2$O maser parameters for G076.38$-$0.62}\\ 
\hline\hline
\multicolumn{1}{c}{Feature} &  \multicolumn{1}{c}{I$_{\rm peak}$} & \multicolumn{1}{c}{$V_{\rm LSR}$} & \multicolumn{1}{c}{$\Delta~x$} & \multicolumn{1}{c}{$\Delta~y$} & \multicolumn{1}{c}{$V_{x}$} & \multicolumn{1}{c}{$V_{y}$} \\
\multicolumn{1}{c}{Number}  &  \multicolumn{1}{c}{(Jy beam$^{-1}$)} & \multicolumn{1}{c}{(km s$^{-1}$)} & \multicolumn{1}{c}{(mas)} & \multicolumn{1}{c}{(mas)} & \multicolumn{1}{c}{(km s$^{-1}$)} & \multicolumn{1}{c}{(km s$^{-1}$)} \\
\hline
\endfirsthead
\caption{continued.}\\
\hline\hline
\multicolumn{1}{c}{Feature} &  \multicolumn{1}{c}{I$_{\rm peak}$} & \multicolumn{1}{c}{$V_{\rm LSR}$} & \multicolumn{1}{c}{$\Delta~x$} & \multicolumn{1}{c}{$\Delta~y$} & \multicolumn{1}{c}{$V_{x}$} & \multicolumn{1}{c}{$V_{y}$} \\
\multicolumn{1}{c}{Number}  &  \multicolumn{1}{c}{(Jy beam$^{-1}$)} & \multicolumn{1}{c}{(km s$^{-1}$)} & \multicolumn{1}{c}{(mas)} & \multicolumn{1}{c}{(mas)} & \multicolumn{1}{c}{(km s$^{-1}$)} & \multicolumn{1}{c}{(km s$^{-1}$)} \\
\hline
\endhead
\hline
\endfoot
\hline
\endlastfoot
    1 &     15.56 &   $-$12.6 &    $-$77.3$\pm$2.0 &     17.2$\pm$2.0 &  $-$32.9$\pm$2.0 &  $-$51.3$\pm$2.1  \\
    2 &      5.88 &   $-$13.1 &    $-$77.4$\pm$2.0 &     16.1$\pm$2.0 & ... $ \; \; \; \, $  &  ... $ \; \; \; \, $  \\
    3 &      2.41 &   $-$13.4 &    $-$75.4$\pm$2.0 &     15.5$\pm$2.0 & ... $ \; \; \; \, $  &  ... $ \; \; \; \, $  \\
    4 &      2.05 &    $-$8.4 &    $-$81.5$\pm$2.0 &      2.4$\pm$2.0 &  $-$43.6$\pm$3.7 &  $-$24.3$\pm$3.8  \\
    5 &      2.05 &    $-$3.2 &   $-$104.7$\pm$2.0 &      5.1$\pm$2.0 & ... $ \; \; \; \, $  &  ... $ \; \; \; \, $  \\
    6 &      1.89 &   $-$14.6 &    $-$78.9$\pm$2.0 &     17.5$\pm$2.0 &  $-$38.6$\pm$0.9 &  $-$23.8$\pm$1.3  \\
    7 &      1.83 &   $-$11.4 &    $-$75.2$\pm$2.0 &     15.9$\pm$2.0 & ... $ \; \; \; \, $  &  ... $ \; \; \; \, $  \\
    8 &      1.60 &   $-$11.4 &    $-$75.9$\pm$2.0 &     15.6$\pm$2.0 &  $-$51.4$\pm$2.5 &  $-$46.8$\pm$2.2  \\
    9 &      1.55 &    $-$9.6 &    $-$80.1$\pm$2.0 &     16.7$\pm$2.0 & ... $ \; \; \; \, $  &  ... $ \; \; \; \, $  \\
   10 &      1.14 &   $-$13.3 &    $-$79.7$\pm$2.0 &     15.4$\pm$2.0 & ... $ \; \; \; \, $  &  ... $ \; \; \; \, $  \\
   11 &      1.00 &    $-$7.9 &    $-$82.9$\pm$2.0 &      0.7$\pm$2.0 & ... $ \; \; \; \, $  &  ... $ \; \; \; \, $  \\
   12 &      0.97 &   $-$10.9 &    $-$76.6$\pm$2.0 &     15.9$\pm$2.0 & ... $ \; \; \; \, $  &  ... $ \; \; \; \, $  \\
   13 &      0.90 &   $-$14.0 &    $-$76.3$\pm$2.0 &     16.0$\pm$2.0 & ... $ \; \; \; \, $  &  ... $ \; \; \; \, $  \\
   14 &      0.90 &     6.2 &     $-$1.9$\pm$2.0 &     39.9$\pm$2.0 &   43.0$\pm$1.6 &   31.5$\pm$0.9  \\
   15 &      0.78 &   $-$11.7 &    $-$77.5$\pm$2.0 &     19.1$\pm$2.0 & ... $ \; \; \; \, $  &  ... $ \; \; \; \, $  \\
   16 &      0.56 &    $-$7.9 &    $-$76.4$\pm$2.0 &     16.9$\pm$2.0 & ... $ \; \; \; \, $  &  ... $ \; \; \; \, $  \\
   17 &      0.47 &   $-$13.9 &    $-$81.1$\pm$2.0 &     16.5$\pm$2.0 & ... $ \; \; \; \, $  &  ... $ \; \; \; \, $  \\
   18 &      0.47 &   $-$11.3 &    $-$85.4$\pm$2.0 &     13.0$\pm$2.0 & ... $ \; \; \; \, $  &  ... $ \; \; \; \, $  \\
   19 &      0.47 &    $-$4.3 &   $-$102.8$\pm$2.0 &      5.8$\pm$2.0 &  $-$35.3$\pm$1.8 &  $-$13.4$\pm$1.9  \\
   20 &      0.45 &    $-$7.9 &    $-$85.8$\pm$2.0 &     10.8$\pm$2.0 & ... $ \; \; \; \, $  &  ... $ \; \; \; \, $  \\
   21 &      0.45 &   $-$10.3 &    $-$76.8$\pm$2.0 &     15.8$\pm$2.0 & ... $ \; \; \; \, $  &  ... $ \; \; \; \, $  \\
   22 &      0.44 &   $-$11.4 &    $-$75.4$\pm$2.0 &     12.1$\pm$2.0 & ... $ \; \; \; \, $  &  ... $ \; \; \; \, $  \\
   23 &      0.43 &     6.8 &     $-$1.3$\pm$2.0 &     39.8$\pm$2.0 &   46.6$\pm$1.0 &   32.9$\pm$1.0  \\
   24 &      0.43 &    $-$8.9 &    $-$77.0$\pm$2.0 &     11.4$\pm$2.0 & ... $ \; \; \; \, $  &  ... $ \; \; \; \, $  \\
   25 &      0.37 &   $-$11.2 &    $-$77.9$\pm$2.0 &     15.1$\pm$2.0 & ... $ \; \; \; \, $  &  ... $ \; \; \; \, $  \\
   26 &      0.34 &   $-$14.9 &    $-$78.1$\pm$2.0 &     17.9$\pm$2.0 & ... $ \; \; \; \, $  &  ... $ \; \; \; \, $  \\
   27 &      0.33 &     8.9 &    $-$10.3$\pm$2.0 &     24.9$\pm$2.0 & ... $ \; \; \; \, $  &  ... $ \; \; \; \, $  \\
   28 &      0.33 &   $-$13.1 &    $-$79.4$\pm$2.0 &     18.0$\pm$2.0 & ... $ \; \; \; \, $  &  ... $ \; \; \; \, $  \\
   29 &      0.31 &     5.4 &     $-$9.8$\pm$2.0 &     41.9$\pm$2.0 & ... $ \; \; \; \, $  &  ... $ \; \; \; \, $  \\
   30 &      0.31 &   $-$10.0 &    $-$96.0$\pm$2.0 &     12.0$\pm$2.0 & ... $ \; \; \; \, $  &  ... $ \; \; \; \, $  \\
   31 &      0.30 &    $-$7.1 &    $-$83.9$\pm$2.0 &     10.6$\pm$2.0 &  $-$38.4$\pm$1.7 &  $-$26.1$\pm$3.3  \\
   32 &      0.29 &    $-$9.1 &    $-$78.3$\pm$2.0 &     13.9$\pm$2.0 & ... $ \; \; \; \, $  &  ... $ \; \; \; \, $  \\
   33 &      0.29 &   $-$12.2 &    $-$85.6$\pm$2.0 &     16.0$\pm$2.0 & ... $ \; \; \; \, $  &  ... $ \; \; \; \, $  \\
   34 &      0.29 &    $-$7.8 &    $-$87.2$\pm$2.0 &     12.2$\pm$2.0 & ... $ \; \; \; \, $  &  ... $ \; \; \; \, $  \\
   35 &      0.28 &     0.2 &    $-$86.9$\pm$2.0 &     13.2$\pm$2.0 & ... $ \; \; \; \, $  &  ... $ \; \; \; \, $  \\
   36 &      0.28 &    $-$8.4 &    $-$78.1$\pm$2.0 &     16.5$\pm$2.0 & ... $ \; \; \; \, $  &  ... $ \; \; \; \, $  \\
   37 &      0.28 &    $-$4.1 &    $-$90.9$\pm$2.0 &      9.4$\pm$2.0 & ... $ \; \; \; \, $  &  ... $ \; \; \; \, $  \\
   38 &      0.24 &   $-$11.8 &    $-$77.4$\pm$2.0 &     13.6$\pm$2.0 & ... $ \; \; \; \, $  &  ... $ \; \; \; \, $  \\
   39 &      0.21 &     6.3 &     $-$2.3$\pm$2.0 &     23.0$\pm$2.0 & ... $ \; \; \; \, $  &  ... $ \; \; \; \, $  \\
   40 &      0.20 &    $-$8.1 &    $-$75.7$\pm$2.0 &     23.2$\pm$2.0 & ... $ \; \; \; \, $  &  ... $ \; \; \; \, $  \\
   41 &      0.20 &    $-$9.6 &    $-$84.8$\pm$2.0 &     12.8$\pm$2.0 & ... $ \; \; \; \, $  &  ... $ \; \; \; \, $  \\
   42 &      0.19 &     7.1 &      0.5$\pm$2.0 &     41.1$\pm$2.0 & ... $ \; \; \; \, $  &  ... $ \; \; \; \, $  \\
   43 &      0.19 &    $-$8.5 &    $-$82.6$\pm$2.0 &      2.3$\pm$2.0 & ... $ \; \; \; \, $  &  ... $ \; \; \; \, $  \\
   44 &      0.19 &   $-$11.2 &    $-$86.0$\pm$2.0 &     14.4$\pm$2.0 & ... $ \; \; \; \, $  &  ... $ \; \; \; \, $  \\
   45 &      0.16 &    $-$7.6 &    $-$87.2$\pm$2.0 &      7.5$\pm$2.0 & ... $ \; \; \; \, $  &  ... $ \; \; \; \, $  \\
   46 &      0.15 &    $-$8.9 &    $-$79.5$\pm$2.0 &     15.2$\pm$2.0 & ... $ \; \; \; \, $  &  ... $ \; \; \; \, $  \\
   47 &      0.14 &    $-$4.1 &    $-$87.5$\pm$2.0 &     13.0$\pm$2.0 & ... $ \; \; \; \, $  &  ... $ \; \; \; \, $  \\
   48 &      0.13 &     3.4 &    $-$14.8$\pm$2.0 &     43.0$\pm$2.0 & ... $ \; \; \; \, $  &  ... $ \; \; \; \, $  \\
   49 &      0.08 &     7.9 &     $-$5.5$\pm$2.0 &     33.6$\pm$2.0 & ... $ \; \; \; \, $  &  ... $ \; \; \; \, $  \\
   50 &      0.08 &     8.0 &    $-$16.3$\pm$2.0 &     30.3$\pm$2.0 & ... $ \; \; \; \, $  &  ... $ \; \; \; \, $  \\
\end{longtable}
\tablefoot{
\\
Column~1 gives the feature label number; 
Cols.~2~and~3 provide the intensity of the strongest spot
and the intensity-weighted LSR velocity, respectively, averaged over the
observing epochs; Cols.~4~and~5 give the offsets (with
the associated errors) in absolute position along the R.A. and Dec. axes; Cols.~6~and~7 give the components of the absolute
proper motion (with the associated errors) along the R.A. and Dec. axes. \\
The absolute position to which the position offsets refer is:
R.A.~(J2000) = 20$^{\rm h}$ 27$^{\rm m}$ 25\fs4816,  
Dec~(J2000) = 37\degree\ 22$^{\prime}$ 48\farcs444. 
The absolute positions are evaluated at the BeSSeL/VLBA observing epoch: April 24, 2010. 
} 
} 

\clearpage

\longtab[11]{ 
\begin{longtable}{rrcrrrr} 
\caption{\label{G079_tab} 22.2~GHz H$_2$O maser parameters for G079.88$+$1.18}\\ 
\hline\hline
\multicolumn{1}{c}{Feature} &  \multicolumn{1}{c}{I$_{\rm peak}$} & \multicolumn{1}{c}{$V_{\rm LSR}$} & \multicolumn{1}{c}{$\Delta~x$} & \multicolumn{1}{c}{$\Delta~y$} & \multicolumn{1}{c}{$V_{x}$} & \multicolumn{1}{c}{$V_{y}$} \\
\multicolumn{1}{c}{Number}  &  \multicolumn{1}{c}{(Jy beam$^{-1}$)} & \multicolumn{1}{c}{(km s$^{-1}$)} & \multicolumn{1}{c}{(mas)} & \multicolumn{1}{c}{(mas)} & \multicolumn{1}{c}{(km s$^{-1}$)} & \multicolumn{1}{c}{(km s$^{-1}$)} \\
\hline
\endfirsthead
\caption{continued.}\\
\hline\hline
\multicolumn{1}{c}{Feature} &  \multicolumn{1}{c}{I$_{\rm peak}$} & \multicolumn{1}{c}{$V_{\rm LSR}$} & \multicolumn{1}{c}{$\Delta~x$} & \multicolumn{1}{c}{$\Delta~y$} & \multicolumn{1}{c}{$V_{x}$} & \multicolumn{1}{c}{$V_{y}$} \\
\multicolumn{1}{c}{Number}  &  \multicolumn{1}{c}{(Jy beam$^{-1}$)} & \multicolumn{1}{c}{(km s$^{-1}$)} & \multicolumn{1}{c}{(mas)} & \multicolumn{1}{c}{(mas)} & \multicolumn{1}{c}{(km s$^{-1}$)} & \multicolumn{1}{c}{(km s$^{-1}$)} \\
\hline
\endhead
\hline
\endfoot
\hline
\endlastfoot
    1 &      8.02 &    $-$4.7 &     31.5$\pm$2.0 &    $-$50.2$\pm$2.0 &  $-$14.3$\pm$2.0 &  $-$13.7$\pm$2.1  \\
    2 &      2.84 &    $-$3.8 &      5.7$\pm$2.0 &    $-$14.7$\pm$2.0 & ... $ \; \; \; \, $  &  ... $ \; \; \; \, $  \\
    3 &      1.50 &    $-$4.2 &     $-$1.7$\pm$2.0 &     $-$2.3$\pm$2.0 & ... $ \; \; \; \, $  &  ... $ \; \; \; \, $  \\
    4 &      1.37 &    $-$2.0 &     $-$9.4$\pm$2.0 &      2.0$\pm$2.0 &  $-$25.3$\pm$2.2 &   53.1$\pm$2.5  \\
    5 &      1.04 &    $-$4.1 &    $-$24.6$\pm$2.0 &   $-$110.0$\pm$2.0 &  $-$21.8$\pm$2.0 &   $-$9.5$\pm$2.4  \\
    6 &      0.15 &    $-$0.3 &    $-$10.5$\pm$2.0 &      2.8$\pm$2.0 & ... $ \; \; \; \, $  &  ... $ \; \; \; \, $  \\
\end{longtable}
\tablefoot{
\\
Column~1 gives the feature label number; 
Cols.~2~and~3 provide the intensity of the strongest spot
and the intensity-weighted LSR velocity, respectively, averaged over the
observing epochs; Cols.~4~and~5 give the offsets (with
the associated errors) in absolute position along the R.A. and Dec. axes; Cols.~6~and~7 give the components of the absolute
proper motion (with the associated errors) along the R.A. and Dec. axes. \\
The absolute position to which the position offsets refer is:
R.A.~(J2000) = 20$^{\rm h}$ 30$^{\rm m}$ 29\fs143,  
Dec~(J2000) = 41\degree\ 15$^{\prime}$ 53\farcs640. 
The absolute positions are evaluated at the BeSSeL/VLBA observing epoch: May 24, 2011. 
} 
} 

\clearpage

\longtab[12]{ 
\begin{longtable}{rrcrrrr} 
\caption{\label{G090_tab} 22.2~GHz H$_2$O maser parameters for G090.21$+$2.32}\\ 
\hline\hline
\multicolumn{1}{c}{Feature} &  \multicolumn{1}{c}{I$_{\rm peak}$} & \multicolumn{1}{c}{$V_{\rm LSR}$} & \multicolumn{1}{c}{$\Delta~x$} & \multicolumn{1}{c}{$\Delta~y$} & \multicolumn{1}{c}{$V_{x}$} & \multicolumn{1}{c}{$V_{y}$} \\
\multicolumn{1}{c}{Number}  &  \multicolumn{1}{c}{(Jy beam$^{-1}$)} & \multicolumn{1}{c}{(km s$^{-1}$)} & \multicolumn{1}{c}{(mas)} & \multicolumn{1}{c}{(mas)} & \multicolumn{1}{c}{(km s$^{-1}$)} & \multicolumn{1}{c}{(km s$^{-1}$)} \\
\hline
\endfirsthead
\caption{continued.}\\
\hline\hline
\multicolumn{1}{c}{Feature} &  \multicolumn{1}{c}{I$_{\rm peak}$} & \multicolumn{1}{c}{$V_{\rm LSR}$} & \multicolumn{1}{c}{$\Delta~x$} & \multicolumn{1}{c}{$\Delta~y$} & \multicolumn{1}{c}{$V_{x}$} & \multicolumn{1}{c}{$V_{y}$} \\
\multicolumn{1}{c}{Number}  &  \multicolumn{1}{c}{(Jy beam$^{-1}$)} & \multicolumn{1}{c}{(km s$^{-1}$)} & \multicolumn{1}{c}{(mas)} & \multicolumn{1}{c}{(mas)} & \multicolumn{1}{c}{(km s$^{-1}$)} & \multicolumn{1}{c}{(km s$^{-1}$)} \\
\hline
\endhead
\hline
\endfoot
\hline
\endlastfoot
    1 &     13.56 &    $-$7.2 &     27.2$\pm$2.0 &    $-$14.4$\pm$2.0 &    0.7$\pm$0.5 &    3.2$\pm$0.4  \\
    2 &     10.20 &    $-$6.6 &     27.2$\pm$2.0 &    $-$17.0$\pm$2.0 &   $-$0.6$\pm$0.3 &    6.0$\pm$0.4  \\
    3 &      3.12 &    $-$6.3 &     27.0$\pm$2.0 &    $-$16.4$\pm$2.0 & ... $ \; \; \; \, $  &  ... $ \; \; \; \, $  \\
    4 &      2.41 &    $-$5.7 &     27.7$\pm$2.0 &    $-$19.3$\pm$2.0 & ... $ \; \; \; \, $  &  ... $ \; \; \; \, $  \\
    5 &      2.11 &    $-$6.6 &     26.6$\pm$2.0 &    $-$17.3$\pm$2.0 & ... $ \; \; \; \, $  &  ... $ \; \; \; \, $  \\
    6 &      1.34 &    $-$6.3 &     27.1$\pm$2.0 &    $-$18.3$\pm$2.0 &   $-$5.0$\pm$1.0 &    1.2$\pm$1.1  \\
    7 &      0.70 &    $-$4.3 &     27.2$\pm$2.0 &    $-$24.7$\pm$2.0 &   $-$1.6$\pm$0.5 &    1.1$\pm$0.8  \\
    8 &      0.64 &    $-$6.7 &     27.2$\pm$2.0 &    $-$14.9$\pm$2.0 & ... $ \; \; \; \, $  &  ... $ \; \; \; \, $  \\
    9 &      0.48 &    $-$5.7 &     26.7$\pm$2.0 &     $-$9.4$\pm$2.0 & ... $ \; \; \; \, $  &  ... $ \; \; \; \, $  \\
   10 &      0.39 &    $-$5.6 &     27.3$\pm$2.0 &     $-$8.0$\pm$2.0 & ... $ \; \; \; \, $  &  ... $ \; \; \; \, $  \\
   11 &      0.26 &    $-$4.9 &     28.0$\pm$2.0 &    $-$21.0$\pm$2.0 & ... $ \; \; \; \, $  &  ... $ \; \; \; \, $  \\
   12 &      0.19 &    $-$4.5 &     27.2$\pm$2.0 &    $-$25.0$\pm$2.0 & ... $ \; \; \; \, $  &  ... $ \; \; \; \, $  \\
   13 &      0.16 &    $-$2.9 &     27.5$\pm$2.0 &    $-$26.1$\pm$2.0 & ... $ \; \; \; \, $  &  ... $ \; \; \; \, $  \\
   14 &      0.09 &    $-$5.4 &     26.1$\pm$2.0 &    $-$23.0$\pm$2.0 & ... $ \; \; \; \, $  &  ... $ \; \; \; \, $  \\
   15 &      0.07 &    $-$5.9 &     28.7$\pm$2.0 &     $-$5.7$\pm$2.0 & ... $ \; \; \; \, $  &  ... $ \; \; \; \, $  \\
\end{longtable}
\tablefoot{
\\
Column~1 gives the feature label number; 
Cols.~2~and~3 provide the intensity of the strongest spot
and the intensity-weighted LSR velocity, respectively, averaged over the
observing epochs; Cols.~4~and~5 give the offsets (with
the associated errors) in absolute position along the R.A. and Dec. axes; Cols.~6~and~7 give the components of the absolute
proper motion (with the associated errors) along the R.A. and Dec. axes. \\
The absolute position to which the position offsets refer is:
R.A.~(J2000) = 21$^{\rm h}$ 2$^{\rm m}$ 22\fs6981,  
Dec~(J2000) = 50\degree\ 3$^{\prime}$ 08\farcs325. 
The absolute positions are evaluated at the BeSSeL/VLBA observing epoch: May 16, 2010. 
} 
} 

\clearpage

\longtab[13]{ 
\begin{longtable}{rrcrrrr} 
\caption{\label{G105_taA} 22.2~GHz H$_2$O maser parameters for G105.42$+$9.88 VLA~3A}\\ 
\hline\hline
\multicolumn{1}{c}{Feature} &  \multicolumn{1}{c}{I$_{\rm peak}$} & \multicolumn{1}{c}{$V_{\rm LSR}$} & \multicolumn{1}{c}{$\Delta~x$} & \multicolumn{1}{c}{$\Delta~y$} & \multicolumn{1}{c}{$V_{x}$} & \multicolumn{1}{c}{$V_{y}$} \\
\multicolumn{1}{c}{Number}  &  \multicolumn{1}{c}{(Jy beam$^{-1}$)} & \multicolumn{1}{c}{(km s$^{-1}$)} & \multicolumn{1}{c}{(mas)} & \multicolumn{1}{c}{(mas)} & \multicolumn{1}{c}{(km s$^{-1}$)} & \multicolumn{1}{c}{(km s$^{-1}$)} \\
\hline
\endfirsthead
\caption{continued.}\\
\hline\hline
\multicolumn{1}{c}{Feature} &  \multicolumn{1}{c}{I$_{\rm peak}$} & \multicolumn{1}{c}{$V_{\rm LSR}$} & \multicolumn{1}{c}{$\Delta~x$} & \multicolumn{1}{c}{$\Delta~y$} & \multicolumn{1}{c}{$V_{x}$} & \multicolumn{1}{c}{$V_{y}$} \\
\multicolumn{1}{c}{Number}  &  \multicolumn{1}{c}{(Jy beam$^{-1}$)} & \multicolumn{1}{c}{(km s$^{-1}$)} & \multicolumn{1}{c}{(mas)} & \multicolumn{1}{c}{(mas)} & \multicolumn{1}{c}{(km s$^{-1}$)} & \multicolumn{1}{c}{(km s$^{-1}$)} \\
\hline
\endhead
\hline
\endfoot
\hline
\endlastfoot
    1 &      0.24 &   $-$20.3 &    $-$25.5$\pm$2.0 &   $-$356.0$\pm$2.0 &   $-$3.1$\pm$1.6 &  $-$17.4$\pm$1.0  \\
    2 &      0.20 &   $-$19.7 &    $-$24.7$\pm$2.0 &   $-$357.5$\pm$2.0 &   $-$3.8$\pm$0.9 &  $-$14.4$\pm$1.0  \\
\end{longtable}
\tablefoot{
\\
Column~1 gives the feature label number; 
Cols.~2~and~3 provide the intensity of the strongest spot
and the intensity-weighted LSR velocity, respectively, averaged over the
observing epochs; Cols.~4~and~5 give the offsets (with
the associated errors) in absolute position along the R.A. and Dec. axes; Cols.~6~and~7 give the components of the absolute
proper motion (with the associated errors) along the R.A. and Dec. axes. \\
The absolute position to which the position offsets refer is:
R.A.~(J2000) = 21$^{\rm h}$ 43$^{\rm m}$ 06\fs4808,  
Dec~(J2000) = 66\degree\ 6$^{\prime}$ 55\farcs345. 
The absolute positions are evaluated at the BeSSeL/VLBA observing epoch: May 24, 2011. 
} 
} 

\clearpage

\longtab[14]{ 
\begin{longtable}{rrcrrrr} 
\caption{\label{G105_taB} 22.2~GHz H$_2$O maser parameters for G105.42$+$9.88 VLA~3B}\\ 
\hline\hline
\multicolumn{1}{c}{Feature} &  \multicolumn{1}{c}{I$_{\rm peak}$} & \multicolumn{1}{c}{$V_{\rm LSR}$} & \multicolumn{1}{c}{$\Delta~x$} & \multicolumn{1}{c}{$\Delta~y$} & \multicolumn{1}{c}{$V_{x}$} & \multicolumn{1}{c}{$V_{y}$} \\
\multicolumn{1}{c}{Number}  &  \multicolumn{1}{c}{(Jy beam$^{-1}$)} & \multicolumn{1}{c}{(km s$^{-1}$)} & \multicolumn{1}{c}{(mas)} & \multicolumn{1}{c}{(mas)} & \multicolumn{1}{c}{(km s$^{-1}$)} & \multicolumn{1}{c}{(km s$^{-1}$)} \\
\hline
\endfirsthead
\caption{continued.}\\
\hline\hline
\multicolumn{1}{c}{Feature} &  \multicolumn{1}{c}{I$_{\rm peak}$} & \multicolumn{1}{c}{$V_{\rm LSR}$} & \multicolumn{1}{c}{$\Delta~x$} & \multicolumn{1}{c}{$\Delta~y$} & \multicolumn{1}{c}{$V_{x}$} & \multicolumn{1}{c}{$V_{y}$} \\
\multicolumn{1}{c}{Number}  &  \multicolumn{1}{c}{(Jy beam$^{-1}$)} & \multicolumn{1}{c}{(km s$^{-1}$)} & \multicolumn{1}{c}{(mas)} & \multicolumn{1}{c}{(mas)} & \multicolumn{1}{c}{(km s$^{-1}$)} & \multicolumn{1}{c}{(km s$^{-1}$)} \\
\hline
\endhead
\hline
\endfoot
\hline
\endlastfoot
    1 &     95.87 &   $-$13.9 &   $-$106.9$\pm$2.0 &   $-$161.4$\pm$2.0 & ... $ \; \; \; \, $  &  ... $ \; \; \; \, $  \\
    2 &     74.51 &   $-$12.7 &   $-$105.9$\pm$2.0 &   $-$159.1$\pm$2.0 & ... $ \; \; \; \, $  &  ... $ \; \; \; \, $  \\
    3 &     31.01 &   $-$13.6 &   $-$111.0$\pm$2.0 &   $-$156.8$\pm$2.0 & ... $ \; \; \; \, $  &  ... $ \; \; \; \, $  \\
    4 &     30.65 &   $-$16.0 &   $-$107.3$\pm$2.0 &   $-$159.7$\pm$2.0 &   $-$3.7$\pm$1.6 &   $-$4.5$\pm$1.7  \\
    5 &     12.49 &   $-$11.9 &   $-$129.0$\pm$2.0 &   $-$154.2$\pm$2.0 & ... $ \; \; \; \, $  &  ... $ \; \; \; \, $  \\
    6 &     10.88 &   $-$14.6 &   $-$102.4$\pm$2.0 &   $-$168.4$\pm$2.0 &   $-$4.9$\pm$0.5 &   $-$3.1$\pm$0.5  \\
    7 &     10.60 &   $-$13.1 &   $-$105.9$\pm$2.0 &   $-$161.9$\pm$2.0 & ... $ \; \; \; \, $  &  ... $ \; \; \; \, $  \\
    8 &     10.47 &   $-$18.1 &   $-$114.6$\pm$2.0 &   $-$156.0$\pm$2.0 &   $-$3.7$\pm$0.4 &   $-$3.7$\pm$0.4  \\
    9 &      6.75 &   $-$13.9 &   $-$103.5$\pm$2.0 &   $-$168.1$\pm$2.0 & ... $ \; \; \; \, $  &  ... $ \; \; \; \, $  \\
   10 &      6.49 &   $-$22.4 &   $-$117.4$\pm$2.0 &   $-$156.0$\pm$2.0 & ... $ \; \; \; \, $  &  ... $ \; \; \; \, $  \\
   11 &      4.76 &   $-$13.1 &   $-$106.5$\pm$2.0 &   $-$159.7$\pm$2.0 &   $-$4.3$\pm$0.6 &   $-$3.7$\pm$0.7  \\
   12 &      3.62 &   $-$10.7 &   $-$103.3$\pm$2.0 &   $-$162.8$\pm$2.0 & ... $ \; \; \; \, $  &  ... $ \; \; \; \, $  \\
   13 &      3.53 &   $-$18.6 &   $-$113.6$\pm$2.0 &   $-$156.8$\pm$2.0 & ... $ \; \; \; \, $  &  ... $ \; \; \; \, $  \\
   14 &      3.37 &   $-$12.1 &   $-$104.2$\pm$2.0 &   $-$162.2$\pm$2.0 & ... $ \; \; \; \, $  &  ... $ \; \; \; \, $  \\
   15 &      2.98 &   $-$11.3 &   $-$101.6$\pm$2.0 &   $-$174.2$\pm$2.0 & ... $ \; \; \; \, $  &  ... $ \; \; \; \, $  \\
   16 &      2.87 &   $-$15.3 &   $-$102.8$\pm$2.0 &   $-$168.8$\pm$2.0 & ... $ \; \; \; \, $  &  ... $ \; \; \; \, $  \\
   17 &      2.87 &   $-$14.4 &   $-$106.7$\pm$2.0 &   $-$160.4$\pm$2.0 & ... $ \; \; \; \, $  &  ... $ \; \; \; \, $  \\
   18 &      2.55 &   $-$17.0 &   $-$104.9$\pm$2.0 &   $-$167.4$\pm$2.0 & ... $ \; \; \; \, $  &  ... $ \; \; \; \, $  \\
   19 &      2.31 &   $-$13.4 &   $-$104.0$\pm$2.0 &   $-$167.1$\pm$2.0 & ... $ \; \; \; \, $  &  ... $ \; \; \; \, $  \\
   20 &      2.24 &   $-$12.9 &   $-$105.4$\pm$2.0 &   $-$161.1$\pm$2.0 &   $-$3.5$\pm$0.4 &   $-$5.7$\pm$0.5  \\
   21 &      2.10 &   $-$13.6 &   $-$104.6$\pm$2.0 &   $-$164.8$\pm$2.0 & ... $ \; \; \; \, $  &  ... $ \; \; \; \, $  \\
   22 &      1.78 &   $-$14.2 &   $-$104.4$\pm$2.0 &   $-$161.7$\pm$2.0 & ... $ \; \; \; \, $  &  ... $ \; \; \; \, $  \\
   23 &      1.54 &   $-$10.2 &   $-$105.2$\pm$2.0 &   $-$160.7$\pm$2.0 & ... $ \; \; \; \, $  &  ... $ \; \; \; \, $  \\
   24 &      1.06 &   $-$12.3 &   $-$129.4$\pm$2.0 &   $-$154.2$\pm$2.0 & ... $ \; \; \; \, $  &  ... $ \; \; \; \, $  \\
   25 &      0.94 &   $-$18.2 &   $-$118.0$\pm$2.0 &   $-$155.1$\pm$2.0 & ... $ \; \; \; \, $  &  ... $ \; \; \; \, $  \\
   26 &      0.66 &   $-$16.3 &   $-$107.7$\pm$2.0 &   $-$160.8$\pm$2.0 & ... $ \; \; \; \, $  &  ... $ \; \; \; \, $  \\
   27 &      0.56 &   $-$12.8 &   $-$104.2$\pm$2.0 &   $-$166.7$\pm$2.0 & ... $ \; \; \; \, $  &  ... $ \; \; \; \, $  \\
   28 &      0.56 &   $-$14.2 &   $-$105.2$\pm$2.0 &   $-$162.0$\pm$2.0 & ... $ \; \; \; \, $  &  ... $ \; \; \; \, $  \\
   29 &      0.55 &   $-$10.6 &   $-$112.9$\pm$2.0 &   $-$157.0$\pm$2.0 & ... $ \; \; \; \, $  &  ... $ \; \; \; \, $  \\
   30 &      0.54 &   $-$17.1 &   $-$104.8$\pm$2.0 &   $-$167.9$\pm$2.0 & ... $ \; \; \; \, $  &  ... $ \; \; \; \, $  \\
   31 &      0.50 &   $-$15.1 &   $-$104.0$\pm$2.0 &   $-$166.4$\pm$2.0 & ... $ \; \; \; \, $  &  ... $ \; \; \; \, $  \\
   32 &      0.46 &   $-$18.4 &   $-$106.8$\pm$2.0 &   $-$163.1$\pm$2.0 & ... $ \; \; \; \, $  &  ... $ \; \; \; \, $  \\
   33 &      0.44 &   $-$12.7 &   $-$104.0$\pm$2.0 &   $-$163.7$\pm$2.0 & ... $ \; \; \; \, $  &  ... $ \; \; \; \, $  \\
   34 &      0.43 &   $-$14.6 &   $-$102.0$\pm$2.0 &   $-$169.1$\pm$2.0 & ... $ \; \; \; \, $  &  ... $ \; \; \; \, $  \\
   35 &      0.37 &   $-$14.9 &   $-$118.6$\pm$2.0 &   $-$153.5$\pm$2.0 & ... $ \; \; \; \, $  &  ... $ \; \; \; \, $  \\
   36 &      0.37 &   $-$10.7 &   $-$129.8$\pm$2.0 &   $-$153.8$\pm$2.0 & ... $ \; \; \; \, $  &  ... $ \; \; \; \, $  \\
   37 &      0.36 &   $-$15.9 &   $-$104.3$\pm$2.0 &   $-$172.0$\pm$2.0 & ... $ \; \; \; \, $  &  ... $ \; \; \; \, $  \\
   38 &      0.35 &   $-$19.7 &   $-$103.2$\pm$2.0 &   $-$171.2$\pm$2.0 & ... $ \; \; \; \, $  &  ... $ \; \; \; \, $  \\
   39 &      0.35 &   $-$16.8 &   $-$114.0$\pm$2.0 &   $-$155.9$\pm$2.0 & ... $ \; \; \; \, $  &  ... $ \; \; \; \, $  \\
   40 &      0.35 &   $-$13.2 &   $-$114.0$\pm$2.0 &   $-$156.5$\pm$2.0 & ... $ \; \; \; \, $  &  ... $ \; \; \; \, $  \\
   41 &      0.33 &   $-$12.9 &   $-$102.6$\pm$2.0 &   $-$168.1$\pm$2.0 & ... $ \; \; \; \, $  &  ... $ \; \; \; \, $  \\
   42 &      0.32 &   $-$14.4 &   $-$103.9$\pm$2.0 &   $-$166.6$\pm$2.0 & ... $ \; \; \; \, $  &  ... $ \; \; \; \, $  \\
   43 &      0.31 &   $-$10.1 &   $-$114.7$\pm$2.0 &   $-$171.5$\pm$2.0 & ... $ \; \; \; \, $  &  ... $ \; \; \; \, $  \\
   44 &      0.30 &   $-$14.6 &   $-$102.0$\pm$2.0 &   $-$168.6$\pm$2.0 & ... $ \; \; \; \, $  &  ... $ \; \; \; \, $  \\
   45 &      0.29 &   $-$13.8 &   $-$111.5$\pm$2.0 &   $-$156.9$\pm$2.0 & ... $ \; \; \; \, $  &  ... $ \; \; \; \, $  \\
   46 &      0.28 &   $-$12.0 &   $-$104.2$\pm$2.0 &   $-$166.0$\pm$2.0 & ... $ \; \; \; \, $  &  ... $ \; \; \; \, $  \\
   47 &      0.26 &   $-$11.6 &   $-$105.6$\pm$2.0 &   $-$162.5$\pm$2.0 & ... $ \; \; \; \, $  &  ... $ \; \; \; \, $  \\
   48 &      0.25 &   $-$14.2 &   $-$102.2$\pm$2.0 &   $-$168.1$\pm$2.0 & ... $ \; \; \; \, $  &  ... $ \; \; \; \, $  \\
   49 &      0.24 &   $-$14.3 &   $-$103.6$\pm$2.0 &   $-$168.1$\pm$2.0 & ... $ \; \; \; \, $  &  ... $ \; \; \; \, $  \\
   50 &      0.23 &   $-$12.7 &   $-$113.7$\pm$2.0 &   $-$155.0$\pm$2.0 & ... $ \; \; \; \, $  &  ... $ \; \; \; \, $  \\
   51 &      0.22 &   $-$22.8 &   $-$117.6$\pm$2.0 &   $-$156.8$\pm$2.0 & ... $ \; \; \; \, $  &  ... $ \; \; \; \, $  \\
   52 &      0.20 &   $-$15.6 &   $-$104.6$\pm$2.0 &   $-$170.4$\pm$2.0 & ... $ \; \; \; \, $  &  ... $ \; \; \; \, $  \\
   53 &      0.19 &   $-$11.4 &   $-$102.3$\pm$2.0 &   $-$168.3$\pm$2.0 & ... $ \; \; \; \, $  &  ... $ \; \; \; \, $  \\
   54 &      0.19 &   $-$23.7 &   $-$104.8$\pm$2.0 &   $-$178.9$\pm$2.0 & ... $ \; \; \; \, $  &  ... $ \; \; \; \, $  \\
   55 &      0.18 &   $-$16.0 &   $-$105.6$\pm$2.0 &   $-$162.9$\pm$2.0 & ... $ \; \; \; \, $  &  ... $ \; \; \; \, $  \\
   56 &      0.17 &   $-$14.4 &   $-$114.4$\pm$2.0 &   $-$156.0$\pm$2.0 & ... $ \; \; \; \, $  &  ... $ \; \; \; \, $  \\
   57 &      0.17 &   $-$17.1 &   $-$105.7$\pm$2.0 &   $-$164.9$\pm$2.0 & ... $ \; \; \; \, $  &  ... $ \; \; \; \, $  \\
   58 &      0.16 &   $-$14.0 &   $-$114.9$\pm$2.0 &   $-$156.1$\pm$2.0 & ... $ \; \; \; \, $  &  ... $ \; \; \; \, $  \\
   59 &      0.15 &   $-$19.9 &   $-$103.1$\pm$2.0 &   $-$169.5$\pm$2.0 & ... $ \; \; \; \, $  &  ... $ \; \; \; \, $  \\
   60 &      0.14 &   $-$24.4 &   $-$104.4$\pm$2.0 &   $-$173.2$\pm$2.0 & ... $ \; \; \; \, $  &  ... $ \; \; \; \, $  \\
   61 &      0.13 &    $-$7.9 &   $-$105.3$\pm$2.0 &   $-$162.4$\pm$2.0 & ... $ \; \; \; \, $  &  ... $ \; \; \; \, $  \\
   62 &      0.13 &   $-$11.5 &   $-$104.6$\pm$2.0 &   $-$162.6$\pm$2.0 & ... $ \; \; \; \, $  &  ... $ \; \; \; \, $  \\
   63 &      0.13 &   $-$25.5 &   $-$105.7$\pm$2.0 &   $-$179.8$\pm$2.0 & ... $ \; \; \; \, $  &  ... $ \; \; \; \, $  \\
   64 &      0.13 &   $-$15.7 &   $-$104.3$\pm$2.0 &   $-$168.1$\pm$2.0 & ... $ \; \; \; \, $  &  ... $ \; \; \; \, $  \\
   65 &      0.12 &   $-$17.6 &   $-$126.1$\pm$2.0 &   $-$185.7$\pm$2.0 & ... $ \; \; \; \, $  &  ... $ \; \; \; \, $  \\
   66 &      0.12 &   $-$12.3 &   $-$103.3$\pm$2.0 &   $-$167.5$\pm$2.0 & ... $ \; \; \; \, $  &  ... $ \; \; \; \, $  \\
   67 &      0.12 &   $-$13.2 &   $-$109.0$\pm$2.0 &   $-$157.7$\pm$2.0 & ... $ \; \; \; \, $  &  ... $ \; \; \; \, $  \\
   68 &      0.11 &   $-$18.2 &   $-$104.9$\pm$2.0 &   $-$169.5$\pm$2.0 & ... $ \; \; \; \, $  &  ... $ \; \; \; \, $  \\
   69 &      0.11 &   $-$21.8 &   $-$118.2$\pm$2.0 &   $-$156.1$\pm$2.0 & ... $ \; \; \; \, $  &  ... $ \; \; \; \, $  \\
   70 &      0.11 &   $-$21.8 &   $-$104.9$\pm$2.0 &   $-$167.1$\pm$2.0 & ... $ \; \; \; \, $  &  ... $ \; \; \; \, $  \\
   71 &      0.10 &   $-$16.5 &   $-$106.5$\pm$2.0 &   $-$162.2$\pm$2.0 & ... $ \; \; \; \, $  &  ... $ \; \; \; \, $  \\
   72 &      0.08 &   $-$26.0 &   $-$117.1$\pm$2.0 &   $-$154.6$\pm$2.0 & ... $ \; \; \; \, $  &  ... $ \; \; \; \, $  \\
\end{longtable}
\tablefoot{
\\
Column~1 gives the feature label number; 
Cols.~2~and~3 provide the intensity of the strongest spot
and the intensity-weighted LSR velocity, respectively, averaged over the
observing epochs; Cols.~4~and~5 give the offsets (with
the associated errors) in absolute position along the R.A. and Dec. axes; Cols.~6~and~7 give the components of the absolute
proper motion (with the associated errors) along the R.A. and Dec. axes. \\
The absolute position to which the position offsets refer is:
R.A.~(J2000) = 21$^{\rm h}$ 43$^{\rm m}$ 06\fs4808,  
Dec~(J2000) = 66\degree\ 6$^{\prime}$ 55\farcs345. 
The absolute positions are evaluated at the BeSSeL/VLBA observing epoch: May 24, 2011. 
} 
} 

\clearpage

\longtab[15]{ 
\begin{longtable}{rrcrrrr} 
\caption{\label{G108_20b} 22.2~GHz H$_2$O maser parameters for G108.20$+$0.59}\\ 
\hline\hline
\multicolumn{1}{c}{Feature} &  \multicolumn{1}{c}{I$_{\rm peak}$} & \multicolumn{1}{c}{$V_{\rm LSR}$} & \multicolumn{1}{c}{$\Delta~x$} & \multicolumn{1}{c}{$\Delta~y$} & \multicolumn{1}{c}{$V_{x}$} & \multicolumn{1}{c}{$V_{y}$} \\
\multicolumn{1}{c}{Number}  &  \multicolumn{1}{c}{(Jy beam$^{-1}$)} & \multicolumn{1}{c}{(km s$^{-1}$)} & \multicolumn{1}{c}{(mas)} & \multicolumn{1}{c}{(mas)} & \multicolumn{1}{c}{(km s$^{-1}$)} & \multicolumn{1}{c}{(km s$^{-1}$)} \\
\hline
\endfirsthead
\caption{continued.}\\
\hline\hline
\multicolumn{1}{c}{Feature} &  \multicolumn{1}{c}{I$_{\rm peak}$} & \multicolumn{1}{c}{$V_{\rm LSR}$} & \multicolumn{1}{c}{$\Delta~x$} & \multicolumn{1}{c}{$\Delta~y$} & \multicolumn{1}{c}{$V_{x}$} & \multicolumn{1}{c}{$V_{y}$} \\
\multicolumn{1}{c}{Number}  &  \multicolumn{1}{c}{(Jy beam$^{-1}$)} & \multicolumn{1}{c}{(km s$^{-1}$)} & \multicolumn{1}{c}{(mas)} & \multicolumn{1}{c}{(mas)} & \multicolumn{1}{c}{(km s$^{-1}$)} & \multicolumn{1}{c}{(km s$^{-1}$)} \\
\hline
\endhead
\hline
\endfoot
\hline
\endlastfoot
    1 &     10.50 &   $-$54.2 &    $-$85.3$\pm$1.0 &   $-$101.2$\pm$1.0 &   15.1$\pm$1.4 &   15.3$\pm$1.4  \\
    2 &      4.17 &   $-$53.8 &    $-$96.3$\pm$1.0 &   $-$114.0$\pm$1.0 & ... $ \; \; \; \, $  &  ... $ \; \; \; \, $  \\
    3 &      2.64 &   $-$52.4 &   $-$148.7$\pm$1.0 &   $-$216.6$\pm$1.0 &   $-$2.0$\pm$3.5 &  $-$34.8$\pm$3.8  \\
    4 &      2.15 &   $-$49.0 &    $-$99.6$\pm$1.0 &   $-$146.0$\pm$1.0 &  $-$20.2$\pm$2.7 &    8.5$\pm$2.9  \\
    5 &      2.00 &   $-$55.8 &    $-$93.4$\pm$1.0 &   $-$110.4$\pm$1.0 & ... $ \; \; \; \, $  &  ... $ \; \; \; \, $  \\
    6 &      1.88 &   $-$55.8 &    $-$80.2$\pm$1.0 &    $-$98.9$\pm$1.0 &   11.7$\pm$1.4 &   15.1$\pm$1.5  \\
    7 &      1.83 &   $-$55.8 &    $-$90.1$\pm$1.0 &   $-$102.4$\pm$1.0 &    7.4$\pm$2.1 &   18.3$\pm$2.0  \\
    8 &      1.26 &   $-$53.7 &    $-$91.5$\pm$1.0 &   $-$110.3$\pm$1.0 &   $-$2.1$\pm$3.8 &   21.2$\pm$3.7  \\
    9 &      0.80 &   $-$49.0 &    $-$99.3$\pm$1.0 &   $-$146.8$\pm$1.0 & ... $ \; \; \; \, $  &  ... $ \; \; \; \, $  \\
   10 &      0.79 &   $-$46.4 &    $-$93.7$\pm$1.0 &   $-$179.2$\pm$1.0 & ... $ \; \; \; \, $  &  ... $ \; \; \; \, $  \\
   11 &      0.74 &   $-$55.6 &    $-$92.8$\pm$1.0 &   $-$110.2$\pm$1.0 & ... $ \; \; \; \, $  &  ... $ \; \; \; \, $  \\
   12 &      0.74 &   $-$57.2 &      4.8$\pm$1.0 &   $-$128.9$\pm$1.0 & ... $ \; \; \; \, $  &  ... $ \; \; \; \, $  \\
   13 &      0.73 &   $-$56.0 &    $-$94.3$\pm$1.0 &   $-$104.6$\pm$1.0 &    2.6$\pm$1.6 &   27.1$\pm$1.8  \\
   14 &      0.67 &   $-$55.4 &    $-$92.9$\pm$1.0 &   $-$110.0$\pm$1.0 & ... $ \; \; \; \, $  &  ... $ \; \; \; \, $  \\
   15 &      0.61 &   $-$56.0 &    $-$88.3$\pm$1.0 &   $-$103.5$\pm$1.0 & ... $ \; \; \; \, $  &  ... $ \; \; \; \, $  \\
   16 &      0.56 &   $-$54.9 &    $-$72.4$\pm$1.0 &   $-$101.7$\pm$1.0 &   11.6$\pm$2.0 &   15.8$\pm$2.0  \\
   17 &      0.48 &   $-$42.1 &    $-$94.9$\pm$1.0 &   $-$152.7$\pm$1.0 &    3.1$\pm$2.3 &  $-$41.7$\pm$3.0  \\
   18 &      0.43 &   $-$55.2 &    $-$90.4$\pm$1.0 &    $-$84.0$\pm$1.0 &    6.2$\pm$1.6 &   15.6$\pm$1.4  \\
   19 &      0.40 &   $-$46.2 &   $-$145.2$\pm$1.0 &   $-$219.9$\pm$1.0 & ... $ \; \; \; \, $  &  ... $ \; \; \; \, $  \\
   20 &      0.34 &   $-$57.2 &    $-$84.8$\pm$1.0 &   $-$104.3$\pm$1.0 & ... $ \; \; \; \, $  &  ... $ \; \; \; \, $  \\
   21 &      0.32 &   $-$52.6 &   $-$148.1$\pm$1.0 &   $-$219.1$\pm$1.0 & ... $ \; \; \; \, $  &  ... $ \; \; \; \, $  \\
   22 &      0.29 &   $-$54.0 &    $-$88.1$\pm$1.0 &   $-$106.7$\pm$1.0 & ... $ \; \; \; \, $  &  ... $ \; \; \; \, $  \\
   23 &      0.29 &   $-$44.2 &   $-$136.9$\pm$1.0 &   $-$215.4$\pm$1.0 & ... $ \; \; \; \, $  &  ... $ \; \; \; \, $  \\
   24 &      0.28 &   $-$55.6 &    $-$79.8$\pm$1.0 &    $-$98.9$\pm$1.0 &   10.1$\pm$2.3 &   14.2$\pm$2.6  \\
   25 &      0.26 &   $-$50.8 &    $-$99.8$\pm$1.0 &   $-$146.9$\pm$1.0 & ... $ \; \; \; \, $  &  ... $ \; \; \; \, $  \\
   26 &      0.24 &   $-$56.0 &    $-$90.1$\pm$1.0 &   $-$102.5$\pm$1.0 & ... $ \; \; \; \, $  &  ... $ \; \; \; \, $  \\
   27 &      0.21 &   $-$53.9 &     $-$0.6$\pm$1.0 &   $-$104.0$\pm$1.0 & ... $ \; \; \; \, $  &  ... $ \; \; \; \, $  \\
   28 &      0.17 &   $-$54.6 &    $-$89.1$\pm$1.0 &    $-$83.8$\pm$1.0 & ... $ \; \; \; \, $  &  ... $ \; \; \; \, $  \\
   29 &      0.16 &   $-$53.3 &      1.2$\pm$1.0 &   $-$113.9$\pm$1.0 & ... $ \; \; \; \, $  &  ... $ \; \; \; \, $  \\
   30 &      0.14 &   $-$53.5 &    $-$85.8$\pm$1.0 &   $-$100.9$\pm$1.0 & ... $ \; \; \; \, $  &  ... $ \; \; \; \, $  \\
   31 &      0.12 &   $-$51.8 &   $-$149.4$\pm$1.0 &   $-$217.8$\pm$1.0 & ... $ \; \; \; \, $  &  ... $ \; \; \; \, $  \\
   32 &      0.10 &   $-$54.2 &    $-$93.5$\pm$1.0 &   $-$105.4$\pm$1.0 & ... $ \; \; \; \, $  &  ... $ \; \; \; \, $  \\
   33 &      0.10 &   $-$56.0 &    $-$94.2$\pm$1.0 &   $-$112.3$\pm$1.0 & ... $ \; \; \; \, $  &  ... $ \; \; \; \, $  \\
   34 &      0.09 &   $-$58.1 &    $-$85.4$\pm$1.0 &   $-$106.1$\pm$1.0 & ... $ \; \; \; \, $  &  ... $ \; \; \; \, $  \\
   35 &      0.09 &   $-$61.5 &    $-$96.5$\pm$1.0 &   $-$135.5$\pm$1.0 & ... $ \; \; \; \, $  &  ... $ \; \; \; \, $  \\
   36 &      0.09 &   $-$54.4 &    $-$91.6$\pm$1.0 &    $-$84.2$\pm$1.0 & ... $ \; \; \; \, $  &  ... $ \; \; \; \, $  \\
   37 &      0.09 &   $-$55.2 &    $-$95.0$\pm$1.0 &   $-$105.2$\pm$1.0 & ... $ \; \; \; \, $  &  ... $ \; \; \; \, $  \\
   38 &      0.09 &   $-$55.2 &    $-$94.9$\pm$1.0 &   $-$105.4$\pm$1.0 & ... $ \; \; \; \, $  &  ... $ \; \; \; \, $  \\
   39 &      0.08 &   $-$51.6 &   $-$149.9$\pm$1.0 &   $-$217.7$\pm$1.0 & ... $ \; \; \; \, $  &  ... $ \; \; \; \, $  \\
\end{longtable}
\tablefoot{
\\
Column~1 gives the feature label number; 
Cols.~2~and~3 provide the intensity of the strongest spot
and the intensity-weighted LSR velocity, respectively, averaged over the
observing epochs; Cols.~4~and~5 give the offsets (with
the associated errors) in absolute position along the R.A. and Dec. axes; Cols.~6~and~7 give the components of the absolute
proper motion (with the associated errors) along the R.A. and Dec. axes. \\
The absolute position to which the position offsets refer is:
R.A.~(J2000) = 22$^{\rm h}$ 49$^{\rm m}$ 31\fs4775,  
Dec~(J2000) = 59\degree\ 55$^{\prime}$ 42\farcs006. 
The absolute positions are evaluated at the BeSSeL/VLBA observing epoch: June 5, 2010. 
} 
} 

\clearpage

\longtab[16]{ 
\begin{longtable}{rrcrrrr} 
\caption{\label{G108_59b} 22.2~GHz H$_2$O maser parameters for G108.59$+$0.49}\\ 
\hline\hline
\multicolumn{1}{c}{Feature} &  \multicolumn{1}{c}{I$_{\rm peak}$} & \multicolumn{1}{c}{$V_{\rm LSR}$} & \multicolumn{1}{c}{$\Delta~x$} & \multicolumn{1}{c}{$\Delta~y$} & \multicolumn{1}{c}{$V_{x}$} & \multicolumn{1}{c}{$V_{y}$} \\
\multicolumn{1}{c}{Number}  &  \multicolumn{1}{c}{(Jy beam$^{-1}$)} & \multicolumn{1}{c}{(km s$^{-1}$)} & \multicolumn{1}{c}{(mas)} & \multicolumn{1}{c}{(mas)} & \multicolumn{1}{c}{(km s$^{-1}$)} & \multicolumn{1}{c}{(km s$^{-1}$)} \\
\hline
\endfirsthead
\caption{continued.}\\
\hline\hline
\multicolumn{1}{c}{Feature} &  \multicolumn{1}{c}{I$_{\rm peak}$} & \multicolumn{1}{c}{$V_{\rm LSR}$} & \multicolumn{1}{c}{$\Delta~x$} & \multicolumn{1}{c}{$\Delta~y$} & \multicolumn{1}{c}{$V_{x}$} & \multicolumn{1}{c}{$V_{y}$} \\
\multicolumn{1}{c}{Number}  &  \multicolumn{1}{c}{(Jy beam$^{-1}$)} & \multicolumn{1}{c}{(km s$^{-1}$)} & \multicolumn{1}{c}{(mas)} & \multicolumn{1}{c}{(mas)} & \multicolumn{1}{c}{(km s$^{-1}$)} & \multicolumn{1}{c}{(km s$^{-1}$)} \\
\hline
\endhead
\hline
\endfoot
\hline
\endlastfoot
    1 &      4.59 &   $-$50.2 &     59.1$\pm$5.0 &     71.9$\pm$5.0 & ... $ \; \; \; \, $  &  ... $ \; \; \; \, $  \\
    2 &      2.55 &   $-$54.0 &      5.0$\pm$5.0 &      2.0$\pm$5.0 &  $-$39.4$\pm$1.5 &  $-$19.6$\pm$2.5  \\
    3 &      0.87 &   $-$52.3 &     57.8$\pm$5.0 &     65.7$\pm$5.0 & ... $ \; \; \; \, $  &  ... $ \; \; \; \, $  \\
    4 &      0.36 &   $-$50.0 &     59.0$\pm$5.0 &     72.7$\pm$5.0 & ... $ \; \; \; \, $  &  ... $ \; \; \; \, $  \\
    5 &      0.25 &   $-$53.7 &     12.5$\pm$5.0 &      2.2$\pm$5.0 & ... $ \; \; \; \, $  &  ... $ \; \; \; \, $  \\
\end{longtable}
\tablefoot{
\\
Column~1 gives the feature label number; 
Cols.~2~and~3 provide the intensity of the strongest spot
and the intensity-weighted LSR velocity, respectively, averaged over the
observing epochs; Cols.~4~and~5 give the offsets (with
the associated errors) in absolute position along the R.A. and Dec. axes; Cols.~6~and~7 give the components of the absolute
proper motion (with the associated errors) along the R.A. and Dec. axes. \\
The absolute position to which the position offsets refer is:
R.A.~(J2000) = 22$^{\rm h}$ 52$^{\rm m}$ 38\fs316,  
Dec~(J2000) = 60\degree\ 0$^{\prime}$ 51\farcs876. 
The absolute positions are evaluated at the BeSSeL/VLBA observing epoch: December 10, 2010. 
} 
} 

\clearpage

\longtab[17]{ 
\begin{longtable}{rrcrrrr} 
\caption{\label{G111_24b} 22.2~GHz H$_2$O maser parameters for G111.24$-$1.24}\\ 
\hline\hline
\multicolumn{1}{c}{Feature} &  \multicolumn{1}{c}{I$_{\rm peak}$} & \multicolumn{1}{c}{$V_{\rm LSR}$} & \multicolumn{1}{c}{$\Delta~x$} & \multicolumn{1}{c}{$\Delta~y$} & \multicolumn{1}{c}{$V_{x}$} & \multicolumn{1}{c}{$V_{y}$} \\
\multicolumn{1}{c}{Number}  &  \multicolumn{1}{c}{(Jy beam$^{-1}$)} & \multicolumn{1}{c}{(km s$^{-1}$)} & \multicolumn{1}{c}{(mas)} & \multicolumn{1}{c}{(mas)} & \multicolumn{1}{c}{(km s$^{-1}$)} & \multicolumn{1}{c}{(km s$^{-1}$)} \\
\hline
\endfirsthead
\caption{continued.}\\
\hline\hline
\multicolumn{1}{c}{Feature} &  \multicolumn{1}{c}{I$_{\rm peak}$} & \multicolumn{1}{c}{$V_{\rm LSR}$} & \multicolumn{1}{c}{$\Delta~x$} & \multicolumn{1}{c}{$\Delta~y$} & \multicolumn{1}{c}{$V_{x}$} & \multicolumn{1}{c}{$V_{y}$} \\
\multicolumn{1}{c}{Number}  &  \multicolumn{1}{c}{(Jy beam$^{-1}$)} & \multicolumn{1}{c}{(km s$^{-1}$)} & \multicolumn{1}{c}{(mas)} & \multicolumn{1}{c}{(mas)} & \multicolumn{1}{c}{(km s$^{-1}$)} & \multicolumn{1}{c}{(km s$^{-1}$)} \\
\hline
\endhead
\hline
\endfoot
\hline
\endlastfoot
    1 &     12.15 &   $-$53.0 &    $-$28.4$\pm$1.0 &      9.9$\pm$1.0 &  $-$34.8$\pm$3.2 &  $-$10.1$\pm$4.2  \\
    2 &      5.86 &   $-$50.8 &    $-$30.7$\pm$1.0 &     16.5$\pm$1.0 &  $-$21.5$\pm$1.2 &   $-$8.0$\pm$1.6  \\
    3 &      4.30 &   $-$49.5 &    $-$30.1$\pm$1.0 &     11.3$\pm$1.0 & ... $ \; \; \; \, $  &  ... $ \; \; \; \, $  \\
    4 &      3.14 &   $-$51.6 &    $-$26.8$\pm$1.0 &      7.5$\pm$1.0 &  $-$24.5$\pm$3.7 &  $-$29.5$\pm$4.1  \\
    5 &      2.74 &   $-$52.6 &    $-$29.6$\pm$1.0 &     12.0$\pm$1.0 &  $-$25.6$\pm$1.5 &  $-$21.2$\pm$2.3  \\
    6 &      1.92 &   $-$49.1 &    $-$29.6$\pm$1.0 &     11.1$\pm$1.0 & ... $ \; \; \; \, $  &  ... $ \; \; \; \, $  \\
    7 &      1.61 &   $-$49.8 &    $-$32.2$\pm$1.0 &     15.2$\pm$1.0 & ... $ \; \; \; \, $  &  ... $ \; \; \; \, $  \\
    8 &      1.59 &   $-$49.3 &    $-$27.5$\pm$1.0 &      9.2$\pm$1.0 & ... $ \; \; \; \, $  &  ... $ \; \; \; \, $  \\
    9 &      1.53 &   $-$51.0 &    $-$32.2$\pm$1.0 &     17.0$\pm$1.0 & ... $ \; \; \; \, $  &  ... $ \; \; \; \, $  \\
   10 &      1.33 &   $-$52.9 &    $-$29.7$\pm$1.0 &     13.0$\pm$1.0 &  $-$28.3$\pm$1.8 &  $-$13.9$\pm$1.7  \\
   11 &      1.26 &   $-$49.0 &    $-$28.3$\pm$1.0 &      9.5$\pm$1.0 & ... $ \; \; \; \, $  &  ... $ \; \; \; \, $  \\
   12 &      0.98 &   $-$50.1 &    $-$31.9$\pm$1.0 &     17.7$\pm$1.0 & ... $ \; \; \; \, $  &  ... $ \; \; \; \, $  \\
   13 &      0.48 &   $-$48.3 &    $-$28.1$\pm$1.0 &     10.8$\pm$1.0 &  $-$38.1$\pm$2.6 &   $-$7.9$\pm$4.3  \\
   14 &      0.44 &   $-$50.3 &    $-$29.9$\pm$1.0 &     12.7$\pm$1.0 &  $-$17.6$\pm$1.6 &  $-$19.2$\pm$2.3  \\
   15 &      0.38 &   $-$52.2 &    $-$30.6$\pm$1.0 &     14.8$\pm$1.0 &  $-$20.1$\pm$1.3 &  $-$12.0$\pm$2.4  \\
   16 &      0.30 &   $-$51.0 &    $-$27.2$\pm$1.0 &      8.1$\pm$1.0 &  $-$34.2$\pm$4.5 &  $-$37.7$\pm$5.0  \\
   17 &      0.30 &   $-$48.7 &    $-$28.8$\pm$1.0 &      9.5$\pm$1.0 & ... $ \; \; \; \, $  &  ... $ \; \; \; \, $  \\
   18 &      0.26 &   $-$53.9 &    $-$27.5$\pm$1.0 &      8.4$\pm$1.0 & ... $ \; \; \; \, $  &  ... $ \; \; \; \, $  \\
   19 &      0.19 &   $-$53.0 &    $-$41.4$\pm$1.0 &      3.1$\pm$1.0 & ... $ \; \; \; \, $  &  ... $ \; \; \; \, $  \\
   20 &      0.12 &   $-$50.8 &    $-$23.7$\pm$1.0 &      4.5$\pm$1.0 & ... $ \; \; \; \, $  &  ... $ \; \; \; \, $  \\
   21 &      0.12 &   $-$49.5 &    $-$30.1$\pm$1.0 &     13.4$\pm$1.0 &  $-$18.2$\pm$5.5 &  $-$16.6$\pm$8.2  \\
\end{longtable}
\tablefoot{
\\
Column~1 gives the feature label number; 
Cols.~2~and~3 provide the intensity of the strongest spot
and the intensity-weighted LSR velocity, respectively, averaged over the
observing epochs; Cols.~4~and~5 give the offsets (with
the associated errors) in absolute position along the R.A. and Dec. axes; Cols.~6~and~7 give the components of the absolute
proper motion (with the associated errors) along the R.A. and Dec. axes. \\
The absolute position to which the position offsets refer is:
R.A.~(J2000) = 23$^{\rm h}$ 17$^{\rm m}$ 20\fs7888,  
Dec~(J2000) = 59\degree\ 28$^{\prime}$ 46\farcs970. 
The absolute positions are evaluated at the BeSSeL/VLBA observing epoch: December 10, 2010. 
} 
} 

\clearpage

\longtab[18]{ 
\begin{longtable}{rrcrrrr} 
\caption{\label{G160_tab} 22.2~GHz H$_2$O maser parameters for G160.14$+$3.16}\\ 
\hline\hline
\multicolumn{1}{c}{Feature} &  \multicolumn{1}{c}{I$_{\rm peak}$} & \multicolumn{1}{c}{$V_{\rm LSR}$} & \multicolumn{1}{c}{$\Delta~x$} & \multicolumn{1}{c}{$\Delta~y$} & \multicolumn{1}{c}{$V_{x}$} & \multicolumn{1}{c}{$V_{y}$} \\
\multicolumn{1}{c}{Number}  &  \multicolumn{1}{c}{(Jy beam$^{-1}$)} & \multicolumn{1}{c}{(km s$^{-1}$)} & \multicolumn{1}{c}{(mas)} & \multicolumn{1}{c}{(mas)} & \multicolumn{1}{c}{(km s$^{-1}$)} & \multicolumn{1}{c}{(km s$^{-1}$)} \\
\hline
\endfirsthead
\caption{continued.}\\
\hline\hline
\multicolumn{1}{c}{Feature} &  \multicolumn{1}{c}{I$_{\rm peak}$} & \multicolumn{1}{c}{$V_{\rm LSR}$} & \multicolumn{1}{c}{$\Delta~x$} & \multicolumn{1}{c}{$\Delta~y$} & \multicolumn{1}{c}{$V_{x}$} & \multicolumn{1}{c}{$V_{y}$} \\
\multicolumn{1}{c}{Number}  &  \multicolumn{1}{c}{(Jy beam$^{-1}$)} & \multicolumn{1}{c}{(km s$^{-1}$)} & \multicolumn{1}{c}{(mas)} & \multicolumn{1}{c}{(mas)} & \multicolumn{1}{c}{(km s$^{-1}$)} & \multicolumn{1}{c}{(km s$^{-1}$)} \\
\hline
\endhead
\hline
\endfoot
\hline
\endlastfoot
    1 &      5.82 &   $-$16.8 &     39.9$\pm$1.0 &    $-$10.2$\pm$1.0 &   10.5$\pm$0.9 &  $-$21.0$\pm$0.8  \\
    2 &      3.85 &   $-$16.0 &     72.0$\pm$1.0 &      2.0$\pm$1.0 &   25.6$\pm$1.4 &   $-$9.8$\pm$1.6  \\
    3 &      0.59 &   $-$17.6 &     62.9$\pm$1.0 &     $-$6.6$\pm$1.0 &   19.5$\pm$1.1 &  $-$15.3$\pm$1.0  \\
    4 &      0.15 &   $-$15.7 &     72.3$\pm$1.0 &      2.5$\pm$1.0 &   27.8$\pm$1.9 &   $-$7.5$\pm$2.9  \\
    5 &      0.13 &   $-$17.4 &     39.4$\pm$1.0 &    $-$10.1$\pm$1.0 & ... $ \; \; \; \, $  &  ... $ \; \; \; \, $  \\
\end{longtable}
\tablefoot{
\\
Column~1 gives the feature label number; 
Cols.~2~and~3 provide the intensity of the strongest spot
and the intensity-weighted LSR velocity, respectively, averaged over the
observing epochs; Cols.~4~and~5 give the offsets (with
the associated errors) in absolute position along the R.A. and Dec. axes; Cols.~6~and~7 give the components of the absolute
proper motion (with the associated errors) along the R.A. and Dec. axes. \\
The absolute position to which the position offsets refer is:
R.A.~(J2000) = 5$^{\rm h}$ 1$^{\rm m}$ 40\fs236,  
Dec~(J2000) = 47\degree\ 7$^{\prime}$ 19\farcs020. 
The absolute positions are evaluated at the BeSSeL/VLBA observing epoch: April 16, 2010. 
} 
} 

\clearpage

\longtab[19]{ 
\begin{longtable}{rrcrrrr} 
\caption{\label{G168_tab} 22.2~GHz H$_2$O maser parameters for G168.06$+$0.82}\\ 
\hline\hline
\multicolumn{1}{c}{Feature} &  \multicolumn{1}{c}{I$_{\rm peak}$} & \multicolumn{1}{c}{$V_{\rm LSR}$} & \multicolumn{1}{c}{$\Delta~x$} & \multicolumn{1}{c}{$\Delta~y$} & \multicolumn{1}{c}{$V_{x}$} & \multicolumn{1}{c}{$V_{y}$} \\
\multicolumn{1}{c}{Number}  &  \multicolumn{1}{c}{(Jy beam$^{-1}$)} & \multicolumn{1}{c}{(km s$^{-1}$)} & \multicolumn{1}{c}{(mas)} & \multicolumn{1}{c}{(mas)} & \multicolumn{1}{c}{(km s$^{-1}$)} & \multicolumn{1}{c}{(km s$^{-1}$)} \\
\hline
\endfirsthead
\caption{continued.}\\
\hline\hline
\multicolumn{1}{c}{Feature} &  \multicolumn{1}{c}{I$_{\rm peak}$} & \multicolumn{1}{c}{$V_{\rm LSR}$} & \multicolumn{1}{c}{$\Delta~x$} & \multicolumn{1}{c}{$\Delta~y$} & \multicolumn{1}{c}{$V_{x}$} & \multicolumn{1}{c}{$V_{y}$} \\
\multicolumn{1}{c}{Number}  &  \multicolumn{1}{c}{(Jy beam$^{-1}$)} & \multicolumn{1}{c}{(km s$^{-1}$)} & \multicolumn{1}{c}{(mas)} & \multicolumn{1}{c}{(mas)} & \multicolumn{1}{c}{(km s$^{-1}$)} & \multicolumn{1}{c}{(km s$^{-1}$)} \\
\hline
\endhead
\hline
\endfoot
\hline
\endlastfoot
    1 &     37.97 &   $-$29.8 &      7.9$\pm$2.0 &     $-$6.9$\pm$2.0 & ... $ \; \; \; \, $  &  ... $ \; \; \; \, $  \\
    2 &     16.34 &   $-$29.5 &      9.8$\pm$2.0 &     $-$5.0$\pm$2.0 &   18.1$\pm$2.8 &  $-$25.8$\pm$3.1  \\
    3 &      7.08 &   $-$29.8 &      6.7$\pm$2.0 &     $-$7.9$\pm$2.0 &   28.5$\pm$3.9 &   $-$8.6$\pm$4.5  \\
    4 &      1.47 &   $-$26.5 &    206.7$\pm$2.0 &   $-$124.4$\pm$2.0 &   10.3$\pm$4.6 &   19.0$\pm$4.5  \\
    5 &      1.07 &   $-$29.4 &     10.4$\pm$2.0 &     $-$4.7$\pm$2.0 & ... $ \; \; \; \, $  &  ... $ \; \; \; \, $  \\
    6 &      1.02 &   $-$26.2 &    229.4$\pm$2.0 &   $-$127.8$\pm$2.0 &   30.4$\pm$2.9 &   14.6$\pm$2.7  \\
    7 &      0.69 &   $-$27.5 &    $-$90.3$\pm$2.0 &   $-$108.1$\pm$2.0 & ... $ \; \; \; \, $  &  ... $ \; \; \; \, $  \\
    8 &      0.66 &   $-$30.2 &    $-$89.4$\pm$2.0 &   $-$102.7$\pm$2.0 &   $-$6.4$\pm$5.9 &  $-$55.1$\pm$18.1  \\
    9 &      0.48 &   $-$25.7 &    211.5$\pm$2.0 &   $-$124.9$\pm$2.0 &   26.0$\pm$9.7 &   19.2$\pm$9.4  \\
   10 &      0.18 &   $-$26.5 &     17.5$\pm$2.0 &     27.9$\pm$2.0 & ... $ \; \; \; \, $  &  ... $ \; \; \; \, $  \\
   11 &      0.14 &   $-$25.1 &     12.5$\pm$2.0 &     34.8$\pm$2.0 &   26.8$\pm$5.3 &   22.3$\pm$5.9  \\
   12 &      0.13 &   $-$27.7 &    $-$90.0$\pm$2.0 &   $-$109.8$\pm$2.0 & ... $ \; \; \; \, $  &  ... $ \; \; \; \, $  \\
   13 &      0.09 &   $-$28.6 &      4.2$\pm$2.0 &    $-$12.6$\pm$2.0 &   10.2$\pm$5.7 &  $-$37.0$\pm$10.0  \\
   14 &      0.06 &   $-$25.6 &    174.5$\pm$2.0 &   $-$128.6$\pm$2.0 & ... $ \; \; \; \, $  &  ... $ \; \; \; \, $  \\
   15 &      0.06 &   $-$20.2 &     83.3$\pm$2.0 &   $-$185.1$\pm$2.0 & ... $ \; \; \; \, $  &  ... $ \; \; \; \, $  \\
\end{longtable}
\tablefoot{
\\
Column~1 gives the feature label number; 
Cols.~2~and~3 provide the intensity of the strongest spot
and the intensity-weighted LSR velocity, respectively, averaged over the
observing epochs; Cols.~4~and~5 give the offsets (with
the associated errors) in absolute position along the R.A. and Dec. axes; Cols.~6~and~7 give the components of the absolute
proper motion (with the associated errors) along the R.A. and Dec. axes. \\
The absolute position to which the position offsets refer is:
R.A.~(J2000) = 5$^{\rm h}$ 17$^{\rm m}$ 13\fs7436,  
Dec~(J2000) = 39\degree\ 22$^{\prime}$ 19\farcs915. 
The absolute positions are evaluated at the BeSSeL/VLBA observing epoch: April 16, 2010. 
} 
} 

\clearpage

\longtab[20]{ 
\begin{longtable}{rrcrrrr} 
\caption{\label{G176_tab} 22.2~GHz H$_2$O maser parameters for G176.52$+$0.20}\\ 
\hline\hline
\multicolumn{1}{c}{Feature} &  \multicolumn{1}{c}{I$_{\rm peak}$} & \multicolumn{1}{c}{$V_{\rm LSR}$} & \multicolumn{1}{c}{$\Delta~x$} & \multicolumn{1}{c}{$\Delta~y$} & \multicolumn{1}{c}{$V_{x}$} & \multicolumn{1}{c}{$V_{y}$} \\
\multicolumn{1}{c}{Number}  &  \multicolumn{1}{c}{(Jy beam$^{-1}$)} & \multicolumn{1}{c}{(km s$^{-1}$)} & \multicolumn{1}{c}{(mas)} & \multicolumn{1}{c}{(mas)} & \multicolumn{1}{c}{(km s$^{-1}$)} & \multicolumn{1}{c}{(km s$^{-1}$)} \\
\hline
\endfirsthead
\caption{continued.}\\
\hline\hline
\multicolumn{1}{c}{Feature} &  \multicolumn{1}{c}{I$_{\rm peak}$} & \multicolumn{1}{c}{$V_{\rm LSR}$} & \multicolumn{1}{c}{$\Delta~x$} & \multicolumn{1}{c}{$\Delta~y$} & \multicolumn{1}{c}{$V_{x}$} & \multicolumn{1}{c}{$V_{y}$} \\
\multicolumn{1}{c}{Number}  &  \multicolumn{1}{c}{(Jy beam$^{-1}$)} & \multicolumn{1}{c}{(km s$^{-1}$)} & \multicolumn{1}{c}{(mas)} & \multicolumn{1}{c}{(mas)} & \multicolumn{1}{c}{(km s$^{-1}$)} & \multicolumn{1}{c}{(km s$^{-1}$)} \\
\hline
\endhead
\hline
\endfoot
\hline
\endlastfoot
\\
 \multicolumn{7}{c}{\bf Maser subset~{\em IG}}  \\
\\
\hline
    1 &     17.28 &   $-$16.6 &     $-$5.3$\pm$5.0 &     51.3$\pm$5.0 &    9.8$\pm$0.5 &  $-$13.6$\pm$0.5  \\
    2 &     13.77 &   $-$16.4 &    $-$11.2$\pm$5.0 &     55.1$\pm$5.0 &    0.9$\pm$0.8 &   $-$2.7$\pm$0.8  \\
    3 &      7.89 &   $-$16.9 &     $-$5.7$\pm$5.0 &     50.1$\pm$5.0 &    8.0$\pm$1.0 &   $-$5.4$\pm$1.0  \\
    4 &      6.98 &   $-$14.6 &    $-$13.8$\pm$5.0 &     67.6$\pm$5.0 &   $-$5.1$\pm$0.5 &    6.8$\pm$0.5  \\
    5 &      6.95 &   $-$14.4 &    $-$13.6$\pm$5.0 &     66.0$\pm$5.0 &   $-$6.5$\pm$0.5 &    2.4$\pm$0.5  \\
    6 &      5.54 &   $-$16.8 &     $-$5.0$\pm$5.0 &     49.0$\pm$5.0 &    8.3$\pm$1.0 &   $-$2.0$\pm$1.0  \\
    7 &      4.06 &   $-$14.4 &    $-$15.5$\pm$5.0 &     65.9$\pm$5.0 & ... $ \; \; \; \, $  &  ... $ \; \; \; \, $  \\
    8 &      3.98 &   $-$16.7 &     $-$3.4$\pm$5.0 &     49.9$\pm$5.0 &    5.3$\pm$0.5 &  $-$10.8$\pm$0.5  \\
    9 &      3.88 &   $-$14.7 &    $-$15.3$\pm$5.0 &     69.5$\pm$5.0 &    4.0$\pm$1.0 &    6.5$\pm$1.0  \\
   10 &      3.30 &   $-$14.4 &    $-$13.9$\pm$5.0 &     66.8$\pm$5.0 &  $-$16.4$\pm$1.0 &    2.6$\pm$1.0  \\
   11 &      2.57 &   $-$14.6 &    $-$13.7$\pm$5.0 &     68.2$\pm$5.0 & ... $ \; \; \; \, $  &  ... $ \; \; \; \, $  \\
   12 &      2.40 &   $-$17.1 &     $-$3.8$\pm$5.0 &     50.2$\pm$5.0 & ... $ \; \; \; \, $  &  ... $ \; \; \; \, $  \\
   13 &      2.38 &   $-$14.5 &    $-$14.1$\pm$5.0 &     66.0$\pm$5.0 & ... $ \; \; \; \, $  &  ... $ \; \; \; \, $  \\
   14 &      2.13 &   $-$16.0 &    $-$13.1$\pm$5.0 &     61.4$\pm$5.0 & ... $ \; \; \; \, $  &  ... $ \; \; \; \, $  \\
   15 &      2.09 &   $-$16.5 &     $-$2.5$\pm$5.0 &     48.3$\pm$5.0 & ... $ \; \; \; \, $  &  ... $ \; \; \; \, $  \\
   16 &      2.00 &   $-$14.1 &    $-$15.2$\pm$5.0 &     67.0$\pm$5.0 & ... $ \; \; \; \, $  &  ... $ \; \; \; \, $  \\
   17 &      1.90 &   $-$16.3 &    $-$11.6$\pm$5.0 &     54.3$\pm$5.0 & ... $ \; \; \; \, $  &  ... $ \; \; \; \, $  \\
   18 &      1.89 &   $-$14.4 &    $-$15.2$\pm$5.0 &     68.5$\pm$5.0 & ... $ \; \; \; \, $  &  ... $ \; \; \; \, $  \\
   19 &      1.75 &   $-$16.8 &     $-$2.1$\pm$5.0 &     49.2$\pm$5.0 & ... $ \; \; \; \, $  &  ... $ \; \; \; \, $  \\
   20 &      1.71 &   $-$14.5 &    $-$14.7$\pm$5.0 &     69.0$\pm$5.0 & ... $ \; \; \; \, $  &  ... $ \; \; \; \, $  \\
   21 &      1.64 &   $-$17.3 &     $-$3.7$\pm$5.0 &     49.6$\pm$5.0 & ... $ \; \; \; \, $  &  ... $ \; \; \; \, $  \\
   22 &      1.62 &   $-$16.6 &     $-$3.3$\pm$5.0 &     48.7$\pm$5.0 & ... $ \; \; \; \, $  &  ... $ \; \; \; \, $  \\
   23 &      1.52 &   $-$14.3 &    $-$15.1$\pm$5.0 &     68.2$\pm$5.0 & ... $ \; \; \; \, $  &  ... $ \; \; \; \, $  \\
   24 &      1.41 &   $-$17.0 &     $-$5.0$\pm$5.0 &     49.8$\pm$5.0 & ... $ \; \; \; \, $  &  ... $ \; \; \; \, $  \\
   25 &      1.34 &   $-$16.9 &    $-$10.9$\pm$5.0 &     55.7$\pm$5.0 &   $-$5.3$\pm$1.8 &   $-$4.5$\pm$1.9  \\
   26 &      1.33 &   $-$17.0 &     $-$6.3$\pm$5.0 &     50.1$\pm$5.0 & ... $ \; \; \; \, $  &  ... $ \; \; \; \, $  \\
   27 &      1.15 &   $-$15.7 &     $-$6.4$\pm$5.0 &     51.5$\pm$5.0 & ... $ \; \; \; \, $  &  ... $ \; \; \; \, $  \\
   28 &      1.01 &   $-$14.3 &    $-$16.4$\pm$5.0 &     65.9$\pm$5.0 & ... $ \; \; \; \, $  &  ... $ \; \; \; \, $  \\
   29 &      0.98 &   $-$14.1 &    $-$13.7$\pm$5.0 &     63.2$\pm$5.0 & ... $ \; \; \; \, $  &  ... $ \; \; \; \, $  \\
   30 &      0.88 &   $-$16.7 &     $-$7.1$\pm$5.0 &     52.3$\pm$5.0 & ... $ \; \; \; \, $  &  ... $ \; \; \; \, $  \\
   31 &      0.75 &   $-$16.0 &     $-$6.7$\pm$5.0 &     51.1$\pm$5.0 & ... $ \; \; \; \, $  &  ... $ \; \; \; \, $  \\
   32 &      0.74 &   $-$14.6 &    $-$16.1$\pm$5.0 &     64.7$\pm$5.0 &    3.8$\pm$1.0 &    3.9$\pm$1.1  \\
   33 &      0.73 &   $-$16.2 &    $-$10.8$\pm$5.0 &     55.6$\pm$5.0 & ... $ \; \; \; \, $  &  ... $ \; \; \; \, $  \\
   34 &      0.71 &   $-$15.8 &     $-$9.3$\pm$5.0 &     53.9$\pm$5.0 & ... $ \; \; \; \, $  &  ... $ \; \; \; \, $  \\
   35 &      0.64 &   $-$16.1 &     $-$8.4$\pm$5.0 &     51.6$\pm$5.0 & ... $ \; \; \; \, $  &  ... $ \; \; \; \, $  \\
   36 &      0.41 &   $-$15.7 &    $-$10.1$\pm$5.0 &     53.3$\pm$5.0 & ... $ \; \; \; \, $  &  ... $ \; \; \; \, $  \\
   37 &      0.39 &   $-$17.1 &     $-$7.4$\pm$5.0 &     52.6$\pm$5.0 & ... $ \; \; \; \, $  &  ... $ \; \; \; \, $  \\
   38 &      0.23 &   $-$15.7 &     $-$6.9$\pm$5.0 &     51.6$\pm$5.0 & ... $ \; \; \; \, $  &  ... $ \; \; \; \, $  \\
   39 &      0.10 &   $-$15.8 &     $-$5.8$\pm$5.0 &     50.1$\pm$5.0 & ... $ \; \; \; \, $  &  ... $ \; \; \; \, $  \\
\hline
   \\
    \\
 \multicolumn{7}{c}{\bf Maser subset~{\em OG}}  \\
\\
\hline
    1 &      3.03 &   $-$12.1 &     14.2$\pm$5.0 &     73.3$\pm$5.0 &   39.7$\pm$1.0 &   14.6$\pm$1.0  \\
    2 &      1.84 &   $-$11.8 &     13.3$\pm$5.0 &     73.1$\pm$5.0 & ... $ \; \; \; \, $  &  ... $ \; \; \; \, $  \\
    3 &      1.31 &   $-$13.6 &    119.8$\pm$5.0 &     92.7$\pm$5.0 & ... $ \; \; \; \, $  &  ... $ \; \; \; \, $  \\
    4 &      0.69 &   $-$13.8 &    116.0$\pm$5.0 &     89.8$\pm$5.0 &   17.9$\pm$1.4 &   $-$9.7$\pm$1.6  \\
    5 &      0.66 &   $-$13.2 &    116.4$\pm$5.0 &     89.6$\pm$5.0 & ... $ \; \; \; \, $  &  ... $ \; \; \; \, $  \\
    6 &      0.64 &   $-$15.0 &    105.9$\pm$5.0 &     93.5$\pm$5.0 & ... $ \; \; \; \, $  &  ... $ \; \; \; \, $  \\
    7 &      0.59 &    $-$9.3 &     22.3$\pm$5.0 &     75.7$\pm$5.0 & ... $ \; \; \; \, $  &  ... $ \; \; \; \, $  \\
    8 &      0.58 &   $-$15.3 &    110.8$\pm$5.0 &     96.2$\pm$5.0 & ... $ \; \; \; \, $  &  ... $ \; \; \; \, $  \\
    9 &      0.52 &   $-$11.2 &     14.0$\pm$5.0 &     72.7$\pm$5.0 & ... $ \; \; \; \, $  &  ... $ \; \; \; \, $  \\
   10 &      0.39 &   $-$19.7 &     51.5$\pm$5.0 &    119.3$\pm$5.0 &    8.5$\pm$0.8 &   10.9$\pm$0.8  \\
   11 &      0.38 &   $-$15.2 &    108.3$\pm$5.0 &     94.4$\pm$5.0 & ... $ \; \; \; \, $  &  ... $ \; \; \; \, $  \\
   12 &      0.38 &    $-$9.8 &     93.0$\pm$5.0 &     87.6$\pm$5.0 & ... $ \; \; \; \, $  &  ... $ \; \; \; \, $  \\
   13 &      0.35 &   $-$15.4 &    107.4$\pm$5.0 &     94.6$\pm$5.0 & ... $ \; \; \; \, $  &  ... $ \; \; \; \, $  \\
   14 &      0.34 &   $-$12.4 &    117.5$\pm$5.0 &     88.7$\pm$5.0 & ... $ \; \; \; \, $  &  ... $ \; \; \; \, $  \\
   15 &      0.34 &   $-$11.8 &     13.3$\pm$5.0 &     73.5$\pm$5.0 & ... $ \; \; \; \, $  &  ... $ \; \; \; \, $  \\
   16 &      0.33 &   $-$15.6 &    103.5$\pm$5.0 &     92.7$\pm$5.0 &   21.0$\pm$1.8 &    9.9$\pm$1.9  \\
   17 &      0.32 &    $-$9.7 &     92.8$\pm$5.0 &     87.2$\pm$5.0 &   21.1$\pm$1.9 &   15.3$\pm$2.0  \\
   18 &      0.32 &   $-$15.7 &    104.8$\pm$5.0 &     93.6$\pm$5.0 & ... $ \; \; \; \, $  &  ... $ \; \; \; \, $  \\
   19 &      0.31 &   $-$15.7 &    105.5$\pm$5.0 &     93.6$\pm$5.0 & ... $ \; \; \; \, $  &  ... $ \; \; \; \, $  \\
   20 &      0.21 &   $-$10.8 &     19.5$\pm$5.0 &     75.6$\pm$5.0 & ... $ \; \; \; \, $  &  ... $ \; \; \; \, $  \\
   21 &      0.21 &   $-$15.7 &    108.7$\pm$5.0 &     93.0$\pm$5.0 & ... $ \; \; \; \, $  &  ... $ \; \; \; \, $  \\
   22 &      0.19 &   $-$19.4 &     52.2$\pm$5.0 &    119.9$\pm$5.0 & ... $ \; \; \; \, $  &  ... $ \; \; \; \, $  \\
   23 &      0.15 &   $-$12.2 &    117.8$\pm$5.0 &     88.1$\pm$5.0 & ... $ \; \; \; \, $  &  ... $ \; \; \; \, $  \\
   24 &      0.14 &   $-$15.2 &    110.5$\pm$5.0 &     96.2$\pm$5.0 & ... $ \; \; \; \, $  &  ... $ \; \; \; \, $  \\
   25 &      0.13 &   $-$20.2 &    128.2$\pm$5.0 &    201.5$\pm$5.0 & ... $ \; \; \; \, $  &  ... $ \; \; \; \, $  \\
   26 &      0.13 &   $-$15.4 &    104.4$\pm$5.0 &     93.9$\pm$5.0 & ... $ \; \; \; \, $  &  ... $ \; \; \; \, $  \\
   27 &      0.13 &   $-$19.6 &     52.5$\pm$5.0 &    120.3$\pm$5.0 & ... $ \; \; \; \, $  &  ... $ \; \; \; \, $  \\
   28 &      0.13 &   $-$15.4 &    111.2$\pm$5.0 &     96.1$\pm$5.0 & ... $ \; \; \; \, $  &  ... $ \; \; \; \, $  \\
   29 &      0.12 &   $-$10.8 &     18.1$\pm$5.0 &     74.9$\pm$5.0 & ... $ \; \; \; \, $  &  ... $ \; \; \; \, $  \\
   30 &      0.11 &   $-$19.6 &     51.3$\pm$5.0 &    118.9$\pm$5.0 &    5.3$\pm$1.9 &   12.7$\pm$2.2  \\
   31 &      0.10 &    $-$9.7 &     94.2$\pm$5.0 &     88.6$\pm$5.0 & ... $ \; \; \; \, $  &  ... $ \; \; \; \, $  \\
   32 &      0.07 &    $-$9.8 &     19.6$\pm$5.0 &     76.8$\pm$5.0 & ... $ \; \; \; \, $  &  ... $ \; \; \; \, $  \\
\end{longtable}
\tablefoot{
\\
Column~1 gives the feature label number; 
Cols.~2~and~3 provide the intensity of the strongest spot
and the intensity-weighted LSR velocity, respectively, averaged over the
observing epochs; Cols.~4~and~5 give the offsets (with
the associated errors) in absolute position along the R.A. and Dec. axes; Cols.~6~and~7 give the components of the absolute
proper motion (with the associated errors) along the R.A. and Dec. axes. \\
The absolute position to which the position offsets refer is:
R.A.~(J2000) = 5$^{\rm h}$ 37$^{\rm m}$ 52\fs1353,  
Dec~(J2000) = 32\degree\ 0$^{\prime}$ 03\farcs878. 
The absolute positions are evaluated at the BeSSeL/VLBA observing epoch: April 16, 2010. 
} 
} 

\clearpage

\longtab[21]{ 
\begin{longtable}{rrcrrrr} 
\caption{\label{G182_tab} 22.2~GHz H$_2$O maser parameters for G182.68$-$3.27}\\ 
\hline\hline
\multicolumn{1}{c}{Feature} &  \multicolumn{1}{c}{I$_{\rm peak}$} & \multicolumn{1}{c}{$V_{\rm LSR}$} & \multicolumn{1}{c}{$\Delta~x$} & \multicolumn{1}{c}{$\Delta~y$} & \multicolumn{1}{c}{$V_{x}$} & \multicolumn{1}{c}{$V_{y}$} \\
\multicolumn{1}{c}{Number}  &  \multicolumn{1}{c}{(Jy beam$^{-1}$)} & \multicolumn{1}{c}{(km s$^{-1}$)} & \multicolumn{1}{c}{(mas)} & \multicolumn{1}{c}{(mas)} & \multicolumn{1}{c}{(km s$^{-1}$)} & \multicolumn{1}{c}{(km s$^{-1}$)} \\
\hline
\endfirsthead
\caption{continued.}\\
\hline\hline
\multicolumn{1}{c}{Feature} &  \multicolumn{1}{c}{I$_{\rm peak}$} & \multicolumn{1}{c}{$V_{\rm LSR}$} & \multicolumn{1}{c}{$\Delta~x$} & \multicolumn{1}{c}{$\Delta~y$} & \multicolumn{1}{c}{$V_{x}$} & \multicolumn{1}{c}{$V_{y}$} \\
\multicolumn{1}{c}{Number}  &  \multicolumn{1}{c}{(Jy beam$^{-1}$)} & \multicolumn{1}{c}{(km s$^{-1}$)} & \multicolumn{1}{c}{(mas)} & \multicolumn{1}{c}{(mas)} & \multicolumn{1}{c}{(km s$^{-1}$)} & \multicolumn{1}{c}{(km s$^{-1}$)} \\
\hline
\endhead
\hline
\endfoot
\hline
\endlastfoot
    1 &      3.16 &    $-$7.3 &    $-$96.9$\pm$1.0 &    152.2$\pm$1.0 & ... $ \; \; \; \, $  &  ... $ \; \; \; \, $  \\
    2 &      0.84 &    $-$5.6 &    $-$89.6$\pm$1.0 &    169.8$\pm$1.0 &   19.0$\pm$5.6 &   20.5$\pm$5.7  \\
    3 &      0.10 &    $-$7.4 &    $-$94.6$\pm$1.0 &    150.4$\pm$1.0 & ... $ \; \; \; \, $  &  ... $ \; \; \; \, $  \\
\end{longtable}
\tablefoot{
\\
Column~1 gives the feature label number; 
Cols.~2~and~3 provide the intensity of the strongest spot
and the intensity-weighted LSR velocity, respectively, averaged over the
observing epochs; Cols.~4~and~5 give the offsets (with
the associated errors) in absolute position along the R.A. and Dec. axes; Cols.~6~and~7 give the components of the absolute
proper motion (with the associated errors) along the R.A. and Dec. axes. \\
The absolute position to which the position offsets refer is:
R.A.~(J2000) = 5$^{\rm h}$ 39$^{\rm m}$ 28\fs426,  
Dec~(J2000) = 24\degree\ 56$^{\prime}$ 31\farcs956. 
The absolute positions are evaluated at the BeSSeL/VLBA observing epoch: June 20, 2010. 
} 
} 

\clearpage

\longtab[22]{ 
\begin{longtable}{rrcrrrr} 
\caption{\label{G183_tab} 22.2~GHz H$_2$O maser parameters for G183.72$-$3.66}\\ 
\hline\hline
\multicolumn{1}{c}{Feature} &  \multicolumn{1}{c}{I$_{\rm peak}$} & \multicolumn{1}{c}{$V_{\rm LSR}$} & \multicolumn{1}{c}{$\Delta~x$} & \multicolumn{1}{c}{$\Delta~y$} & \multicolumn{1}{c}{$V_{x}$} & \multicolumn{1}{c}{$V_{y}$} \\
\multicolumn{1}{c}{Number}  &  \multicolumn{1}{c}{(Jy beam$^{-1}$)} & \multicolumn{1}{c}{(km s$^{-1}$)} & \multicolumn{1}{c}{(mas)} & \multicolumn{1}{c}{(mas)} & \multicolumn{1}{c}{(km s$^{-1}$)} & \multicolumn{1}{c}{(km s$^{-1}$)} \\
\hline
\endfirsthead
\caption{continued.}\\
\hline\hline
\multicolumn{1}{c}{Feature} &  \multicolumn{1}{c}{I$_{\rm peak}$} & \multicolumn{1}{c}{$V_{\rm LSR}$} & \multicolumn{1}{c}{$\Delta~x$} & \multicolumn{1}{c}{$\Delta~y$} & \multicolumn{1}{c}{$V_{x}$} & \multicolumn{1}{c}{$V_{y}$} \\
\multicolumn{1}{c}{Number}  &  \multicolumn{1}{c}{(Jy beam$^{-1}$)} & \multicolumn{1}{c}{(km s$^{-1}$)} & \multicolumn{1}{c}{(mas)} & \multicolumn{1}{c}{(mas)} & \multicolumn{1}{c}{(km s$^{-1}$)} & \multicolumn{1}{c}{(km s$^{-1}$)} \\
\hline
\endhead
\hline
\endfoot
\hline
\endlastfoot
    1 &     81.29 &     4.4 &     21.2$\pm$1.0 &     $-$5.9$\pm$1.0 & ... $ \; \; \; \, $  &  ... $ \; \; \; \, $  \\
    2 &     75.12 &     3.9 &     19.5$\pm$1.0 &     $-$5.8$\pm$1.0 &   $-$5.4$\pm$1.0 &   $-$1.0$\pm$0.4  \\
    3 &     49.05 &     3.7 &     21.2$\pm$1.0 &     $-$5.9$\pm$1.0 & ... $ \; \; \; \, $  &  ... $ \; \; \; \, $  \\
    4 &     48.79 &    $-$1.8 &     14.7$\pm$1.0 &     $-$9.1$\pm$1.0 & ... $ \; \; \; \, $  &  ... $ \; \; \; \, $  \\
    5 &     26.06 &     2.1 &     $-$9.5$\pm$1.0 &    $-$16.7$\pm$1.0 & ... $ \; \; \; \, $  &  ... $ \; \; \; \, $  \\
    6 &     24.85 &     1.4 &    $-$10.1$\pm$1.0 &    $-$17.2$\pm$1.0 & ... $ \; \; \; \, $  &  ... $ \; \; \; \, $  \\
    7 &     23.48 &     1.3 &    $-$13.0$\pm$1.0 &    $-$20.4$\pm$1.0 &    3.8$\pm$1.1 &   11.7$\pm$0.6  \\
    8 &      9.44 &     4.1 &     24.6$\pm$1.0 &     $-$6.3$\pm$1.0 & ... $ \; \; \; \, $  &  ... $ \; \; \; \, $  \\
    9 &      8.58 &     3.6 &     $-$9.1$\pm$1.0 &    $-$13.7$\pm$1.0 & ... $ \; \; \; \, $  &  ... $ \; \; \; \, $  \\
   10 &      7.47 &     4.5 &     22.8$\pm$1.0 &     $-$5.7$\pm$1.0 & ... $ \; \; \; \, $  &  ... $ \; \; \; \, $  \\
   11 &      6.50 &     2.2 &     33.7$\pm$1.0 &    $-$50.2$\pm$1.0 & ... $ \; \; \; \, $  &  ... $ \; \; \; \, $  \\
   12 &      6.04 &     4.6 &     22.9$\pm$1.0 &     $-$5.7$\pm$1.0 & ... $ \; \; \; \, $  &  ... $ \; \; \; \, $  \\
   13 &      3.42 &    $-$5.6 &     19.6$\pm$1.0 &    $-$57.7$\pm$1.0 & ... $ \; \; \; \, $  &  ... $ \; \; \; \, $  \\
   14 &      3.14 &     2.5 &    $-$44.8$\pm$1.0 &    $-$50.3$\pm$1.0 & ... $ \; \; \; \, $  &  ... $ \; \; \; \, $  \\
   15 &      2.97 &     3.5 &      2.6$\pm$1.0 &    $-$10.5$\pm$1.0 & ... $ \; \; \; \, $  &  ... $ \; \; \; \, $  \\
   16 &      0.90 &     1.8 &      9.3$\pm$1.0 &   $-$121.9$\pm$1.1 & ... $ \; \; \; \, $  &  ... $ \; \; \; \, $  \\
   17 &      0.53 &    $-$1.9 &    $-$11.4$\pm$1.0 &    $-$19.1$\pm$1.0 & ... $ \; \; \; \, $  &  ... $ \; \; \; \, $  \\
   18 &      0.38 &    $-$7.8 &     17.2$\pm$1.0 &    $-$58.2$\pm$1.0 & ... $ \; \; \; \, $  &  ... $ \; \; \; \, $  \\
   19 &      0.34 &   $-$16.9 &      3.5$\pm$1.0 &    $-$10.0$\pm$1.0 & ... $ \; \; \; \, $  &  ... $ \; \; \; \, $  \\
   20 &      0.21 &    $-$0.4 &    $-$45.4$\pm$1.0 &    $-$55.6$\pm$1.0 & ... $ \; \; \; \, $  &  ... $ \; \; \; \, $  \\
   21 &      0.20 &   $-$14.4 &      7.0$\pm$1.0 &     $-$8.4$\pm$1.0 & ... $ \; \; \; \, $  &  ... $ \; \; \; \, $  \\
   22 &      0.17 &    $-$3.2 &     $-$9.7$\pm$1.0 &    $-$16.8$\pm$1.0 & ... $ \; \; \; \, $  &  ... $ \; \; \; \, $  \\
\end{longtable}
\tablefoot{
\\
Column~1 gives the feature label number; 
Cols.~2~and~3 provide the intensity of the strongest spot
and the intensity-weighted LSR velocity, respectively, averaged over the
observing epochs; Cols.~4~and~5 give the offsets (with
the associated errors) in absolute position along the R.A. and Dec. axes; Cols.~6~and~7 give the components of the absolute
proper motion (with the associated errors) along the R.A. and Dec. axes. \\
The absolute position to which the position offsets refer is:
R.A.~(J2000) = 5$^{\rm h}$ 40$^{\rm m}$ 24\fs228,  
Dec~(J2000) = 23\degree\ 50$^{\prime}$ 54\farcs744. 
The absolute positions are evaluated at the BeSSeL/VLBA observing epoch: June 20, 2010. 
} 
} 

\clearpage

\longtab[23]{ 
\begin{longtable}{rrcrrrr} 
\caption{\label{G229_01b} 22.2~GHz H$_2$O maser parameters for G229.57$+$0.15 VLA-1}\\ 
\hline\hline
\multicolumn{1}{c}{Feature} &  \multicolumn{1}{c}{I$_{\rm peak}$} & \multicolumn{1}{c}{$V_{\rm LSR}$} & \multicolumn{1}{c}{$\Delta~x$} & \multicolumn{1}{c}{$\Delta~y$} & \multicolumn{1}{c}{$V_{x}$} & \multicolumn{1}{c}{$V_{y}$} \\
\multicolumn{1}{c}{Number}  &  \multicolumn{1}{c}{(Jy beam$^{-1}$)} & \multicolumn{1}{c}{(km s$^{-1}$)} & \multicolumn{1}{c}{(mas)} & \multicolumn{1}{c}{(mas)} & \multicolumn{1}{c}{(km s$^{-1}$)} & \multicolumn{1}{c}{(km s$^{-1}$)} \\
\hline
\endfirsthead
\caption{continued.}\\
\hline\hline
\multicolumn{1}{c}{Feature} &  \multicolumn{1}{c}{I$_{\rm peak}$} & \multicolumn{1}{c}{$V_{\rm LSR}$} & \multicolumn{1}{c}{$\Delta~x$} & \multicolumn{1}{c}{$\Delta~y$} & \multicolumn{1}{c}{$V_{x}$} & \multicolumn{1}{c}{$V_{y}$} \\
\multicolumn{1}{c}{Number}  &  \multicolumn{1}{c}{(Jy beam$^{-1}$)} & \multicolumn{1}{c}{(km s$^{-1}$)} & \multicolumn{1}{c}{(mas)} & \multicolumn{1}{c}{(mas)} & \multicolumn{1}{c}{(km s$^{-1}$)} & \multicolumn{1}{c}{(km s$^{-1}$)} \\
\hline
\endhead
\hline
\endfoot
\hline
\endlastfoot
    1 &      7.97 &    57.2 &   1039.7$\pm$2.0 &   1529.0$\pm$2.0 &   $-$0.7$\pm$1.5 &  $-$18.9$\pm$1.6  \\
    2 &      0.11 &    55.7 &   1040.1$\pm$2.0 &   1529.9$\pm$2.0 & ... $ \; \; \; \, $  &  ... $ \; \; \; \, $  \\
\end{longtable}
\tablefoot{
\\
Column~1 gives the feature label number; 
Cols.~2~and~3 provide the intensity of the strongest spot
and the intensity-weighted LSR velocity, respectively, averaged over the
observing epochs; Cols.~4~and~5 give the offsets (with
the associated errors) in absolute position along the R.A. and Dec. axes; Cols.~6~and~7 give the components of the absolute
proper motion (with the associated errors) along the R.A. and Dec. axes. \\
The absolute position to which the position offsets refer is:
R.A.~(J2000) = 7$^{\rm h}$ 23$^{\rm m}$ 01\fs7718,  
Dec~(J2000) = $-$14\degree\ 41$^{\prime}$ 34\farcs339. 
The absolute positions are evaluated at the BeSSeL/VLBA observing epoch: October 7, 2010. 
} 
} 

\clearpage

\longtab[24]{ 
\begin{longtable}{rrcrrrr} 
\caption{\label{G229_02b} 22.2~GHz H$_2$O maser parameters for G229.57$+$0.15 VLA-2}\\ 
\hline\hline
\multicolumn{1}{c}{Feature} &  \multicolumn{1}{c}{I$_{\rm peak}$} & \multicolumn{1}{c}{$V_{\rm LSR}$} & \multicolumn{1}{c}{$\Delta~x$} & \multicolumn{1}{c}{$\Delta~y$} & \multicolumn{1}{c}{$V_{x}$} & \multicolumn{1}{c}{$V_{y}$} \\
\multicolumn{1}{c}{Number}  &  \multicolumn{1}{c}{(Jy beam$^{-1}$)} & \multicolumn{1}{c}{(km s$^{-1}$)} & \multicolumn{1}{c}{(mas)} & \multicolumn{1}{c}{(mas)} & \multicolumn{1}{c}{(km s$^{-1}$)} & \multicolumn{1}{c}{(km s$^{-1}$)} \\
\hline
\endfirsthead
\caption{continued.}\\
\hline\hline
\multicolumn{1}{c}{Feature} &  \multicolumn{1}{c}{I$_{\rm peak}$} & \multicolumn{1}{c}{$V_{\rm LSR}$} & \multicolumn{1}{c}{$\Delta~x$} & \multicolumn{1}{c}{$\Delta~y$} & \multicolumn{1}{c}{$V_{x}$} & \multicolumn{1}{c}{$V_{y}$} \\
\multicolumn{1}{c}{Number}  &  \multicolumn{1}{c}{(Jy beam$^{-1}$)} & \multicolumn{1}{c}{(km s$^{-1}$)} & \multicolumn{1}{c}{(mas)} & \multicolumn{1}{c}{(mas)} & \multicolumn{1}{c}{(km s$^{-1}$)} & \multicolumn{1}{c}{(km s$^{-1}$)} \\
\hline
\endhead
\hline
\endfoot
\hline
\endlastfoot
    1 &      4.05 &    46.7 &    536.7$\pm$2.0 &   1419.6$\pm$2.0 & ... $ \; \; \; \, $  &  ... $ \; \; \; \, $  \\
    2 &      3.45 &    46.8 &    534.8$\pm$2.0 &   1419.8$\pm$2.0 &   17.3$\pm$1.5 &    1.1$\pm$1.6  \\
    3 &      2.52 &    46.7 &    536.4$\pm$2.0 &   1419.4$\pm$2.0 &    9.2$\pm$1.6 &   $-$0.1$\pm$1.8  \\
    4 &      2.46 &    46.9 &    539.6$\pm$2.0 &   1420.5$\pm$2.0 &   13.7$\pm$1.4 &    0.7$\pm$1.5  \\
    5 &      2.12 &    54.9 &    477.8$\pm$2.0 &   1348.1$\pm$2.0 & ... $ \; \; \; \, $  &  ... $ \; \; \; \, $  \\
    6 &      1.16 &    47.0 &    544.6$\pm$2.0 &   1420.8$\pm$2.0 &   18.2$\pm$1.8 &   $-$2.5$\pm$2.0  \\
    7 &      0.89 &    46.2 &    520.4$\pm$2.0 &   1412.0$\pm$2.0 &   11.7$\pm$1.9 &   $-$2.5$\pm$1.5  \\
    8 &      0.80 &    59.7 &    478.9$\pm$2.0 &   1349.7$\pm$2.0 & ... $ \; \; \; \, $  &  ... $ \; \; \; \, $  \\
    9 &      0.77 &    46.8 &    536.9$\pm$2.0 &   1420.3$\pm$2.0 &   12.4$\pm$1.7 &    0.8$\pm$3.5  \\
   10 &      0.73 &    56.1 &    498.0$\pm$2.0 &   1369.5$\pm$2.0 &   19.7$\pm$2.9 &  $-$22.3$\pm$3.3  \\
   11 &      0.64 &    43.0 &    485.8$\pm$2.0 &   1390.2$\pm$2.0 &   39.3$\pm$2.9 &   11.3$\pm$5.0  \\
   12 &      0.46 &    55.0 &    437.1$\pm$2.0 &   1339.3$\pm$2.0 &   12.5$\pm$2.3 &  $-$21.1$\pm$2.5  \\
   13 &      0.42 &    46.8 &    535.4$\pm$2.0 &   1420.1$\pm$2.0 & ... $ \; \; \; \, $  &  ... $ \; \; \; \, $  \\
   14 &      0.40 &    59.2 &    479.6$\pm$2.0 &   1350.4$\pm$2.0 &   11.5$\pm$2.8 &  $-$32.8$\pm$2.3  \\
   15 &      0.34 &    46.7 &    534.2$\pm$2.0 &   1419.3$\pm$2.0 &   22.9$\pm$4.4 &   $-$5.1$\pm$4.0  \\
   16 &      0.30 &    47.1 &    545.4$\pm$2.0 &   1420.8$\pm$2.0 &   16.2$\pm$3.3 &   $-$0.1$\pm$4.1  \\
   17 &      0.21 &    47.0 &    546.0$\pm$2.0 &   1421.0$\pm$2.0 & ... $ \; \; \; \, $  &  ... $ \; \; \; \, $  \\
   18 &      0.19 &    48.1 &    519.7$\pm$2.0 &   1366.6$\pm$2.0 & ... $ \; \; \; \, $  &  ... $ \; \; \; \, $  \\
   19 &      0.18 &    61.0 &    519.0$\pm$2.0 &   1375.5$\pm$2.0 & ... $ \; \; \; \, $  &  ... $ \; \; \; \, $  \\
   20 &      0.17 &    43.4 &    483.3$\pm$2.0 &   1393.4$\pm$2.0 & ... $ \; \; \; \, $  &  ... $ \; \; \; \, $  \\
   21 &      0.16 &    46.8 &    540.3$\pm$2.0 &   1421.2$\pm$2.0 & ... $ \; \; \; \, $  &  ... $ \; \; \; \, $  \\
   22 &      0.15 &    42.3 &    481.7$\pm$2.0 &   1391.7$\pm$2.0 & ... $ \; \; \; \, $  &  ... $ \; \; \; \, $  \\
   23 &      0.13 &    52.8 &    408.6$\pm$2.0 &   1430.5$\pm$2.0 & ... $ \; \; \; \, $  &  ... $ \; \; \; \, $  \\
   24 &      0.12 &    55.0 &    437.5$\pm$2.0 &   1338.8$\pm$2.0 & ... $ \; \; \; \, $  &  ... $ \; \; \; \, $  \\
   25 &      0.10 &    61.9 &    470.3$\pm$2.0 &   1354.4$\pm$2.0 &   $-$0.6$\pm$2.5 &  $-$20.5$\pm$4.0  \\
   26 &      0.08 &    63.9 &    473.0$\pm$2.0 &   1352.7$\pm$2.0 &  $-$19.5$\pm$5.3 &  $-$50.9$\pm$9.4  \\
\end{longtable}
\tablefoot{
\\
Column~1 gives the feature label number; 
Cols.~2~and~3 provide the intensity of the strongest spot
and the intensity-weighted LSR velocity, respectively, averaged over the
observing epochs; Cols.~4~and~5 give the offsets (with
the associated errors) in absolute position along the R.A. and Dec. axes; Cols.~6~and~7 give the components of the absolute
proper motion (with the associated errors) along the R.A. and Dec. axes. \\
The absolute position to which the position offsets refer is:
R.A.~(J2000) = 7$^{\rm h}$ 23$^{\rm m}$ 01\fs7718,  
Dec~(J2000) = $-$14\degree\ 41$^{\prime}$ 34\farcs339. 
The absolute positions are evaluated at the BeSSeL/VLBA observing epoch: October 7, 2010. 
} 
} 

\clearpage

\longtab[25]{ 
\begin{longtable}{rrcrrrr} 
\caption{\label{G236_tab} 22.2~GHz H$_2$O maser parameters for G236.82$+$1.98}\\ 
\hline\hline
\multicolumn{1}{c}{Feature} &  \multicolumn{1}{c}{I$_{\rm peak}$} & \multicolumn{1}{c}{$V_{\rm LSR}$} & \multicolumn{1}{c}{$\Delta~x$} & \multicolumn{1}{c}{$\Delta~y$} & \multicolumn{1}{c}{$V_{x}$} & \multicolumn{1}{c}{$V_{y}$} \\
\multicolumn{1}{c}{Number}  &  \multicolumn{1}{c}{(Jy beam$^{-1}$)} & \multicolumn{1}{c}{(km s$^{-1}$)} & \multicolumn{1}{c}{(mas)} & \multicolumn{1}{c}{(mas)} & \multicolumn{1}{c}{(km s$^{-1}$)} & \multicolumn{1}{c}{(km s$^{-1}$)} \\
\hline
\endfirsthead
\caption{continued.}\\
\hline\hline
\multicolumn{1}{c}{Feature} &  \multicolumn{1}{c}{I$_{\rm peak}$} & \multicolumn{1}{c}{$V_{\rm LSR}$} & \multicolumn{1}{c}{$\Delta~x$} & \multicolumn{1}{c}{$\Delta~y$} & \multicolumn{1}{c}{$V_{x}$} & \multicolumn{1}{c}{$V_{y}$} \\
\multicolumn{1}{c}{Number}  &  \multicolumn{1}{c}{(Jy beam$^{-1}$)} & \multicolumn{1}{c}{(km s$^{-1}$)} & \multicolumn{1}{c}{(mas)} & \multicolumn{1}{c}{(mas)} & \multicolumn{1}{c}{(km s$^{-1}$)} & \multicolumn{1}{c}{(km s$^{-1}$)} \\
\hline
\endhead
\hline
\endfoot
\hline
\endlastfoot
\\
 \multicolumn{7}{c}{\bf Maser subset~{\em IG}}  \\
\\
\hline
    1 &     11.18 &    42.1 &     17.6$\pm$3.0 &    366.3$\pm$3.0 &  $-$16.6$\pm$1.4 &  $-$20.3$\pm$2.5  \\
    2 &      2.72 &    41.7 &     16.9$\pm$3.0 &    363.9$\pm$3.0 &  $-$11.7$\pm$1.2 &   $-$5.9$\pm$1.5  \\
    3 &      1.95 &    37.8 &     23.5$\pm$3.0 &    361.0$\pm$3.0 &   $-$4.6$\pm$1.4 &  $-$13.4$\pm$1.5  \\
    4 &      1.78 &    59.8 &     67.7$\pm$3.0 &    411.7$\pm$3.0 &   17.0$\pm$1.7 &    7.6$\pm$2.5  \\
    5 &      1.28 &    60.3 &     68.2$\pm$2.0 &    413.2$\pm$2.0 &   10.4$\pm$2.5 &   26.7$\pm$3.2  \\
    6 &      1.22 &    59.8 &     68.5$\pm$2.0 &    412.9$\pm$2.3 & ... $ \; \; \; \, $  &  ... $ \; \; \; \, $  \\
    7 &      1.09 &    61.1 &     67.8$\pm$3.0 &    414.5$\pm$3.0 &   10.3$\pm$1.6 &   16.4$\pm$1.5  \\
    8 &      0.94 &    39.0 &     20.1$\pm$3.0 &    372.8$\pm$3.0 & ... $ \; \; \; \, $  &  ... $ \; \; \; \, $  \\
    9 &      0.78 &    41.1 &     17.7$\pm$3.0 &    367.9$\pm$3.0 &  $-$12.1$\pm$1.1 &   $-$1.9$\pm$1.9  \\
   10 &      0.46 &    38.7 &     20.5$\pm$2.0 &    373.5$\pm$2.0 & ... $ \; \; \; \, $  &  ... $ \; \; \; \, $  \\
   11 &      0.43 &    49.2 &     29.3$\pm$2.0 &    381.5$\pm$2.0 & ... $ \; \; \; \, $  &  ... $ \; \; \; \, $  \\
   12 &      0.37 &    61.7 &     49.2$\pm$3.0 &    391.9$\pm$3.0 & ... $ \; \; \; \, $  &  ... $ \; \; \; \, $  \\
   13 &      0.34 &    39.9 &     23.0$\pm$3.0 &    361.2$\pm$3.0 & ... $ \; \; \; \, $  &  ... $ \; \; \; \, $  \\
   14 &      0.29 &    45.4 &     22.8$\pm$3.0 &    361.4$\pm$3.0 & ... $ \; \; \; \, $  &  ... $ \; \; \; \, $  \\
   15 &      0.19 &    56.0 &     36.8$\pm$3.0 &    357.7$\pm$3.0 & ... $ \; \; \; \, $  &  ... $ \; \; \; \, $  \\
   16 &      0.18 &    40.5 &     31.8$\pm$2.0 &    384.3$\pm$2.0 & ... $ \; \; \; \, $  &  ... $ \; \; \; \, $  \\
   17 &      0.17 &    64.8 &     46.7$\pm$2.0 &    378.6$\pm$2.0 & ... $ \; \; \; \, $  &  ... $ \; \; \; \, $  \\
   18 &      0.13 &    59.4 &     67.6$\pm$2.0 &    409.0$\pm$2.0 & ... $ \; \; \; \, $  &  ... $ \; \; \; \, $  \\
   19 &      0.09 &    49.8 &     34.4$\pm$2.0 &    384.2$\pm$2.0 & ... $ \; \; \; \, $  &  ... $ \; \; \; \, $  \\
   20 &      0.09 &    63.4 &     52.6$\pm$2.0 &    395.9$\pm$2.0 & ... $ \; \; \; \, $  &  ... $ \; \; \; \, $  \\
   21 &      0.09 &    48.8 &     29.1$\pm$2.0 &    382.2$\pm$2.0 & ... $ \; \; \; \, $  &  ... $ \; \; \; \, $  \\
   22 &      0.08 &    64.8 &     33.8$\pm$2.0 &    356.8$\pm$2.0 & ... $ \; \; \; \, $  &  ... $ \; \; \; \, $  \\
   23 &      0.06 &    66.9 &     34.2$\pm$2.0 &    356.8$\pm$2.0 & ... $ \; \; \; \, $  &  ... $ \; \; \; \, $  \\
\hline
   \\
    \\
 \multicolumn{7}{c}{\bf Maser subset~{\em OG}}  \\
\\
\hline
    1 &      4.49 &    52.0 &     75.1$\pm$3.0 &    365.1$\pm$3.0 &    2.1$\pm$1.6 &   $-$1.7$\pm$2.4  \\
    2 &      3.29 &    52.4 &     99.2$\pm$3.0 &    296.3$\pm$3.0 &    7.9$\pm$1.6 &   $-$2.2$\pm$3.2  \\
    3 &      2.24 &    52.1 &     75.1$\pm$2.0 &    365.3$\pm$2.0 & ... $ \; \; \; \, $  &  ... $ \; \; \; \, $  \\
    4 &      1.62 &    52.4 &     99.7$\pm$2.0 &    296.4$\pm$2.0 & ... $ \; \; \; \, $  &  ... $ \; \; \; \, $  \\
    5 &      0.85 &    52.3 &    102.4$\pm$3.0 &    298.7$\pm$3.0 &    7.8$\pm$3.4 &   $-$9.6$\pm$8.8  \\
    6 &      0.70 &    52.3 &    102.8$\pm$2.0 &    298.2$\pm$2.0 & ... $ \; \; \; \, $  &  ... $ \; \; \; \, $  \\
    7 &      0.65 &    53.1 &    101.9$\pm$3.0 &    303.1$\pm$3.0 &    3.3$\pm$1.7 &   $-$0.7$\pm$5.2  \\
    8 &      0.64 &    50.5 &     93.3$\pm$3.0 &    265.6$\pm$3.0 &    5.9$\pm$1.7 &  $-$15.7$\pm$2.0  \\
    9 &      0.62 &    52.5 &     74.8$\pm$3.0 &    367.2$\pm$3.0 &    1.8$\pm$1.7 &   $-$1.0$\pm$2.1  \\
   10 &      0.58 &    53.1 &    102.1$\pm$2.0 &    303.5$\pm$2.0 & ... $ \; \; \; \, $  &  ... $ \; \; \; \, $  \\
   11 &      0.55 &    53.5 &     98.2$\pm$3.0 &    293.3$\pm$3.0 &    3.3$\pm$3.6 &   $-$8.4$\pm$2.0  \\
   12 &      0.52 &    47.1 &     69.3$\pm$2.0 &    362.7$\pm$2.0 & ... $ \; \; \; \, $  &  ... $ \; \; \; \, $  \\
   13 &      0.41 &    52.5 &     74.9$\pm$2.0 &    367.3$\pm$2.0 & ... $ \; \; \; \, $  &  ... $ \; \; \; \, $  \\
   14 &      0.27 &    52.8 &    102.2$\pm$2.0 &    299.9$\pm$2.0 & ... $ \; \; \; \, $  &  ... $ \; \; \; \, $  \\
   15 &      0.25 &    53.5 &     98.4$\pm$2.0 &    292.9$\pm$2.0 & ... $ \; \; \; \, $  &  ... $ \; \; \; \, $  \\
   16 &      0.24 &    54.1 &     97.7$\pm$3.0 &    284.0$\pm$3.0 &    6.2$\pm$3.3 &   $-$2.4$\pm$4.4  \\
   17 &      0.22 &    53.0 &     97.1$\pm$3.0 &    282.0$\pm$3.0 &    6.3$\pm$4.1 &   $-$5.1$\pm$4.4  \\
   18 &      0.17 &    52.8 &    102.7$\pm$2.0 &    298.9$\pm$2.0 & ... $ \; \; \; \, $  &  ... $ \; \; \; \, $  \\
   19 &      0.16 &    50.8 &     93.6$\pm$2.0 &    265.0$\pm$2.0 & ... $ \; \; \; \, $  &  ... $ \; \; \; \, $  \\
   20 &      0.10 &    50.3 &     94.8$\pm$2.0 &    266.6$\pm$2.0 & ... $ \; \; \; \, $  &  ... $ \; \; \; \, $  \\
   21 &      0.08 &    47.8 &     65.9$\pm$2.0 &    364.0$\pm$2.0 & ... $ \; \; \; \, $  &  ... $ \; \; \; \, $  \\
\end{longtable}
\tablefoot{
\\
Column~1 gives the feature label number; 
Cols.~2~and~3 provide the intensity of the strongest spot
and the intensity-weighted LSR velocity, respectively, averaged over the
observing epochs; Cols.~4~and~5 give the offsets (with
the associated errors) in absolute position along the R.A. and Dec. axes; Cols.~6~and~7 give the components of the absolute
proper motion (with the associated errors) along the R.A. and Dec. axes. \\
The absolute position to which the position offsets refer is:
R.A.~(J2000) = 7$^{\rm h}$ 44$^{\rm m}$ 28\fs2367,  
Dec~(J2000) = $-$20\degree\ 8$^{\prime}$ 30\farcs606. 
The absolute positions are evaluated at the BeSSeL/VLBA observing epoch: October 7, 2010. 
} 
} 

\clearpage

\longtab[26]{ 
\begin{longtable}{rrcrrrr} 
\caption{\label{G240_tab} 22.2~GHz H$_2$O maser parameters for G240.32$+$0.07}\\ 
\hline\hline
\multicolumn{1}{c}{Feature} &  \multicolumn{1}{c}{I$_{\rm peak}$} & \multicolumn{1}{c}{$V_{\rm LSR}$} & \multicolumn{1}{c}{$\Delta~x$} & \multicolumn{1}{c}{$\Delta~y$} & \multicolumn{1}{c}{$V_{x}$} & \multicolumn{1}{c}{$V_{y}$} \\
\multicolumn{1}{c}{Number}  &  \multicolumn{1}{c}{(Jy beam$^{-1}$)} & \multicolumn{1}{c}{(km s$^{-1}$)} & \multicolumn{1}{c}{(mas)} & \multicolumn{1}{c}{(mas)} & \multicolumn{1}{c}{(km s$^{-1}$)} & \multicolumn{1}{c}{(km s$^{-1}$)} \\
\hline
\endfirsthead
\caption{continued.}\\
\hline\hline
\multicolumn{1}{c}{Feature} &  \multicolumn{1}{c}{I$_{\rm peak}$} & \multicolumn{1}{c}{$V_{\rm LSR}$} & \multicolumn{1}{c}{$\Delta~x$} & \multicolumn{1}{c}{$\Delta~y$} & \multicolumn{1}{c}{$V_{x}$} & \multicolumn{1}{c}{$V_{y}$} \\
\multicolumn{1}{c}{Number}  &  \multicolumn{1}{c}{(Jy beam$^{-1}$)} & \multicolumn{1}{c}{(km s$^{-1}$)} & \multicolumn{1}{c}{(mas)} & \multicolumn{1}{c}{(mas)} & \multicolumn{1}{c}{(km s$^{-1}$)} & \multicolumn{1}{c}{(km s$^{-1}$)} \\
\hline
\endhead
\hline
\endfoot
\hline
\endlastfoot
    1 &     12.39 &    67.6 &   $-$632.7$\pm$1.0 &    893.4$\pm$1.0 &  $-$27.2$\pm$2.7 &   14.2$\pm$3.1  \\
    2 &      5.41 &    68.1 &   $-$636.4$\pm$1.0 &    902.7$\pm$1.0 & ... $ \; \; \; \, $  &  ... $ \; \; \; \, $  \\
    3 &      3.92 &    65.7 &   $-$631.7$\pm$1.0 &    897.0$\pm$1.0 &  $-$31.7$\pm$4.9 &   10.3$\pm$4.4  \\
    4 &      2.57 &    68.9 &   $-$635.2$\pm$1.0 &    902.7$\pm$1.0 &  $-$34.8$\pm$3.5 &    7.8$\pm$2.6  \\
    5 &      1.23 &    69.7 &   $-$630.1$\pm$1.0 &    912.6$\pm$1.0 & ... $ \; \; \; \, $  &  ... $ \; \; \; \, $  \\
    6 &      1.09 &    63.4 &   $-$632.5$\pm$1.0 &    898.9$\pm$1.0 & ... $ \; \; \; \, $  &  ... $ \; \; \; \, $  \\
    7 &      0.64 &    66.9 &   $-$636.5$\pm$1.0 &    905.7$\pm$1.0 & ... $ \; \; \; \, $  &  ... $ \; \; \; \, $  \\
    8 &      0.38 &    62.9 &   $-$632.0$\pm$1.0 &    898.7$\pm$1.0 & ... $ \; \; \; \, $  &  ... $ \; \; \; \, $  \\
    9 &      0.33 &    63.0 &   $-$633.7$\pm$1.0 &    917.3$\pm$1.0 & ... $ \; \; \; \, $  &  ... $ \; \; \; \, $  \\
   10 &      0.30 &    63.5 &   $-$635.9$\pm$1.0 &    896.7$\pm$1.0 & ... $ \; \; \; \, $  &  ... $ \; \; \; \, $  \\
\end{longtable}
\tablefoot{
\\
Column~1 gives the feature label number; 
Cols.~2~and~3 provide the intensity of the strongest spot
and the intensity-weighted LSR velocity, respectively, averaged over the
observing epochs; Cols.~4~and~5 give the offsets (with
the associated errors) in absolute position along the R.A. and Dec. axes; Cols.~6~and~7 give the components of the absolute
proper motion (with the associated errors) along the R.A. and Dec. axes. \\
The absolute position to which the position offsets refer is:
R.A.~(J2000) = 7$^{\rm h}$ 44$^{\rm m}$ 51\fs967,  
Dec~(J2000) = $-$24\degree\ 7$^{\prime}$ 42\farcs384. 
The absolute positions are evaluated at the BeSSeL/VLBA observing epoch: March 9, 2011. 
} 
} 

\clearpage
\twocolumn

\longtab[27]{ 
\begin{longtable}{rrcrrrr} 
\caption{\label{G359_97b} 22.2~GHz H$_2$O maser parameters for G359.97$-$0.46}\\ 
\hline\hline
\multicolumn{1}{c}{Feature} &  \multicolumn{1}{c}{I$_{\rm peak}$} & \multicolumn{1}{c}{$V_{\rm LSR}$} & \multicolumn{1}{c}{$\Delta~x$} & \multicolumn{1}{c}{$\Delta~y$} & \multicolumn{1}{c}{$V_{x}$} & \multicolumn{1}{c}{$V_{y}$} \\
\multicolumn{1}{c}{Number}  &  \multicolumn{1}{c}{(Jy beam$^{-1}$)} & \multicolumn{1}{c}{(km s$^{-1}$)} & \multicolumn{1}{c}{(mas)} & \multicolumn{1}{c}{(mas)} & \multicolumn{1}{c}{(km s$^{-1}$)} & \multicolumn{1}{c}{(km s$^{-1}$)} \\
\hline
\endfirsthead
\caption{continued.}\\
\hline\hline
\multicolumn{1}{c}{Feature} &  \multicolumn{1}{c}{I$_{\rm peak}$} & \multicolumn{1}{c}{$V_{\rm LSR}$} & \multicolumn{1}{c}{$\Delta~x$} & \multicolumn{1}{c}{$\Delta~y$} & \multicolumn{1}{c}{$V_{x}$} & \multicolumn{1}{c}{$V_{y}$} \\
\multicolumn{1}{c}{Number}  &  \multicolumn{1}{c}{(Jy beam$^{-1}$)} & \multicolumn{1}{c}{(km s$^{-1}$)} & \multicolumn{1}{c}{(mas)} & \multicolumn{1}{c}{(mas)} & \multicolumn{1}{c}{(km s$^{-1}$)} & \multicolumn{1}{c}{(km s$^{-1}$)} \\
\hline
\endhead
\hline
\endfoot
\hline
\endlastfoot
    1 &     13.19 &     4.8 &    $-$21.5$\pm$15.0 &     $-$7.3$\pm$15.0 & ... $ \; \; \; \, $  &  ... $ \; \; \; \, $  \\
    2 &      5.30 &     4.3 &    $-$29.6$\pm$15.0 &    $-$17.7$\pm$15.0 & ... $ \; \; \; \, $  &  ... $ \; \; \; \, $  \\
    3 &      0.54 &    10.8 &    $-$39.2$\pm$15.0 &    $-$16.9$\pm$15.0 & ... $ \; \; \; \, $  &  ... $ \; \; \; \, $  \\
    4 &      0.34 &    10.6 &    $-$36.1$\pm$15.0 &    $-$13.4$\pm$15.0 & ... $ \; \; \; \, $  &  ... $ \; \; \; \, $  \\
\end{longtable}
\tablefoot{
\\
Column~1 gives the feature label number; 
Cols.~2~and~3 provide the intensity of the strongest spot
and the intensity-weighted LSR velocity, respectively, averaged over the
observing epochs; Cols.~4~and~5 give the offsets (with
the associated errors) in absolute position along the R.A. and Dec. axes; Cols.~6~and~7 give the components of the absolute
proper motion (with the associated errors) along the R.A. and Dec. axes. \\
The absolute position to which the position offsets refer is:
R.A.~(J2000) = 17$^{\rm h}$ 47$^{\rm m}$ 20\fs1851,  
Dec~(J2000) = $-$29\degree\ 11$^{\prime}$ 59\farcs255. 
The absolute positions are evaluated at the BeSSeL/VLBA observing epoch: September 10, 2010. 
} 
} 

\clearpage

\section{Analysis of the PM distributions in each POETS target}
\label{app_pm}

\begin{figure*}
\centering
\includegraphics[width=\textwidth]{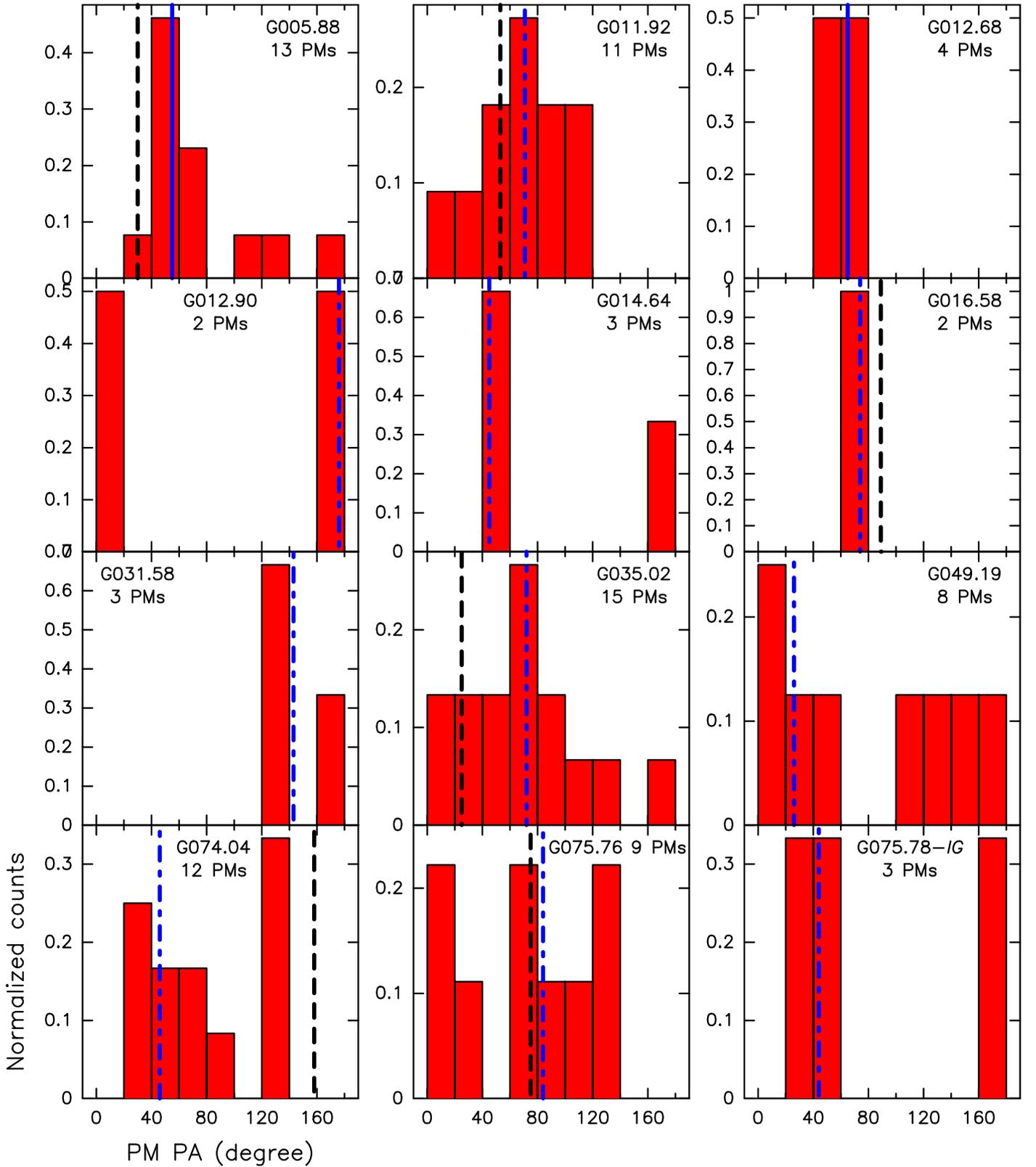}
\caption{Each plot shows the histogram of the PM~PA (0\degree\ $\le$ PA $\le$ 180\degree) for a specific POETS target, whose name is reported on the top together with the number of PMs employed to produce the histogram. The histogram bin is \ 20\degree\ and the histogram values are normalized by the number of PMs. The average error on the PM~PA is \ $\le$10\degree \ for most of the sources. The vertical {\it blue} line marks the PA of the preferential direction of the water maser PMs, PD$_{\rm PM}$ (see Sect.~\ref{obs_pm}), using {\it filled} or {\it dot-dashed lines} for more or less collimated cases, respectively. In sources where the presence of a jet has been established through observations of thermal line or continuum tracers, the jet PA from the literature, A$_{\rm LJ}$, is denoted by a vertical {\it black dashed} line.}
\label{histo_PA_1}
\end{figure*}

\begin{figure*}
\centering
\includegraphics[width=\textwidth]{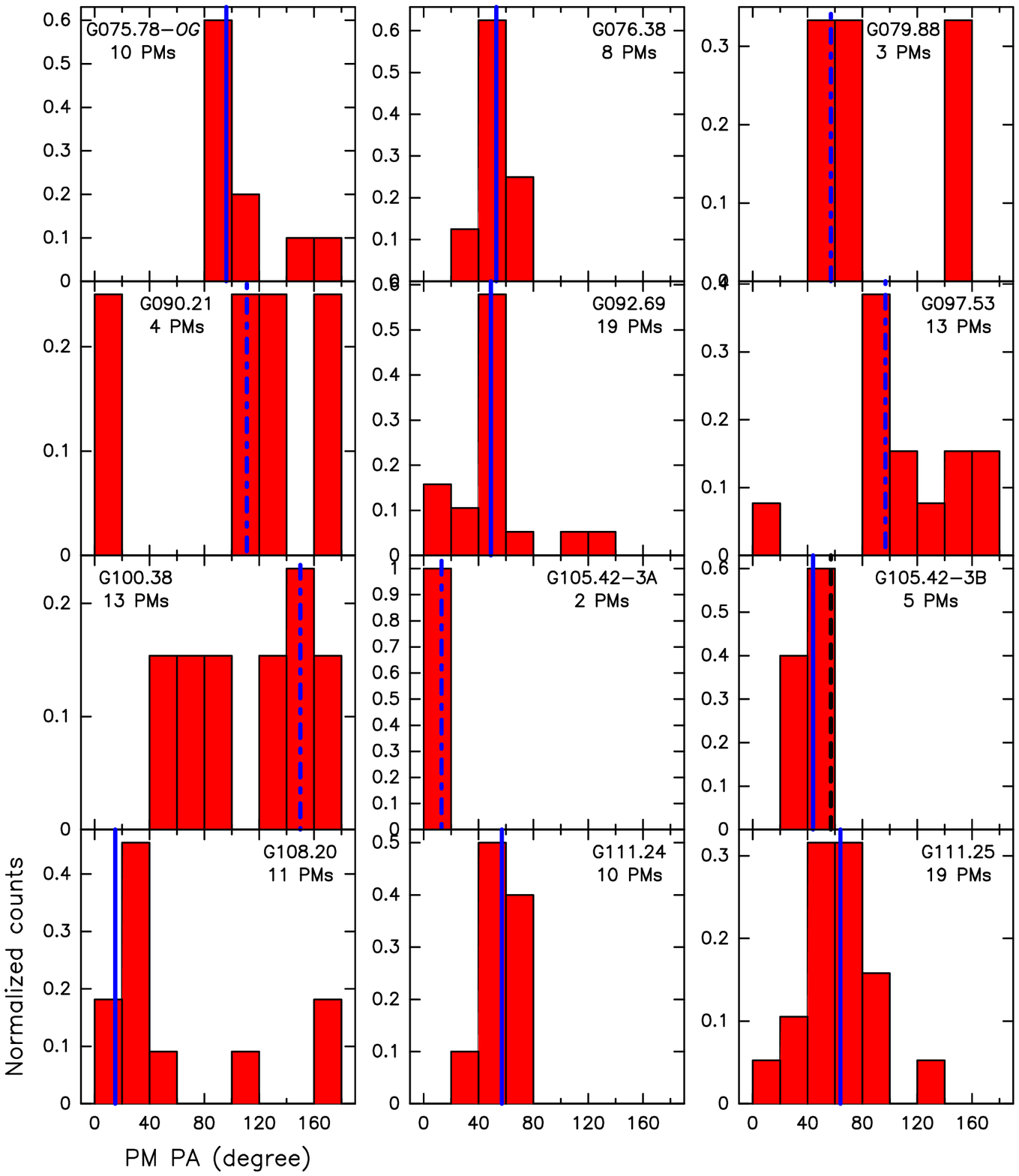}
\caption{As in Fig.~\ref{histo_PA_1}}
\label{histo_PA_2}
\end{figure*}

\begin{figure*}
\centering
\includegraphics[width=\textwidth]{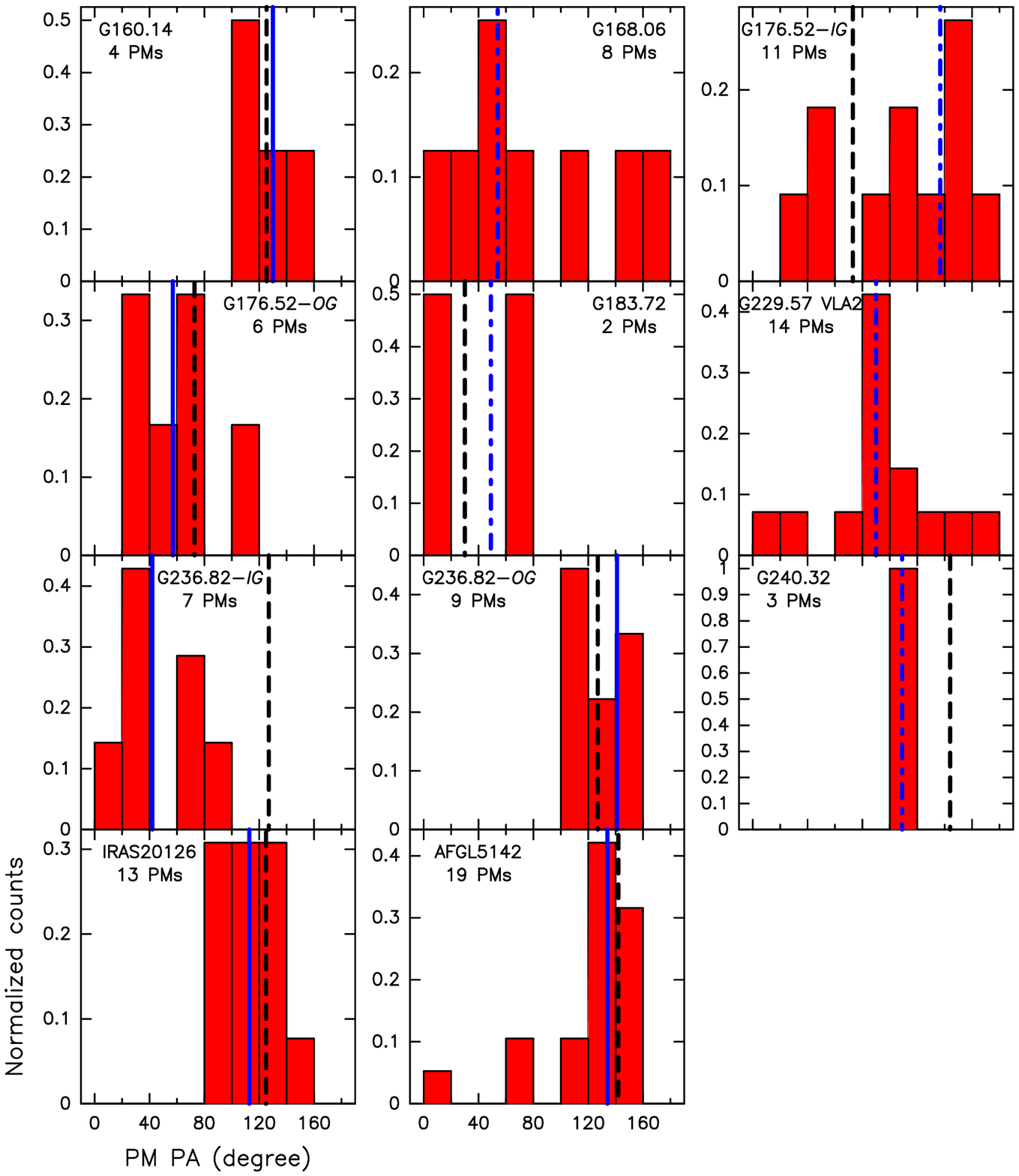}
\caption{As in Fig.~\ref{histo_PA_1}}
\label{histo_PA_3}
\end{figure*}

Figs.~\ref{histo_PA_1}-\ref{histo_PA_3} show the histograms of the PM directions for all the targets with at least two measured PMs. We have not considered the two sources \ G012.91$-$0.26 \ and \ G108.59$+$0.49, where the maser emission is likely not associated with the radio continuum.

\end{appendix}

\end{document}